\documentclass[a4paper,12pt]{article}
\usepackage{expl3}
\usepackage{amsmath,amssymb,mathrsfs,amsthm,tikz,shuffle,paralist}
\usepackage{dsfont}

\usepackage{xcolor}
\usepackage{relsize}
\usepackage{subfig}

\usepackage[font={small},labelfont=bf]{caption}

\usepackage[compat=1.1.0]{tikz-feynman}
\usetikzlibrary{shapes.misc}
\usetikzlibrary{cd}
\usepackage{todonotes}
\usepackage{stackengine}
\usepackage[bottom]{footmisc}
\usepackage{booktabs} 

\usetikzlibrary{arrows}
\newcommand{\midarrow}{\tikz \draw [-{Stealth[length=2.2mm]}](0,0) -- (.1,0);}

\newcommand{\nint}{{n_{\text{int}}}}
\newcommand{\nex}{{n_{\text{ext}}}}
\newcommand{\nISP}{{n_{\text{ISP}}}}
\newcommand{\LS}{{\text{LS}}}

\newcommand{\RationalizeRoots}{\texttt{\textup{RationalizeRoots}}}
\newcommand{\DlogBasis}{\texttt{\textup{DlogBasis}}}
\newcommand{\Maple}{\texttt{\textup{Maple}}}
\newcommand{\FIRE}{\texttt{\textup{FIRE}}}

\usepackage{url}
\def\UrlBreaks{\do\/\do-}
\usepackage{breakurl}
\usepackage[colorlinks=true
,urlcolor=blue
,anchorcolor=blue
,citecolor=blue
,filecolor=blue
,linkcolor=blue
,menucolor=blue
,pagecolor=blue
,linktocpage=true
,pdfproducer=medialab
,pdfa=true
]{hyperref}
\hypersetup{breaklinks=true}

\usepackage[utf8]{inputenc}
\usepackage{graphicx}

\usepackage{jheppub}

\begin{document}

\title{Classifying post-Minkowskian geometries for gravitational waves via loop-by-loop Baikov}

\author[]{Hjalte Frellesvig,}
\emailAdd{hjalte.frellesvig@nbi.ku.dk}
\author[]{Roger Morales,}
\emailAdd{roger.morales@nbi.ku.dk}
\author[]{and Matthias Wilhelm}
\emailAdd{matthias.wilhelm@nbi.ku.dk}

\affiliation[]{%
Niels Bohr International Academy, Niels Bohr Institute, Copenhagen University, Blegdamsvej 17, 2100 Copenhagen \O{}, Denmark}

\date{\today}%

\abstract{We use the loop-by-loop Baikov representation to investigate the geometries in Feynman integrals contributing to the classical dynamics of a black-hole two-body system in the post-Minkowskian expansion of general relativity. These geometries determine the spaces of functions to which the corresponding Feynman diagrams evaluate. As a proof of principle, we provide a full classification of the geometries appearing up to three loops, i.e. fourth post-Minkowskian order, for all diagrams relevant to the conservative as well as the dissipative dynamics, finding full agreement with the literature. Moreover, we show that the non-planar top topology at four loops, which is the most complicated sector with respect to integration-by-parts identities, has an algebraic leading singularity and thus can only depend on non-trivial geometries through its subsectors.}

\maketitle

\section{Introduction}
\label{sec:intro}

The groundbreaking discovery of gravitational waves has prompted a revolution in many branches of physics~\cite{LIGOScientific:2016aoc,LIGOScientific:2017vwq}, including particle physics, astronomy and cosmology, and their study has become a central topic of attention. A cornerstone of this novel field is the study of gravitational waves emitted during the inspiral and merger of two astronomical compact objects, such as black holes and neutron stars, bound in a gravitational two-body system. The forthcoming launch of third-generation gravitational-wave detectors will provide an unprecedented window of high-precision data for such events, which demands comparable precision of theoretical predictions for the interpretation of the results~\cite{Berti:2022wzk,Buonanno:2022pgc}.

To approach this theoretically rich problem, numerous complementary techniques have been developed, ranging from numerical to analytical. Among others, these include the effective-one-body (EOB) formalism~\cite{Buonanno:1998gg,Buonanno:2000ef} as well as the post-Newtonian~\cite{Goldberger:2004jt,Blanchet:2013haa,Porto:2016pyg,Levi:2018nxp}, post-Minkowskian~\cite{Damour:2016gwp,Buonanno:2022pgc} and self-force~\cite{Mino:1996nk,Quinn:1996am,Poisson:2011nh,Barack:2018yvs} expansions.

In particular, the post-Minkowskian (PM) expansion allows us to describe the inspiral phase perturbatively in Newton's constant $G$. Notably, it also accounts for relativistic effects, making it valid through all orders in the velocity. Since it is possible to relate the bound and unbound observables through analytic continuation~\cite{Damour:2016gwp,Kalin:2019rwq,Kalin:2019inp,Cho:2021arx,Dlapa:2024cje} as well as the EOB, this enables the use of Quantum Field Theory (QFT) techniques, specifically Feynman diagrams and scattering amplitudes, for the PM expansion~\cite{Bjerrum-Bohr:2018xdl,Cheung:2018wkq,Kosower:2018adc,Bern:2019nnu,KoemansCollado:2019ggb,Cristofoli:2019neg,Bern:2019crd,Kalin:2019rwq,Kalin:2019inp,Parra-Martinez:2020dzs,Kalin:2020mvi,Kalin:2020fhe,Mogull:2020sak,DiVecchia:2021bdo,Herrmann:2021tct,Bern:2021dqo,Dlapa:2021npj,Bern:2021yeh,Dlapa:2021vgp,Cho:2021arx,Kalin:2022hph,Dlapa:2022lmu,Dlapa:2023hsl,Jakobsen:2023ndj,Damgaard:2023ttc,Jakobsen:2023hig,Jakobsen:2023pvx,Dlapa:2024cje,Driesse:2024xad}; see refs.~\cite{Bjerrum-Bohr:2022blt,Buonanno:2022pgc} for an overview. Therefore, the extraction of higher-precision theoretical predictions for the inspiral phase is closely linked to the computation of higher-loop Feynman integrals, where an $L$-loop correction is of order $G^{L+1}$ and contributes to the $(L+1)$PM order. The complete state of the art currently stands at three loops, which corresponds to a 4PM correction~\cite{Bern:2021dqo,Dlapa:2021npj,Bern:2021yeh,Dlapa:2021vgp,Dlapa:2022lmu,Dlapa:2023hsl,Jakobsen:2023ndj,Damgaard:2023ttc,Jakobsen:2023hig,Jakobsen:2023pvx}, while at four loops only partial results in the conservative sector are currently known~\cite{Driesse:2024xad}.

While most known Feynman integrals in QFT can be written in terms of multiple polylogarithms~\cite{Chen:1977oja,Goncharov:1995ifj}, new special functions and geometries also start to appear, especially at high loop orders and in cases with many physical scales. The simplest non-polylogarithmic cases involve elliptic curves and K3 surfaces, but in general more intricate algebraic varieties occur, such as higher-genus curves and Calabi-Yau manifolds; see ref.~\cite{Bourjaily:2022bwx} for a recent review. This has already been observed in several $L$-loop families of Feynman integrals, such as the banana, traintrack and tardigrade diagrams~\cite{Broadhurst:1993mw,Bourjaily:2018ycu,Bourjaily:2018yfy,Bonisch:2021yfw,Broedel:2021zij,Duhr:2022pch,Lairez:2022zkj,Pogel:2022vat,Duhr:2022dxb,Cao:2023tpx,Duhr:2024hjf}, which involve Calabi-Yau manifolds of dimensions growing linearly with the loop order. The calculation of these Feynman integrals can become highly challenging. While the necessary calculational tools for some of these geometries currently are in the initial process of development~\cite{Bonisch:2021yfw,Frellesvig:2021hkr,Muller:2022gec,Duhr:2022pch,Lairez:2022zkj,Dlapa:2022wdu,Pogel:2022vat,Duhr:2022dxb,Gorges:2023zgv,DHoker:2023vax,Marzucca:2023gto,delaCruz:2024xit,Duhr:2024hjf,Jockers:2024tpc}, a systematic understanding of the full scope of geometries that can occur in Feynman integrals is still lacking. In this paper, we address this gap by initiating a systematic exploration of the space of underlying geometries in the PM expansion, which subsequently allows us to characterize the space of functions that PM integrals evaluate to, guiding the development of the respective tools.

Explicit calculations have shown that the PM expansion is expressible in terms of polylogarithms up to two loops (3PM order). However, at three loops products of complete elliptic integrals appear in the result, whose source has been identified as a K3 surface in the corresponding Feynman integrals~\cite{Bern:2021dqo,Dlapa:2021npj,Dlapa:2022wdu}. Moreover, the present authors have shown in ref.~\cite{Frellesvig:2023bbf} that a novel Calabi-Yau three-fold arises in a Feynman diagram at four loops in the conservative sector. This later was confirmed in ref.~\cite{Klemm:2024wtd}, where also a second Calabi-Yau three-fold was identified at four-loop order in the dissipative sector. In this work, we provide the details of the methodology used in ref.~\cite{Frellesvig:2023bbf} to identify the first Calabi-Yau three-fold, and provide a systematic analysis of the geometries appearing at lower loops.

There are various techniques available in the literature that can be used to characterize the geometries in Feynman integrals; see ref.~\cite{Bourjaily:2022bwx} for an overview. One option is to calculate the linear system of differential equations~\cite{Kotikov:1990kg} for the master integrals, and derive from it a higher-order differential operator -- known as the Picard-Fuchs operator -- for the individual master integrals. Studying the properties of the Picard-Fuchs operator 
characterizes the geometry at hand. However, calculating the differential equation and solving the necessary integration-by-part (IBP) identities can become a bottleneck in the computation. To avoid this complication, in this work we will use instead an approach based on the leading singularity (LS)~\cite{Cachazo:2008vp}. The leading singularity, which is related to the maximally iterated discontinuity,\footnote{
Strictly speaking, our notion of leading singularity is related to \textit{generalized} unitarity cuts, which are deformations of the integration contour and have no direct interpretation as a discontinuity. See ref.~\cite{Britto:2024mna} for a detailed discussion.
} can be calculated using any Feynman integral representation. 

In this work, we propose the loop-by-loop Baikov representation~\cite{Baikov:1996iu,Frellesvig:2017aai} as a particularly advantageous choice. Using this representation, we find that many geometries occurring in Feynman integrals in the post-Minkowskian expansion can be related across different diagram topologies and loop orders, which substantially reduces the number of diagrams that we need to analyze, as we gather in tab.~\ref{tab:indep_diagrams}. As a proof of principle, we classify the geometries appearing in all diagrams up to three loops, i.e.\ the current complete state of the art, for both the conservative and the dissipative sector, finding full agreement with all known results in the literature. 
Moreover, we demonstrate the potential of our approach by analyzing a particular non-planar four-loop diagram beyond the current state of the art, namely the most complicated four-loop diagram from the perspective of IBPs.
While we use the scattering-amplitudes-based approach in this work, our results also hold in alternative formulations such as the worldline formalism. In addition to classifying geometries, the leading singularity is also helpful for bringing the differential equation into canonical form~\cite{Henn:2013pwa}, which is typically obtained in polylogarithmic cases by dividing the integrals by their leading singularities~\cite{Henn:2014qga, Primo:2016ebd, WasserMSc,Dlapa:2021qsl}.

The remainder of this paper is organized as follows. In sec.~\ref{sec:PM} we review the post-Minkowskian expansion and discuss how to restrict the analysis to Feynman diagrams contributing to the classical dynamics. Sec.~\ref{sec:Methodology} introduces the notion of leading singularity and its relation to the geometry in Feynman diagrams. Specifically, we introduce the loop-by-loop Baikov representation and the methodology used to calculate the leading singularity, as well as a compendium of simplifying relations found among PM geometries. In sec.~\ref{sec:one_loop} and sec.~\ref{sec:two_loop}, we respectively carry out the analysis at one and two loops. We turn in sec.~\ref{sec:three_loop} to three-loops, where we also study the first non-trivial geometry -- a K3 surface -- appearing in the PM expansion.
In sec.~\ref{sec:four_loop_nonplanar} we then study the four-loop non-planar top topology.
We finally conclude and discuss further directions in sec.~\ref{sec:conclusion}.

We include four appendices. App.~\ref{app: Nonplanar_unraveling} shows that for the purpose of the geometry, vertices at the matter lines become orderless, which allows us to relate the geometry of numerous non-planar diagrams to planar counterparts. App.~\ref{app: IBPs_bubbles_triangles} showcases that diagrams containing certain bubble and triangle subdiagrams with cubic vertices are expressible in terms of lower sectors; thus, they can be dropped from the analysis of geometries. Similarly, app.~\ref{app: Superclassical_reduction} shows that the leading singularity of superclassical diagrams with a box iteration can be related to that of lower-loop graphs, allowing us to relate geometries across different loop orders. Lastly, in app.~\ref{app:parity} we prove that the differential equations for post-Minkowskian Feynman integrals in general split into two decoupled blocks according to a certain parity, to be explained in sec.~\ref{sec:soft_expansion}.

\section{Feynman diagrams in the post-Minkowskian expansion}
\label{sec:PM}

In this section, we review the formalism used to perturbatively calculate corrections to the classical dynamics of two inspiraling, non-spinning black holes.\footnote{Spin effects only modify the numerator of the integrals once the vertices are dressed; thus, they do not increase the complexity of our analysis. Similarly, the analogous description for neutron stars only affects
the numerators.}
To calculate such corrections, we will use the modern scattering-amplitudes-based approach, which exploits the relation between the bound problem and the scattering problem through analytic continuation~\cite{Damour:2016gwp,Kalin:2019rwq,Kalin:2019inp,Cho:2021arx,Dlapa:2024cje}. In particular, we will closely follow the conventions of refs.~\cite{Parra-Martinez:2020dzs,Bern:2021dqo,Herrmann:2021tct,Bern:2021yeh}; however, our results are also valid for alternative formulations such as the worldline formalism.
With that aim, we assume that the internal structure of each object can be neglected; thus, the size of the bodies is much smaller than the distance between them. For the gravitational two-body problem, this point-particle approximation implies that the Schwarzschild radius $r_s \sim G m$ is much smaller than the impact parameter $|b|$, given in momentum space by $|b|\sim 1/|q|$. This naturally defines the small expansion parameter $r_s/|b| \sim G m |q| \ll 1$, valid for the long-distance regime, which in turn is compatible with the post-Minkowskian (PM) expansion of general relativity, such that an $n$PM correction corresponds to an order $G^n$ in the expansion. 

In sec.~\ref{sec:PM_kinematics} we provide the kinematics for the scattering process, and in sec.~\ref{sec:soft_expansion} we describe the soft expansion inherent to the classical regime. Then, in sec.~\ref{sec:classical_diagrams} we review which Feynman diagrams are relevant to the classical dynamics of the scattering process.

\subsection{Kinematics}
\label{sec:PM_kinematics}
The setup we consider is the two-to-two scattering of massive scalars, which model two non-spinning black holes when matched to the effective field theory~\cite{Cheung:2018wkq}, mediated by the gravitational interaction.

Throughout this paper, 
the arrows in Feynman diagrams indicate the direction of the corresponding momenta, thick lines denote the massive scalars and thin lines denote the gravitons.
For example, we have at tree level
\begin{equation}
\label{eq: diag_initial_param}
\hspace*{-0.4cm} 
\begin{tikzpicture}[baseline={([yshift=-0.1cm]current bounding box.center)}] 
	\node[] (a) at (0,0) {};
	\node[] (b) at (0,-1) {};
	\node[label=left:{$p_1$}] (p1) at ($(a)+(-1,0)$) {};
	\node[label=left:{$p_2$}] (p2) at ($(b)+(-1,0)$) {};
	\node[label=right:{$p_3={-}p_2{-}q$}] (p3) at ($(b)+(1,0)$) {};
	\node[label=right:{$p_4={-}p_1{+}q$}] (p4) at ($(a)+(1,0)$) {};
	\draw[line width=0.15mm, postaction={decorate}] (b.center) -- node[sloped, allow upside down, label={[xshift=0.75cm, yshift=0cm]$q$}] {\midarrow} (a.center);
	\draw[line width=0.5mm, postaction={decorate}] (a.center) -- node[sloped, allow upside down] {\midarrow} (p1.center);
	\draw[line width=0.5mm, postaction={decorate}] (a.center) -- node[sloped, allow upside down] {\midarrow} (p4.center);
	\draw[line width=0.5mm, postaction={decorate}] (b.center) -- node[sloped, allow upside down] {\midarrow} (p2.center);
	\draw[line width=0.5mm, postaction={decorate}] (b.center) -- node[sloped, allow upside down] {\midarrow} (p3.center);
\end{tikzpicture}.
\end{equation}
The scattering is characterized by the following kinematic invariants: 
\begin{equation}
s=(p_1+p_2)^2, \qquad t=(p_1+p_4)^2=q^2, \qquad u=(p_1+p_3)^2,
\end{equation}
\begin{equation}
p_1^2=p_4^2=m_1^2, \qquad p_2^2=p_3^2=m_2^2,
\end{equation}
where $q$ is the momentum transfer, and $m_i$ are the masses of the scalar particles. The physical region relevant to the scattering process corresponds to $s>(m_1+m_2)^2$, $t=q^2<0$ and $u<0$. Due to the relation $s+t+u=2m_1^2+2m_2^2$, a priori we would need four parameters (including the masses) to describe the scattering. Nonetheless, as we will see below, in the soft expansion the dependence on the masses fully factors out, as they become part of the prefactor. Therefore, the kinematic dependence is reduced to two parameters, which can be further reduced to a single parameter by rescaling and dimensional analysis. Anticipating this, we introduce
\begin{equation}
\sigma = \frac{p_1 \cdot p_2}{m_1 m_2} = \frac{s-m_1^2-m_2^2}{2\, m_1 m_2}
\end{equation}
to uniquely characterize the process, which is the relativistic Lorentz factor of particle~1 in the rest frame of particle 2, where $\sigma>1$ in the physical region.

\subsection{Soft expansion}
\label{sec:soft_expansion}
In addition to the long-distance dynamics, a classical two-body system inherently has a large angular momentum $J \gg \hbar$~\cite{Damour:2016gwp,Damour:2017zjx,Cheung:2018wkq,Bern:2019crd}, which in natural units defines a hierarchy of scales
\begin{equation}
m_1^2, \, m_2^2, \, s, \, u \sim J^2 \, |t| \gg |t|=|q|^2.
\end{equation}
As a consequence, the classical limit naturally translates to keeping the momentum transfer $|q|$ small~\cite{Neill:2013wsa}, the so-called soft expansion. In particular, this limit is implemented in the perturbative expansion via the method of regions~\cite{Beneke:1997zp}, which allows us to systematically discard quantum contributions by truncating hard (quantum) loop momenta $\sim m$ in favor of soft loop momenta $\sim |q|$~\cite{Neill:2013wsa}.

To facilitate the soft expansion, it is customary to decompose the momenta onto components along and perpendicular to the momentum transfer~\cite{Landshoff:1969yyn,Parra-Martinez:2020dzs},
\begin{equation}
\label{eq: diag_param}
\begin{tikzpicture}[baseline=(current bounding box.center)] 
	\node[] (a) at (0,0) {};
	\node[] (b) at (0,-1) {};
	\node[label=left:{$\overline{p}_1{-}\frac{q}{2}$}] (p1) at ($(a)+(-1,0)$) {};
	\node[label=left:{$\overline{p}_2{+}\frac{q}{2}$}] (p2) at ($(b)+(-1,0)$) {};
	\node[label=right:{$\overline{p}_2{-}\frac{q}{2}$}] (p3) at ($(b)+(1,0)$) {};
	\node[label=right:{$\overline{p}_1{+}\frac{q}{2}$}] (p4) at ($(a)+(1,0)$) {};
	\draw[line width=0.15mm, postaction={decorate}] (b.center) -- node[sloped, allow upside down, label={[xshift=0.75cm, yshift=0cm]$q$}] {\midarrow} (a.center);
	\draw[line width=0.5mm, postaction={decorate}] (p1.center) -- node[sloped, allow upside down] {\midarrow} (a.center);
	\draw[line width=0.5mm, postaction={decorate}] (a.center) -- node[sloped, allow upside down] {\midarrow} (p4.center);
	\draw[line width=0.5mm, postaction={decorate}] (p2.center) -- node[sloped, allow upside down] {\midarrow} (b.center);
	\draw[line width=0.5mm, postaction={decorate}] (b.center) -- node[sloped, allow upside down] {\midarrow} (p3.center);
\end{tikzpicture}.
\end{equation}
To see that $\overline{p}_i \cdot q=0$, one can for example compute $0=p_1^2-p_4^2= - 2 \, \overline{p}_1 \cdot q$, and similarly for $\overline{p}_2$. Using these kinematics, graviton propagators with loop momentum $k$ within the soft region scale like 
\begin{equation}
\begin{tikzpicture}[baseline={([yshift=-0.1cm]current bounding box.center)}] 
	\node[] (a) at (0,0) {};
	\node[] (p4) at ($(a)+(2,0)$) {};
	\draw[line width=0.15mm, postaction={decorate}] (a.center) -- node[sloped, allow upside down] {\midarrow} (p4.center);
\end{tikzpicture} = \frac{1}{k^2} \sim \frac{1}{|q|^2}.
\end{equation}
In contrast, matter propagators can be expanded for small $|q|$ as follows:
\begin{equation}
\frac{1}{(k+\overline{p}_i \pm \frac{q}{2})^2-m_i^2}=\frac{1}{k^2+2k \cdot (\overline{p}_i \pm \frac{q}{2})} = \frac{1}{2k\cdot \overline{p}_i} \Big(1-\frac{k^2 \pm k \cdot q}{2k\cdot \overline{p}_i} + \mathcal{O}(q^2) \Big).
\end{equation}
Therefore, they become linearized (also known as eikonal) propagators to first approximation, while receiving higher power corrections at later stages in the expansion. We furthermore introduce the soft four-velocities $u_i^\mu=\overline{p}_i^\mu /\overline{m}_i$, with the soft masses being $\overline{m}_i^2=\overline{p}_i^2=m_i^2-\frac{q^2}{4}$, so that $u_i^2=1$ and $u_i \cdot q=0$. This way
\begin{equation}
\frac{1}{2k\cdot \overline{p}_i} = \frac{1}{2 u_i \cdot k \sqrt{m_i^2-\frac{q^2}{4}}} = \frac{1}{2u_i \cdot k} \Big(\frac{1}{m_i}+ \frac{q^2}{8 m_i^3} + \mathcal{O}(q^4)\Big),
\end{equation}
and the matter propagators become
\begin{equation}
\begin{tikzpicture}[baseline={([yshift=-0.1cm]current bounding box.center)}] 
	\node[] (a) at (0,0) {};
	\node[] (p4) at ($(a)+(2,0)$) {};
	\draw[line width=0.5mm, postaction={decorate}] (a.center) -- node[sloped, allow upside down] {\midarrow} (p4.center);
\end{tikzpicture} = \frac{1}{m_i}\frac{1}{2u_i\cdot k}+ \mathcal{O}(q^2) \sim \frac{1}{|q|},
\end{equation}
which leads to simpler linearized propagators at leading order in $|q|$ and a factorization of the mass dependence. Since $q^2<0$ becomes the only dimensionful scale, its dependence can be fixed by dimensional analysis and it can be set to $q^2=-1$, although we will keep the full dependence since it does not affect the complexity of our analysis. Thus, the only relevant scale for the process is
\begin{align}
y &= u_1 \cdot u_2 = \frac{\overline{p}_1 \cdot \overline{p}_2}{\overline{m}_1 \overline{m}_2} = \frac{p_1 \cdot p_2 + \frac{q^2}{4}}{\overline{m}_1 \overline{m}_2}= \sigma + q^2 \, \frac{\sigma (m_1^2+m_2^2)+2m_1m_2}{8 m_1^2 m_2^2} + \mathcal{O}(q^4).
\end{align}
As we will see, the square root $\sqrt{y^2-1}$ will regularly appear; hence, the change of variables $y=\frac{x^2+1}{2x}$ will be used to rationalize it.

The existence of the linearized propagators leads to an additional symmetry compared to quadratic propagators: taking $u_1 \rightarrow -u_1$ and $u_2 \rightarrow -u_2$ as well as conjugating the corresponding Feynman $i\epsilon$, the matter propagators change by a sign. We refer to this transformation as \emph{parity}, and we can organize the Feynman integrals into contributions that are \emph{even} under parity, and those that are \emph{odd} under parity; see app.~\ref{app:parity} for further details.

One of the main observables is the impulse $\Delta p_i^\mu$, i.e.\ the (loop) corrected change in momentum of particle $i$. Integrals with different parity occur in different contributions to the impulse $\Delta^{(n)} p_i^\mu$ at $n$PM order, that is, at $n-1$ loops. Schematically, it can be calculated by the formalism of ref.\ \cite{Kosower:2018adc}, or similarly by a Fourier transform~\cite{Dlapa:2023hsl}
\begin{equation}
\label{eq: Fourier_impulse}
\Delta^{(n)} p_i^\mu \sim G^n M^{n+1} \int d^D q \, e^{i q \cdot b} \delta(u_1 \cdot q) \delta(u_2 \cdot q) \left[ q^\mu \, \mathcal{I}^{(n)}_{q} + \sum_{i=1}^2 u_i^\mu \, \mathcal{I}^{(n)}_{u_i} \right],
\end{equation}
where $M$ denotes a generic mass, and $\mathcal{I}^{(n)}_{q,u_i}$ are $n$PM integrals. By dimensional analysis, it can be seen that $[\mathcal{I}^{(n)}_{q}]=n-3$, while $[\mathcal{I}^{(n)}_{u_i}]=n-2$. The only dimensionful dependence in the integrals comes from $|q|$, and odd powers of $|q|$ can only arise from linearized propagators, which are odd under parity, while all other factors in the integral are even. The previous scaling means that $\mathcal{I}^{(n)}_{q}$ and $\mathcal{I}^{(n)}_{u_i}$ respectively involve odd and even-parity PM integrals at even PM order, and the reverse for odd PM orders.

The result can then be organized as
\begin{equation}
 \Delta^{(n)} p_i^\mu=c^{(n)}_{b} \, b^\mu+ \sum_{i=1}^2 c^{(n)}_{u_i} \, u_i^\mu\,;
\end{equation}
cf.\ ref.~\cite{Kalin:2020mvi}. By the argument above, PM integrals of odd (even) parity contribute to $c^{(n)}_{b}$ for $n$ even (odd), while the integrals of even (odd) parity contribute to $c^{(n)}_{u_i}$~\cite{Dlapa:2023hsl}. For conservative dynamics, $c^{(n)}_{u_i}$ can be bootstrapped from lower loops~\cite{Kalin:2020fhe,Dlapa:2023hsl,Mogull:2020sak}, so at even (odd) PM order, only the odd (even) integrals can contribute with new functions. For dissipative dynamics, by contrast, even (odd) integrals can also develop new functions for $n$ even (odd). Since our objective is to equally study the Feynman integrals relevant to the conservative and the dissipative dynamics, for each classical diagram we will thus analyze the corresponding integrals of even as well as odd parity; see also sec.~\ref{sec:methodology_systematics} for more details.

\subsection{Classical diagrams}
\label{sec:classical_diagrams}
In order to identify the classical contributions to the gravitational potential, we first perform a power counting in $G$ and in $|q|$ to single out the relevant diagrams.  As discussed above, graviton propagators scale as $|q|^{-2}$ whereas matter propagators linearize and scale as $|q|^{-1}$. Additionally, the self-interaction vertex of $n$ gravitons scales like
\begin{equation}
\begin{tikzpicture}[baseline={([yshift=-0.1cm]current bounding box.center)}] 
	\node[] (a) at (0,0) {};
	\node[label=left:{$1$}] (a1) at (-1,0) {};
	\node[label={[xshift=0.3cm, yshift=-0.6cm]$2$}] (a2) at (0.71,-0.71) {};
	\node[label={[xshift=0.3cm, yshift=-0.2cm]$n$}] (an) at (0.71,0.71) {};
	\draw[line width=0.15mm] (a.center) -- (a1.center);
	\draw[line width=0.15mm] (a.center) -- (a2.center);
	\draw[line width=0.15mm] (a.center) -- (an.center);
	\node at (0.6,-0.2)[circle,fill,inner sep=0.6pt]{};
	\node at (0.6,0)[circle,fill,inner sep=0.6pt]{};
	\node at (0.6,0.2)[circle,fill,inner sep=0.6pt]{};
\end{tikzpicture} \sim |q|^{2} G^{\frac{n}{2}-1},
\end{equation}
and the interaction of a matter line with $n$ gravitons scales as \begin{equation}
\begin{tikzpicture}[baseline={([yshift=-0.1cm]current bounding box.center)}] 
	\node[] (a) at (0,0) {};
	\node[] (a0) at (-1,0) {};
	\node[] (a1) at (1,0) {};
	\node[label={[xshift=-0.2cm, yshift=-0.6cm]$1$}] (a2) at (-0.71,-0.71) {};
	\node[label={[xshift=0.2cm, yshift=-0.6cm]$n$}] (an) at (0.71,-0.71) {};
	\draw[line width=0.5mm] (a.center) -- (a0.center);
	\draw[line width=0.5mm] (a.center) -- (a1.center);
	\draw[line width=0.15mm] (a.center) -- (a2.center);
	\draw[line width=0.15mm] (a.center) -- (an.center);
	\node at (-0.2,-0.6)[circle,fill,inner sep=0.6pt]{};
	\node at (0,-0.6)[circle,fill,inner sep=0.6pt]{};
	\node at (0.2,-0.6)[circle,fill,inner sep=0.6pt]{};
\end{tikzpicture} \sim G^{\frac{n}{2}}.
\end{equation}
Lastly, the integration measure for each loop scales as $\int d^4k \sim |q|^4$.

As shown above, the long-distance regime establishes an expansion in terms of the small parameter $r_s/|b| \sim G m |q| \ll 1$, supplemented with the soft-$|q|$ expansion to extract the classical dynamics. Since the tree-level contribution contains Newtonian physics $\sim |q|^{-2} G$, each loop subsequently adds a $|q| \, G$ correction. Thus, diagrams contributing to the classical potential between the two black holes scale as $|q|^{L-2} G^{L+1}$ at $L$-loops~\cite{Neill:2013wsa}, corresponding to an $(L+1)$-PM correction. 

For a given PM order, diagrams subleading in $|q|$ are quantum corrections and can be discarded. There are also diagrams with fewer powers of $|q|$ relative to the classical scaling, which correspond to superclassical (or iteration) contributions and cancel order-by-order when matching the full theory with the effective field theory~\cite{Bern:2019crd}; thus, in principle they can also be dropped for the purpose of identifying geometries, but we will keep them in our discussion for illustrative purposes and since their leading singularities are useful for bringing the corresponding differential equation into canonical form.
\begin{figure}[tb]
\centering
\subfloat[]{\begin{tikzpicture}[baseline={([yshift=-0.1cm]current bounding box.center)}, scale=0.85] 
	\node[] (a) at (0,0) {};
	\node[] (d1) at (0,-0.4) {};
	\node[] (d2) at (0,-1.2) {};
	\node[] (b) at (0,-1.6) {};
	\node[] (p1) at ($(a)+(-0.7,0)$) {};
	\node[] (p2) at ($(b)+(-0.7,0)$) {};
	\node[] (p3) at ($(b)+(0.7,0)$) {};
	\node[] (p4) at ($(a)+(0.7,0)$) {};
	\draw[line width=0.15mm] (d1.center) -- (a.center);
	\draw[line width=0.15mm] (d2.center) -- (b.center);
	\draw[line width=0.5mm] (p1.center) -- (p4.center);
	\draw[line width=0.5mm] (p2.center) -- (p3.center);
	\draw[line width=0.15mm] (0,-0.8) circle (0.4);
\end{tikzpicture}}
\qquad
\subfloat[]{
\begin{tikzpicture}[baseline={([yshift=-0.1cm]current bounding box.center)}, scale=1.1] 
	\node[] (a) at (0,0) {};
	\node[] (b) at (0,-1) {};
	\node[] (c) at (0.5,-0.5) {};
	\node[] (p1) at ($(a)+(-0.85,0)$) {};
	\node[] (p2) at ($(b)+(-0.85,0)$) {};
	\node[] (p3) at ($(b)+(1,0)$) {};
	\node[] (p4) at ($(a)+(1,0)$) {};
	\draw[line width=0.15mm] (a.center) -- (b.center);
	\draw[line width=0.5mm] (p1.center) -- (a.center);
	\draw[line width=0.5mm] (a.center) -- (c.center);
	\draw[line width=0.5mm] (c.center) -- (p4.center);
	\draw[line width=0.5mm] (p2.center) -- (b.center);
	\draw[line width=0.5mm] (b.center) -- (c.center);
	\draw[line width=0.5mm] (c.center) -- (p3.center);
\end{tikzpicture}}
\qquad
\subfloat[]{\begin{tikzpicture}[baseline={([yshift=-0.1cm]current bounding box.center)}, scale=0.9] 
	\node[] (a) at (0,0) {};
	\node[] (b) at (0,-1) {};
	\node[] (p1) at ($(a)+(-0.5,0)$) {};
	\node[] (p2) at ($(b)+(-0.5,0)$) {};
	\node[] (p3) at ($(b)+(1.5,0)$) {};
	\node[] (p4) at ($(a)+(1.5,0)$) {};
	\draw[line width=0.15mm] (a.center) -- (b.center);
	\draw[line width=0.5mm] (p1.center) -- (p4.center);
	\draw[line width=0.5mm] (p2.center) -- (p3.center);
	\draw[line width=0.15mm] (1,0) arc (0:180:0.5);
\end{tikzpicture}}
\caption{(a): One-loop diagram with quantum power counting that does not contribute to the classical dynamics; (b) and (c): Examples of one-loop diagrams that vanish in dimensional regularisation.}
\label{fig: diagrams_discarded}
\end{figure}

As explained in refs.~\cite{Bern:2019crd,Herrmann:2021tct}, the classical limit allows us to discard several diagrams from the beginning; see fig.~\ref{fig: diagrams_discarded} for three examples. Concretely, we will discard diagrams containing pure graviton loops, see fig.~\ref{fig: diagrams_discarded}(a), as they have a quantum power counting and do not contribute to the classical dynamics. To avoid such quantum contributions, we will require that all closed loops contain at least one matter line. In particular, $|q|$ scaling  implies that classical diagrams have exactly $L$ matter lines at $L$ loops.\footnote{For any connected graph, we have $L=I-V+1$, where $I$ is the number of internal edges and $V$ is the number of vertices. We can split the latter numbers into graviton and scalar contributions: $V=V_g+V_s$ and $I=I_g+I_s$. Noticing that $V_s=2+I_s$ as well as recalling the $|q|$ scaling of propagators, vertices and the loop integration measure, we find $I_s=L$ for a classical diagram.}${}^{,}$\footnote{In particular, the worldline formulation makes this manifest by having $L$ delta-function matter propagators at $L$ loops.} Furthermore, we will also discard diagrams where none of the loop integrals depend on the momentum transfer $q^2$ (the only dimensionful scale), see figs.~\ref{fig: diagrams_discarded}(b) and~\ref{fig: diagrams_discarded}(c), since they are dimensionless and thus vanish in dimensional regularisation.

\section{Classification of geometries via leading singularities}
\label{sec:Methodology}
A number of methods have been developed and used in the literature to detect and identify the non-trivial geometries appearing in Feynman integrals; see ref.~\cite{Bourjaily:2022bwx} for a review. In essence, they are mainly based on two different approaches: differential equations~\cite{Kotikov:1990kg} and leading singularities~\cite{Cachazo:2008vp}.

On the one hand, one can investigate the differential equations that the Feynman integrals satisfy~\cite{Kotikov:1990kg}. To do so, one first finds a basis of master integrals $\vec{\mathcal{I}}$ for the problem of interest, a minimal set to which all integrals can be reduced via integration-by-parts identities (IBPs)~\cite{Chetyrkin:1981qh}. Then, one derives a linear system of coupled first-order differential equations for $\vec{\mathcal{I}}$ with respect to a kinematic variable $x$: $\partial_x \vec{\mathcal{I}}=A \, \vec{\mathcal{I}}$, where $A$ is a matrix with entries depending on the kinematics and the dimension. Due to the presence of linearized propagators, the differential equations of PM integrals decouple into two blocks, see e.g.\ refs.~\cite{Parra-Martinez:2020dzs,DiVecchia:2021bdo,Herrmann:2021tct}. As we prove in app.~\ref{app:parity}, this decoupling is due to a splitting of the IBP identities and the differential equations under the parity transformation. We assume that there is no further decoupling.
 
By taking further derivatives, one can then derive a single higher-order differential equation
\begin{equation}
\mathcal{L}_n \mathcal{I}_i = \left( \frac{d^n}{d x^n} + \sum_{j=0}^{n-1} c_j(x) \frac{d^j}{d x^j} \right) \mathcal{I}_i = \text{inhomogeniety}
\end{equation}
for the master integral $\mathcal{I}_i$ of interest, where the operator $\mathcal{L}_n$ is known as the Picard-Fuchs operator and $c_j(x)$ are rational functions. The inhomogeneity is given by a linear combination of master integrals from subsectors, which are integrals where a number of propagators in $\mathcal{I}_i$ are absent, corresponding to a diagram where the corresponding lines have been pinched. 
Each subsector in the inhomogeneity (iteratively) has a Picard-Fuchs operator associated to it. When solving the differential equation, one works upwards from bottom sectors to top sectors.  

The rational factorization of the Picard-Fuchs operator into a product of irreducible differential operators $\mathcal{L}_{n_1} \cdots \mathcal{L}_{n_m}$ allows us to detect non-trivial geometries. If $\mathcal{L}_n$ factorizes into a product $\mathcal{L}_1 \cdots \mathcal{L}_1$ of operators of order one, and the same is true (iteratively) for the Picard-Fuchs operators of the inhomogeneity, the differential equation for the respective integral can be brought into dlog form. In most known cases, a dlog form implies that the integral is polylogarithmic; see, however, ref.~\cite{Duhr:2020gdd} for an example of an integral in dlog form that cannot be expressed in terms of polylogarithms. In the following, we will focus on whether the integrals have the dlog property and not address the question of whether or not they can be written in terms of polylogarithms, which lies beyond the scope of this work. In particular, a sector is only expressible in dlog form if all its relevant subsectors are also expressible in that form; note that some subsectors only have quantum corrections and are thus not relevant in the classical limit, though. However, the presence of any $\mathcal{L}_{n_i>1}$ indicates a non-trivial geometry. Then, the solutions to the homogeneous differential equation are periods of these geometries, such that the solutions of the inhomogeneous differential equation lie beyond the space of polylogarithms. In all known examples, such a geometry was found to be either a Calabi-Yau $(n_i-1)$-fold or a higher-genus curve~\cite{Huang:2013kh,Marzucca:2023gto}, which we can distinguish by a further analysis of the leading singularity and of the Picard-Fuchs operator.\footnote{Very recently, it was found that the homogeneous differential equation for the four-loop banana integral, which contains a Calabi-Yau three-fold, can also be solved via a genus-two curve, albeit with transcendental coefficients~\cite{Jockers:2024tpc}.}
 In particular, the homogeneous differential equation allows us to study the geometries in a Feynman integral subsector by subsector.

The drawback of using differential equations to detect geometries is that one first needs to solve the IBPs, which can become extremely challenging. Even though algorithmic methods for solving IBPs exist, such as ref.~\cite{Laporta:2000dsw}, with several publicly available computer code implementations~\cite{Smirnov:2023yhb,vonManteuffel:2012np,Lee:2013mka,Klappert:2020nbg}, solving them usually represents the main bottleneck in state-of-the-art calculations. See however refs.~\cite{Mastrolia:2018uzb,Frellesvig:2019uqt} for an alternative approach to identifying integral relations, which was recently applied to the PM expansion up to one loop in refs.~\cite{Brunello:2023fef,Frellesvig:2024swj}, as well as refs.~\cite{Lairez:2022zkj,delaCruz:2024xit} for a method of obtaining Picard-Fuchs operators via the polynomial equations that define the algebraic variety, thus avoiding standard IBPs.

As an alternative approach, one can compute the leading singularity (LS) of the diagram, which is related to the maximally iterated discontinuity of the integral~\cite{Cachazo:2008vp}. Discontinuities can be calculated via cuts, which are generally computed as the propagators go on-shell, meaning that they are replaced by delta functions $\frac{i}{Q_i^2 - m_i^2} \to \delta(Q_i^2 - m_i^2)$. In our approach, we use generalized cuts~\cite{Britto:2024mna}, in which we deform the integration contour such that it encircles the pole at $Q_i^2=m_i^2$, effectively taking the residue. The leading singularity corresponds to taking the maximal cut, i.e.\ all propagators on-shell, as well as deforming all further integrals to closed contours, which can be calculated via further residues in cases of poles.
Note that in the case of branch cuts, the corresponding contour integrals generically cannot be calculated via residues, though. Further note that for multiple poles and/or branch cuts, several different leading singularities exist; only in simple cases do we have exactly two leading singularities, which add up to zero via the global residue theorem and are thus equal up to a sign~\cite{Bosma:2017ens, Bourjaily:2020hjv, Frellesvig:2021hkr, Weinzierl:2022eaz}.\footnote{Note that we are treating propagators and irreducible scalar products (ISPs, to be defined below) differently here.
We can always take the residues corresponding to propagators since the diagrams in which they are absent are accounted for by subsectors. By contrast, there are master integrals with powers of the ISPs in the numerator, for which the poles are absent. Therefore, for ISPs we also need to take into account contours that do not encircle these poles; see sec.~\ref{sec:methodology_systematics} for further details.}

Importantly, the leading singularity provides an indication of the geometry of the Feynman integral. Specifically, if all leading singularities of a diagram and all its subsectors are algebraic, then the diagram has a dlog form. Otherwise, the diagram depends on a geometry that goes beyond polylogarithms. Moreover, if a sector has only vanishing leading singularities, it has no master integrals~\cite{Lee:2013hzt, Bosma:2017ens}.%
\footnote{The opposite direction is more subtle. A sector can have a non-vanishing leading singularity in $D=4$ when it is reducible to lower sectors via IBP relations that vanish in $D=4$. An example of this is the one-loop one-mass triangle integral, which is reducible to the one-loop bubble in $D\neq4$ dimensions and has a leading singularity proportional to $0^{(D-4)/2}$, with leading singularity $1/s$ in $D=4$.
}

The advantage of using the leading singularity to identify geometries is that it localizes many of the integration variables, making it simpler to calculate than the differential equation. Moreover, we can use our results to obtain a basis of pure master integrals, which is typically obtained in polylogarithmic cases by dividing each master integral by its leading singularity~\cite{Henn:2014qga, Primo:2016ebd, WasserMSc, Dlapa:2021qsl}.

Lastly, throughout this paper we will assume that there are no total cancellations of geometries in the sum of the diagrams. Unlike in $\mathcal{N}=4$ super Yang-Mills theory, where supersymmetry imposes numerous cancellations between boson and fermion contributions to the amplitude, we do not expect mechanisms ensuring such total cancellations in the PM expansion of general relativity. Thus, we will assume that the geometries appearing in individual Feynman diagrams are also present in the amplitude, which will be indicative of the space of functions it evaluates to. At four loops and at first self-force order, i.e.\ for diagrams containing only one linearized propagator on one of the two matter lines,
however, it was found in ref.~\cite{Driesse:2024xad} that the complete elliptic integrals arising in the diagrams for the conservative sector do not appear in their final result for the momentum impulse. The absence of these elliptic integrals is not due to a cancellation of geometries between diagrams, though. Instead, it is due to the fact that the Fourier integral in eq.~\eqref{eq: Fourier_impulse} introduces a factor of $\varepsilon$ for odd $n$PM orders and that the particular combination of master integrals in general relativity that contains those elliptic integrals is finite, i.e.\ it does not have $\varepsilon^{-1}$ poles. 
The elliptic integrals thus appear at higher orders in $\varepsilon$ in dimensional regularisation, and thus do not contribute to the result in strict $D=4$.\footnote{We would like to thank Gustav Jakobsen and Zhengwen Liu for pointing this out.} 
It will be interesting to study whether something similar occurs for the diagrams that contain Calabi-Yau integrals at four loops~\cite{Frellesvig:2023bbf,Klemm:2024wtd}.
However, this requires the full integrand and is thus beyond the scope of this work.

In sec.~\ref{sec:Baikov}, we will present the loop-by-loop Baikov representation of Feynman integrals~\cite{Baikov:1996iu,Frellesvig:2017aai} as a convenient choice when computing leading singularities, since cutting propagators can be done straightforwardly in this representation. In sec.~\ref{sec:methodology_systematics}, we then present the systematics of our analysis, such as the loop-by-loop integration order followed and relevant changes of variables that expose further poles. By decomposing the leading singularity into a sequence of lower-loop calculations, we will reveal powerful identities between the geometries of different diagrams in sec.~\ref{sec:reduction_relations}. In particular, we will be able to recycle lower-loop results and easily classify the geometries in the PM integrals by means e.g.\ of the reduction of box iterations and the unraveling of matter propagators for non-planar graphs. Moreover, we find simple visual rules that identify certain sectors that have no master integrals. 

\subsection{The Baikov representation}
\label{sec:Baikov}

In this section, we go over the  details of the Baikov representation for Feynman integrals, and emphasize its usefulness when computing the leading singularity, which we use to detect and identify the geometry behind the integrals.

Let us consider a scalar Feynman diagram in $D$ dimensions with $\nint$ propagators of power $\nu_i$, mass $m_i$ and generic momenta $Q_i$ in the loop-momentum representation:
\begin{equation}
\label{eq: loop momentum Feynman integral}
\mathcal{I}= \int \prod_{j=1}^L d^D k_j \prod_{i=1}^\nint \frac{1}{{(Q_i^2-m_i^2)}^{\nu_i}},
\end{equation}
where $k_j$ denote the $L$ loop momenta. 

Let us change variables to the Baikov variables $z_i=Q_i^2-m_i^2$, assuming for now that this change of variables is bijective; see below for a discussion of the non-bijective case. We then obtain the standard Baikov representation~\cite{Baikov:1996iu}, see also ref.~\cite{Weinzierl:2022eaz} for an introduction:
\begin{equation}
\label{eq: standard_Baikov}
\mathcal{I}=\mathcal{J}\, \frac{{(\det G(p_1,\dots,p_E))}^{\frac{-D+E+1}{2}}}{\prod_{j=1}^L \Gamma{\Big( \frac{D-E+1-j}{2} \Big)}} \int_{\mathfrak{C}} \frac{d^\nint z}{z_1^{\nu_1} \cdots z_{\nint}^{\nu_\nint}} \, {\mathcal{B}(z_1,\dots,z_{\nint})}^{\frac{D-L-E-1}{2}},
\end{equation}
where $\mathcal{J}$ is the Jacobian for the change of variables, which depends on the exact conventions for the propagators, but is commonly given by 
$\mathcal{J}=\pm 2^{L-\nint}$.
Here, $\mathcal{B}(z_1,\dots,z_{\nint})=\det G(k_1,\dots,k_L,p_1,\dots,p_E)$ is the Baikov polynomial, the roots of which specify the domain of integration $\mathfrak{C}$. We also introduce $G$, which denotes the Gram matrix with entries $G_{ij}(Q_1,\dots,Q_n)=Q_i \cdot Q_j$. Moreover, for a case of $n_{\text{ext}}$ external particles, $E=\text{dim} \langle p_1, \dots, p_{n_{\text{ext}}} \rangle$ is the number of independent external momenta, which for generic kinematics and $D \geq E$ is equal to $n_{\text{ext}}-1$. 

For the Baikov representation to be attainable, we need to express the Baikov polynomial in terms of the Baikov variables $z_i=Q_i^2-m_i^2$. Thus, we first need to invert the change of variables; in other words, it needs to be bijective. This requires that the number of independent scalar products $N_{\text{V}}$ involving the loop momenta is equal to the number of Baikov variables. Concretely, we have $L(L+1)/2$ products of loop momenta $k_i \cdot k_j$, and $EL$ products $p_i \cdot k_j$ with the independent external momenta at $L$-loops. Hence, we need $\nint=N_{\text{V}}$ for 
\begin{equation}
\label{eq: number_scalar_prods}
N_{\text{V}} = \frac{1}{2} L(L+1) + E L.
\end{equation}
This is in general not guaranteed. On the one hand, if $\nint>N_{\text{V}}$ the Baikov representation becomes overcomplete -- since the propagators are not all independent, and thus the Baikov variables can be related; see also ref.~\cite{Weinzierl:2022eaz} for details. This case is rather special; however, we will use it to relate the leading singularity of certain non-planar diagrams to their planar counterparts, see sec.~\ref{sec:two_loop} and app.~\ref{app: Nonplanar_unraveling}. 

On the other hand, in the most common scenario $\nint<N_{\text{V}}$. In this case, we need to artificially add extra Baikov variables $z_j$ such that the remaining $\nISP=N_{\text{V}}-\nint$ irreducible scalar products (ISPs) are introduced. This can be interpreted as finding a bigger set of Baikov variables where $\nint=N_{\text{V}}$ is satisfied by construction, but not including their propagators for the induced representation since ISPs can only appear in the numerator; see ref.~\cite{Weinzierl:2022eaz} for details. In particular, we have in this case
\begin{equation}
\label{eq: standard_Baikov_ISP}
\mathcal{I}=\mathcal{J}\, \frac{{(\det G(p_1,\dots,p_E))}^{\frac{-D+E+1}{2}}}{\prod_{j=1}^L \Gamma{\Big( \frac{D-E+1-j}{2} \Big)}} \int_{\mathfrak{C}} \frac{d^{N_{\text{V}}} \! z\  \mathcal{N}(z_{\nint+1},\dots,z_{N_{\text{V}}}) }{z_1^{\nu_1} \cdots z_{\nint}^{\nu_\nint}} \ {\mathcal{B}(z_1,\dots,z_{N_{\text{V}}})}^{\frac{D-L-E-1}{2}},
\end{equation}
where we are now also allowing for a generic numerator $\mathcal{N}$ that only depends on the extra Baikov variables.

The Baikov representation is especially useful when calculating cuts of Feynman integrals, for which the propagators go on-shell. Under the cut, we deform the contour of the integral to encircle the pole $Q_i^2 = m_i^2$, which is equivalent to computing the residue at $z_i=0$. To obtain the integral on the maximal cut, we thus compute the residues at $z_i=0$ for $i=1,\dots,\nint$, which fully localizes those integrals. For instance, for all $\nu_i=1$ we obtain
\begin{align}
& \mathcal{I}_{\text{max-cut}}^{\nu_i=1}= \mathcal{J}\, \frac{{(\det G(p_1,\dots,p_E))}^{\frac{-D+E+1}{2}}}{\prod_{j=1}^L \Gamma{\Big( \frac{D-E+1-j}{2} \Big)}} \nonumber \\
& \qquad \qquad \times \int d^{\nISP} z \ \mathcal{N}(z_{\nint+1},\dots,z_{N_{\text{V}}}) \ {\mathcal{B}(\underbrace{0,\dots,0}_{\nint},z_{\nint+1},\dots,z_{N_{\text{V}}})}^{\frac{D-L-E-1}{2}}.
\label{eq: cut_Baikov}
\end{align}

At the maximal cut, $\nISP$ integrals remain. Yet often further poles can be revealed, allowing for further residues to be taken. This is sometimes apparent only after a change of variables, as we will illustrate with the examples in sec.~\ref{sec:two_loop}. We will then generalize the notion of leading singularity to be the integral associated to the integration contours around those poles too. However, we want to avoid as many of these integrals and changes of variables as possible, since they can be non-trivial to find.

In order to reduce the number of remaining integrals, we will furthermore use the so-called loop-by-loop Baikov representation~\cite{Frellesvig:2017aai}, where the standard Baikov representation is employed for each loop separately.\footnote{Equivalently, the loop-by-loop Baikov representation can be obtained from the standard Baikov representation by integrating out variables, see e.g.\ ref.~\cite{Jiang:2023qnl}.} In this case, $N_{\text{V}} = L + \sum_{i=1}^L E_i$; thus, we frequently have many fewer ISPs than when using the standard Baikov representation. The drawback is that we now have $2L-1$ different Baikov polynomials that depend on loop momenta, whereas with standard Baikov we have only one. Therefore, the expressions become much more intricate, and the complexity of the Baikov polynomials is highly dependent on the loop-by-loop order followed. Furthermore, one can encounter sub-loops where the kinematics are not generic due to $E_i>D$, keeping in mind that in $D$ dimensions we can only have $E_i \leq D$ independent momenta. While a Baikov representation strictly in $D$ dimensions is possible~\cite{Frellesvig:2021vdl}, we are able to avoid such cases up to four loops with an educated choice of the integration order; see sec.~\ref{sec:methodology_systematics}.

In our case, we will consider diagrams within the post-Minkowskian expansion in $D=4-2\varepsilon$ dimensions, with the main subtlety being the presence of linearized matter propagators; see sec.~\ref{sec:soft_expansion}. Nonetheless, the Baikov representation is well-suited for this task. This is because when inverting the change of variables $z_j=Q_j^2-m_j^2$ in order to express all scalar products in terms of Baikov variables, the product $u_i \cdot k$ that appears in the linearized propagators can be immediately mapped to the Baikov variable $z= 2 u_i \cdot k$ denoting the corresponding propagator. Hence, dealing with PM diagrams does not cause any problems in this representation; in fact, linearized propagators simplify the calculation.

\subsection{Modus operandi and further details}
\label{sec:methodology_systematics}

In this subsection, we provide the systematics of the analysis used to classify the geometries present in the Feynman integrals order-by-order in the PM expansion.  We also provide the changes of variables used to expose further poles to take residues at, as well as the methodology employed for integrals with higher-order poles.

First of all, the overall aim is to generate as few ISPs as possible by choosing an educated diagram parametrization and loop-by-loop integration order, which consequently reduces the number of integration variables in the leading singularity. As such, we will normally parametrize the diagrams such that the momentum transfer $q$ appears in the last loop. This way, we avoid the proliferation of products $k_i \cdot q$ with lower loop momenta, which increase the number of ISPs. Additionally, we will start the integration order with the simplest sub-loops (bubbles, triangles) and progress toward boxes, avoiding the appearance of pentagons (or higher-sided polygons) with $E_i \nless D=4-2\varepsilon$. While this turns out to always be possible up to three loops, starting at four loops we unavoidably encounter diagrams with pentagon sub-loops; see sec.~\ref{sec:four_loop_nonplanar} for an example. In such cases, we have a vanishing prefactor 
\begin{equation}
\frac{1}{\Gamma (\frac{D-E_i}{2})} = \frac{1}{\Gamma (-\varepsilon)} = -\varepsilon + \mathcal{O}(\varepsilon^2)
\end{equation}
in the Baikov representation. This is an artifact of dimensional regularisation, and the $\varepsilon$-dependence is compensated by a $\varepsilon^{-1}$-divergence when integrating over the Baikov contour $\mathfrak{C}$. As shown in ref.~\cite{Frellesvig:2021vdl} for the elliptic double-box diagram, the resulting geometry is nonetheless in agreement with that from a Baikov representation strictly in $D=4$. To avoid this explicit $\varepsilon$ dependence, we can study such diagrams in $D=6-2\varepsilon$, since the four dimensional result can be related to it through dimension-shift identities~\cite{Tarasov:1996br, Lee:2012te}, and thus they depend on the same geometry.

Starting at two loops in sec.~\ref{sec:two_loop}, we will avoid repeating calculations and we will compute the leading singularities by recycling lower-loop results, which exploits the advantage of using a loop-by-loop Baikov representation. For instance, one can calculate
\begin{equation}
\LS \left( \begin{tikzpicture}[baseline={([yshift=-0.1cm]current bounding box.center)}] 
	\node[] (a) at (0,0) {};
	\node[] (a1) at (0.8,0) {};
	\node[] (b) at (0.4,-0.8) {};
	\node[] (b1) at (1.2,-0.8) {};
	\node[] (p1) at ($(a)+(-0.2,0)$) {};
	\node[] (p2) at ($(b)+(-0.6,0)$) {};
	\node[] (p3) at ($(b1)+(0.2,0)$) {};
	\node[] (p4) at ($(a1)+(0.6,0)$) {};
	\draw[line width=0.15mm] (b.center) -- (a.center);
	\draw[line width=0.15mm] (b.center) -- (a1.center);
	\draw[line width=0.15mm] (b1.center) -- (a1.center);
	\draw[line width=0.5mm] (p1.center) -- (p4.center);
	\draw[line width=0.5mm] (p2.center) -- (p3.center);
\end{tikzpicture} \right) = \LS \left(
\begin{tikzpicture}[baseline={([yshift=-0.1cm]current bounding box.center)}] 
	\node[] (a) at (0,0) {};
	\node[] (a1) at (0.8,0) {};
	\node[] (b) at (0.4,-0.8) {};
	\node[] (p1) at ($(a)+(-0.2,0)$) {};
	\node[] (p2) at ($(b)+(-0.6,0)$) {};
	\node[] (p3) at ($(b)+(0.6,0)$) {};
	\node[] (p4) at ($(a1)+(0.2,0)$) {};
	\draw[line width=0.15mm] (b.center) -- (a.center);
	\draw[line width=0.15mm] (b.center) -- (a1.center);
	\draw[line width=0.15mm,-{Latex[length=2.2mm]}](1.1,-0.75) -- (1.1,-0.05);
	\node[label={[xshift=0.35cm, yshift=-0.45cm]$k_2$}] (k) at (1.1,-0.4) {};
	\draw[line width=0.5mm] (p1.center) -- (p4.center);
	\draw[line width=0.5mm] (p2.center) -- (p3.center);
\end{tikzpicture} \times
\begin{tikzpicture}[baseline={([yshift=-0.1cm]current bounding box.center)}]
	\node[] (a) at (0,0) {};
	\node[] (a1) at (0.8,0) {};
	\node[] (b) at (0,-0.8) {};
	\node[] (b1) at (0.8,-0.8) {};
	\node[] (p1) at ($(a)+(-0.2,0)$) {};
	\node[] (p2) at ($(b)+(-0.2,0)$) {};
	\node[] (p3) at ($(b1)+(0.2,0)$) {};
	\node[] (p4) at ($(a1)+(0.2,0)$) {};
	\draw[line width=0.15mm, dashed, postaction={decorate}] (b.center) -- node[sloped, allow upside down, label={[xshift=0.2cm, yshift=0cm]$k_2$}] {\midarrow} (a.center);
	\draw[line width=0.15mm] (b1.center) -- (a1.center);
	\draw[line width=0.5mm] (p1.center) -- (a.center);
	\draw[line width=0.5mm, dashed, dash phase=1pt, postaction={decorate}] (a.center) -- node[sloped, allow upside down, label={[xshift=0cm, yshift=-0.2cm]$2u_1{\cdot}k_2$}] {\midarrow} (a1.center);
	\draw[line width=0.5mm] (a1.center) -- (p4.center);
	\draw[line width=0.5mm] (p2.center) -- node[label={[xshift=0cm, yshift=-0.85cm]\textcolor{white}{$2u_2{\cdot}k_2$}}] {} (p3.center);
\end{tikzpicture}
\right)
\end{equation}
by first computing the leading singularity for the first triangle, with $k_2$ as momentum transfer. Afterwards, we can calculate the contribution from the $k_2$-loop, which requires adding the extra Baikov variables $z_6=k_2^2$ and $z_7=2u_1 \cdot k_2$ to match the ISPs, which are indicated with dashed propagators. The equation above is a schematic representation of the loop-by-loop decomposition chosen, and should be understood as calculating the leading singularity for the loop with the extra (dashed) propagators in the last place.

Also starting at two loops, we see the appearance of further poles once the maximal cut has been computed, which allow for further residues. In general, a convenient way to calculate them is by using the {\texttt{\textup{LeadingSingularities}}} command implemented in the {\DlogBasis} package~\cite{Henn:2020lye}. To expose those further poles, one usually needs to rationalize square roots depending on the extra Baikov variables. Conveniently, the package {\RationalizeRoots}~\cite{Besier:2019kco} can in many cases provide such rationalizations. For instance, for square roots of quadratic polynomials $\sqrt{(z-r_1)(z-r_2)}$, where $r_i$ are the roots, we found the change of variables~\cite{Besier:2018jen} to~$t$
\vspace*{-0.2cm}
\begin{equation}
\label{eq: change_of_variables_(z-r1)(z-r2)}
z = r_1-\frac{(r_2-r_1)(1-t)^2}{4t},
\end{equation}
to be very useful to expose further poles. Similarly, in integrals such as
\begin{equation}
\int \frac{d z \, d \vec{z}_j}{\sqrt{z-{P(\vec{z}_j)}^2} \sqrt{\widetilde{P}(z,\vec{z}_j)}},
\end{equation}
where $P(\vec{z}_j)$ is a polynomial in the remaining Baikov variables $\vec{z}_j$ and $\widetilde{P}(z,\vec{z}_j)$ depends on all variables, we found it useful to rationalize the first square root with the change of variables
\begin{equation}
\label{eq: change_of_variables_(z-c_squared)}
z = \frac{1-2i t \, P(\vec{z}_j)}{t^2}.
\end{equation}

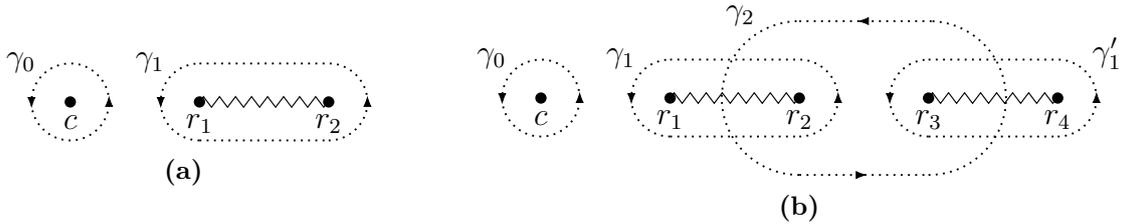
\begin{figure}[tb]
\centering
\subfloat[]{\begin{tikzpicture}[baseline={([yshift=-0.1cm]current bounding box.center)}, scale=0.85] 
	\node[] (p) at (0,0) {};
	\node[] (r1) at (2,0) {};
	\node[] (r2) at (4,0) {};
	\draw[line width=0.15mm, decorate, decoration={zigzag, segment length=6, amplitude=2}] (r1.center) -- (r2.center);
	\node at (p.center) [circle,fill,inner sep=1.5pt]{};
	\draw[line width=0.25mm, dotted] (p.center) circle (0.6);
	\draw[line width=0.15mm,-{Latex[length=1.5mm]}]($(p)+(-0.6,-0.05)$) -- ($(p)+(-0.6,-0.1)$);
	\draw[line width=0.15mm,-{Latex[length=1.5mm]}]($(p)+(0.6,0.05)$) -- ($(p)+(0.6,0.1)$);
	\node[label={[xshift=0cm, yshift=-0.65cm]$c$}] (lp) at (p.center) {};
	\node[label={[xshift=0.2cm, yshift=-0.75cm]$\gamma_0$}] (lg0) at ($(p)+(-1,1)$) {};
	\node at (r1.center) [circle,fill,inner sep=1.5pt]{};
	\node at (r2.center) [circle,fill,inner sep=1.5pt]{};
	\draw[line width=0.25mm, dotted] ($(r2)+(0,-0.6)$) arc (270:360:0.6);
	\draw[line width=0.25mm, dotted] ($(r2)+(0.6,0)$) arc (0:90:0.6);
	\draw[line width=0.25mm, dotted] ($(r1)+(0,-0.6)$) arc (270:90:0.6);
	\draw[line width=0.25mm, dotted] ($(r1)+(0,-0.6)$) -- ($(r2)+(0,-0.6)$);
	\draw[line width=0.25mm, dotted] ($(r1)+(0,0.6)$) -- ($(r2)+(0,0.6)$);
	\draw[line width=0.15mm,-{Latex[length=1.5mm]}]($(r1)+(-0.6,-0.05)$) -- ($(r1)+(-0.6,-0.1)$);
	\draw[line width=0.15mm,-{Latex[length=1.5mm]}]($(r2)+(0.6,0.05)$) -- ($(r2)+(0.6,0.1)$);
	\node[label={[xshift=0cm, yshift=-0.7cm]$r_1$}] (lr1) at (r1.center) {};
	\node[label={[xshift=0cm, yshift=-0.7cm]$r_2$}] (lr2) at (r2.center) {};
	\node[label={[xshift=0.2cm, yshift=-0.75cm]$\gamma_1$}] (lg1) at ($(r1)+(-1,1)$) {};
\end{tikzpicture}}
\qquad \quad
\subfloat[]{\begin{tikzpicture}[baseline={([yshift=-0.1cm]current bounding box.center)}, scale=0.85] 
	\node[] (p) at (0,0) {};
	\node[] (r1) at (2,0) {};
	\node[] (r2) at (4,0) {};
	\node[] (r3) at (6,0) {};
	\node[] (r4) at (8,0) {};
	\draw[line width=0.15mm, decorate, decoration={zigzag, segment length=6, amplitude=2}] (r1.center) -- (r2.center);
	\draw[line width=0.15mm, decorate, decoration={zigzag, segment length=6, amplitude=2}] (r3.center) -- (r4.center);
	\node at (p.center) [circle,fill,inner sep=1.5pt]{};
	\draw[line width=0.25mm, dotted] (p.center) circle (0.6);
	\draw[line width=0.15mm,-{Latex[length=1.5mm]}]($(p)+(-0.6,-0.05)$) -- ($(p)+(-0.6,-0.1)$);
	\draw[line width=0.15mm,-{Latex[length=1.5mm]}]($(p)+(0.6,0.05)$) -- ($(p)+(0.6,0.1)$);
	\node[label={[xshift=0cm, yshift=-0.65cm]$c$}] (lp) at (p.center) {};
	\node[label={[xshift=0.2cm, yshift=-0.75cm]$\gamma_0$}] (lg0) at ($(p)+(-1,1)$) {};
	\node at (r1.center) [circle,fill,inner sep=1.5pt]{};
	\node at (r2.center) [circle,fill,inner sep=1.5pt]{};
	\draw[line width=0.25mm, dotted] ($(r2)+(0,-0.6)$) arc (270:360:0.6);
	\draw[line width=0.25mm, dotted] ($(r2)+(0.6,0)$) arc (0:90:0.6);
	\draw[line width=0.25mm, dotted] ($(r1)+(0,-0.6)$) arc (270:90:0.6);
	\draw[line width=0.25mm, dotted] ($(r1)+(0,-0.6)$) -- ($(r2)+(0,-0.6)$);
	\draw[line width=0.25mm, dotted] ($(r1)+(0,0.6)$) -- ($(r2)+(0,0.6)$);
	\draw[line width=0.15mm,-{Latex[length=1.5mm]}]($(r1)+(-0.6,-0.05)$) -- ($(r1)+(-0.6,-0.1)$);
	\draw[line width=0.15mm,-{Latex[length=1.5mm]}]($(r2)+(0.6,0.05)$) -- ($(r2)+(0.6,0.1)$);
	\node[label={[xshift=0cm, yshift=-0.7cm]$r_1$}] (lr1) at (r1.center) {};
	\node[label={[xshift=0cm, yshift=-0.7cm]$r_2$}] (lr2) at (r2.center) {};
	\node[label={[xshift=0.2cm, yshift=-0.75cm]$\gamma_1$}] (lg1) at ($(r1)+(-1,1)$) {};
	\node at (r3.center) [circle,fill,inner sep=1.5pt]{};
	\node at (r4.center) [circle,fill,inner sep=1.5pt]{};
	\draw[line width=0.25mm, dotted] ($(r4)+(0,-0.6)$) arc (270:360:0.6);
	\draw[line width=0.25mm, dotted] ($(r4)+(0.6,0)$) arc (0:90:0.6);
	\draw[line width=0.25mm, dotted] ($(r3)+(0,-0.6)$) arc (270:90:0.6);
	\draw[line width=0.25mm, dotted] ($(r3)+(0,-0.6)$) -- ($(r4)+(0,-0.6)$);
	\draw[line width=0.25mm, dotted] ($(r3)+(0,0.6)$) -- ($(r4)+(0,0.6)$);
	\draw[line width=0.15mm,-{Latex[length=1.5mm]}]($(r3)+(-0.6,-0.05)$) -- ($(r3)+(-0.6,-0.1)$);
	\draw[line width=0.15mm,-{Latex[length=1.5mm]}]($(r4)+(0.6,0.05)$) -- ($(r4)+(0.6,0.1)$);
	\node[label={[xshift=0cm, yshift=-0.7cm]$r_3$}] (lr3) at (r3.center) {};
	\node[label={[xshift=0cm, yshift=-0.7cm]$r_4$}] (lr4) at (r4.center) {};
	\node[label={[xshift=-0.2cm, yshift=-0.75cm]$\gamma'_1$}] (lg1p) at ($(r4)+(1,1)$) {};
	\draw[line width=0.25mm, dotted] ($(r3)+(0,-1.2)$) arc (270:360:1.2);
	\draw[line width=0.25mm, dotted] ($(r3)+(1.2,0)$) arc (0:90:1.2);
	\draw[line width=0.25mm, dotted] ($(r2)+(0,-1.2)$) arc (270:90:1.2);
	\draw[line width=0.25mm, dotted] ($(r2)+(0,-1.2)$) -- ($(r3)+(0,-1.2)$);
	\draw[line width=0.25mm, dotted] ($(r2)+(0,1.2)$) -- ($(r3)+(0.1,1.2)$);
	\draw[line width=0.15mm,-{Latex[length=1.5mm]}](5,-1.2) -- (5.05,-1.2);
	\draw[line width=0.15mm,-{Latex[length=1.5mm]}](4.95,1.2) -- (4.9,1.2);
	\node[label={[xshift=0.5cm, yshift=-0.6cm]$\gamma_2$}] (lg2) at ($(r2)+(-1.5,1.5)$) {};
\end{tikzpicture}}
\caption{Examples of closed contours for the integral in eq.~\eqref{eq: residues_square_root_example} for a quadratic polynomial (a), and a quartic polynomial (b).}
\label{fig: contours_branch_cuts}
\end{figure}
One should nonetheless be careful about taking further residues when there are square roots involved. For example, let us consider
\begin{equation}
\LS \left( \int \frac{dx}{(x-c) \sqrt{P_n(x)}} \right),
\label{eq: residues_square_root_example}
\end{equation}
where $P_n(x)=(x-r_1)\cdots(x-r_n)$ is a polynomial of degree $n$ in $x$ with non-coinciding roots. For a quadratic polynomial we have two closed contours $\gamma_0$ and $\gamma_1$, respectively around the pole or around the branch cut, see fig.~\ref{fig: contours_branch_cuts}(a). Since there are no more poles or branch cuts (including at infinity), the corresponding closed integrals are therefore equal up to a sign. However, for higher-degree polynomials more contours can be taken. For instance, let us consider a polynomial of degree four, such that the integral above has a pole and two branch cuts; see fig.~\ref{fig: contours_branch_cuts}(b). In that case, while the closed integral $\gamma_0$ around the simple pole yields the sum of the closed integrals over $\gamma_1$ and $\gamma'_1$, it is not possible to individually obtain the integrals over $\gamma_1$ and $\gamma'_1$ that way. Moreover, there is a new contour $\gamma_2$. In fact, for a quartic polynomial the integral in eq.~\eqref{eq: residues_square_root_example} evaluates to elliptic integrals, which is manifest by taking the leading singularity as the closed integral over $\gamma_2$. To avoid the subtlety of choosing a contour in our calculations, we will generally avoid taking simple residues if there are square roots involved, and instead rationalize first (if possible) the accompanying square roots.

In some cases, see sec.~\ref{sec:two_loop_abelian}, the maximal cut leads to an integral without a discontinuity. Consequently, deforming the Baikov contour $\mathfrak{C}$ to a circle around a pole to compute the residue would yield zero. In these cases, we instead have to integrate the leading singularity along a contour between the roots of the Baikov polynomial of the diagram.\footnote{For Feynman integrals in dimensional regularization, the Baikov contour is a multivariate generalization of the Pochhammer contour~\cite{aomoto2011theory, yoshida2013hypergeometric, Matsubara-Heo:2023ylc}. In $D=4$, contributions from the branch-points correspond to the residues discussed above. 
In cases where the $D \rightarrow 4$ limit of the contour gives no contribution from those branch-points, 
another contribution will have to be chosen, and in cases where the integral between those branch points converges, that is a valid choice and it will be the same as the integral between the zeros of the Baikov polynomial.}

Lastly, as explained in sec.~\ref{sec:soft_expansion}, for each Feynman diagram we need to analyze both the integrals of even and odd parity. Since we assume all master integrals in a given parity sector to be coupled in the differential equations, meaning that there are no decoupled blocks within each parity and thus all master integrals give rise to the same geometry, for each parity sector we choose one master integral that yields a leading singularity that is simple to compute. For the parity specified by the number of matter propagators in the diagram we can just consider the scalar integral. Starting at three loops in sec.~\ref{sec:three_loop_non-abelian}, however, we find higher-order poles after taking the maximal cut. As explained in ref.~\cite{Henn:2020lye}, this can be remedied by adding numerator factors. In particular, we will add minimal numerators containing ISPs such as $k^2$ and $(2 u_i \cdot k)^2$, where the latter is squared so that it does not modify the parity of the integral. By contrast, we may obtain a master integral of opposite parity by adding one power of a linearized ISP $2u_i \cdot k$ in the numerator, or by doubling (adding a dot to) a matter propagator.

\subsection{Relating geometries in different Feynman integrals}
\label{sec:reduction_relations}

Using the Baikov representation we also discovered a number of relations among leading singularities as well as further simplifications coming from the IBP identities that the integrals satisfy, which we gather in this subsection. These relations vastly simplify the analysis, since they significantly reduce the number of independent diagrams that we need to consider, and they allow us to immediately relate geometries across different sectors and loop orders.

First of all, we note that diagrams that can be related by vertical and horizontal reflections trivially depend on the same geometry. Similarly, since the matter propagators are linearized, if we flip either of the matter lines, the leading singularity at most changes by a sign. 
Therefore, the geometry is also the same. For example, by flipping the matter line on top, we easily find that
\begin{align}
\LS \left( \begin{tikzpicture}[baseline={([yshift=-0.1cm]current bounding box.center)}] 
	\node[] (a) at (0,0) {};
	\node[] (b) at (0,-0.5) {};
	\node[] (a1) at (0.5,0) {};
	\node[] (b1) at (0.5,-0.5) {};
	\node[] (c) at (0,-1) {};
	\node[] (c1) at (0.5,-1) {};
	\node[] (p1) at ($(a)+(-0.2,0)$) {};
	\node[] (p2) at ($(c)+(-0.2,0)$) {};
	\node[] (p3) at ($(c1)+(0.2,0)$) {};
	\node[] (p4) at ($(a1)+(0.2,0)$) {};
	\draw[line width=0.15mm] (b.center) -- (a1.center);
	\draw[line width=0.15mm] (a.center) -- (0.175,-0.175);
	\draw[line width=0.15mm] (0.325,-0.325) -- (b1.center);
	\draw[line width=0.15mm] (b1.center) -- (b.center);
	\draw[line width=0.15mm] (b.center) -- (c.center);
	\draw[line width=0.15mm] (c1.center) -- (b1.center);
	\draw[line width=0.5mm] (a.center) -- (a1.center);
	\draw[line width=0.5mm] (c.center) -- (c1.center);
	\draw[line width=0.5mm] (p1.center) -- (a.center);
	\draw[line width=0.5mm] (a1.center) -- (p4.center);
	\draw[line width=0.5mm] (p2.center) -- (c.center);
	\draw[line width=0.5mm] (c1.center) -- (p3.center);
\end{tikzpicture} \right) & \, \propto \LS \left( \begin{tikzpicture}[baseline={([yshift=-0.1cm]current bounding box.center)}] 
	\node[] (a) at (0,0) {};
	\node[] (b) at (0,-0.5) {};
	\node[] (a1) at (0.5,0) {};
	\node[] (b1) at (0.5,-0.5) {};
	\node[] (c) at (0,-1) {};
	\node[] (c1) at (0.5,-1) {};
	\node[] (p1) at ($(a)+(-0.2,0)$) {};
	\node[] (p2) at ($(c)+(-0.2,0)$) {};
	\node[] (p3) at ($(c1)+(0.2,0)$) {};
	\node[] (p4) at ($(a1)+(0.2,0)$) {};
	\draw[line width=0.15mm] (b.center) -- (a.center);
	\draw[line width=0.15mm] (a1.center) -- (b1.center);
	\draw[line width=0.15mm] (b1.center) -- (b.center);
	\draw[line width=0.15mm] (b.center) -- (c.center);
	\draw[line width=0.15mm] (c1.center) -- (b1.center);
	\draw[line width=0.5mm] (a.center) -- (a1.center);
	\draw[line width=0.5mm] (c.center) -- (c1.center);
	\draw[line width=0.5mm] (p1.center) -- (a.center);
	\draw[line width=0.5mm] (a1.center) -- (p4.center);
	\draw[line width=0.5mm] (p2.center) -- (c.center);
	\draw[line width=0.5mm] (c1.center) -- (p3.center);
\end{tikzpicture} \right), \\[0.1cm]
\LS \left( \begin{tikzpicture}[baseline={([yshift=-0.1cm]current bounding box.center)}, scale=0.8] 
	\node[] (a) at (0,0) {};
	\node[] (b) at (0,-1) {};
	\node[] (a1) at (1,0) {};
	\node[] (b1) at (1,-1) {};
	\node[] (p1) at ($(a)+(-0.25,0)$) {};
	\node[] (p2) at ($(b)+(-0.25,0)$) {};
	\node[] (p3) at ($(b1)+(0.25,0)$) {};
	\node[] (p4) at ($(a1)+(0.25,0)$) {};
	\draw[line width=0.15mm] (b1.center) -- (a.center);
	\draw[line width=0.15mm] (b.center) -- (a.center);
		\draw[line width=0.15mm] (a1.center) -- (0.6,-0.4);
		\draw[line width=0.15mm] (0.4,-0.6) -- (b.center);
	\draw[line width=0.5mm] (a.center) -- (a1.center);
	\draw[line width=0.5mm] (b.center) -- (b1.center);
	\draw[line width=0.5mm] (p1.center) -- (a.center);
	\draw[line width=0.5mm] (a1.center) -- (p4.center);
	\draw[line width=0.5mm] (p2.center) -- (b.center);
	\draw[line width=0.5mm] (b1.center) -- (p3.center);
\end{tikzpicture} \right) & \, \propto \LS \left( \begin{tikzpicture}[baseline={([yshift=-0.1cm]current bounding box.center)}] 
	\node[] (a) at (0,0) {};
	\node[] (a1) at (0.8,0) {};
	\node[] (b) at (0.4,-0.8) {};
	\node[] (b1) at (1.2,-0.8) {};
	\node[] (p1) at ($(a)+(-0.2,0)$) {};
	\node[] (p2) at ($(b)+(-0.6,0)$) {};
	\node[] (p3) at ($(b1)+(0.2,0)$) {};
	\node[] (p4) at ($(a1)+(0.6,0)$) {};
	\draw[line width=0.15mm] (b.center) -- (a.center);
	\draw[line width=0.15mm] (b.center) -- (a1.center);
		\draw[line width=0.15mm] (b1.center) -- (a1.center);
	\draw[line width=0.5mm] (p1.center) -- (p4.center);
	\draw[line width=0.5mm] (p2.center) -- (p3.center);
\end{tikzpicture} \right),
\end{align}
\begin{align}
\LS \left( \begin{tikzpicture}[baseline={([yshift=-0.1cm]current bounding box.center)}] 
	\node[] (a) at (0,0) {};
	\node[] (a1) at (0.8,0) {};
	\node[] (a2) at (1.6,0) {};
	\node[] (b) at (0,-0.8) {};
	\node[] (b1) at (0.8,-0.8) {};
	\node[] (b2) at (1.6,-0.8) {};
	\node[] (p1) at ($(a)+(-0.2,0)$) {};
	\node[] (p2) at ($(b)+(-0.2,0)$) {};
	\node[] (p3) at ($(b2)+(0.2,0)$) {};
	\node[] (p4) at ($(a2)+(0.2,0)$) {};
	\draw[line width=0.15mm] (b.center) -- (a.center);
	\draw[line width=0.15mm] (b1.center) -- (a1.center);
	\draw[line width=0.15mm] (b2.center) -- (a2.center);
	\draw[line width=0.5mm] (p1.center) -- (p4.center);
	\draw[line width=0.5mm] (p2.center) -- (p3.center);
\end{tikzpicture} \right) & \, \propto \LS \left( \begin{tikzpicture}[baseline={([yshift=-0.1cm]current bounding box.center)}] 
	\node[] (a) at (0,0) {};
	\node[] (a1) at (0.8,0) {};
	\node[] (a2) at (1.6,0) {};
	\node[] (b) at (0,-0.8) {};
	\node[] (b1) at (0.8,-0.8) {};
	\node[] (b2) at (1.6,-0.8) {};
	\node[] (p1) at ($(a)+(-0.2,0)$) {};
	\node[] (p2) at ($(b)+(-0.2,0)$) {};
	\node[] (p3) at ($(b2)+(0.2,0)$) {};
	\node[] (p4) at ($(a2)+(0.2,0)$) {};
	\draw[line width=0.15mm] (a.center) -- (0.7,-0.35);
	\draw[line width=0.15mm] (0.9,-0.45) -- (b2.center);
	\draw[line width=0.15mm] (b1.center) -- (a1.center);
	\draw[line width=0.15mm] (b.center) -- (0.7,-0.45);
	\draw[line width=0.15mm] (0.9,-0.35) -- (a2.center);
	\draw[line width=0.5mm] (p1.center) -- (p4.center);
	\draw[line width=0.5mm] (p2.center) -- (p3.center);
\end{tikzpicture} \right).
\end{align}

In fact, as shown in app.~\ref{app: Nonplanar_unraveling}, we can exploit a redundancy in the leading singularity of non-planar loops to find that the linearized propagators can be unraveled, essentially meaning that all vertices at the matter lines are orderless:
\begin{equation}
\label{eq: unraveling_matter_props}
\LS \left( \begin{tikzpicture}[baseline={([yshift=-0.1cm]current bounding box.center)}, scale=0.8]
	\node[] (a) at (0,0.3) {};
	\node[] (a1) at (1,0.3) {};
	\node[] (a2) at (3,0.3) {};
	\node[] (b) at (0,-1) {};
	\node[] (bmod1) at (0.87,-1.03) {};
	\node[] (bmod2) at (1.28,-1.4) {};
	\node[] (bmod3) at (2.72,-1.4) {};
	\node[] (bmod4) at (3.13,-1.03) {};
	\node[] (aend) at (4,0.3) {};
	\node[] (bend) at (4,-1) {};
	\node[] (p1) at (0.69,0.3) {};
	\node[] (p4) at (3.31,0.3) {};
	\fill[gray!20] (2,-0.2) circle (2.1);
	\fill[white] (2,-0.2) circle (1.4);
	\draw[line width=0.15mm] (bmod1.center) -- (a2.center);
	\draw[line width=0.15mm] (bmod2.center) -- (a2.center);
	\node at (1.125,-1)[circle,fill,inner sep=0.6pt]{};
	\node at (1.3,-1)[circle,fill,inner sep=0.6pt]{};
	\node at (1.475,-1)[circle,fill,inner sep=0.6pt]{};
	\draw[line width=0.15mm] (bmod3.center) -- (2.11,-0.8);
	\draw[line width=0.15mm] (1.67,-0.37) -- (a1.center);
	\draw[line width=0.15mm] (bmod4.center) -- (2.37,-0.55);
	\draw[line width=0.15mm] (1.85,-0.23) -- (a1.center);
	\node at (2.525,-1)[circle,fill,inner sep=0.6pt]{};
	\node at (2.7,-1)[circle,fill,inner sep=0.6pt]{};
	\node at (2.875,-1)[circle,fill,inner sep=0.6pt]{};
	\draw[line width=0.5mm] (p1.center) -- (p4.center);
\end{tikzpicture} \right) \propto \LS \left( \begin{tikzpicture}[baseline={([yshift=-0.1cm]current bounding box.center)}, scale=0.8]
	\node[] (a) at (0,0.3) {};
	\node[] (a1) at (1.2,0.3) {};
	\node[] (a2) at (2.8,0.3) {};
	\node[] (b) at (0,-1) {};
	\node[] (bmod1) at (0.87,-1.03) {};
	\node[] (bmod2) at (1.5,-1.51) {};
	\node[] (bmod3) at (2.5,-1.51) {};
	\node[] (bmod4) at (3.13,-1.03) {};
	\node[] (aend) at (4,0.3) {};
	\node[] (bend) at (4,-1) {};
	\node[] (p1) at (0.69,0.3) {};
	\node[] (p4) at (3.31,0.3) {};
	\fill[gray!20] (2,-0.2) circle (2.1);
	\fill[white] (2,-0.2) circle (1.4);
	\draw[line width=0.15mm] (bmod1.center) -- (a1.center);
	\draw[line width=0.15mm] (bmod2.center) -- (a1.center);
	\node at (1,-0.9)[circle,fill,inner sep=0.6pt]{};
	\node at (1.15,-0.9)[circle,fill,inner sep=0.6pt]{};
	\node at (1.3,-0.9)[circle,fill,inner sep=0.6pt]{};
	\draw[line width=0.15mm] (bmod3.center) -- (a2.center);
	\draw[line width=0.15mm] (bmod4.center) -- (a2.center);
	\node at (2.7,-0.9)[circle,fill,inner sep=0.6pt]{};
	\node at (2.85,-0.9)[circle,fill,inner sep=0.6pt]{};
	\node at (3,-0.9)[circle,fill,inner sep=0.6pt]{};
	\draw[line width=0.5mm] (p1.center) -- (p4.center);
\end{tikzpicture} \right).
\end{equation}
The dots in the graviton lines account for the fact that the identity is valid beyond cubic vertices in the matter lines, and the grey zone indicates that this identity is also valid as part of a bigger diagram. In general, we will exploit the unraveling of matter propagators to relate the leading singularity of non-planar diagrams to their planar counterparts, which are easier to calculate. This identity is therefore especially useful for diagrams whose non-planarity arises solely from the order of the vertices in the matter lines. While this is sufficient to relate the leading singularity of all non-planar diagrams for $L \leq 3$, starting at four loops one also finds truly non-planar diagrams, see for example sec.~\ref{sec:four_loop_nonplanar}.

For instance, let us consider the three- and five-loop amoeba diagrams, which at the level of the leading singularity respectively involve a Calabi-Yau three-fold and a Calabi-Yau eight-fold for quadratic propagators and generic kinematics~\cite{Bourjaily:2018yfy}. In the PM expansion, we can easily find planar counterparts using the unraveling transformation for the propagators highlighted in red:
\begin{equation}
\LS \left(
\begin{tikzpicture}[baseline={([yshift=-0.1cm]current bounding box.center)}, scale=0.8] 
	\node[] (a) at (0,0) {};
	\node[] (b) at (0,-1) {};
	\node[] (a1) at (1,0) {};
	\node[] (b1) at (1,-1) {};
	\node[] (a2) at (2,0) {};
	\node[] (b2) at (2,-1) {};
	\node[] (p1) at ($(a)+(-0.3,0)$) {};
	\node[] (p2) at ($(b)+(-0.3,0)$) {};
	\node[] (p3) at ($(b2)+(0.3,0)$) {};
	\node[] (p4) at ($(a2)+(0.3,0)$) {};
	\draw[line width=0.15mm] (b1.center) -- (a.center);
	\draw[line width=0.15mm] (a1.center) -- (0.6,-0.4);
	\draw[line width=0.15mm] (0.4,-0.6) -- (b.center);
	\draw[line width=0.15mm] (b2.center) -- (a1.center);
	\draw[line width=0.15mm] (a2.center) -- (1.6,-0.4);
	\draw[line width=0.15mm] (1.4,-0.6) -- (b1.center);
	\draw[line width=0.5mm, red] (a.center) -- (a1.center);
	\draw[line width=0.5mm] (p1.center) -- (a.center);
	\draw[line width=0.5mm] (a1.center) -- (p4.center);
	\draw[line width=0.5mm] (p2.center) -- (p3.center);
\end{tikzpicture} \right)
\propto
\LS \left(
\begin{tikzpicture}[baseline={([yshift=-0.1cm]current bounding box.center)}, scale=0.8] 
	\node[] (a) at (0,0) {};
	\node[] (b) at (0,-1) {};
	\node[] (a1) at (1,0) {};
	\node[] (b1) at (1,-1) {};
	\node[] (a2) at (2,0) {};
	\node[] (b2) at (2,-1) {};
	\node[] (p1) at ($(a)+(-0.3,0)$) {};
	\node[] (p2) at ($(b)+(-0.3,0)$) {};
	\node[] (p3) at ($(b2)+(0.3,0)$) {};
	\node[] (p4) at ($(a2)+(0.3,0)$) {};
	\draw[line width=0.15mm] (b.center) -- (a.center);
	\draw[line width=0.15mm] (b1.center) -- (a1.center);
	\draw[line width=0.15mm] (b1.center) -- (a2.center);
	\draw[line width=0.15mm] (a.center) -- (0.8,-0.4);
	\draw[line width=0.15mm] (1.06,-0.53) -- (1.27,-0.64);
	\draw[line width=0.15mm] (1.53,-0.77) -- (b2.center);
	\draw[line width=0.5mm] (p1.center) -- (p4.center);
	\draw[line width=0.5mm, red] (b1.center) -- (b2.center);
	\draw[line width=0.5mm] (p2.center) -- (b1.center);
	\draw[line width=0.5mm] (b2.center) -- (p3.center);
\end{tikzpicture} \right) \propto
\LS \left(
\begin{tikzpicture}[baseline={([yshift=-0.1cm]current bounding box.center)}] 
	\node[] (a) at (0.4,0) {};
	\node[] (a1) at (1.2,0) {};
	\node[] (a2) at (2,0) {};
	\node[] (b) at (0,-0.8) {};
	\node[] (b1) at (0.8,-0.8) {};
	\node[] (b2) at (1.6,-0.8) {};
	\node[] (p1) at ($(a)+(-0.6,0)$) {};
	\node[] (p2) at ($(b)+(-0.2,0)$) {};
	\node[] (p3) at ($(b2)+(0.6,0)$) {};
	\node[] (p4) at ($(a2)+(0.2,0)$) {};
	\draw[line width=0.15mm] (b.center) -- (a.center);
	\draw[line width=0.15mm] (b1.center) -- (a.center);
		\draw[line width=0.15mm] (b2.center) -- (a1.center);
	\draw[line width=0.15mm] (b2.center) -- (a2.center);
	\draw[line width=0.5mm] (p1.center) -- (p4.center);
	\draw[line width=0.5mm] (p2.center) -- (p3.center);
\end{tikzpicture}
\right) \propto 
\frac{x}{x^2-1},
\end{equation}
\begin{align}
\LS \left(
\begin{tikzpicture}[baseline={([yshift=-0.1cm]current bounding box.center)}, scale=0.8] 
	\node[] (a) at (0,0) {};
	\node[] (b) at (0,-1) {};
	\node[] (a1) at (1,0) {};
	\node[] (b1) at (1,-1) {};
	\node[] (a2) at (2,0) {};
	\node[] (b2) at (2,-1) {};
	\node[] (a3) at (3,0) {};
	\node[] (b3) at (3,-1) {};
	\node[] (p1) at ($(a)+(-0.3,0)$) {};
	\node[] (p2) at ($(b)+(-0.3,0)$) {};
	\node[] (p3) at ($(b3)+(0.3,0)$) {};
	\node[] (p4) at ($(a3)+(0.3,0)$) {};
	\draw[line width=0.15mm] (b1.center) -- (a.center);
	\draw[line width=0.15mm] (a1.center) -- (0.6,-0.4);
	\draw[line width=0.15mm] (0.4,-0.6) -- (b.center);
	\draw[line width=0.15mm] (b2.center) -- (a1.center);
	\draw[line width=0.15mm] (a2.center) -- (1.6,-0.4);
	\draw[line width=0.15mm] (1.4,-0.6) -- (b1.center);
	\draw[line width=0.15mm] (b3.center) -- (a2.center);
	\draw[line width=0.15mm] (a3.center) -- (2.6,-0.4);
	\draw[line width=0.15mm] (2.4,-0.6) -- (b2.center);
	\draw[line width=0.5mm, red] (a.center) -- (a1.center);
	\draw[line width=0.5mm] (p1.center) -- (a.center);
	\draw[line width=0.5mm] (a1.center) -- (p4.center);
	\draw[line width=0.5mm] (p2.center) -- (p3.center);
\end{tikzpicture} \right) & \propto \LS \left(
\begin{tikzpicture}[baseline={([yshift=-0.1cm]current bounding box.center)}, scale=0.8] 
	\node[] (a) at (0,0) {};
	\node[] (b) at (0,-1) {};
	\node[] (a1) at (1,0) {};
	\node[] (b1) at (1,-1) {};
	\node[] (a2) at (2,0) {};
	\node[] (b2) at (2,-1) {};
	\node[] (a3) at (3,0) {};
	\node[] (b3) at (3,-1) {};
	\node[] (p1) at ($(a)+(-0.3,0)$) {};
	\node[] (p2) at ($(b)+(-0.3,0)$) {};
	\node[] (p3) at ($(b3)+(0.3,0)$) {};
	\node[] (p4) at ($(a3)+(0.3,0)$) {};
	\draw[line width=0.15mm] (b.center) -- (a.center);
	\draw[line width=0.15mm] (b1.center) -- (a1.center);
	\draw[line width=0.15mm] (b1.center) -- (a2.center);
	\draw[line width=0.15mm] (a.center) -- (0.8,-0.4);
	\draw[line width=0.15mm] (1.06,-0.53) -- (1.27,-0.64);
	\draw[line width=0.15mm] (1.53,-0.77) -- (b2.center);
	\draw[line width=0.15mm] (b3.center) -- (a2.center);
	\draw[line width=0.15mm] (a3.center) -- (2.6,-0.4);
	\draw[line width=0.15mm] (2.4,-0.6) -- (b2.center);
	\draw[line width=0.5mm] (p1.center) -- (p4.center);
	\draw[line width=0.5mm, red] (b1.center) -- (b2.center);
	\draw[line width=0.5mm] (p2.center) -- (b1.center);
	\draw[line width=0.5mm] (b2.center) -- (p3.center);
\end{tikzpicture} \right) \propto
\LS \left(
\begin{tikzpicture}[baseline={([yshift=-0.1cm]current bounding box.center)}, scale=0.8] 
	\node[] (a) at (0,0) {};
	\node[] (b) at (0,-1) {};
	\node[] (a1) at (1,0) {};
	\node[] (b1) at (1,-1) {};
	\node[] (a2) at (2,0) {};
	\node[] (b2) at (2,-1) {};
	\node[] (a3) at (3,0) {};
	\node[] (b3) at (3,-1) {};
	\node[] (p1) at ($(a)+(-0.3,0)$) {};
	\node[] (p2) at ($(b)+(-0.3,0)$) {};
	\node[] (p3) at ($(b3)+(0.3,0)$) {};
	\node[] (p4) at ($(a3)+(0.3,0)$) {};
	\draw[line width=0.15mm] (b.center) -- (a.center);
	\draw[line width=0.15mm] (b1.center) -- (a.center);
	\draw[line width=0.15mm] (b2.center) -- (a1.center);
	\draw[line width=0.15mm] (b2.center) -- (a2.center);
	\draw[line width=0.15mm] (b3.center) -- (a2.center);
	\draw[line width=0.15mm] (b1.center) -- (1.467,-0.77);
	\draw[line width=0.15mm] (1.727,-0.637) -- (1.94,-0.53);
	\draw[line width=0.15mm] (2.06,-0.47) -- (2.273,-0.363);
	\draw[line width=0.15mm] (2.53,-0.23) -- (a3.center);
	\draw[line width=0.5mm, red] (a2.center) -- (a3.center);
	\draw[line width=0.5mm] (p1.center) -- (a2.center);
	\draw[line width=0.5mm] (a3.center) -- (p4.center);
	\draw[line width=0.5mm] (p2.center) -- (p3.center);
\end{tikzpicture} \right) \nonumber \\[0.1cm]
& \propto
\LS \left(
\begin{tikzpicture}[baseline={([yshift=-0.1cm]current bounding box.center)}, scale=0.8] 
	\node[] (a) at (0,0) {};
	\node[] (b) at (0,-1) {};
	\node[] (a1) at (1,0) {};
	\node[] (b1) at (1,-1) {};
	\node[] (a2) at (2,0) {};
	\node[] (b2) at (2,-1) {};
	\node[] (a3) at (3,0) {};
	\node[] (b3) at (3,-1) {};
	\node[] (p1) at ($(a)+(-0.3,0)$) {};
	\node[] (p2) at ($(b)+(-0.3,0)$) {};
	\node[] (p3) at ($(b3)+(0.3,0)$) {};
	\node[] (p4) at ($(a3)+(0.3,0)$) {};
	\draw[line width=0.15mm] (b.center) -- (a.center);
	\draw[line width=0.15mm] (b1.center) -- (a.center);
	\draw[line width=0.15mm] (b2.center) -- (a1.center);
	\draw[line width=0.15mm] (a2.center) -- (1.6,-0.4);
	\draw[line width=0.15mm] (1.4,-0.6) -- (b1.center);
	\draw[line width=0.15mm] (b2.center) -- (a3.center);
	\draw[line width=0.15mm] (b3.center) -- (a3.center);
	\draw[line width=0.5mm, red] (a1.center) -- (a2.center);
	\draw[line width=0.5mm] (p1.center) -- (a1.center);
	\draw[line width=0.5mm] (a2.center) -- (p4.center);
	\draw[line width=0.5mm] (p2.center) -- (p3.center);
\end{tikzpicture} \right) \propto
\LS \left(
\begin{tikzpicture}[baseline={([yshift=-0.1cm]current bounding box.center)}, scale=0.8] 
	\node[] (a) at (0,0) {};
	\node[] (b) at (0,-1) {};
	\node[] (a1) at (1,0) {};
	\node[] (b1) at (1,-1) {};
	\node[] (a2) at (2,0) {};
	\node[] (b2) at (2,-1) {};
	\node[] (a3) at (3,0) {};
	\node[] (b3) at (3,-1) {};
	\node[] (p1) at ($(a)+(-0.3,0)$) {};
	\node[] (p2) at ($(b)+(-0.3,0)$) {};
	\node[] (p3) at ($(b3)+(0.3,0)$) {};
	\node[] (p4) at ($(a3)+(0.3,0)$) {};
	\draw[line width=0.15mm] (b.center) -- (a.center);
	\draw[line width=0.15mm] (b1.center) -- (a.center);
	\draw[line width=0.15mm] (a2.center) -- (b2.center);
	\draw[line width=0.15mm] (a1.center) -- (b1.center);
	\draw[line width=0.15mm] (b2.center) -- (a3.center);
	\draw[line width=0.15mm] (b3.center) -- (a3.center);
	\draw[line width=0.5mm] (p1.center) -- (p4.center);
	\draw[line width=0.5mm] (p2.center) -- (p3.center);
\end{tikzpicture} \right) \nonumber \\[0.1cm]
& \propto \frac{q^2 \, x^3}{(x^2-1)^3}.
\end{align}
Since we obtain algebraic leading singularities, we can conclude that if these PM diagrams were to depend on a non-trivial geometry, it would arise solely from subsectors.

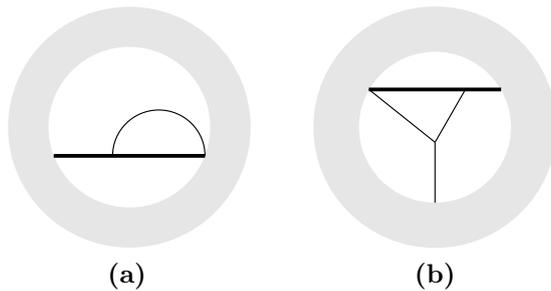
\begin{figure}[t]
\centering
\subfloat[]{\begin{tikzpicture}[baseline={([yshift=-0.1cm]current bounding box.center)}, scale=0.76]
	\node[] (a) at (0,-0.3) {};
	\node[] (a1) at (1.1,-0.3) {};
	\node[] (a2) at (2.7,-0.3) {};
	\node[] (p1) at (0.69,-0.3) {};
	\node[] (p4) at (3.31,-0.3) {};
	\fill[gray!20] (2,0.2) circle (2.1);
	\fill[white] (2,0.2) circle (1.4);
	\draw[line width=0.15mm] (p4.center) arc (0:180:0.8);
	\draw[line width=0.5mm] (p1.center) -- (p4.center);
\end{tikzpicture}} \qquad
\subfloat[]{\begin{tikzpicture}[baseline={([yshift=-0.1cm]current bounding box.center)}]
	\node[] (a) at (1.6,0.3) {};
	\node[] (a1) at (2.4,0.3) {};
	\node[] (b) at (2,-1.2) {};
	\node[] (p1) at (1.13,0.3) {};
	\node[] (p4) at (2.87,0.3) {};
	\fill[gray!20] (2,-0.2) circle (1.6);
	\fill[white] (2,-0.2) circle (1);
	\draw[line width=0.15mm] (p1.center) -- (2,-0.4);
	\draw[line width=0.15mm] (a1.center) -- (2,-0.4);
	\draw[line width=0.15mm] (b.center) -- (2,-0.4);
	\draw[line width=0.5mm] (p1.center) -- (p4.center);
\end{tikzpicture}}
\caption{Diagrams with zero master integrals in their sector: (a) A bubble correction with at least one cubic vertex; (b) A triangle attached to a matter line with a cubic graviton self-interaction and at least one cubic vertex on the matter line. The grey zone indicates that this diagrams should be thought of as being part of a bigger diagram.}
\label{fig: diagrams_zero_masters}
\end{figure}
Secondly, diagrams containing bubble corrections with at least one cubic vertex, as represented in fig.~\ref{fig: diagrams_zero_masters}(a), have zero master integrals in their sector; see app.~\ref{app: IBPs_bubbles_triangles} for details. Therefore, these diagrams are fully expressible in terms of lower sectors, and thus do not need to be considered in our analysis. As we also show in app.~\ref{app: IBPs_bubbles_triangles}, in conjunction with the unraveling of matter propagators in eq.~\eqref{eq: unraveling_matter_props}, we can generalize the previous argument to bubble corrections that go over more than one matter propagator, and also omit them from our analysis of geometries. For instance, we find that the following two-loop diagrams have zero master integrals in their respective sectors:
\begin{equation}
\raisebox{0.17cm}{\begin{tikzpicture}[baseline={([yshift=-0.1cm]current bounding box.center)}, scale=0.8] 
	\node[] (b) at (0.5,-1) {};
	\node[] (c) at (0,0) {};
	\node[] (c1) at (1,0) {};
	\node[] (p1) at ($(b)+(-0.95,0)$) {};
	\node[] (p2) at ($(c)+(-0.45,0)$) {};
	\node[] (p3) at ($(c1)+(0.45,0)$) {};
	\node[] (p4) at ($(b)+(0.95,0)$) {};
	\draw[line width=0.15mm] (b.center) -- (c.center);
	\draw[line width=0.15mm] (b.center) -- (c1.center);
	\draw[line width=0.5mm] (p1.center) -- (p4.center);
	\draw[line width=0.5mm] (p2.center) -- (p3.center);
	\draw[line width=0.15mm] (1.2,0) arc (0:180:0.7);
\end{tikzpicture}}, \qquad
\begin{tikzpicture}[baseline={([yshift=-0.1cm]current bounding box.center)}, scale=0.8] 
	\node[] (a) at (0,0) {};
	\node[] (a1) at (1,0) {};
	\node[] (b) at (0,-1) {};
	\node[] (b1) at (1,-1) {};
	\node[] (p1) at ($(a)+(-0.25,0)$) {};
	\node[] (p2) at ($(b)+(-0.25,0)$) {};
	\node[] (p3) at ($(b1)+(0.25,0)$) {};
	\node[] (p4) at ($(a1)+(0.25,0)$) {};
	\draw[line width=0.15mm] (b.center) -- (a.center);
	\draw[line width=0.15mm] (b1.center) -- (a1.center);
	\draw[line width=0.5mm] (p1.center) -- (p4.center);
	\draw[line width=0.5mm] (p2.center) -- (p3.center);
	\draw[line width=0.15mm] (0.8,0) arc (0:180:0.3);
\end{tikzpicture}, \qquad
\raisebox{0.175cm}{\begin{tikzpicture}[baseline={([yshift=-0.1cm]current bounding box.center)}, scale=0.8] 
	\node[] (a) at (0,0) {};
	\node[] (a1) at (1,0) {};
	\node[] (b) at (0,-1) {};
	\node[] (b1) at (1,-1) {};
	\node[] (p1) at ($(a)+(-0.85,0)$) {};
	\node[] (p2) at ($(b)+(-0.85,0)$) {};
	\node[] (p3) at ($(b1)+(0.25,0)$) {};
	\node[] (p4) at ($(a1)+(0.25,0)$) {};
	\draw[line width=0.15mm] (b.center) -- (a.center);
	\draw[line width=0.15mm] (a1.center) -- (b1.center);
	\draw[line width=0.5mm] (p1.center) -- (p4.center);
	\draw[line width=0.5mm] (p2.center) -- (p3.center);
	\draw[line width=0.15mm] (1,0) arc (0:180:0.8);
\end{tikzpicture}}, \qquad
\begin{tikzpicture}[baseline={([yshift=-0.1cm]current bounding box.center)}, scale=0.8] 
	\node[] (a) at (0.5,-1) {};
	\node[] (b) at (0.5,-0.5) {};
	\node[] (c) at (0,0) {};
	\node[] (c1) at (1,0) {};
	\node[] (p1) at ($(a)+(-1.05,0)$) {};
	\node[] (p2) at ($(c)+(-0.55,0)$) {};
	\node[] (p3) at ($(c1)+(0.25,0)$) {};
	\node[] (p4) at ($(a)+(0.75,0)$) {};
	\draw[line width=0.15mm] (b.center) -- (a.center);
	\draw[line width=0.15mm] (b.center) -- (c.center);
	\draw[line width=0.15mm] (b.center) -- (c1.center);
	\draw[line width=0.5mm] (p1.center) -- (p4.center);
	\draw[line width=0.5mm] (p2.center) -- (p3.center);
	\draw[line width=0.15mm] (0.3,0) arc (0:180:0.3);
\end{tikzpicture}.
\end{equation}

Thirdly, as we also show in app.~\ref{app: IBPs_bubbles_triangles}, diagrams containing triangles on a matter line with at least two cubic vertices and one of them being the graviton self-interaction vertex (as shown in fig.~\ref{fig: diagrams_zero_masters}(b)) also have zero master integrals in their sector. This applies, for example, to the following three-loop diagrams, where the triangles of the form of fig.~\ref{fig: diagrams_zero_masters}(b) are highlighted in red:
\begin{equation}
\begin{tikzpicture}[baseline={([yshift=-0.1cm]current bounding box.center)}] 
	\node[] (a) at (0,0) {};
	\node[] (a1) at (0.5,0) {};
	\node[] (a2) at (1,0) {};
	\node[] (b) at (0,-0.5) {};
	\node[] (c) at (0,-1) {};
	\node[] (c1) at (1,-1) {};
	\node[] (p1) at ($(a)+(-0.2,0)$) {};
	\node[] (p2) at ($(c)+(-0.2,0)$) {};
	\node[] (p3) at ($(c1)+(0.2,0)$) {};
	\node[] (p4) at ($(a2)+(0.2,0)$) {};
	\draw[line width=0.15mm, red] (b.center) -- (a.center);
	\draw[line width=0.15mm, red] (b.center) -- (c.center);
	\draw[line width=0.15mm, red] (b.center) -- (c1.center);
	\draw[line width=0.15mm] (c1.center) -- (a1.center);
	\draw[line width=0.15mm] (c1.center) -- (a2.center);
	\draw[line width=0.5mm] (p1.center) -- (p4.center);
	\draw[line width=0.5mm] (p2.center) -- (c.center);
	\draw[line width=0.5mm, red] (c.center) -- (c1.center);
	\draw[line width=0.5mm] (c1.center) -- (p3.center);
\end{tikzpicture}, \quad
\raisebox{0.17cm}{\begin{tikzpicture}[baseline={([yshift=-0.1cm]current bounding box.center)}] 
	\node[] (a) at (0,0) {};
	\node[] (a1) at (1,0) {};
	\node[] (b) at (0,-1) {};
	\node[] (b1) at (1,-1) {};
	\node[] (p1) at ($(a)+(-0.2,0)$) {};
	\node[] (p2) at ($(b)+(-0.2,0)$) {};
	\node[] (p3) at ($(b1)+(0.2,0)$) {};
	\node[] (p4) at ($(a1)+(0.2,0)$) {};
	\draw[line width=0.15mm] (b.center) -- (a.center);
	\draw[line width=0.15mm] (b1.center) -- (a1.center);
	\draw[line width=0.15mm, red] (0.5,0) -- (0.5,0.5);
	\draw[line width=0.5mm] (p1.center) -- (0.5,0);
	\draw[line width=0.5mm, red] (0.5,0) -- (a1.center);
	\draw[line width=0.5mm] (a1.center) -- (p4.center);
	\draw[line width=0.5mm] (p2.center) -- (p3.center);
	\draw[line width=0.15mm, red] (1,0) arc (0:180:0.5);
\end{tikzpicture}}, \quad 
\begin{tikzpicture}[baseline={([yshift=-0.1cm]current bounding box.center)}] 
	\node[] (a) at (0,0) {};
	\node[] (a1) at (0.5,0) {};
	\node[] (a2) at (1,0) {};
	\node[] (b) at (0,-0.5) {};
	\node[] (b1) at (0.5,-0.5) {};
	\node[] (c) at (0,-1) {};
	\node[] (c1) at (1,-1) {};
	\node[] (p1) at ($(a)+(-0.2,0)$) {};
	\node[] (p2) at ($(c)+(-0.2,0)$) {};
	\node[] (p3) at ($(c1)+(0.2,0)$) {};
	\node[] (p4) at ($(a2)+(0.2,0)$) {};
	\draw[line width=0.15mm] (b.center) -- (a.center);
	\draw[line width=0.15mm, red] (b1.center) -- (a1.center);
	\draw[line width=0.15mm] (c1.center) -- (a2.center);
	\draw[line width=0.15mm, red] (b1.center) -- (b.center);
	\draw[line width=0.15mm] (b.center) -- (c.center);
	\draw[line width=0.15mm, red] (a2.center) -- (b1.center);
	\draw[line width=0.5mm] (p1.center) -- (a1.center);
	\draw[line width=0.5mm, red] (a1.center) -- (a2.center);
	\draw[line width=0.5mm] (a2.center) -- (p4.center);
	\draw[line width=0.5mm] (p2.center) -- (p3.center);
\end{tikzpicture}, \quad \raisebox{0.05cm}{\begin{tikzpicture}[baseline={([yshift=-0.1cm]current bounding box.center)}] 
	\node[] (a) at (0,0) {};
	\node[] (a1) at (1,0) {};
	\node[] (b) at (0,-1) {};
	\node[] (b1) at (1,-1) {};
	\node[] (p1) at ($(a)+(-0.2,0)$) {};
	\node[] (p2) at ($(b)+(-0.2,0)$) {};
	\node[] (p3) at ($(b1)+(0.2,0)$) {};
	\node[] (p4) at ($(a1)+(0.2,0)$) {};
	\draw[line width=0.15mm, red] (b.center) -- (a.center);
	\draw[line width=0.15mm] (b1.center) -- (a1.center);
	\draw[line width=0.15mm, red] (0,-0.5) -- (0.5,0);
	\draw[line width=0.5mm] (p1.center) -- (a.center);
	\draw[line width=0.5mm, red] (a.center) -- (0.5,0);
	\draw[line width=0.5mm] (0.5,0) -- (p4.center);
	\draw[line width=0.5mm] (p2.center) -- (p3.center);
	\draw[line width=0.15mm] (1,0) arc (0:180:0.25);
\end{tikzpicture}}.
\end{equation}

Lastly, as shown in app.~\ref{app: Superclassical_reduction}, we have a reduction for superclassical diagrams containing a box iteration at one end,
\begin{equation}
\label{eq: reduction_superclassical}
\LS \left(
\begin{tikzpicture}[baseline={([yshift=-0.1cm]current bounding box.center)}] 
	\node[] (a) at (0,0) {};
	\node[] (a1) at (1.5,0) {};
	\node[] (b) at (0,-1.2) {};
	\node[] (b1) at (1.5,-1.2) {};
	\node[] (p1) at ($(a)+(-0.3,0)$) {};
	\node[] (p2) at ($(b)+(-0.3,0)$) {};
	\node[] (p3) at ($(b1)+(0.3,0)$) {};
	\node[] (p4) at ($(a1)+(0.3,0)$) {};
	\draw[line width=0.15mm] (b1.center) -- (a1.center);
	\draw[line width=0.5mm] (p1.center) -- (0.4,0);
	\draw[line width=0.5mm] (0.4,0) -- (p4.center);
	\draw[line width=0.5mm] (p2.center) -- (0.4,-1.2);
	\draw[line width=0.5mm] (0.4,-1.2) -- (p3.center);
	\fill[gray!50] (0.4,-0.6) ellipse (0.4 and 0.7);
	\draw (0.4,-0.6) ellipse (0.4 and 0.7);
\end{tikzpicture} \right) \propto \frac{x}{x^2-1} \, \LS \left(
\begin{tikzpicture}[baseline={([yshift=-0.1cm]current bounding box.center)}] 
	\node[] (a) at (0,0) {};
	\node[] (a1) at (0.8,0) {};
	\node[] (b) at (0,-1.2) {};
	\node[] (b1) at (0.8,-1.2) {};
	\node[] (p1) at ($(a)+(-0.3,0)$) {};
	\node[] (p2) at ($(b)+(-0.3,0)$) {};
	\node[] (p3) at ($(b1)+(0.3,0)$) {};
	\node[] (p4) at ($(a1)+(0.3,0)$) {};
	\draw[line width=0.5mm] (p1.center) -- (0.4,0);
	\draw[line width=0.5mm] (0.4,0) -- (p4.center);
	\draw[line width=0.5mm] (p2.center) -- (0.4,-1.2);
	\draw[line width=0.5mm] (0.4,-1.2) -- (p3.center);
	\fill[gray!50] (0.4,-0.6) ellipse (0.4 and 0.7);
	\draw (0.4,-0.6) ellipse (0.4 and 0.7);
\end{tikzpicture} \right),
\end{equation}
which can also be embedded in a larger reduction of box iterations
\begin{equation}
\label{eq: reduction_superclassical_generic}
\LS \left(
\begin{tikzpicture}[baseline={([yshift=-0.1cm]current bounding box.center)}] 
	\node[] (a) at (0,0) {};
	\node[] (a1) at (0.4,0) {};
	\node[] (a2) at (1.5,0) {};
	\node[] (a3) at (2.6,0) {};
	\node[] (a4) at (3,0) {};
	\node[] (b) at (0,-1.2) {};
	\node[] (b1) at (0.4,-1.2) {};
	\node[] (b2) at (1.5,-1.2) {};
	\node[] (b3) at (2.6,-1.2) {};
	\node[] (b4) at (3,-1.2) {};
	\node[] (p1) at ($(a)+(-0.3,0)$) {};
	\node[] (p2) at ($(b)+(-0.3,0)$) {};
	\node[] (p3) at ($(b4)+(0.3,0)$) {};
	\node[] (p4) at ($(a4)+(0.3,0)$) {};
	\draw[line width=0.15mm] (a2.center) -- (b2.center);
	\draw[line width=0.5mm] (p1.center) -- (0.4,0);
	\draw[line width=0.5mm] (0.4,0) -- (2.6,0);
	\draw[line width=0.5mm] (2.6,0) -- (p4.center);
	\draw[line width=0.5mm] (p2.center) -- (0.4,-1.2);
	\draw[line width=0.5mm] (0.4,-1.2) -- (2.6,-1.2);
	\draw[line width=0.5mm] (2.6,-1.2) -- (p3.center);
	\fill[gray!50] ($(a1)+(0,-0.6)$) ellipse (0.4 and 0.7);
	\draw ($(a1)+(0,-0.6)$) ellipse (0.4 and 0.7);
	\fill[gray!50] ($(a3)+(0,-0.6)$) ellipse (0.4 and 0.7);
	\draw ($(a3)+(0,-0.6)$) ellipse (0.4 and 0.7);
\end{tikzpicture} \right) \propto \frac{x}{x^2-1} \, \LS \left(
\begin{tikzpicture}[baseline={([yshift=-0.1cm]current bounding box.center)}] 
	\node[] (a) at (0,0) {};
	\node[] (a1) at (0.4,0) {};
	\node[] (a2) at (2.4,0) {};
	\node[] (a3) at (2.8,0) {};
	\node[] (b) at (0,-1.2) {};
	\node[] (b1) at (0.4,-1.2) {};
	\node[] (b2) at (2.4,-1.2) {};
	\node[] (b3) at (2.8,-1.2) {};
	\node[] (p1) at ($(a)+(-0.3,0)$) {};
	\node[] (p2) at ($(b)+(-0.3,0)$) {};
	\node[] (p3) at ($(b3)+(0.3,0)$) {};
	\node[] (p4) at ($(a3)+(0.3,0)$) {};
	\draw[line width=0.5mm] (p1.center) -- (0.4,0);
	\draw[line width=0.5mm] (0.4,0) -- (2.6,0);
	\draw[line width=0.5mm] (2.6,0) -- (p4.center);
	\draw[line width=0.5mm] (p2.center) -- (0.4,-1.2);
	\draw[line width=0.5mm] (0.4,-1.2) -- (2.6,-1.2);
	\draw[line width=0.5mm] (2.6,-1.2) -- (p3.center);
	\fill[gray!50] ($(a1)+(0,-0.6)$) ellipse (0.4 and 0.7);
	\draw ($(a1)+(0,-0.6)$) ellipse (0.4 and 0.7);
	\fill[gray!50] ($(a2)+(0,-0.6)$) ellipse (0.4 and 0.7);
	\draw ($(a2)+(0,-0.6)$) ellipse (0.4 and 0.7);
\end{tikzpicture} \right).
\end{equation}
The last reduction is completely homologous to the result for dual-conformal invariant ladder integrals with quadratic propagators given in refs.~\cite{McLeod:2023qdf,Cao:2023tpx}.

While these superclassical reductions are valid regardless of the parity of the blob, if we dot one of the matter propagators of the superclassical loop, thus changing its parity, the corresponding leading singularity vanishes:
\begin{equation}
\label{eq: reduction_superclassical_dot_vanish}
\LS \left(
\begin{tikzpicture}[baseline={([yshift=-0.1cm]current bounding box.center)}] 
	\node[] (a) at (0,0) {};
	\node[] (a1) at (1.5,0) {};
	\node[] (b) at (0,-1.2) {};
	\node[] (b1) at (1.5,-1.2) {};
	\node[] (p1) at ($(a)+(-0.3,0)$) {};
	\node[] (p2) at ($(b)+(-0.3,0)$) {};
	\node[] (p3) at ($(b1)+(0.3,0)$) {};
	\node[] (p4) at ($(a1)+(0.3,0)$) {};
	\draw[line width=0.15mm] (b1.center) -- (a1.center);
	\draw[line width=0.5mm] (p1.center) -- (0.4,0);
	\draw[line width=0.5mm] (0.4,0) -- (p4.center);
	\draw[line width=0.5mm] (p2.center) -- (0.4,-1.2);
	\draw[line width=0.5mm] (0.4,-1.2) -- (p3.center);
	\fill[gray!50] (0.4,-0.6) ellipse (0.4 and 0.7);
	\draw (0.4,-0.6) ellipse (0.4 and 0.7);
	\node at (1,0) [circle,fill,inner sep=1.5pt]{};
\end{tikzpicture} \right) = \LS \left(
\begin{tikzpicture}[baseline={([yshift=-0.1cm]current bounding box.center)}] 
	\node[] (a) at (0,0) {};
	\node[] (a1) at (0.4,0) {};
	\node[] (a2) at (1.5,0) {};
	\node[] (a3) at (2.6,0) {};
	\node[] (a4) at (3,0) {};
	\node[] (b) at (0,-1.2) {};
	\node[] (b1) at (0.4,-1.2) {};
	\node[] (b2) at (1.5,-1.2) {};
	\node[] (b3) at (2.6,-1.2) {};
	\node[] (b4) at (3,-1.2) {};
	\node[] (p1) at ($(a)+(-0.3,0)$) {};
	\node[] (p2) at ($(b)+(-0.3,0)$) {};
	\node[] (p3) at ($(b4)+(0.3,0)$) {};
	\node[] (p4) at ($(a4)+(0.3,0)$) {};
	\draw[line width=0.15mm] (a2.center) -- (b2.center);
	\draw[line width=0.5mm] (p1.center) -- (0.4,0);
	\draw[line width=0.5mm] (0.4,0) -- (2.6,0);
	\draw[line width=0.5mm] (2.6,0) -- (p4.center);
	\draw[line width=0.5mm] (p2.center) -- (0.4,-1.2);
	\draw[line width=0.5mm] (0.4,-1.2) -- (2.6,-1.2);
	\draw[line width=0.5mm] (2.6,-1.2) -- (p3.center);
	\fill[gray!50] ($(a1)+(0,-0.6)$) ellipse (0.4 and 0.7);
	\draw ($(a1)+(0,-0.6)$) ellipse (0.4 and 0.7);
	\fill[gray!50] ($(a3)+(0,-0.6)$) ellipse (0.4 and 0.7);
	\draw ($(a3)+(0,-0.6)$) ellipse (0.4 and 0.7);
	\node at (1,0) [circle,fill,inner sep=1.5pt]{};
\end{tikzpicture} \right) = 0.
\end{equation}
This implies that these diagrams with opposite parity have zero master integrals in their sector, and therefore do not appear as subsectors.

With these identities, we can immediately reduce the leading singularity of all superclassical diagrams containing a box iteration at one end into lower loops and lower sectors. Starting at three loops, however, we also find superclassical diagrams the leading singularity of which cannot be computed straightforwardly with these reductions, see sec.~\ref{sec:three_loop_abelian}.

All in all, the previous relations allow us to significantly reduce the number of independent integrals that we need to analyze. In particular, as we gather in tab.~\ref{tab:indep_diagrams}, we obtain a reduction of one order of magnitude in the number of integrals at both two and three loops, thus allowing for a systematic study of the corresponding geometries.

\begin{table}[t]
\begin{center}
\begin{tabular}{lccc}
\toprule
$L$ & 1 & 2 & 3 \\[0.1cm] \toprule
Initial number & 2 & 23 & 531 \\ \midrule
Unraveling of matter propagators & 2 & 19 & 317 \\ \midrule
Reduction of bubbles & 2 & 9 & 54 \\ \midrule
Reduction of triangles & 2 & 6 & 23 \\ \midrule
Reduction of superclassical diagrams & 1 & 4 & 15 \\[0.1cm] \bottomrule
\end{tabular}
\end{center}
\caption{Number of remaining independent integrals at $L$-loop order after imposing the simplifying relations that are gathered in this section. The initial number already excludes quantum contributions, reflections, flips of one matter line, dimensionless integrals and integrals that factorize into products of lower-loop integrals.}
\label{tab:indep_diagrams}
\end{table}

\section{One-loop diagrams}
\label{sec:one_loop}

\begin{figure}[t]
\centering
\begin{tikzpicture}[baseline={([yshift=-0.1cm]current bounding box.center)}, scale=1] 
	\node[] (a) at (0,0) {};
	\node[] (a1) at (1,0) {};
	\node[] (b) at (0,-1) {};
	\node[] (b1) at (1,-1) {};
	\node[] (p1) at ($(a)+(-0.25,0)$) {};
	\node[] (p2) at ($(b)+(-0.25,0)$) {};
	\node[] (p3) at ($(b1)+(0.25,0)$) {};
	\node[] (p4) at ($(a1)+(0.25,0)$) {};
	\draw[line width=0.15mm] (b.center) -- (a.center);
	\draw[line width=0.15mm] (b1.center) -- (a1.center);
	\draw[line width=0.5mm] (p1.center) -- (p4.center);
	\draw[line width=0.5mm] (p2.center) -- (p3.center);
\end{tikzpicture} \quad \begin{tikzpicture}[baseline={([yshift=-0.1cm]current bounding box.center)}, scale=1] 
	\node[] (a) at (0,0) {};
	\node[] (a1) at (1,0) {};
	\node[] (b) at (0,-1) {};
	\node[] (b1) at (1,-1) {};
	\node[] (p1) at ($(a)+(-0.25,0)$) {};
	\node[] (p2) at ($(b)+(-0.25,0)$) {};
	\node[] (p3) at ($(b1)+(0.25,0)$) {};
	\node[] (p4) at ($(a1)+(0.25,0)$) {};
	\draw[line width=0.15mm] (b.center) -- (a1.center);
	\draw[line width=0.15mm] (a.center) -- (0.375,-0.375);
	\draw[line width=0.15mm] (0.625,-0.625) -- (b1.center);
	\draw[line width=0.5mm] (p1.center) -- (p4.center);
	\draw[line width=0.5mm] (p2.center) -- (p3.center);
\end{tikzpicture} \quad \begin{tikzpicture}[baseline={([yshift=-0.1cm]current bounding box.center)}, scale=1] 
	\node[] (a) at (0.5,0) {};
	\node[] (b) at (0.5,-0.5) {};
	\node[] (c) at (0,-1) {};
	\node[] (c1) at (1,-1) {};
	\node[] (p1) at ($(a)+(-0.75,0)$) {};
	\node[] (p2) at ($(c)+(-0.25,0)$) {};
	\node[] (p3) at ($(c1)+(0.25,0)$) {};
	\node[] (p4) at ($(a)+(0.75,0)$) {};
	\draw[line width=0.15mm] (b.center) -- (a.center);
	\draw[line width=0.15mm] (b.center) -- (c.center);
	\draw[line width=0.15mm] (b.center) -- (c1.center);
	\draw[line width=0.5mm] (p1.center) -- (p4.center);
	\draw[line width=0.5mm] (p2.center) -- (p3.center);
\end{tikzpicture} \quad  \begin{tikzpicture}[baseline={([yshift=-0.1cm]current bounding box.center)}, scale=1] 
	\node[] (a) at (0.5,0) {};
	\node[] (b) at (0,-1) {};
	\node[] (b1) at (1,-1) {};
	\node[] (p1) at ($(a)+(-0.75,0)$) {};
	\node[] (p2) at ($(b)+(-0.25,0)$) {};
	\node[] (p3) at ($(b1)+(0.25,0)$) {};
	\node[] (p4) at ($(a)+(0.75,0)$) {};
	\draw[line width=0.15mm] (b.center) -- (a.center);
	\draw[line width=0.15mm] (b1.center) -- (a.center);
	\draw[line width=0.5mm] (p1.center) -- (p4.center);
	\draw[line width=0.5mm] (p2.center) -- (p3.center);
\end{tikzpicture}
\caption{One-loop classical and superclassical diagrams.}
\label{fig: diagrams_1_loop}
\end{figure}
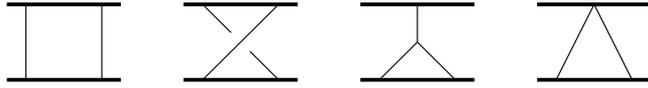

In this section, we begin the systematic analysis of the leading singularities of the diagrams relevant to the classical dynamics, starting with the one-loop corrections. As we gather in fig.~\ref{fig: diagrams_1_loop}, we a priori have three diagrams with four propagators, one subsector with three propagators, as well as their vertical and horizontal reflections.

However, using the relations gathered in sec.~\ref{sec:reduction_relations}, we can see that the leading singularity of the crossed-box diagram can be immediately related to that of the box diagram by flipping one matter line, and that the latter is straightforward to compute:
\begin{equation}
\LS \left( \begin{tikzpicture}[baseline={([yshift=-0.1cm]current bounding box.center)}] 
	\node[] (a) at (0,0) {};
	\node[] (a1) at (0.8,0) {};
	\node[] (b) at (0,-0.8) {};
	\node[] (b1) at (0.8,-0.8) {};
	\node[] (p1) at ($(a)+(-0.2,0)$) {};
	\node[] (p2) at ($(b)+(-0.2,0)$) {};
	\node[] (p3) at ($(b1)+(0.2,0)$) {};
	\node[] (p4) at ($(a1)+(0.2,0)$) {};
	\draw[line width=0.15mm] (b.center) -- (a1.center);
	\draw[line width=0.15mm] (a.center) -- (0.3,-0.3);
	\draw[line width=0.15mm] (0.5,-0.5) -- (b1.center);
	\draw[line width=0.5mm] (p1.center) -- (p4.center);
	\draw[line width=0.5mm] (p2.center) -- (p3.center);
\end{tikzpicture} \right) \propto \LS \left( \begin{tikzpicture}[baseline={([yshift=-0.1cm]current bounding box.center)}] 
	\node[] (a) at (0,0) {};
	\node[] (a1) at (0.8,0) {};
	\node[] (b) at (0,-0.8) {};
	\node[] (b1) at (0.8,-0.8) {};
	\node[] (p1) at ($(a)+(-0.2,0)$) {};
	\node[] (p2) at ($(b)+(-0.2,0)$) {};
	\node[] (p3) at ($(b1)+(0.2,0)$) {};
	\node[] (p4) at ($(a1)+(0.2,0)$) {};
	\draw[line width=0.15mm] (b.center) -- (a.center);
	\draw[line width=0.15mm] (b1.center) -- (a1.center);
	\draw[line width=0.5mm] (p1.center) -- (p4.center);
	\draw[line width=0.5mm] (p2.center) -- (p3.center);
\end{tikzpicture} \right) \propto \frac{x}{q^2 (x^2-1)}.
\end{equation}
In general, for each diagram we also find a version where one of the matter lines is flipped, which up to a minus sign depends on the same Feynman integral and geometry. Therefore, in the following we will omit them and consider just one representative.

In addition, we can trivially relate the last two integrals of fig.~\ref{fig: diagrams_1_loop},
\begin{equation}
\begin{tikzpicture}[baseline={([yshift=-0.1cm]current bounding box.center)}] 
	\node[] (a) at (0.4,0) {};
	\node[] (b) at (0.4,-0.4) {};
	\node[] (c) at (0,-0.8) {};
	\node[] (c1) at (0.8,-0.8) {};
	\node[] (p1) at ($(a)+(-0.6,0)$) {};
	\node[] (p2) at ($(c)+(-0.2,0)$) {};
	\node[] (p3) at ($(c1)+(0.2,0)$) {};
	\node[] (p4) at ($(a)+(0.6,0)$) {};
	\draw[line width=0.15mm] (b.center) -- (a.center);
	\draw[line width=0.15mm] (b.center) -- (c.center);
	\draw[line width=0.15mm] (b.center) -- (c1.center);
	\draw[line width=0.5mm] (p1.center) -- (p4.center);
	\draw[line width=0.5mm] (p2.center) -- (p3.center);
\end{tikzpicture} = \frac{1}{q^2} \, \begin{tikzpicture}[baseline={([yshift=-0.1cm]current bounding box.center)}] 
	\node[] (a) at (0.4,0) {};
	\node[] (b) at (0,-0.8) {};
	\node[] (b1) at (0.8,-0.8) {};
	\node[] (p1) at ($(a)+(-0.6,0)$) {};
	\node[] (p2) at ($(b)+(-0.2,0)$) {};
	\node[] (p3) at ($(b1)+(0.2,0)$) {};
	\node[] (p4) at ($(a)+(0.6,0)$) {};
	\draw[line width=0.15mm] (b.center) -- (a.center);
	\draw[line width=0.15mm] (b1.center) -- (a.center);
	\draw[line width=0.5mm] (p1.center) -- (p4.center);
	\draw[line width=0.5mm] (p2.center) -- (p3.center);
\end{tikzpicture}.
\end{equation}
In general, there are integrals at all loops which contain trivial $\frac{1}{q^2}$ graviton propagators. In the following, we will thus also omit them and consider only one representative where all the trivial propagators are pinched. 

Therefore, in practice we only need to study the triangle diagram at one loop. Let us nonetheless first go over the details of the computation for the box diagram for illustrative purposes.

\subsection{Box diagram}
\label{sec:one_loop_box}

The first example we consider is the box diagram with propagators of power $\nu_i=1$, with a superclassical power counting of $|q|^{-2} G^2$ and a parametrization given in fig.~\ref{fig: params_box_and_triangle}(a). We have $\nex=4$, thus $E=\text{dim} \langle p_1, \dots, p_{n_{\text{ext}}} \rangle=3$ and the diagram automatically satisfies $N_{\text{V}}=4=\nint$. Therefore, the Baikov representation in $D=4$ is straightforward by eq.~\eqref{eq: standard_Baikov}:
\begin{equation}
\begin{tikzpicture}[baseline={([yshift=-0.1cm]current bounding box.center)}] 
	\node[] (a) at (0,0) {};
	\node[] (a1) at (0.8,0) {};
	\node[] (b) at (0,-0.8) {};
	\node[] (b1) at (0.8,-0.8) {};
	\node[] (p1) at ($(a)+(-0.2,0)$) {};
	\node[] (p2) at ($(b)+(-0.2,0)$) {};
	\node[] (p3) at ($(b1)+(0.2,0)$) {};
	\node[] (p4) at ($(a1)+(0.2,0)$) {};
	\draw[line width=0.15mm] (b.center) -- (a.center);
	\draw[line width=0.15mm] (b1.center) -- (a1.center);
	\draw[line width=0.5mm] (p1.center) -- (p4.center);
	\draw[line width=0.5mm] (p2.center) -- (p3.center);
\end{tikzpicture} \propto \int \frac{d z_1 d z_2 d z_3 d z_4}{z_1 z_2 z_3 z_4} \frac{1}{\sqrt{\det G(k,u_1,u_2,q)}},
\end{equation}
where we omit constant prefactors, and where 
\begin{align}
\det G(k,u_1,u_2,q) = & \, (y^2-1) \, \lambda(k^2, (k-q)^2,q^2) \nonumber \\
& -4 q^2 ((k \cdot u_1)^2 + (k \cdot u_2)^2 - 2 y (k \cdot u_1) (k \cdot u_2)),
\end{align}
for $\lambda(x,y,z)=x^2+y^2+z^2-2xy-2xz-2yz$ being the Källén function. This Gram determinant becomes a quadratic polynomial $P_2(z_1,\dots,z_4)$ in the Baikov variables, with coefficients depending on $q^2$ and $y$, where we define 
\begin{equation}
z_1=k^2, \qquad z_2=(k-q)^2, \qquad z_3=2u_1 \cdot k, \qquad z_4=-2u_2 \cdot k.
\end{equation}

Deforming the contour to encircle the poles at $z_i=0$ and taking the corresponding residues, we obtain by eq.~\eqref{eq: cut_Baikov} that
\begin{equation}
\label{eq: LS_box}
\LS \left( \begin{tikzpicture}[baseline={([yshift=-0.1cm]current bounding box.center)}] 
	\node[] (a) at (0,0) {};
	\node[] (a1) at (0.8,0) {};
	\node[] (b) at (0,-0.8) {};
	\node[] (b1) at (0.8,-0.8) {};
	\node[] (p1) at ($(a)+(-0.2,0)$) {};
	\node[] (p2) at ($(b)+(-0.2,0)$) {};
	\node[] (p3) at ($(b1)+(0.2,0)$) {};
	\node[] (p4) at ($(a1)+(0.2,0)$) {};
	\draw[line width=0.15mm] (b.center) -- (a.center);
	\draw[line width=0.15mm] (b1.center) -- (a1.center);
	\draw[line width=0.5mm] (p1.center) -- (p4.center);
	\draw[line width=0.5mm] (p2.center) -- (p3.center);
\end{tikzpicture} \right) \propto \frac{1}{\sqrt{P_2(0,\dots,0)}} \propto \frac{1}{\sqrt{(y^2-1) \, \lambda(0,0,q^2)}} = \frac{1}{q^2 \sqrt{y^2-1}} \propto \frac{x}{q^2 (x^2-1)},
\end{equation}
where we used the change of variables $y=\frac{x^2+1}{2x}$ to rationalize the square root in the kinematics. We see that the result correctly scales as $|q|^{-2}$, and as $G^2$ once the vertices are dressed. Importantly, this result also holds if we relax the condition $u_i \cdot q=0$, which will be relevant when considering box sub-loops appearing inside bigger diagrams. The result for the leading singularity is in agreement with eq.~\eqref{eq: reduction_superclassical} and with refs.~\cite{Parra-Martinez:2020dzs,Herrmann:2021tct,DiVecchia:2021bdo}, where a basis for pure master integrals is provided up to two loops.\footnote{Moreover, it agrees with the well-known result $1/(st)$ for the leading singularity of the box integral with quadratic propagators \cite{Britto:2004nc} upon expanding the result in the classical limit.}

So far, we have considered the contribution that is even under the parity transformation, see sec.~\ref{sec:soft_expansion} and app.~\ref{app:parity} for more details. 
Since at one loop we do not have ISPs, the odd contribution can be calculated instead by adding a dot to a matter propagator,
\begin{equation}
\LS \left( \begin{tikzpicture}[baseline={([yshift=-0.1cm]current bounding box.center)}] 
	\node[] (a) at (0,0) {};
	\node[] (a1) at (0.8,0) {};
	\node[] (b) at (0,-0.8) {};
	\node[] (b1) at (0.8,-0.8) {};
	\node[] (p1) at ($(a)+(-0.2,0)$) {};
	\node[] (p2) at ($(b)+(-0.2,0)$) {};
	\node[] (p3) at ($(b1)+(0.2,0)$) {};
	\node[] (p4) at ($(a1)+(0.2,0)$) {};
	\draw[line width=0.15mm] (b.center) -- (a.center);
	\draw[line width=0.15mm] (b1.center) -- (a1.center);
	\draw[line width=0.5mm] (p1.center) -- (p4.center);
	\draw[line width=0.5mm] (p2.center) -- (p3.center);
	\node at ($(a1)-(0.4,0)$) [circle,fill,inner sep=1.5pt]{};
have \end{tikzpicture} \right) = 0.
\end{equation}
Since it vanishes, the odd-parity box has zero master integrals in its sector.

\subsection{Triangle diagram}
\label{sec:one_loop_triangle}
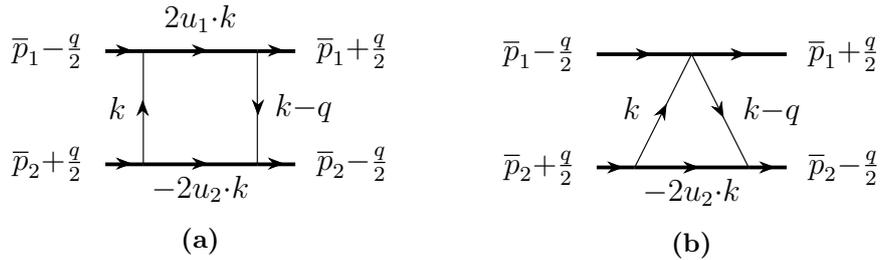
\begin{figure}[t]
\centering
\subfloat[]{\begin{tikzpicture}[baseline={([yshift=-0.1cm]current bounding box.center)}] 
	\node[] (a) at (0,0) {};
	\node[] (b) at (0,-1.5) {};
	\node[] (a1) at (1.5,0) {};
	\node[] (b1) at (1.5,-1.5) {};
	\node[label=left:{$\overline{p}_1{-}\frac{q}{2}$}] (p1) at ($(a)+(-0.5,0)$) {};
	\node[label=left:{$\overline{p}_2{+}\frac{q}{2}$}] (p2) at ($(b)+(-0.5,0)$) {};
	\node[label=right:{$\overline{p}_2{-}\frac{q}{2}$}] (p3) at ($(b1)+(0.5,0)$) {};
	\node[label=right:{$\overline{p}_1{+}\frac{q}{2}$}] (p4) at ($(a1)+(0.5,0)$) {};
	\draw[line width=0.15mm, postaction={decorate}] (b.center) -- node[sloped, allow upside down, label={[xshift=0.15cm, yshift=0cm]$k$}] {\midarrow} (a.center);
	\draw[line width=0.15mm, postaction={decorate}] (a1.center) -- node[sloped, allow upside down, label={[xshift=-0.15cm, yshift=0cm]$k{-}q$}] {\midarrow} (b1.center);
	\draw[line width=0.5mm, postaction={decorate}] (a.center) -- node[sloped, allow upside down, label={[xshift=0cm, yshift=-0.15cm]$2u_1{\cdot} k$}] {\midarrow} (a1.center);
	\draw[line width=0.5mm, postaction={decorate}] (b.center) -- node[sloped, allow upside down, label={[xshift=0cm, yshift=-0.9cm]$-2u_2{\cdot} k$}] {\midarrow} (b1.center);
	\draw[line width=0.5mm, postaction={decorate}] (p1.center) -- node[sloped, allow upside down] {\midarrow} (a.center);
	\draw[line width=0.5mm, postaction={decorate}] (a1.center) -- node[sloped, allow upside down] {\midarrow} (p4.center);
	\draw[line width=0.5mm, postaction={decorate}] (p2.center) -- node[sloped, allow upside down] {\midarrow} (b.center);
	\draw[line width=0.5mm, postaction={decorate}] (b1.center) -- node[sloped, allow upside down] {\midarrow} (p3.center);
\end{tikzpicture}} \quad \qquad \subfloat[]{\begin{tikzpicture}[baseline=(current bounding box.center)] 
	\node[] (a) at (0.75,0) {};
	\node[] (b) at (0,-1.5) {};
	\node[] (b1) at (1.5,-1.5) {};
	\node[label=left:{$\overline{p}_1{-}\frac{q}{2}$}] (p1) at ($(a)+(-1.25,0)$) {};
	\node[label=left:{$\overline{p}_2{+}\frac{q}{2}$}] (p2) at ($(b)+(-0.5,0)$) {};
	\node[label=right:{$\overline{p}_2{-}\frac{q}{2}$}] (p3) at ($(b1)+(0.5,0)$) {};
	\node[label=right:{$\overline{p}_1{+}\frac{q}{2}$}] (p4) at ($(a)+(1.25,0)$) {};
	\draw[line width=0.15mm, postaction={decorate}] (b.center) -- node[sloped, allow upside down, label={[xshift=0.05cm, yshift=-0.35cm]$k$}] {\midarrow} (a.center);
	\draw[line width=0.15mm, postaction={decorate}] (a.center) -- node[sloped, allow upside down, label={[xshift=-0.05cm, yshift=-0.44cm]$k{-}q$}] {\midarrow} (b1.center);
	\node[label={[xshift=0cm, yshift=-0.15cm]\textcolor{white}{$2u_1{\cdot} k$}}] (lwhite) at (0.75,0) {};
	\draw[line width=0.5mm, postaction={decorate}] (b.center) -- node[sloped, allow upside down, label={[xshift=0cm, yshift=-0.9cm]$-2u_2{\cdot} k$}] {\midarrow} (b1.center);
	\draw[line width=0.5mm, postaction={decorate}] (p1.center) -- node[sloped, allow upside down] {\midarrow} (a.center);
	\draw[line width=0.5mm, postaction={decorate}] (a.center) -- node[sloped, allow upside down] {\midarrow} (p4.center);
	\draw[line width=0.5mm, postaction={decorate}] (p2.center) -- node[sloped, allow upside down] {\midarrow} (b.center);
	\draw[line width=0.5mm, postaction={decorate}] (b1.center) -- node[sloped, allow upside down] {\midarrow} (p3.center);
\end{tikzpicture}}
\caption{Parametrization of the loop momenta for the box diagram (a), and the triangle diagram (b).}
\label{fig: params_box_and_triangle}
\end{figure}

Next, we consider the triangle diagram, which is a subsector of the box diagram and has a classical power counting of $|q|^{-1} G^2$. Following the parametrization of fig.~\ref{fig: params_box_and_triangle}(b), we observe that the momentum of the first scalar particle does not appear in the loop. Hence, we have $E=2$ and it automatically satisfies $N_{\text{V}}=3=\nint$. Directly applying eqs.~\eqref{eq: standard_Baikov} and~\eqref{eq: cut_Baikov}, we obtain
\begin{align}
\label{eq: Baikov_triangle}
\hspace{-0.3cm} \begin{tikzpicture}[baseline={([yshift=-0.1cm]current bounding box.center)}] 
	\node[] (a) at (0.4,0) {};
	\node[] (b) at (0,-0.8) {};
	\node[] (b1) at (0.8,-0.8) {};
	\node[] (p1) at ($(a)+(-0.6,0)$) {};
	\node[] (p2) at ($(b)+(-0.2,0)$) {};
	\node[] (p3) at ($(b1)+(0.2,0)$) {};
	\node[] (p4) at ($(a)+(0.6,0)$) {};
	\draw[line width=0.15mm] (b.center) -- (a.center);
	\draw[line width=0.15mm] (b1.center) -- (a.center);
	\draw[line width=0.5mm] (p1.center) -- (p4.center);
	\draw[line width=0.5mm] (p2.center) -- (p3.center);
\end{tikzpicture} & \propto \int \frac{d z_1 d z_2 d z_3}{z_1 z_2 z_3} \frac{1}{\sqrt{\det G(u_2,q)}} = \int \frac{d z_1 d z_2 d z_3}{z_1 z_2 z_3} \frac{1}{\sqrt{q^2-(u_2\cdot q)^2}}, \\[0.2cm]
\hspace{-0.275cm} \LS \left( \begin{tikzpicture}[baseline={([yshift=-0.1cm]current bounding box.center)}] 
	\node[] (a) at (0.4,0) {};
	\node[] (b) at (0,-0.8) {};
	\node[] (b1) at (0.8,-0.8) {};
	\node[] (p1) at ($(a)+(-0.6,0)$) {};
	\node[] (p2) at ($(b)+(-0.2,0)$) {};
	\node[] (p3) at ($(b1)+(0.2,0)$) {};
	\node[] (p4) at ($(a)+(0.6,0)$) {};
	\draw[line width=0.15mm] (b.center) -- (a.center);
	\draw[line width=0.15mm] (b1.center) -- (a.center);
	\draw[line width=0.5mm] (p1.center) -- (p4.center);
	\draw[line width=0.5mm] (p2.center) -- (p3.center);
\end{tikzpicture} \right) & \propto \frac{1}{\sqrt{q^2-(u_2 \cdot q)^2}} = \frac{1}{|q|}.
\label{eq: LS_triangle}
\end{align}
The simplification comes from the fact that for triangle loops only the simpler Gram determinant $\det G(p_1,\dots,p_E)=\det G(u_2,q)$ appears in eq.~\eqref{eq: standard_Baikov}, which does not depend on the loop momentum. The result has the expected $|q|^{-1}$ scaling, and is in agreement with the one-loop pure basis provided in refs.~\cite{Parra-Martinez:2020dzs,Herrmann:2021tct,DiVecchia:2021bdo}. Notice that we also give the result for $u_i \cdot q \neq 0$, which will be relevant for diagrams containing triangle sub-loops.

In this case, the even-parity contribution can be calculated again with a dotted matter propagator,
\begin{equation}
\label{eq: LS_triangle_dotted_matter}
\LS \left( \begin{tikzpicture}[baseline={([yshift=-0.1cm]current bounding box.center)}] 
	\node[] (a) at (0.4,0) {};
	\node[] (b) at (0,-0.8) {};
	\node[] (b1) at (0.8,-0.8) {};
	\node[] (p1) at ($(a)+(-0.6,0)$) {};
	\node[] (p2) at ($(b)+(-0.2,0)$) {};
	\node[] (p3) at ($(b1)+(0.2,0)$) {};
	\node[] (p4) at ($(a)+(0.6,0)$) {};
	\draw[line width=0.15mm] (b.center) -- (a.center);
	\draw[line width=0.15mm] (b1.center) -- (a.center);
	\draw[line width=0.5mm] (p1.center) -- (p4.center);
	\draw[line width=0.5mm] (p2.center) -- (p3.center);
	\node at ($(b1)-(0.4,0)$) [circle,fill,inner sep=1.5pt]{};
\end{tikzpicture} \right) \propto \frac{\varepsilon \, (u_2 \cdot q)}{q^2 \sqrt{q^2-(u_2 \cdot q)^2}} = 0.
\end{equation}
It thus has zero master integrals in its sector. The $\varepsilon$-dependence arises during the Laurent expansion of the integrand when calculating the residue at the double pole.

\section{Two-loop diagrams}
\label{sec:two_loop}
\begin{figure}[t]
\centering
\parbox{\textwidth}{\begin{align*}
& \begin{tikzpicture}[baseline={([yshift=-0.1cm]current bounding box.center)}] 
	\node[] (a) at (0,0) {};
	\node[] (a1) at (0.9,0) {};
	\node[] (a2) at (1.2,0) {};
	\node[] (b) at (0.45,-0.6) {};
	\node[] (c) at (0.45,-1) {};
	\node[] (c1) at (1.2,-1) {};
	\node[] (p1) at ($(a)+(-0.2,0)$) {};
	\node[] (p2) at ($(c)+(-0.65,0)$) {};
	\node[] (p3) at ($(c1)+(0.2,0)$) {};
	\node[] (p4) at ($(a2)+(0.2,0)$) {};
	\draw[line width=0.15mm] (b.center) -- (a.center);
	\draw[line width=0.15mm] (b.center) -- (a1.center);
	\draw[line width=0.15mm] (b.center) -- (c.center);
	\draw[line width=0.15mm] (a2.center) -- (c1.center);
	\draw[line width=0.5mm] (p1.center) -- (p4.center);
	\draw[line width=0.5mm] (p2.center) -- (p3.center);
\end{tikzpicture} \quad \begin{tikzpicture}[baseline={([yshift=-0.1cm]current bounding box.center)}] 
	\node[] (a) at (0,0) {};
	\node[] (a1) at (0.8,0) {};
	\node[] (a2) at (1.6,0) {};
	\node[] (b) at (0,-0.8) {};
	\node[] (b1) at (0.8,-0.8) {};
	\node[] (b2) at (1.6,-0.8) {};
	\node[] (p1) at ($(a)+(-0.2,0)$) {};
	\node[] (p2) at ($(b)+(-0.2,0)$) {};
	\node[] (p3) at ($(b2)+(0.2,0)$) {};
	\node[] (p4) at ($(a2)+(0.2,0)$) {};
	\draw[line width=0.15mm] (b.center) -- (a.center);
	\draw[line width=0.15mm] (b1.center) -- (a1.center);
	\draw[line width=0.15mm] (b2.center) -- (a2.center);
	\draw[line width=0.5mm] (p1.center) -- (p4.center);
	\draw[line width=0.5mm] (p2.center) -- (p3.center);
\end{tikzpicture} \quad \raisebox{0.15cm}{\begin{tikzpicture}[baseline={([yshift=-0.1cm]current bounding box.center)}, scale=0.8] 
	\node[] (a) at (0,0) {};
	\node[] (a1) at (1,0) {};
	\node[] (b) at (0,-1) {};
	\node[] (b1) at (1,-1) {};
	\node[] (p1) at ($(a)+(-0.45,0)$) {};
	\node[] (p2) at ($(b)+(-0.45,0)$) {};
	\node[] (p3) at ($(b1)+(0.45,0)$) {};
	\node[] (p4) at ($(a1)+(0.45,0)$) {};
	\draw[line width=0.15mm] (b.center) -- (a.center);
	\draw[line width=0.15mm] (a1.center) -- (b1.center);
	\draw[line width=0.5mm] (p1.center) -- (p4.center);
	\draw[line width=0.5mm] (p2.center) -- (p3.center);
	\draw[line width=0.15mm] (1.2,0) arc (0:180:0.7);
\end{tikzpicture}} \quad \begin{tikzpicture}[baseline={([yshift=-0.1cm]current bounding box.center)}, scale=0.8] 
	\node[] (a) at (0,0) {};
	\node[] (a1) at (1,0) {};
	\node[] (b) at (0,-1) {};
	\node[] (b1) at (1,-1) {};
	\node[] (p1) at ($(a)+(-0.25,0)$) {};
	\node[] (p2) at ($(b)+(-0.25,0)$) {};
	\node[] (p3) at ($(b1)+(0.25,0)$) {};
	\node[] (p4) at ($(a1)+(0.25,0)$) {};
	\draw[line width=0.15mm] (b.center) -- (a.center);
	\draw[line width=0.15mm] (b1.center) -- (a1.center);
	\draw[line width=0.5mm] (p1.center) -- (p4.center);
	\draw[line width=0.5mm] (p2.center) -- (p3.center);
	\draw[line width=0.15mm] (0.8,0) arc (0:180:0.3);
\end{tikzpicture} \quad
\begin{tikzpicture}[baseline={([yshift=-0.1cm]current bounding box.center)}, scale=0.8] 
	\node[] (a) at (0,0) {};
	\node[] (a1) at (1,0) {};
	\node[] (b) at (0,-1) {};
	\node[] (b1) at (1,-1) {};
	\node[] (p1) at ($(a)+(-0.55,0)$) {};
	\node[] (p2) at ($(b)+(-0.55,0)$) {};
	\node[] (p3) at ($(b1)+(0.25,0)$) {};
	\node[] (p4) at ($(a1)+(0.25,0)$) {};
	\draw[line width=0.15mm] (b.center) -- (a.center);
	\draw[line width=0.15mm] (b1.center) -- (a1.center);
	\draw[line width=0.5mm] (p1.center) -- (p4.center);
	\draw[line width=0.5mm] (p2.center) -- (p3.center);
	\draw[line width=0.15mm] (0.3,0) arc (0:180:0.3);
\end{tikzpicture} \quad \begin{tikzpicture}[baseline={([yshift=-0.1cm]current bounding box.center)}] 
	\node[] (a) at (0,0) {};
	\node[] (b) at (0,-0.5) {};
	\node[] (a1) at (0.5,0) {};
	\node[] (b1) at (0.5,-0.5) {};
	\node[] (c) at (0,-1) {};
	\node[] (c1) at (0.5,-1) {};
	\node[] (p1) at ($(a)+(-0.2,0)$) {};
	\node[] (p2) at ($(c)+(-0.2,0)$) {};
	\node[] (p3) at ($(c1)+(0.2,0)$) {};
	\node[] (p4) at ($(a1)+(0.2,0)$) {};
	\draw[line width=0.15mm] (a.center) -- (b.center);
	\draw[line width=0.15mm] (a1.center) -- (b1.center);
	\draw[line width=0.15mm] (b1.center) -- (b.center);
	\draw[line width=0.15mm] (b.center) -- (c.center);
	\draw[line width=0.15mm] (c1.center) -- (b1.center);
	\draw[line width=0.5mm] (a.center) -- (a1.center);
	\draw[line width=0.5mm] (c.center) -- (c1.center);
	\draw[line width=0.5mm] (p1.center) -- (a.center);
	\draw[line width=0.5mm] (a1.center) -- (p4.center);
	\draw[line width=0.5mm] (p2.center) -- (c.center);
	\draw[line width=0.5mm] (c1.center) -- (p3.center);
\end{tikzpicture} \\[0.2cm]
& \begin{tikzpicture}[baseline={([yshift=-0.1cm]current bounding box.center)}] 
	\node[] (a) at (0,0) {};
	\node[] (a1) at (0.5,0) {};
	\node[] (a2) at (0.9,0) {};
	\node[] (b) at (0.7,-0.27) {};
	\node[] (c) at (0.45,-0.6) {};
	\node[] (d) at (0.45,-1) {};
	\node[] (d1) at (1.2,-1) {};
	\node[] (p1) at ($(a)+(-0.2,0)$) {};
	\node[] (p2) at ($(d)+(-0.65,0)$) {};
	\node[] (p3) at ($(d1)+(0.2,0)$) {};
	\node[] (p4) at ($(a2)+(0.5,0)$) {};
	\draw[line width=0.15mm] (c.center) -- (a.center);
	\draw[line width=0.15mm] (0.65,-0.22) -- (a1.center);
	\draw[line width=0.15mm] (c.center) -- (a2.center);
	\draw[line width=0.15mm] (c.center) -- (d.center);
	\draw[line width=0.15mm] (0.75,-0.35) -- (d1.center);
	\draw[line width=0.5mm] (p1.center) -- (p4.center);
	\draw[line width=0.5mm] (p2.center) -- (p3.center);
\end{tikzpicture} \quad \begin{tikzpicture}[baseline={([yshift=-0.1cm]current bounding box.center)}] 
	\node[] (a) at (0,0) {};
	\node[] (a1) at (0.8,0) {};
	\node[] (a2) at (1.6,0) {};
	\node[] (b) at (0,-0.8) {};
	\node[] (b1) at (0.8,-0.8) {};
	\node[] (b2) at (1.6,-0.8) {};
	\node[] (p1) at ($(a)+(-0.2,0)$) {};
	\node[] (p2) at ($(b)+(-0.2,0)$) {};
	\node[] (p3) at ($(b2)+(0.2,0)$) {};
	\node[] (p4) at ($(a2)+(0.2,0)$) {};
	\draw[line width=0.15mm] (b.center) -- (a.center);
	\draw[line width=0.15mm] (b1.center) -- (a2.center);
	\draw[line width=0.15mm] (b2.center) -- (1.3,-0.5);
	\draw[line width=0.15mm] (1.1,-0.3) -- (a1.center);
	\draw[line width=0.5mm] (p1.center) -- (p4.center);
	\draw[line width=0.5mm] (p2.center) -- (p3.center);
\end{tikzpicture}
\end{align*}}
\caption{Two-loop classical and superclassical top topology diagrams. The second row contains non-planar topologies, the leading singularity of which can be related by unraveling~\eqref{eq: unraveling_matter_props} to their planar counterparts above.}
\label{fig: top_topologies_two_loop}
\vspace{0.5cm}
\parbox{\textwidth}{\begin{align*}
& \raisebox{0.17cm}{\begin{tikzpicture}[baseline={([yshift=-0.1cm]current bounding box.center)}, scale=0.8] 
	\node[] (b) at (0.5,-1) {};
	\node[] (c) at (0,0) {};
	\node[] (c1) at (1,0) {};
	\node[] (p1) at ($(b)+(-0.95,0)$) {};
	\node[] (p2) at ($(c)+(-0.45,0)$) {};
	\node[] (p3) at ($(c1)+(0.45,0)$) {};
	\node[] (p4) at ($(b)+(0.95,0)$) {};
	\draw[line width=0.15mm] (b.center) -- (c.center);
	\draw[line width=0.15mm] (b.center) -- (c1.center);
	\draw[line width=0.5mm] (p1.center) -- (p4.center);
	\draw[line width=0.5mm] (p2.center) -- (p3.center);
	\draw[line width=0.15mm] (1.2,0) arc (0:180:0.7);
\end{tikzpicture}} \enspace \, \quad
\raisebox{0.025cm}{\begin{tikzpicture}[baseline={([yshift=-0.1cm]current bounding box.center)}, scale=0.8] 
	\node[] (b) at (0.5,-1) {};
	\node[] (c) at (0,0) {};
	\node[] (c1) at (1,0) {};
	\node[] (p1) at ($(b)+(-0.75,0)$) {};
	\node[] (p2) at ($(c)+(-0.25,0)$) {};
	\node[] (p3) at ($(c1)+(0.25,0)$) {};
	\node[] (p4) at ($(b)+(0.75,0)$) {};
	\draw[line width=0.15mm] (b.center) -- (c.center);
	\draw[line width=0.15mm] (b.center) -- (c1.center);
	\draw[line width=0.5mm] (p1.center) -- (p4.center);
	\draw[line width=0.5mm] (p2.center) -- (p3.center);
	\draw[line width=0.15mm] (0.8,0) arc (0:180:0.3);
\end{tikzpicture}} \quad \enspace
\raisebox{0.025cm}{\begin{tikzpicture}[baseline={([yshift=-0.1cm]current bounding box.center)}, scale=0.8] 
	\node[] (b) at (0.5,-1) {};
	\node[] (c) at (0,0) {};
	\node[] (c1) at (1,0) {};
	\node[] (p1) at ($(b)+(-1.05,0)$) {};
	\node[] (p2) at ($(c)+(-0.55,0)$) {};
	\node[] (p3) at ($(c1)+(0.25,0)$) {};
	\node[] (p4) at ($(b)+(0.75,0)$) {};
	\draw[line width=0.15mm] (b.center) -- (c.center);
	\draw[line width=0.15mm] (b.center) -- (c1.center);
	\draw[line width=0.5mm] (p1.center) -- (p4.center);
	\draw[line width=0.5mm] (p2.center) -- (p3.center);
	\draw[line width=0.15mm] (0.3,0) arc (0:180:0.3);
\end{tikzpicture}} \quad
\raisebox{0.17cm}{\begin{tikzpicture}[baseline={([yshift=-0.1cm]current bounding box.center)}, scale=0.8] 
	\node[] (b) at (0.5,-1) {};
	\node[] (c) at (0,0) {};
	\node[] (c1) at (1,0) {};
	\node[] (p1) at ($(b)+(-1.15,0)$) {};
	\node[] (p2) at ($(c)+(-0.65,0)$) {};
	\node[] (p3) at ($(c1)+(0.25,0)$) {};
	\node[] (p4) at ($(b)+(0.75,0)$) {};
	\draw[line width=0.15mm] (b.center) -- (c.center);
	\draw[line width=0.15mm] (b.center) -- (c1.center);
	\draw[line width=0.5mm] (p1.center) -- (p4.center);
	\draw[line width=0.5mm] (p2.center) -- (p3.center);
	\draw[line width=0.15mm] (c1.center) arc (0:180:0.7);
\end{tikzpicture}} \ \quad \raisebox{0.025cm}{\begin{tikzpicture}[baseline={([yshift=-0.1cm]current bounding box.center)}] 
	\node[] (a) at (0,0) {};
	\node[] (a1) at (0.8,0) {};
	\node[] (b) at (0,-0.8) {};
	\node[] (b1) at (0.4,-0.8) {};
	\node[] (b2) at (0.8,-0.8) {};
	\node[] (p1) at ($(a)+(-0.2,0)$) {};
	\node[] (p2) at ($(b)+(-0.2,0)$) {};
	\node[] (p3) at ($(b2)+(0.2,0)$) {};
	\node[] (p4) at ($(a1)+(0.2,0)$) {};
	\draw[line width=0.15mm] (b1.center) -- (a.center);
	\draw[line width=0.15mm] (b1.center) -- (a1.center);
	\draw[line width=0.5mm] (p1.center) -- (p4.center);
	\draw[line width=0.5mm] (p2.center) -- (p3.center);
	\draw[line width=0.15mm] (0.4,0) arc (0:180:0.2);
\end{tikzpicture}} \quad \, \enspace \begin{tikzpicture}[baseline={([yshift=-0.1cm]current bounding box.center)}] 
	\node[] (a) at (0,0) {};
	\node[] (b) at (0.25,-0.5) {};
	\node[] (a1) at (0.5,0) {};
	\node[] (d) at (0,-1) {};
	\node[] (d1) at (0.5,-1) {};
	\node[] (p1) at ($(a)+(-0.2,0)$) {};
	\node[] (p2) at ($(d)+(-0.2,0)$) {};
	\node[] (p3) at ($(d1)+(0.2,0)$) {};
	\node[] (p4) at ($(a1)+(0.2,0)$) {};
	\draw[line width=0.15mm] (b.center) -- (a1.center);
	\draw[line width=0.15mm] (b.center) -- (a.center);
	\draw[line width=0.15mm] (b.center) -- (d.center);
	\draw[line width=0.15mm] (d1.center) -- (b.center);
	\draw[line width=0.5mm] (p1.center) -- (p4.center);
	\draw[line width=0.5mm] (p2.center) -- (p3.center);
\end{tikzpicture} \quad \enspace \, \begin{tikzpicture}[baseline={([yshift=-0.1cm]current bounding box.center)}] 
	\node[] (a) at (0,0) {};
	\node[] (b) at (0.5,-0.5) {};
	\node[] (a1) at (0.5,0) {};
	\node[] (c) at (0,-1) {};
	\node[] (c1) at (0.5,-1) {};
	\node[] (p1) at ($(a)+(-0.2,0)$) {};
	\node[] (p2) at ($(c)+(-0.2,0)$) {};
	\node[] (p3) at ($(c1)+(0.2,0)$) {};
	\node[] (p4) at ($(a1)+(0.2,0)$) {};
	\draw[line width=0.15mm] (b.center) -- (a1.center);
	\draw[line width=0.15mm] (b.center) -- (a.center);
	\draw[line width=0.15mm] (a.center) -- (c.center);
	\draw[line width=0.15mm] (c1.center) -- (b.center);
	\draw[line width=0.5mm] (p1.center) -- (p4.center);
	\draw[line width=0.5mm] (p2.center) -- (p3.center);
\end{tikzpicture} \\[0.2cm]
& \begin{tikzpicture}[baseline={([yshift=-0.1cm]current bounding box.center)}] 
	\node[] (a) at (0,0) {};
	\node[] (a1) at (0.75,0) {};
	\node[] (a2) at (1.2,0) {};
	\node[] (b) at (0.38,-0.5) {};
	\node[] (d) at (0.6,-0.8) {};
	\node[] (p1) at ($(a)+(-0.2,0)$) {};
	\node[] (p2) at ($(d)+(-0.8,0)$) {};
	\node[] (p3) at ($(d)+(0.8,0)$) {};
	\node[] (p4) at ($(a2)+(0.2,0)$) {};
	\draw[line width=0.15mm] (d.center) -- (a.center);
	\draw[line width=0.15mm] (b.center) -- (a1.center);
	\draw[line width=0.15mm] (d.center) -- (a2.center);
	\draw[line width=0.5mm] (p1.center) -- (p4.center);
	\draw[line width=0.5mm] (p2.center) -- (p3.center);
\end{tikzpicture} \quad \begin{tikzpicture}[baseline={([yshift=-0.1cm]current bounding box.center)}] 
	\node[] (a) at (0,0) {};
	\node[] (a1) at (0.8,0) {};
	\node[] (a2) at (1.2,0) {};
	\node[] (b) at (0.4,-0.8) {};
	\node[] (b1) at (1.2,-0.8) {};
	\node[] (p1) at ($(a)+(-0.2,0)$) {};
	\node[] (p2) at ($(b)+(-0.6,0)$) {};
	\node[] (p3) at ($(b1)+(0.2,0)$) {};
	\node[] (p4) at ($(a1)+(0.6,0)$) {};
	\draw[line width=0.15mm] (b.center) -- (a.center);
	\draw[line width=0.15mm] (b.center) -- (a1.center);
		\draw[line width=0.15mm] (b1.center) -- (a2.center);
	\draw[line width=0.5mm] (p1.center) -- (p4.center);
	\draw[line width=0.5mm] (p2.center) -- (p3.center);
\end{tikzpicture} \quad \  \begin{tikzpicture}[baseline={([yshift=-0.1cm]current bounding box.center)}] 
	\node[] (a) at (0,0) {};
	\node[] (a1) at (0.4,0) {};
	\node[] (a2) at (0.8,0) {};
	\node[] (b) at (0.4,-0.8) {};
	\node[] (p1) at ($(a)+(-0.2,0)$) {};
	\node[] (p2) at ($(b)+(-0.6,0)$) {};
	\node[] (p3) at ($(b)+(0.6,0)$) {};
	\node[] (p4) at ($(a2)+(0.2,0)$) {};
	\draw[line width=0.15mm] (b.center) -- (a.center);
	\draw[line width=0.15mm] (b.center) -- (a1.center);
	\draw[line width=0.15mm] (b.center) -- (a2.center);
	\draw[line width=0.5mm] (p1.center) -- (p4.center);
	\draw[line width=0.5mm] (p2.center) -- (p3.center);
\end{tikzpicture} \quad
\begin{tikzpicture}[baseline={([yshift=-0.1cm]current bounding box.center)}] 
	\node[] (a) at (0,0) {};
	\node[] (a1) at (0.8,0) {};
	\node[] (b) at (0.4,-0.8) {};
	\node[] (b1) at (1.2,-0.8) {};
	\node[] (p1) at ($(a)+(-0.2,0)$) {};
	\node[] (p2) at ($(b)+(-0.6,0)$) {};
	\node[] (p3) at ($(b1)+(0.2,0)$) {};
	\node[] (p4) at ($(a1)+(0.6,0)$) {};
	\draw[line width=0.15mm] (b.center) -- (a.center);
	\draw[line width=0.15mm] (b.center) -- (a1.center);
		\draw[line width=0.15mm] (b1.center) -- (a1.center);
	\draw[line width=0.5mm] (p1.center) -- (p4.center);
	\draw[line width=0.5mm] (p2.center) -- (p3.center);
\end{tikzpicture} \quad \raisebox{0.17cm}{\begin{tikzpicture}[baseline={([yshift=-0.1cm]current bounding box.center)}, scale=0.8] 
	\node[] (b) at (0,-1) {};
	\node[] (b1) at (1,-1) {};
	\node[] (c) at (0,0) {};
	\node[] (c1) at (1,0) {};
	\node[] (p1) at ($(b)+(-0.65,0)$) {};
	\node[] (p2) at ($(c)+(-0.65,0)$) {};
	\node[] (p3) at ($(c1)+(0.25,0)$) {};
	\node[] (p4) at ($(b1)+(0.25,0)$) {};
	\draw[line width=0.15mm] (b.center) -- (c.center);
	\draw[line width=0.15mm] (b1.center) -- (c1.center);
	\draw[line width=0.5mm] (p1.center) -- (p4.center);
	\draw[line width=0.5mm] (p2.center) -- (p3.center);
	\draw[line width=0.15mm] (c1.center) arc (0:180:0.7);
\end{tikzpicture}} \quad
\raisebox{0.025cm}{\begin{tikzpicture}[baseline={([yshift=-0.1cm]current bounding box.center)}] 
	\node[] (a) at (0,0) {};
	\node[] (a1) at (0.8,0) {};
	\node[] (b) at (0,-0.8) {};
	\node[] (b1) at (0.8,-0.8) {};
	\node[] (p1) at ($(a)+(-0.2,0)$) {};
	\node[] (p2) at ($(b)+(-0.2,0)$) {};
	\node[] (p3) at ($(b1)+(0.2,0)$) {};
	\node[] (p4) at ($(a1)+(0.2,0)$) {};
	\draw[line width=0.15mm] (b.center) -- (a.center);
	\draw[line width=0.15mm] (b1.center) -- (a1.center);
	\draw[line width=0.5mm] (p1.center) -- (p4.center);
	\draw[line width=0.5mm] (p2.center) -- (p3.center);
	\draw[line width=0.15mm] (0.4,0) arc (0:180:0.2);
\end{tikzpicture}} \quad
\raisebox{0.125cm}{\begin{tikzpicture}[baseline={([yshift=-0.1cm]current bounding box.center)}] 
	\node[] (a) at (0,0) {};
	\node[] (a1) at (0.8,0) {};
	\node[] (b) at (0,-0.8) {};
	\node[] (b1) at (0.8,-0.8) {};
	\node[] (p1) at ($(a)+(-0.2,0)$) {};
	\node[] (p2) at ($(b)+(-0.2,0)$) {};
	\node[] (p3) at ($(b1)+(0.2,0)$) {};
	\node[] (p4) at ($(a1)+(0.2,0)$) {};
	\draw[line width=0.15mm] (b.center) -- (a.center);
	\draw[line width=0.15mm] (b1.center) -- (a1.center);
	\draw[line width=0.5mm] (p1.center) -- (p4.center);
	\draw[line width=0.5mm] (p2.center) -- (p3.center);
	\draw[line width=0.15mm] (0.8,0) arc (0:180:0.4);
\end{tikzpicture}} \\[0.3cm]
& \begin{tikzpicture}[baseline={([yshift=-0.1cm]current bounding box.center)}] 
	\node[] (a) at (0,0) {};
	\node[] (a1) at (0.5,0) {};
	\node[] (a2) at (0.9,0) {};
	\node[] (b) at (0.7,-0.27) {};
	\node[] (c) at (0.45,-0.6) {};
	\node[] (d) at (0.45,-0.8) {};
	\node[] (p1) at ($(a)+(-0.35,0)$) {};
	\node[] (p2) at ($(d)+(-0.8,0)$) {};
	\node[] (p3) at ($(d)+(0.8,0)$) {};
	\node[] (p4) at ($(a2)+(0.35,0)$) {};
	\draw[line width=0.15mm] (c.center) -- (a.center);
	\draw[line width=0.15mm] (0.65,-0.22) -- (a1.center);
	\draw[line width=0.15mm] (c.center) -- (a2.center);
	\draw[line width=0.15mm] (c.center) -- (d.center);
	\draw[line width=0.15mm] plot[smooth, tension=0.7] coordinates {(d.center) (0.67,-0.67) (0.77,-0.53) (0.75,-0.37)};
	\draw[line width=0.5mm] (p1.center) -- (p4.center);
	\draw[line width=0.5mm] (p2.center) -- (p3.center);
\end{tikzpicture} \quad \begin{tikzpicture}[baseline={([yshift=-0.1cm]current bounding box.center)}] 
	\node[] (a) at (0,0) {};
	\node[] (a1) at (0.6,0) {};
	\node[] (a2) at (1.2,0) {};
	\node[] (b) at (0.4,-0.8) {};
	\node[] (b1) at (1.2,-0.8) {};
	\node[] (p1) at ($(a)+(-0.2,0)$) {};
	\node[] (p2) at ($(b)+(-0.6,0)$) {};
	\node[] (p3) at ($(b1)+(0.2,0)$) {};
	\node[] (p4) at ($(a2)+(0.2,0)$) {};
	\draw[line width=0.15mm] (b.center) -- (a.center);
	\draw[line width=0.15mm] (b.center) -- (0.78,-0.42);
	\draw[line width=0.15mm] (0.94,-0.26) -- (a2.center);
		\draw[line width=0.15mm] (b1.center) -- (a1.center);
	\draw[line width=0.5mm] (p1.center) -- (p4.center);
	\draw[line width=0.5mm] (p2.center) -- (p3.center);
\end{tikzpicture}
\end{align*}}
\caption{Two-loop classical and superclassical subsectors of the top topologies in fig.~\ref{fig: top_topologies_two_loop}. The third row contains non-planar topologies, the leading singularity of which can be related by unraveling~\eqref{eq: unraveling_matter_props} to their planar counterparts above.}
\label{fig: subsectors_two_loop}
\end{figure}
At two loops, we initially have the 8 top topology diagrams drawn in fig.~\ref{fig: top_topologies_two_loop}, as well as their reflections; and the 16 subsectors of fig.~\ref{fig: subsectors_two_loop} and their reflections.

First of all, we note that the kissing-triangles subsector simply factorizes into a product of two triangles. Thus, we have \cite{Parra-Martinez:2020dzs,Herrmann:2021tct,DiVecchia:2021bdo}
\begin{equation}
\LS \left(
\begin{tikzpicture}[baseline={([yshift=-0.1cm]current bounding box.center)}] 
	\node[] (a) at (0,0) {};
	\node[] (b) at (0.25,-0.5) {};
	\node[] (a1) at (0.5,0) {};
	\node[] (c) at (0,-1) {};
	\node[] (c1) at (0.5,-1) {};
	\node[] (p1) at ($(a)+(-0.2,0)$) {};
	\node[] (p2) at ($(c)+(-0.2,0)$) {};
	\node[] (p3) at ($(c1)+(0.2,0)$) {};
	\node[] (p4) at ($(a1)+(0.2,0)$) {};
	\draw[line width=0.15mm] (b.center) -- (a1.center);
	\draw[line width=0.15mm] (b.center) -- (a.center);
	\draw[line width=0.15mm] (b.center) -- (c.center);
	\draw[line width=0.15mm] (c1.center) -- (b.center);
	\draw[line width=0.5mm] (p1.center) -- (p4.center);
	\draw[line width=0.5mm] (p2.center) -- (p3.center);
\end{tikzpicture} \right) \propto \frac{1}{\sqrt{q^2-(u_1\cdot q)^2} \sqrt{q^2-(u_2\cdot q)^2}} = \frac{1}{q^2}.
\end{equation}
In the following, we will thus omit diagrams containing integrals that fully factorize into products of lower-loop integrals.

By virtue of the unraveling identity \eqref{eq: unraveling_matter_props}, we can immediately relate the leading singularity of the 4 non-planar diagrams of both figures -- as well as their opposite-parity contributions -- to their planar counterparts above. Moreover, diagrams containing bubble corrections with at least one cubic vertex have zero master integrals in their sector for both parity contributions, see fig.~\ref{fig: diagrams_zero_masters}(a) and the discussion below it as well as app.~\ref{app: IBPs_bubbles_triangles}. Therefore, the following diagrams have zero master integrals in their sectors:
\begin{align}
&\raisebox{0.15cm}{\begin{tikzpicture}[baseline={([yshift=-0.1cm]current bounding box.center)}, scale=0.8] 
	\node[] (a) at (0,0) {};
	\node[] (a1) at (1,0) {};
	\node[] (b) at (0,-1) {};
	\node[] (b1) at (1,-1) {};
	\node[] (p1) at ($(a)+(-0.45,0)$) {};
	\node[] (p2) at ($(b)+(-0.45,0)$) {};
	\node[] (p3) at ($(b1)+(0.45,0)$) {};
	\node[] (p4) at ($(a1)+(0.45,0)$) {};
	\draw[line width=0.15mm] (b.center) -- (a.center);
	\draw[line width=0.15mm] (a1.center) -- (b1.center);
	\draw[line width=0.5mm] (p1.center) -- (p4.center);
	\draw[line width=0.5mm] (p2.center) -- (p3.center);
	\draw[line width=0.15mm] (1.2,0) arc (0:180:0.7);
\end{tikzpicture}}, \quad \begin{tikzpicture}[baseline={([yshift=-0.1cm]current bounding box.center)}, scale=0.8] 
	\node[] (a) at (0,0) {};
	\node[] (a1) at (1,0) {};
	\node[] (b) at (0,-1) {};
	\node[] (b1) at (1,-1) {};
	\node[] (p1) at ($(a)+(-0.25,0)$) {};
	\node[] (p2) at ($(b)+(-0.25,0)$) {};
	\node[] (p3) at ($(b1)+(0.25,0)$) {};
	\node[] (p4) at ($(a1)+(0.25,0)$) {};
	\draw[line width=0.15mm] (b.center) -- (a.center);
	\draw[line width=0.15mm] (b1.center) -- (a1.center);
	\draw[line width=0.5mm] (p1.center) -- (p4.center);
	\draw[line width=0.5mm] (p2.center) -- (p3.center);
	\draw[line width=0.15mm] (0.8,0) arc (0:180:0.3);
\end{tikzpicture}, \quad \begin{tikzpicture}[baseline={([yshift=-0.1cm]current bounding box.center)}, scale=0.8] 
	\node[] (a) at (0,0) {};
	\node[] (a1) at (1,0) {};
	\node[] (b) at (0,-1) {};
	\node[] (b1) at (1,-1) {};
	\node[] (p1) at ($(a)+(-0.55,0)$) {};
	\node[] (p2) at ($(b)+(-0.55,0)$) {};
	\node[] (p3) at ($(b1)+(0.25,0)$) {};
	\node[] (p4) at ($(a1)+(0.25,0)$) {};
	\draw[line width=0.15mm] (b.center) -- (a.center);
	\draw[line width=0.15mm] (b1.center) -- (a1.center);
	\draw[line width=0.5mm] (p1.center) -- (p4.center);
	\draw[line width=0.5mm] (p2.center) -- (p3.center);
	\draw[line width=0.15mm] (0.3,0) arc (0:180:0.3);
\end{tikzpicture}, \quad \raisebox{0.17cm}{\begin{tikzpicture}[baseline={([yshift=-0.1cm]current bounding box.center)}, scale=0.8] 
	\node[] (b) at (0,-1) {};
	\node[] (b1) at (1,-1) {};
	\node[] (c) at (0,0) {};
	\node[] (c1) at (1,0) {};
	\node[] (p1) at ($(b)+(-0.65,0)$) {};
	\node[] (p2) at ($(c)+(-0.65,0)$) {};
	\node[] (p3) at ($(c1)+(0.25,0)$) {};
	\node[] (p4) at ($(b1)+(0.25,0)$) {};
	\draw[line width=0.15mm] (b.center) -- (c.center);
	\draw[line width=0.15mm] (b1.center) -- (c1.center);
	\draw[line width=0.5mm] (p1.center) -- (p4.center);
	\draw[line width=0.5mm] (p2.center) -- (p3.center);
	\draw[line width=0.15mm] (c1.center) arc (0:180:0.7);
\end{tikzpicture}}, \quad \raisebox{0.025cm}{\begin{tikzpicture}[baseline={([yshift=-0.1cm]current bounding box.center)}] 
	\node[] (a) at (0,0) {};
	\node[] (a1) at (0.8,0) {};
	\node[] (b) at (0,-0.8) {};
	\node[] (b1) at (0.8,-0.8) {};
	\node[] (p1) at ($(a)+(-0.2,0)$) {};
	\node[] (p2) at ($(b)+(-0.2,0)$) {};
	\node[] (p3) at ($(b1)+(0.2,0)$) {};
	\node[] (p4) at ($(a1)+(0.2,0)$) {};
	\draw[line width=0.15mm] (b.center) -- (a.center);
	\draw[line width=0.15mm] (b1.center) -- (a1.center);
	\draw[line width=0.5mm] (p1.center) -- (p4.center);
	\draw[line width=0.5mm] (p2.center) -- (p3.center);
	\draw[line width=0.15mm] (0.4,0) arc (0:180:0.2);
\end{tikzpicture}}, \\[0.1cm] &\raisebox{0.17cm}{\begin{tikzpicture}[baseline={([yshift=-0.1cm]current bounding box.center)}, scale=0.8] 
	\node[] (b) at (0.5,-1) {};
	\node[] (c) at (0,0) {};
	\node[] (c1) at (1,0) {};
	\node[] (p1) at ($(b)+(-0.95,0)$) {};
	\node[] (p2) at ($(c)+(-0.45,0)$) {};
	\node[] (p3) at ($(c1)+(0.45,0)$) {};
	\node[] (p4) at ($(b)+(0.95,0)$) {};
	\draw[line width=0.15mm] (b.center) -- (c.center);
	\draw[line width=0.15mm] (b.center) -- (c1.center);
	\draw[line width=0.5mm] (p1.center) -- (p4.center);
	\draw[line width=0.5mm] (p2.center) -- (p3.center);
	\draw[line width=0.15mm] (1.2,0) arc (0:180:0.7);
\end{tikzpicture}}, \quad \raisebox{0.025cm}{\begin{tikzpicture}[baseline={([yshift=-0.1cm]current bounding box.center)}, scale=0.8] 
	\node[] (b) at (0.5,-1) {};
	\node[] (c) at (0,0) {};
	\node[] (c1) at (1,0) {};
	\node[] (p1) at ($(b)+(-0.75,0)$) {};
	\node[] (p2) at ($(c)+(-0.25,0)$) {};
	\node[] (p3) at ($(c1)+(0.25,0)$) {};
	\node[] (p4) at ($(b)+(0.75,0)$) {};
	\draw[line width=0.15mm] (b.center) -- (c.center);
	\draw[line width=0.15mm] (b.center) -- (c1.center);
	\draw[line width=0.5mm] (p1.center) -- (p4.center);
	\draw[line width=0.5mm] (p2.center) -- (p3.center);
	\draw[line width=0.15mm] (0.8,0) arc (0:180:0.3);
\end{tikzpicture}}, \quad \raisebox{0.025cm}{\begin{tikzpicture}[baseline={([yshift=-0.1cm]current bounding box.center)}, scale=0.8] 
	\node[] (b) at (0.5,-1) {};
	\node[] (c) at (0,0) {};
	\node[] (c1) at (1,0) {};
	\node[] (p1) at ($(b)+(-1.05,0)$) {};
	\node[] (p2) at ($(c)+(-0.55,0)$) {};
	\node[] (p3) at ($(c1)+(0.25,0)$) {};
	\node[] (p4) at ($(b)+(0.75,0)$) {};
	\draw[line width=0.15mm] (b.center) -- (c.center);
	\draw[line width=0.15mm] (b.center) -- (c1.center);
	\draw[line width=0.5mm] (p1.center) -- (p4.center);
	\draw[line width=0.5mm] (p2.center) -- (p3.center);
	\draw[line width=0.15mm] (0.3,0) arc (0:180:0.3);
\end{tikzpicture}}, \quad \raisebox{0.17cm}{\begin{tikzpicture}[baseline={([yshift=-0.1cm]current bounding box.center)}, scale=0.8] 
	\node[] (b) at (0.5,-1) {};
	\node[] (c) at (0,0) {};
	\node[] (c1) at (1,0) {};
	\node[] (p1) at ($(b)+(-1.15,0)$) {};
	\node[] (p2) at ($(c)+(-0.65,0)$) {};
	\node[] (p3) at ($(c1)+(0.25,0)$) {};
	\node[] (p4) at ($(b)+(0.75,0)$) {};
	\draw[line width=0.15mm] (b.center) -- (c.center);
	\draw[line width=0.15mm] (b.center) -- (c1.center);
	\draw[line width=0.5mm] (p1.center) -- (p4.center);
	\draw[line width=0.5mm] (p2.center) -- (p3.center);
	\draw[line width=0.15mm] (c1.center) arc (0:180:0.7);
\end{tikzpicture}}, \quad \raisebox{0.025cm}{\begin{tikzpicture}[baseline={([yshift=-0.1cm]current bounding box.center)}] 
	\node[] (a) at (0,0) {};
	\node[] (a1) at (0.8,0) {};
	\node[] (b) at (0,-0.8) {};
	\node[] (b1) at (0.4,-0.8) {};
	\node[] (b2) at (0.8,-0.8) {};
	\node[] (p1) at ($(a)+(-0.2,0)$) {};
	\node[] (p2) at ($(b)+(-0.2,0)$) {};
	\node[] (p3) at ($(b2)+(0.2,0)$) {};
	\node[] (p4) at ($(a1)+(0.2,0)$) {};
	\draw[line width=0.15mm] (b1.center) -- (a.center);
	\draw[line width=0.15mm] (b1.center) -- (a1.center);
	\draw[line width=0.5mm] (p1.center) -- (p4.center);
	\draw[line width=0.5mm] (p2.center) -- (p3.center);
	\draw[line width=0.15mm] (0.4,0) arc (0:180:0.2);
\end{tikzpicture}}.
\end{align}

Additionally, there are three diagrams that contain triangles on a matter line, where the graviton self-interaction is cubic and at least one other vertex of the triangle is cubic:
\begin{equation}
\begin{tikzpicture}[baseline={([yshift=-0.1cm]current bounding box.center)}] 
	\node[] (a) at (0,0) {};
	\node[] (a1) at (0.9,0) {};
	\node[] (a2) at (1.2,0) {};
	\node[] (b) at (0.45,-0.6) {};
	\node[] (c) at (0.45,-1) {};
	\node[] (c1) at (1.2,-1) {};
	\node[] (p1) at ($(a)+(-0.2,0)$) {};
	\node[] (p2) at ($(c)+(-0.65,0)$) {};
	\node[] (p3) at ($(c1)+(0.2,0)$) {};
	\node[] (p4) at ($(a2)+(0.2,0)$) {};
	\draw[line width=0.15mm] (b.center) -- (a.center);
	\draw[line width=0.15mm] (b.center) -- (a1.center);
	\draw[line width=0.15mm] (b.center) -- (c.center);
	\draw[line width=0.15mm] (a2.center) -- (c1.center);
	\draw[line width=0.5mm] (p1.center) -- (p4.center);
	\draw[line width=0.5mm] (p2.center) -- (p3.center);
\end{tikzpicture}, \quad \begin{tikzpicture}[baseline={([yshift=-0.1cm]current bounding box.center)}] 
	\node[] (a) at (0,0) {};
	\node[] (a1) at (0.75,0) {};
	\node[] (a2) at (1.2,0) {};
	\node[] (b) at (0.38,-0.5) {};
	\node[] (d) at (0.6,-0.8) {};
	\node[] (p1) at ($(a)+(-0.2,0)$) {};
	\node[] (p2) at ($(d)+(-0.8,0)$) {};
	\node[] (p3) at ($(d)+(0.8,0)$) {};
	\node[] (p4) at ($(a2)+(0.2,0)$) {};
	\draw[line width=0.15mm] (d.center) -- (a.center);
	\draw[line width=0.15mm] (b.center) -- (a1.center);
	\draw[line width=0.15mm] (d.center) -- (a2.center);
	\draw[line width=0.5mm] (p1.center) -- (p4.center);
	\draw[line width=0.5mm] (p2.center) -- (p3.center);
\end{tikzpicture}, \quad
\begin{tikzpicture}[baseline={([yshift=-0.1cm]current bounding box.center)}] 
	\node[] (a) at (0,0) {};
	\node[] (b) at (0.5,-0.5) {};
	\node[] (a1) at (0.5,0) {};
	\node[] (c) at (0,-1) {};
	\node[] (c1) at (0.5,-1) {};
	\node[] (p1) at ($(a)+(-0.2,0)$) {};
	\node[] (p2) at ($(c)+(-0.2,0)$) {};
	\node[] (p3) at ($(c1)+(0.2,0)$) {};
	\node[] (p4) at ($(a1)+(0.2,0)$) {};
	\draw[line width=0.15mm] (b.center) -- (a1.center);
	\draw[line width=0.15mm] (b.center) -- (a.center);
	\draw[line width=0.15mm] (a.center) -- (c.center);
	\draw[line width=0.15mm] (c1.center) -- (b.center);
	\draw[line width=0.5mm] (p1.center) -- (p4.center);
	\draw[line width=0.5mm] (p2.center) -- (p3.center);
\end{tikzpicture}.
\end{equation}
As explained in sec.~\ref{sec:reduction_relations} and app.~\ref{app: IBPs_bubbles_triangles}, they have zero master integrals in their sector; thus, they are fully expressible in terms of subsectors and can be dropped from the analysis.

Furthermore, using the reduction of box iterations \eqref{eq: reduction_superclassical}  we immediately have
\begin{align}
\LS \left( \begin{tikzpicture}[baseline={([yshift=-0.1cm]current bounding box.center)}] 
	\node[] (a) at (0,0) {};
	\node[] (a1) at (0.8,0) {};
	\node[] (a2) at (1.6,0) {};
	\node[] (b) at (0,-0.8) {};
	\node[] (b1) at (0.8,-0.8) {};
	\node[] (b2) at (1.6,-0.8) {};
	\node[] (p1) at ($(a)+(-0.2,0)$) {};
	\node[] (p2) at ($(b)+(-0.2,0)$) {};
	\node[] (p3) at ($(b2)+(0.2,0)$) {};
	\node[] (p4) at ($(a2)+(0.2,0)$) {};
	\draw[line width=0.15mm] (b.center) -- (a.center);
	\draw[line width=0.15mm] (b1.center) -- (a1.center);
	\draw[line width=0.15mm] (b2.center) -- (a2.center);
	\draw[line width=0.5mm] (p1.center) -- (p4.center);
	\draw[line width=0.5mm] (p2.center) -- (p3.center);
\end{tikzpicture} \right) & \propto \frac{x}{x^2-1} \, \LS \left( \begin{tikzpicture}[baseline={([yshift=-0.1cm]current bounding box.center)}] 
	\node[] (a) at (0,0) {};
	\node[] (a1) at (0.8,0) {};
	\node[] (b) at (0,-0.8) {};
	\node[] (b1) at (0.8,-0.8) {};
	\node[] (p1) at ($(a)+(-0.2,0)$) {};
	\node[] (p2) at ($(b)+(-0.2,0)$) {};
	\node[] (p3) at ($(b1)+(0.2,0)$) {};
	\node[] (p4) at ($(a1)+(0.2,0)$) {};
	\draw[line width=0.15mm] (b.center) -- (a.center);
	\draw[line width=0.15mm] (b1.center) -- (a1.center);
	\draw[line width=0.5mm] (p1.center) -- (p4.center);
	\draw[line width=0.5mm] (p2.center) -- (p3.center);
\end{tikzpicture} \right) \propto \frac{x^2}{q^2 (x^2-1)^2}, \\[0.1cm]
\LS \left( \begin{tikzpicture}[baseline={([yshift=-0.1cm]current bounding box.center)}] 
	\node[] (a) at (0,0) {};
	\node[] (a1) at (0.8,0) {};
	\node[] (a2) at (1.2,0) {};
	\node[] (b) at (0.4,-0.8) {};
	\node[] (b1) at (1.2,-0.8) {};
	\node[] (p1) at ($(a)+(-0.2,0)$) {};
	\node[] (p2) at ($(b)+(-0.6,0)$) {};
	\node[] (p3) at ($(b1)+(0.2,0)$) {};
	\node[] (p4) at ($(a1)+(0.6,0)$) {};
	\draw[line width=0.15mm] (b.center) -- (a.center);
	\draw[line width=0.15mm] (b.center) -- (a1.center);
		\draw[line width=0.15mm] (b1.center) -- (a2.center);
	\draw[line width=0.5mm] (p1.center) -- (p4.center);
	\draw[line width=0.5mm] (p2.center) -- (p3.center);
\end{tikzpicture} \right) & \propto \frac{x}{x^2-1} \, \LS \left( \begin{tikzpicture}[baseline={([yshift=-0.1cm]current bounding box.center)}] 
	\node[] (a) at (0,0) {};
	\node[] (a1) at (0.8,0) {};
	\node[] (b) at (0.4,-0.8) {};
	\node[] (p1) at ($(a)+(-0.2,0)$) {};
	\node[] (p2) at ($(b)+(-0.6,0)$) {};
	\node[] (p3) at ($(b)+(0.6,0)$) {};
	\node[] (p4) at ($(a1)+(0.2,0)$) {};
	\draw[line width=0.15mm] (b.center) -- (a.center);
	\draw[line width=0.15mm] (b.center) -- (a1.center);
	\draw[line width=0.5mm] (p1.center) -- (p4.center);
	\draw[line width=0.5mm] (p2.center) -- (p3.center);
\end{tikzpicture} \right) \propto \frac{x}{|q| \, (x^2-1)},
\end{align}
which is consistent with refs.~\cite{Parra-Martinez:2020dzs,Herrmann:2021tct,DiVecchia:2021bdo}, while the opposite-parity geometries can also be related to those of lower sectors using eqs.~\eqref{eq: reduction_superclassical}--\eqref{eq: reduction_superclassical_dot_vanish}.

\begin{figure}[t]
\centering
\parbox{\textwidth}{\begin{equation*}
\begin{tikzpicture}[baseline={([yshift=-0.1cm]current bounding box.center)}] 
	\node[] (a) at (0,0) {};
	\node[] (a1) at (0.4,0) {};
	\node[] (a2) at (0.8,0) {};
	\node[] (b) at (0.4,-0.8) {};
	\node[] (p1) at ($(a)+(-0.2,0)$) {};
	\node[] (p2) at ($(b)+(-0.6,0)$) {};
	\node[] (p3) at ($(b)+(0.6,0)$) {};
	\node[] (p4) at ($(a2)+(0.2,0)$) {};
	\draw[line width=0.15mm] (b.center) -- (a.center);
	\draw[line width=0.15mm] (b.center) -- (a1.center);
	\draw[line width=0.15mm] (b.center) -- (a2.center);
	\draw[line width=0.5mm] (p1.center) -- (p4.center);
	\draw[line width=0.5mm] (p2.center) -- (p3.center);
\end{tikzpicture} \qquad \begin{tikzpicture}[baseline={([yshift=-0.1cm]current bounding box.center)}] 
	\node[] (a) at (0,0) {};
	\node[] (a1) at (0.8,0) {};
	\node[] (b) at (0.4,-0.8) {};
	\node[] (b1) at (1.2,-0.8) {};
	\node[] (p1) at ($(a)+(-0.2,0)$) {};
	\node[] (p2) at ($(b)+(-0.6,0)$) {};
	\node[] (p3) at ($(b1)+(0.2,0)$) {};
	\node[] (p4) at ($(a1)+(0.6,0)$) {};
	\draw[line width=0.15mm] (b.center) -- (a.center);
	\draw[line width=0.15mm] (b.center) -- (a1.center);
		\draw[line width=0.15mm] (b1.center) -- (a1.center);
	\draw[line width=0.5mm] (p1.center) -- (p4.center);
	\draw[line width=0.5mm] (p2.center) -- (p3.center);
\end{tikzpicture} \qquad \raisebox{0.05cm}{\begin{tikzpicture}[baseline={([yshift=-0.1cm]current bounding box.center)}] 
	\node[] (a) at (0,0) {};
	\node[] (a1) at (0.8,0) {};
	\node[] (b) at (0,-0.8) {};
	\node[] (b1) at (0.8,-0.8) {};
	\node[] (p1) at ($(a)+(-0.2,0)$) {};
	\node[] (p2) at ($(b)+(-0.2,0)$) {};
	\node[] (p3) at ($(b1)+(0.2,0)$) {};
	\node[] (p4) at ($(a1)+(0.2,0)$) {};
	\draw[line width=0.15mm] (b.center) -- (a.center);
	\draw[line width=0.15mm] (b1.center) -- (a1.center);
	\draw[line width=0.5mm] (p1.center) -- (p4.center);
	\draw[line width=0.5mm] (p2.center) -- (p3.center);
	\draw[line width=0.15mm] (0.8,0) arc (0:180:0.4);
\end{tikzpicture}} \qquad \begin{tikzpicture}[baseline={([yshift=-0.1cm]current bounding box.center)}] 
	\node[] (a) at (0,0) {};
	\node[] (b) at (0,-0.5) {};
	\node[] (a1) at (0.5,0) {};
	\node[] (b1) at (0.5,-0.5) {};
	\node[] (c) at (0,-1) {};
	\node[] (c1) at (0.5,-1) {};
	\node[] (p1) at ($(a)+(-0.2,0)$) {};
	\node[] (p2) at ($(c)+(-0.2,0)$) {};
	\node[] (p3) at ($(c1)+(0.2,0)$) {};
	\node[] (p4) at ($(a1)+(0.2,0)$) {};
	\draw[line width=0.15mm] (a.center) -- (b.center);
	\draw[line width=0.15mm] (a1.center) -- (b1.center);
	\draw[line width=0.15mm] (b1.center) -- (b.center);
	\draw[line width=0.15mm] (b.center) -- (c.center);
	\draw[line width=0.15mm] (c1.center) -- (b1.center);
	\draw[line width=0.5mm] (a.center) -- (a1.center);
	\draw[line width=0.5mm] (c.center) -- (c1.center);
	\draw[line width=0.5mm] (p1.center) -- (a.center);
	\draw[line width=0.5mm] (a1.center) -- (p4.center);
	\draw[line width=0.5mm] (p2.center) -- (c.center);
	\draw[line width=0.5mm] (c1.center) -- (p3.center);
\end{tikzpicture}
\end{equation*}}
\caption{The 4 remaining two-loop diagrams after all diagrams with leading singularities that are trivial to calculate or that can be related to lower sectors have been discarded.}
\label{fig: diagrams_indep_two_loop}
\end{figure}
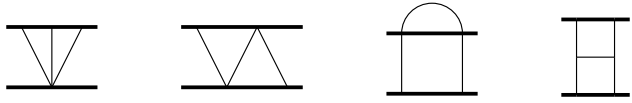
As we show in tab.~\ref{tab:indep_diagrams}, out of the 23 initial non-trivial topologies at two loops, we only need to consider the 4 diagrams of fig.~\ref{fig: diagrams_indep_two_loop}, which already exemplifies the usefulness of the identities among leading singularities gathered in sec.~\ref{sec:reduction_relations}. In sec.~\ref{sec:two_loop_abelian}, we will focus on the first three diagrams, which do not contain graviton self-interactions -- also referred to as Abelian diagrams. We will similarly study the remaining non-Abelian diagram in sec.~\ref{sec:two_loop_non-abelian}.

\subsection{Two-loop Abelian diagrams}
\label{sec:two_loop_abelian}

Let us start analyzing the Abelian subsectors at two loops, that is, the first three diagrams in fig.~\ref{fig: diagrams_indep_two_loop}.

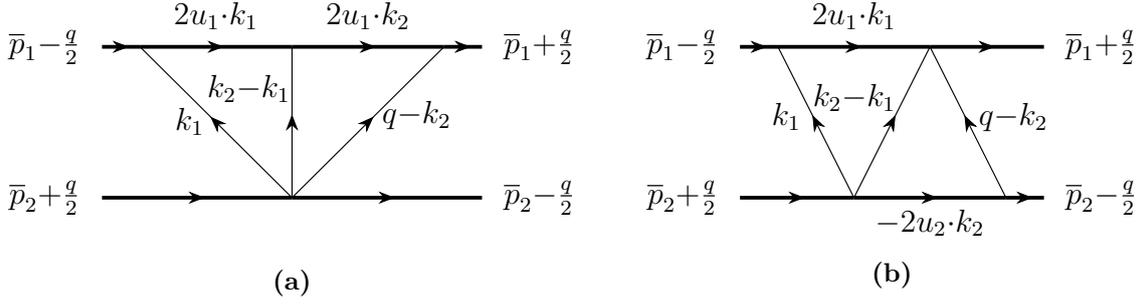
\begin{figure}[t]
\centering
\subfloat[]{\begin{tikzpicture}[baseline=(current bounding box.center)] 
	\node[] (a) at (0,0) {};
	\node[] (a1) at (2,0) {};
	\node[] (b) at (2,-2) {};
	\node[] (a2) at (4,0) {};
	\node[label=left:{$\overline{p}_1{-}\frac{q}{2}$}] (p1) at ($(a)+(-0.5,0)$) {};
	\node[label=left:{$\overline{p}_2{+}\frac{q}{2}$}] (p2) at ($(b)+(-2.5,0)$) {};
	\node[label=right:{$\overline{p}_2{-}\frac{q}{2}$}] (p3) at ($(b)+(2.5,0)$) {};
	\node[label=right:{$\overline{p}_1{+}\frac{q}{2}$}] (p4) at ($(a2)+(0.5,0)$) {};
	\draw[line width=0.15mm, postaction={decorate}] (b.center) -- node[sloped, allow upside down, label={[xshift=0.15cm, yshift=0.5cm]$k_1$}] {\midarrow} (a.center);
	\draw[line width=0.15mm, postaction={decorate}] (b.center) -- node[sloped, allow upside down, label={[xshift=0.35cm, yshift=0.45cm]$k_2{-}k_1$}] {\midarrow} (a1.center);
		\draw[line width=0.15mm, postaction={decorate}] (b.center) -- node[sloped, allow upside down, label={[xshift=1.4cm, yshift=-0.4cm]$q{-}k_2$}] {\midarrow} (a2.center);
	\draw[line width=0.5mm, postaction={decorate}] (a.center) -- node[sloped, allow upside down, label={[xshift=0cm, yshift=-0.15cm]$2u_1{\cdot} k_1$}] {\midarrow} (a1.center);
		\draw[line width=0.5mm, postaction={decorate}] (a1.center) -- node[sloped, allow upside down, label={[xshift=0cm, yshift=-0.15cm]$2u_1{\cdot} k_2$}] {\midarrow} (a2.center);
		\node[label={[xshift=0cm, yshift=-0.9cm]\textcolor{white}{$2u_2{\cdot} k$}}] (lwhite) at (2,-2) {};
	\draw[line width=0.5mm, postaction={decorate}] (p1.center) -- node[sloped, allow upside down] {\midarrow} (a.center);
	\draw[line width=0.5mm, postaction={decorate}] (a2.center) -- node[sloped, allow upside down] {\midarrow} (p4.center);
	\draw[line width=0.5mm, postaction={decorate}] (p2.center) -- node[sloped, allow upside down] {\midarrow} (b.center);
	\draw[line width=0.5mm, postaction={decorate}] (b.center) -- node[sloped, allow upside down] {\midarrow} (p3.center);
\end{tikzpicture}} \hfill \subfloat[]{\raisebox{0.05cm}{\begin{tikzpicture}[baseline=(current bounding box.center)] 
	\node[] (a) at (0,0) {};
	\node[] (a1) at (2,0) {};
	\node[] (b) at (1,-2) {};
	\node[] (b1) at (3,-2) {};
	\node[label=left:{$\overline{p}_1{-}\frac{q}{2}$}] (p1) at ($(a)+(-0.5,0)$) {};
	\node[label=left:{$\overline{p}_2{+}\frac{q}{2}$}] (p2) at ($(b)+(-1.5,0)$) {};
	\node[label=right:{$\overline{p}_2{-}\frac{q}{2}$}] (p3) at ($(b1)+(0.5,0)$) {};
	\node[label=right:{$\overline{p}_1{+}\frac{q}{2}$}] (p4) at ($(a1)+(1.5,0)$) {};
	\draw[line width=0.15mm, postaction={decorate}] (b.center) -- node[sloped, allow upside down, label={[xshift=0.15cm, yshift=0.5cm]$k_1$}] {\midarrow} (a.center);
	\draw[line width=0.15mm, postaction={decorate}] (b.center) -- node[sloped, allow upside down, label={[xshift=0.4cm, yshift=-0.1cm]$k_2{-}k_1$}] {\midarrow} (a1.center);
		\draw[line width=0.15mm, postaction={decorate}] (b1.center) -- node[sloped, allow upside down, label={[xshift=1.4cm, yshift=0.5cm]$q{-}k_2$}] {\midarrow} (a1.center);
	\draw[line width=0.5mm, postaction={decorate}] (a.center) -- node[sloped, allow upside down, label={[xshift=0cm, yshift=-0.15cm]$2u_1{\cdot} k_1$}] {\midarrow} (a1.center);
		\draw[line width=0.5mm, postaction={decorate}] (b.center) -- node[sloped, allow upside down, label={[xshift=0cm, yshift=-0.9cm]$-2u_2{\cdot} k_2$}] {\midarrow} (b1.center);
	\draw[line width=0.5mm, postaction={decorate}] (p1.center) -- node[sloped, allow upside down] {\midarrow} (a.center);
	\draw[line width=0.5mm, postaction={decorate}] (a1.center) -- node[sloped, allow upside down] {\midarrow} (p4.center);
	\draw[line width=0.5mm, postaction={decorate}] (p2.center) -- node[sloped, allow upside down] {\midarrow} (b.center);
	\draw[line width=0.5mm, postaction={decorate}] (b1.center) -- node[sloped, allow upside down] {\midarrow} (p3.center);
\end{tikzpicture}}}
\caption{Parametrization of the loop momenta for the double-triangle diagram (a), and the two-loop zig-zag diagram (b).}
\label{fig: params_double_triangle_and_zigzag}
\end{figure}

The first Abelian subsector is the double-triangle diagram, parametrized as in fig.~\ref{fig: params_double_triangle_and_zigzag}(a). It is the first example where we use the loop-by-loop Baikov representation of eq.~\eqref{eq: standard_Baikov_ISP}, in this case in the integration order $k_1 \to k_2$  to avoid mixing $k_1$ with $q$, which would generate more ISPs. With this parametrization, it has $\nISP = L+\sum_{i=1}^L E_i - \nint = 1$, which requires adding only one extra Baikov variable $z_6=k_2^2$. This can be observed from the Baikov polynomial $\det G(u_1,k_2)$ for the $k_1$-loop, which introduces the scalar products $k_2^2$ and $u_1 \cdot k_2$, where only the latter can be identified with a propagator in the diagram. In practice, this parametrization amounts to recycling the one-loop triangle as follows:
\begin{equation}
\begin{tikzpicture}[baseline={([yshift=-0.1cm]current bounding box.center)}] 
	\node[] (a) at (0,0) {};
	\node[] (a1) at (0.4,0) {};
	\node[] (a2) at (0.8,0) {};
	\node[] (b) at (0.4,-0.8) {};
	\node[] (p1) at ($(a)+(-0.2,0)$) {};
	\node[] (p2) at ($(b)+(-0.6,0)$) {};
	\node[] (p3) at ($(b)+(0.6,0)$) {};
	\node[] (p4) at ($(a2)+(0.2,0)$) {};
	\draw[line width=0.15mm] (b.center) -- (a.center);
	\draw[line width=0.15mm] (b.center) -- (a1.center);
	\draw[line width=0.15mm] (b.center) -- (a2.center);
	\draw[line width=0.5mm] (p1.center) -- (p4.center);
	\draw[line width=0.5mm] (p2.center) -- (p3.center);
\end{tikzpicture} = \begin{tikzpicture}[baseline={([yshift=-0.1cm]current bounding box.center)}] 
	\node[] (a) at (0,0) {};
	\node[] (a1) at (0.8,0) {};
	\node[] (b) at (0.4,-0.8) {};
	\node[] (p1) at ($(a)+(-0.2,0)$) {};
	\node[] (p2) at ($(b)+(-0.6,0)$) {};
	\node[] (p3) at ($(b)+(0.6,0)$) {};
	\node[] (p4) at ($(a1)+(0.2,0)$) {};
	\draw[line width=0.15mm] (b.center) -- (a.center);
	\draw[line width=0.15mm] (a1.center) -- (b.center);
	\draw[line width=0.15mm,-{Latex[length=2.2mm]}](1.1,-0.75) -- (1.1,-0.05);
	\node[label={[xshift=0.35cm, yshift=-0.45cm]$k_2$}] (k) at (1.1,-0.4) {};
	\draw[line width=0.5mm] (p1.center) -- (p4.center);
	\draw[line width=0.5mm] (p2.center) -- (p3.center);
\end{tikzpicture} \times
\begin{tikzpicture}[baseline={([yshift=-0.1cm]current bounding box.center)}] 
	\node[] (a) at (0,0) {};
	\node[] (a1) at (0.8,0) {};
	\node[] (b) at (0.4,-0.8) {};
	\node[] (p1) at ($(a)+(-0.2,0)$) {};
	\node[] (p2) at ($(b)+(-0.6,0)$) {};
	\node[] (p3) at ($(b)+(0.6,0)$) {};
	\node[] (p4) at ($(a1)+(0.2,0)$) {};
	\draw[line width=0.15mm, dashed, postaction={decorate}] (b.center) -- node[sloped, allow upside down, label={[xshift=0.2cm, yshift=0.45cm]$k_2$}] {\midarrow} (a.center);
	\draw[line width=0.15mm] (b.center) -- (a1.center);
	\draw[line width=0.5mm] (p1.center) -- (p4.center);
	\draw[line width=0.5mm] (p2.center) -- (p3.center);
\end{tikzpicture} \propto \int \frac{d z_1 \cdots d z_5}{z_1 \cdots z_5} \frac{d z_6}{\sqrt{\det G(k_2,u_1)} \sqrt{\det G(u_1,q)}},
\label{eq: double_triangle_Baikov_intermediate}
\end{equation}
where we can first compute the $k_1$-triangle as in eq.~\eqref{eq: Baikov_triangle}, but here with momentum transfer $k_2$, and afterwards calculate the contribution from the $k_2$-loop, where we indicate the presence of an ISP by a dashed propagator. Note that two further loop-momentum-dependent Baikov polynomials, namely $\det G(k_1,k_2,u_1)$ and $\det G(k_2,u_1,q)$, occur with vanishing exponent in $D=4$.

The first Gram determinant in eq.~\eqref{eq: double_triangle_Baikov_intermediate} becomes a quadratic polynomial 
\begin{equation}
P_2(z_5,z_6) = z_6 - \frac{z_5^2}{4}
\end{equation}
in the Baikov variables, where $z_5=2u_1 \cdot k_2$, while the second Gram determinant is independent of the loop momenta. Taking the maximal cut, only the integral over the extra variable $z_6$ remains:
\begin{equation}
\LS \left( \begin{tikzpicture}[baseline={([yshift=-0.1cm]current bounding box.center)}] 
	\node[] (a) at (0,0) {};
	\node[] (a1) at (0.4,0) {};
	\node[] (a2) at (0.8,0) {};
	\node[] (b) at (0.4,-0.8) {};
	\node[] (p1) at ($(a)+(-0.2,0)$) {};
	\node[] (p2) at ($(b)+(-0.6,0)$) {};
	\node[] (p3) at ($(b)+(0.6,0)$) {};
	\node[] (p4) at ($(a2)+(0.2,0)$) {};
	\draw[line width=0.15mm] (b.center) -- (a.center);
	\draw[line width=0.15mm] (b.center) -- (a1.center);
	\draw[line width=0.15mm] (b.center) -- (a2.center);
	\draw[line width=0.5mm] (p1.center) -- (p4.center);
	\draw[line width=0.5mm] (p2.center) -- (p3.center);
\end{tikzpicture} \right) \propto \frac{1}{|q|} \, \LS \left( \int \frac{d z_6}{\sqrt{P_2(0,z_6)}} \right) =  \frac{1}{|q|} \, \LS \left( \int \frac{d z_6}{\sqrt{z_6}} \right).
\label{eq: LS_double_triangle_inf}
\end{equation}
First of all, it would seem as if the scaling is not correct, since we expect $|q|^0$; but recall that $z_6 =k_2^2 \sim |q|^2$, which restores the correct scaling. Secondly, the integral contains a higher-order pole. To see this, we can change variables to $z_6=1/t^2$,
\begin{equation}
\LS \left( \begin{tikzpicture}[baseline={([yshift=-0.1cm]current bounding box.center)}] 
	\node[] (a) at (0,0) {};
	\node[] (a1) at (0.4,0) {};
	\node[] (a2) at (0.8,0) {};
	\node[] (b) at (0.4,-0.8) {};
	\node[] (p1) at ($(a)+(-0.2,0)$) {};
	\node[] (p2) at ($(b)+(-0.6,0)$) {};
	\node[] (p3) at ($(b)+(0.6,0)$) {};
	\node[] (p4) at ($(a2)+(0.2,0)$) {};
	\draw[line width=0.15mm] (b.center) -- (a.center);
	\draw[line width=0.15mm] (b.center) -- (a1.center);
	\draw[line width=0.15mm] (b.center) -- (a2.center);
	\draw[line width=0.5mm] (p1.center) -- (p4.center);
	\draw[line width=0.5mm] (p2.center) -- (p3.center);
\end{tikzpicture} \right) \propto \frac{1}{|q|} \, \LS \left( \int \frac{d t}{t^2} \right).
\end{equation}
The same result is obtained using the standard Baikov representation~\eqref{eq: standard_Baikov}.

Following the discussion in sec.~\ref{sec:methodology_systematics}, we instead integrate the leading singularity along a more non-trivial limit of the Baikov contour $\mathfrak{C}$, specified by the roots of the Baikov polynomials,
\begin{equation}
\begin{aligned}
\det G(k_1,k_2,u_1)\Big|_{z_1=\dots=z_5=0}& =-\frac{z_6^2}{4}, \qquad \qquad \enspace \ \det G(k_2,u_1)\Big|_{z_1=\dots=z_5=0}=z_6, \\[0.1cm]
\det G(k_2,u_1,q)\Big|_{z_1=\dots=z_5=0}& =-\frac{(z_6-q^2)^2}{4}, \qquad \det G(u_1,q)\Big|_{z_1=\dots=z_5=0}=q^2.
\end{aligned}
\end{equation}
In this case, we find the roots $z_6=0$ and $z_6=q^2$; thus, we integrate from one to the other, leading to
\begin{equation}
\LS \left( \begin{tikzpicture}[baseline={([yshift=-0.1cm]current bounding box.center)}] 
	\node[] (a) at (0,0) {};
	\node[] (a1) at (0.4,0) {};
	\node[] (a2) at (0.8,0) {};
	\node[] (b) at (0.4,-0.8) {};
	\node[] (p1) at ($(a)+(-0.2,0)$) {};
	\node[] (p2) at ($(b)+(-0.6,0)$) {};
	\node[] (p3) at ($(b)+(0.6,0)$) {};
	\node[] (p4) at ($(a2)+(0.2,0)$) {};
	\draw[line width=0.15mm] (b.center) -- (a.center);
	\draw[line width=0.15mm] (b.center) -- (a1.center);
	\draw[line width=0.15mm] (b.center) -- (a2.center);
	\draw[line width=0.5mm] (p1.center) -- (p4.center);
	\draw[line width=0.5mm] (p2.center) -- (p3.center);
\end{tikzpicture} \right) \propto \frac{1}{|q|} \, \int_{q^2}^0 \frac{d z_6}{\sqrt{z_6}} \propto 1.
\label{eq: LS_double_triangle_u1q0}
\end{equation}
Since the result is a number, we could not have taken a further discontinuity in eq.~\eqref{eq: LS_double_triangle_inf}, thus confirming that we should integrate instead the leading singularity along a more non-trivial limit of the Baikov contour $\mathfrak{C}$. Therefore, we have a unit leading singularity, in agreement with refs.~\cite{Parra-Martinez:2020dzs,Herrmann:2021tct,DiVecchia:2021bdo}. 

Relaxing the condition that $u_i \cdot q=0$, which becomes relevant when the double-triangle diagram appears as a sub-loop, we find instead
\begin{align}
\LS \left( \begin{tikzpicture}[baseline={([yshift=-0.1cm]current bounding box.center)}] 
	\node[] (a) at (0,0) {};
	\node[] (a1) at (0.4,0) {};
	\node[] (a2) at (0.8,0) {};
	\node[] (b) at (0.4,-0.8) {};
	\node[] (p1) at ($(a)+(-0.2,0)$) {};
	\node[] (p2) at ($(b)+(-0.6,0)$) {};
	\node[] (p3) at ($(b)+(0.6,0)$) {};
	\node[] (p4) at ($(a2)+(0.2,0)$) {};
	\draw[line width=0.15mm] (b.center) -- (a.center);
	\draw[line width=0.15mm] (b.center) -- (a1.center);
	\draw[line width=0.15mm] (b.center) -- (a2.center);
	\draw[line width=0.5mm] (p1.center) -- (p4.center);
	\draw[line width=0.5mm] (p2.center) -- (p3.center);
\end{tikzpicture} \right) & \propto \frac{1}{\sqrt{q^2-(u_1 \cdot q)^2}} \, \int_{-\big( (u_1 \cdot q) + \sqrt{(u_1 \cdot q)^2-q^2} \big)^2}^{0} \, \frac{d z_6}{\sqrt{z_6}} \nonumber \\[0.2cm]
& \propto \frac{(u_1 \cdot q) + \sqrt{(u_1 \cdot q)^2-q^2}}{\sqrt{q^2-(u_1 \cdot q)^2}},
\label{eq: LS_double_triangle_u1q_general}
\end{align}
which yields a unit leading singularity if either $u_1 \cdot q=0$ or $q^2=0$. This means that for diagrams containing a double-triangle sub-loop where the respective momentum transfer $k$ satisfies $u_1 \cdot k=0$ or $k^2=0$ at the maximal cut, the resulting leading singularity greatly simplifies.

Regarding the integrals of odd parity, we can add a dot to one of the matter propagators:
\begin{align}
\LS \left( \begin{tikzpicture}[baseline={([yshift=-0.1cm]current bounding box.center)}] 
	\node[] (a) at (0,0) {};
	\node[] (a1) at (0.4,0) {};
	\node[] (a2) at (0.8,0) {};
	\node[] (b) at (0.4,-0.8) {};
	\node[] (p1) at ($(a)+(-0.2,0)$) {};
	\node[] (p2) at ($(b)+(-0.6,0)$) {};
	\node[] (p3) at ($(b)+(0.6,0)$) {};
	\node[] (p4) at ($(a2)+(0.2,0)$) {};
	\draw[line width=0.15mm] (b.center) -- (a.center);
	\draw[line width=0.15mm] (b.center) -- (a1.center);
	\draw[line width=0.15mm] (b.center) -- (a2.center);
	\draw[line width=0.5mm] (p1.center) -- (p4.center);
	\draw[line width=0.5mm] (p2.center) -- (p3.center);
	\node at (0.2,0) [circle,fill,inner sep=1.5pt]{};
\end{tikzpicture} \right) & = \LS \left(
\begin{tikzpicture}[baseline={([yshift=-0.1cm]current bounding box.center)}] 
	\node[] (a) at (0,0) {};
	\node[] (a1) at (0.8,0) {};
	\node[] (b) at (0.4,-0.8) {};
	\node[] (p1) at ($(a)+(-0.2,0)$) {};
	\node[] (p2) at ($(b)+(-0.6,0)$) {};
	\node[] (p3) at ($(b)+(0.6,0)$) {};
	\node[] (p4) at ($(a1)+(0.2,0)$) {};
	\draw[line width=0.15mm] (b.center) -- (a.center);
	\draw[line width=0.15mm] (b.center) -- (a1.center);
	\draw[line width=0.15mm,-{Latex[length=2.2mm]}](1.1,-0.75) -- (1.1,-0.05);
	\node[label={[xshift=0.35cm, yshift=-0.45cm]$k_2$}] (k) at (1.1,-0.4) {};
	\draw[line width=0.5mm] (p1.center) -- (p4.center);
	\draw[line width=0.5mm] (p2.center) -- (p3.center);
	\node at (0.4,0) [circle,fill,inner sep=1.5pt]{};
\end{tikzpicture} \times
\begin{tikzpicture}[baseline={([yshift=-0.1cm]current bounding box.center)}] 
	\node[] (a) at (0,0) {};
	\node[] (a1) at (0.8,0) {};
	\node[] (b) at (0.4,-0.8) {};
	\node[] (p1) at ($(a)+(-0.2,0)$) {};
	\node[] (p2) at ($(b)+(-0.6,0)$) {};
	\node[] (p3) at ($(b)+(0.6,0)$) {};
	\node[] (p4) at ($(a1)+(0.2,0)$) {};
	\draw[line width=0.15mm, dashed, postaction={decorate}] (b.center) -- node[sloped, allow upside down, label={[xshift=0.2cm, yshift=0.45cm]$k_2$}] {\midarrow} (a.center);
	\draw[line width=0.15mm] (b.center) -- (a1.center);
	\draw[line width=0.5mm] (p1.center) -- (p4.center);
	\draw[line width=0.5mm] (p2.center) -- (p3.center);
\end{tikzpicture}
\right) \nonumber \\
& \propto \LS \left( \int d z_6 \frac{\varepsilon \, (u_1 \cdot k_2)}{z_6 \sqrt{z_6-(u_1 \cdot k_2)^2}} \frac{1}{\sqrt{\det G(u_1,q)}} \right) =  0,
\end{align}
which vanishes since $u_1 \cdot k_2=0$ at the maximal cut, and where we used the result~\eqref{eq: LS_triangle_dotted_matter} for the one-loop triangle diagram with a dotted matter propagator.

The second Abelian subsector, which we call the two-loop zig-zag, is parametrized in fig.~\ref{fig: params_double_triangle_and_zigzag}(b). For this diagram we use the integration order $k_1 \to k_2$, and in this instance we have $\nISP=2$, which requires adding $z_6=k_2^2$ and $z_7=2u_1 \cdot k_2$ to match the two ISPs. This can be traced to the Baikov polynomial $\det G(k_2,u_1)$ for the $k_1$-loop, which introduces the scalar products $k_2^2$ and $u_1 \cdot k_2$ of which none are already present in the propagators of the diagram. Schematically, we obtain the same as doing
\begin{align}
\LS \left( \begin{tikzpicture}[baseline={([yshift=-0.1cm]current bounding box.center)}] 
	\node[] (a) at (0,0) {};
	\node[] (a1) at (0.8,0) {};
	\node[] (b) at (0.4,-0.8) {};
	\node[] (b1) at (1.2,-0.8) {};
	\node[] (p1) at ($(a)+(-0.2,0)$) {};
	\node[] (p2) at ($(b)+(-0.6,0)$) {};
	\node[] (p3) at ($(b1)+(0.2,0)$) {};
	\node[] (p4) at ($(a1)+(0.6,0)$) {};
	\draw[line width=0.15mm] (b.center) -- (a.center);
	\draw[line width=0.15mm] (b.center) -- (a1.center);
		\draw[line width=0.15mm] (b1.center) -- (a1.center);
	\draw[line width=0.5mm] (p1.center) -- (p4.center);
	\draw[line width=0.5mm] (p2.center) -- (p3.center);
\end{tikzpicture} \right) & = \LS \left(
\begin{tikzpicture}[baseline={([yshift=-0.1cm]current bounding box.center)}] 
	\node[] (a) at (0,0) {};
	\node[] (a1) at (0.8,0) {};
	\node[] (b) at (0.4,-0.8) {};
	\node[] (p1) at ($(a)+(-0.2,0)$) {};
	\node[] (p2) at ($(b)+(-0.6,0)$) {};
	\node[] (p3) at ($(b)+(0.6,0)$) {};
	\node[] (p4) at ($(a1)+(0.2,0)$) {};
	\draw[line width=0.15mm] (b.center) -- (a.center);
	\draw[line width=0.15mm] (a1.center) -- (b.center);
	\draw[line width=0.15mm,-{Latex[length=2.2mm]}](1.1,-0.75) -- (1.1,-0.05);
	\node[label={[xshift=0.35cm, yshift=-0.45cm]$k_2$}] (k) at (1.1,-0.4) {};
	\draw[line width=0.5mm] (p1.center) -- (p4.center);
	\draw[line width=0.5mm] (p2.center) -- (p3.center);
\end{tikzpicture} \times
\begin{tikzpicture}[baseline={([yshift=-0.1cm]current bounding box.center)}]
	\node[] (a) at (0,0) {};
	\node[] (a1) at (0.8,0) {};
	\node[] (b) at (0,-0.8) {};
	\node[] (b1) at (0.8,-0.8) {};
	\node[] (p1) at ($(a)+(-0.2,0)$) {};
	\node[] (p2) at ($(b)+(-0.2,0)$) {};
	\node[] (p3) at ($(b1)+(0.2,0)$) {};
	\node[] (p4) at ($(a1)+(0.2,0)$) {};
	\draw[line width=0.15mm, dashed, postaction={decorate}] (b.center) -- node[sloped, allow upside down, label={[xshift=0.2cm, yshift=0cm]$k_2$}] {\midarrow} (a.center);
	\draw[line width=0.15mm] (b1.center) -- (a1.center);
	\draw[line width=0.5mm] (p1.center) -- (a.center);
	\draw[line width=0.5mm, dashed, dash phase=1pt, postaction={decorate}] (a.center) -- node[sloped, allow upside down, label={[xshift=0cm, yshift=-0.2cm]$2u_1 {\cdot} k_2$}] {\midarrow} (a1.center);
	\draw[line width=0.5mm] (a1.center) -- (p4.center);
	\draw[line width=0.5mm] (p2.center) -- node[label={[xshift=0cm, yshift=-0.85cm]\textcolor{white}{$2u_2 {\cdot} k_2$}}] {} (p3.center);
\end{tikzpicture}
\right) \nonumber \\
& \propto \LS \left( \int \frac{x \, d z_6 d z_7}{\sqrt{4 z_6 - z_7^2} \sqrt{(x^2-1)^2(q^2-z_6)^2-4q^2x^2z_7^2}} \right).
\label{eq: LS_double-triangle_prime_step_pre}
\end{align}
In this case, we encounter a new obstacle: multiple square roots in the extra Baikov variables. However, the second square root can be easily rationalized with the change of variables~\eqref{eq: change_of_variables_(z-r1)(z-r2)} since the roots $r_i$ with respect to $z_7$ do not contain further square roots, yielding
\begin{equation}
\label{eq: LS_double-triangle_prime_step}
\LS \left( \begin{tikzpicture}[baseline={([yshift=-0.1cm]current bounding box.center)}] 
	\node[] (a) at (0,0) {};
	\node[] (a1) at (0.8,0) {};
	\node[] (b) at (0.4,-0.8) {};
	\node[] (b1) at (1.2,-0.8) {};
	\node[] (p1) at ($(a)+(-0.2,0)$) {};
	\node[] (p2) at ($(b)+(-0.6,0)$) {};
	\node[] (p3) at ($(b1)+(0.2,0)$) {};
	\node[] (p4) at ($(a1)+(0.6,0)$) {};
	\draw[line width=0.15mm] (b.center) -- (a.center);
	\draw[line width=0.15mm] (b.center) -- (a1.center);
		\draw[line width=0.15mm] (b1.center) -- (a1.center);
	\draw[line width=0.5mm] (p1.center) -- (p4.center);
	\draw[line width=0.5mm] (p2.center) -- (p3.center);
\end{tikzpicture} \right) \propto  \LS \left( \int \frac{x \, d z_6 d t_7}{\sqrt{P_6(z_6,t_7)}} \right).
\end{equation}
The polynomial $P_6$ is of overall degree 6, but is only quadratic in $z_6$, $$P_6(z_6,t_7)=-(x^2-1)^2(1+t_7^2)^2(z_6-r_1(t_7))(z_6-r_2(t_7))$$ for $r_i(t_7)$ being the roots with respect to $z_6$. Therefore, it can be rationalized again with respect to $z_6$ with the change of variables~\eqref{eq: change_of_variables_(z-r1)(z-r2)}:
\begin{equation}
\label{eq: LS_double_triangle_prime}
\LS \left( \begin{tikzpicture}[baseline={([yshift=-0.1cm]current bounding box.center)}] 
	\node[] (a) at (0,0) {};
	\node[] (a1) at (0.8,0) {};
	\node[] (b) at (0.4,-0.8) {};
	\node[] (b1) at (1.2,-0.8) {};
	\node[] (p1) at ($(a)+(-0.2,0)$) {};
	\node[] (p2) at ($(b)+(-0.6,0)$) {};
	\node[] (p3) at ($(b1)+(0.2,0)$) {};
	\node[] (p4) at ($(a1)+(0.6,0)$) {};
	\draw[line width=0.15mm] (b.center) -- (a.center);
	\draw[line width=0.15mm] (b.center) -- (a1.center);
		\draw[line width=0.15mm] (b1.center) -- (a1.center);
	\draw[line width=0.5mm] (p1.center) -- (p4.center);
	\draw[line width=0.5mm] (p2.center) -- (p3.center);
\end{tikzpicture} \right) \propto \frac{x}{x^2-1} \, \LS \left( \int \frac{d t_6 d t_7}{t_6 (1+t_7^2)} \right) \propto \frac{x}{x^2-1},
\end{equation}
which is independent of $q$ and also holds if we relax $u_i \cdot q=0$. The result is in agreement with the two-loop pure basis in refs.~\cite{Parra-Martinez:2020dzs,Herrmann:2021tct,DiVecchia:2021bdo}.

For this diagram, while we could obtain a contribution of odd parity by adding one factor of the ISP $z_7=2u_1 \cdot k_2$ in the numerator, we find it simpler to alternatively add a dot:
\begin{align}
\LS \left( \begin{tikzpicture}[baseline={([yshift=-0.1cm]current bounding box.center)}] 
	\node[] (a) at (0,0) {};
	\node[] (a1) at (0.8,0) {};
	\node[] (b) at (0.4,-0.8) {};
	\node[] (b1) at (1.2,-0.8) {};
	\node[] (p1) at ($(a)+(-0.2,0)$) {};
	\node[] (p2) at ($(b)+(-0.6,0)$) {};
	\node[] (p3) at ($(b1)+(0.2,0)$) {};
	\node[] (p4) at ($(a1)+(0.6,0)$) {};
	\draw[line width=0.15mm] (b.center) -- (a.center);
	\draw[line width=0.15mm] (b.center) -- (a1.center);
		\draw[line width=0.15mm] (b1.center) -- (a1.center);
	\draw[line width=0.5mm] (p1.center) -- (p4.center);
	\draw[line width=0.5mm] (p2.center) -- (p3.center);
	\node at (0.4,0) [circle,fill,inner sep=1.5pt]{};
\end{tikzpicture} \right) & = \LS \left(
\begin{tikzpicture}[baseline={([yshift=-0.1cm]current bounding box.center)}] 
	\node[] (a) at (0,0) {};
	\node[] (a1) at (0.8,0) {};
	\node[] (b) at (0.4,-0.8) {};
	\node[] (p1) at ($(a)+(-0.2,0)$) {};
	\node[] (p2) at ($(b)+(-0.6,0)$) {};
	\node[] (p3) at ($(b)+(0.6,0)$) {};
	\node[] (p4) at ($(a1)+(0.2,0)$) {};
	\draw[line width=0.15mm] (b.center) -- (a.center);
	\draw[line width=0.15mm] (a1.center) -- (b.center);
	\draw[line width=0.15mm,-{Latex[length=2.2mm]}](1.1,-0.75) -- (1.1,-0.05);
	\node[label={[xshift=0.35cm, yshift=-0.45cm]$k_2$}] (k) at (1.1,-0.4) {};
	\draw[line width=0.5mm] (p1.center) -- (p4.center);
	\draw[line width=0.5mm] (p2.center) -- (p3.center);
	\node at (0.4,0) [circle,fill,inner sep=1.5pt]{};
\end{tikzpicture} \times
\begin{tikzpicture}[baseline={([yshift=-0.1cm]current bounding box.center)}]
	\node[] (a) at (0,0) {};
	\node[] (a1) at (0.8,0) {};
	\node[] (b) at (0,-0.8) {};
	\node[] (b1) at (0.8,-0.8) {};
	\node[] (p1) at ($(a)+(-0.2,0)$) {};
	\node[] (p2) at ($(b)+(-0.2,0)$) {};
	\node[] (p3) at ($(b1)+(0.2,0)$) {};
	\node[] (p4) at ($(a1)+(0.2,0)$) {};
	\draw[line width=0.15mm, dashed, postaction={decorate}] (b.center) -- node[sloped, allow upside down, label={[xshift=0.2cm, yshift=0cm]$k_2$}] {\midarrow} (a.center);
	\draw[line width=0.15mm] (b1.center) -- (a1.center);
	\draw[line width=0.5mm] (p1.center) -- (a.center);
	\draw[line width=0.5mm, dashed, dash phase=1pt, postaction={decorate}] (a.center) -- node[sloped, allow upside down, label={[xshift=0cm, yshift=-0.2cm]$2u_1 {\cdot} k_2$}] {\midarrow} (a1.center);
	\draw[line width=0.5mm] (a1.center) -- (p4.center);
	\draw[line width=0.5mm] (p2.center) -- node[label={[xshift=0cm, yshift=-0.85cm]\textcolor{white}{$2u_2 {\cdot} k_2$}}] {} (p3.center);
\end{tikzpicture}
\right) \nonumber \\
& \propto \LS \left( \int \frac{\varepsilon \, x \, z_7 \, d z_6 d z_7}{z_6 \sqrt{4 z_6 - z_7^2} \sqrt{(x^2-1)^2(q^2-z_6)^2-4q^2x^2z_7^2}} \right) \nonumber \\
& \propto \frac{1}{|q|} \, \LS \left( \int \frac{d z_6 d t_7}{z_6 t_7} \right) \propto \frac{1}{|q|},
\end{align}
in agreement with refs.~\cite{Parra-Martinez:2020dzs,Herrmann:2021tct,DiVecchia:2021bdo}. To obtain the result, we used eq.~\eqref{eq: LS_triangle_dotted_matter} for the one-loop triangle diagram with a dotted matter propagator, and then changed variables to $z_7 \to \sqrt{z'_7}$, such that the Jacobian cancels the extra factor of $z_7$ in the numerator. Subsequently, we can simultaneously rationalize the two square roots, now linear in $z'_7$, with the change of variables~\eqref{eq: change_of_variables_(z-r1)(z-r2)}.

For the last Abelian subsector, we need to add one extra Baikov variable $z_6=2u_1 \cdot k_2$ under the integration order $k_1 \to k_2$, and use that the leading singularity of the bubble yields $2 u_1 \cdot k_2$,
\begin{align}
\LS \left( \begin{tikzpicture}[baseline={([yshift=-0.1cm]current bounding box.center)}] 
	\node[] (a) at (0,0) {};
	\node[] (a1) at (0.8,0) {};
	\node[] (b) at (0,-0.8) {};
	\node[] (b1) at (0.8,-0.8) {};
	\node[] (p1) at ($(a)+(-0.2,0)$) {};
	\node[] (p2) at ($(b)+(-0.2,0)$) {};
	\node[] (p3) at ($(b1)+(0.2,0)$) {};
	\node[] (p4) at ($(a1)+(0.2,0)$) {};
	\draw[line width=0.15mm] (b.center) -- (a.center);
	\draw[line width=0.15mm] (b1.center) -- (a1.center);
	\draw[line width=0.5mm] (p1.center) -- (p4.center);
	\draw[line width=0.5mm] (p2.center) -- (p3.center);
	\draw[line width=0.15mm] (0.8,0) arc (0:180:0.4);
\end{tikzpicture} \right) & =
\LS \left( \begin{tikzpicture}[baseline={([yshift=-0.1cm]current bounding box.center)}] 
	\node[] (a) at (0,0) {};
	\node[] (a1) at (1.2,0) {};
	\node[] (p1) at ($(a)+(-0.3,0)$) {};
	\node[] (p4) at ($(a1)+(0.3,0)$) {};
	\draw[line width=0.5mm, postaction={decorate}] (p1.center) -- node[sloped, allow upside down, label={[xshift=0cm, yshift=-0.95cm]$2u_1 {\cdot} (k_1{+}k_2)$}] {\midarrow} (p4.center);
	\draw[line width=0.15mm] (a1.center) arc (0:180:0.6);
	\draw[line width=0.15mm, postaction={decorate}] (0.6,0.6) -- node[sloped, allow upside down, label={[xshift=0cm, yshift=0.85cm]$k_1$}] {\midarrow} (0.595,0.6);
\end{tikzpicture} \times
\begin{tikzpicture}[baseline={([yshift=-0.1cm]current bounding box.center)}] 
	\node[] (a) at (0,0) {};
	\node[] (a1) at (0.8,0) {};
	\node[] (b) at (0,-0.8) {};
	\node[] (b1) at (0.8,-0.8) {};
	\node[] (p1) at ($(a)+(-0.2,0)$) {};
	\node[] (p2) at ($(b)+(-0.2,0)$) {};
	\node[] (p3) at ($(b1)+(0.2,0)$) {};
	\node[] (p4) at ($(a1)+(0.2,0)$) {};
	\draw[line width=0.15mm] (b.center) -- (a.center);
	\draw[line width=0.15mm] (b1.center) -- (a1.center);
	\draw[line width=0.5mm] (p1.center) -- (a.center);
	\draw[line width=0.5mm, dashed, dash phase=1pt, postaction={decorate}] (a.center) -- node[sloped, allow upside down, label={[xshift=0cm, yshift=-0.2cm]$2u_1 {\cdot} k_2$}] {\midarrow} (a1.center);
	\draw[line width=0.5mm] (a1.center) -- (p4.center);
	\draw[line width=0.5mm] (p2.center) -- node[sloped, allow upside down, label={[xshift=0cm, yshift=-0.85cm]\textcolor{white}{$2u_2 {\cdot} k_2$}}] {} (p3.center);
\end{tikzpicture} \right) \propto \LS \left( \begin{tikzpicture}[baseline={([yshift=-0.1cm]current bounding box.center)}] 
	\node[] (a) at (0,0) {};
	\node[] (a1) at (0.8,0) {};
	\node[] (b) at (0,-0.8) {};
	\node[] (b1) at (0.8,-0.8) {};
	\node[] (p1) at ($(a)+(-0.2,0)$) {};
	\node[] (p2) at ($(b)+(-0.2,0)$) {};
	\node[] (p3) at ($(b1)+(0.2,0)$) {};
	\node[] (p4) at ($(a1)+(0.2,0)$) {};
	\draw[line width=0.15mm] (b.center) -- (a.center);
	\draw[line width=0.15mm] (b1.center) -- (a1.center);
	\draw[line width=0.5mm] (p1.center) -- (a.center);
	\draw[line width=0.5mm, dashed, dash phase=1pt, postaction={decorate}] (a.center) -- node[sloped, allow upside down, label={[xshift=0cm, yshift=-0.2cm]$2u_1 {\cdot} k_2$}] {\midarrow} (a1.center);
	\draw[line width=0.5mm] (a1.center) -- (p4.center);
	\draw[line width=0.5mm] (p2.center) -- node[sloped, allow upside down, label={[xshift=0cm, yshift=-0.85cm]\textcolor{white}{$2u_2 {\cdot} k_2$}}] {} (p3.center);
\end{tikzpicture} {\times} (2u_1 {\cdot} k_2) \right) \nonumber \\
& \propto \LS \left( \int \frac{x \, z_6 \, d z_6}{|q| \sqrt{q^2(x^2-1)^2-4x^2 z_6^2}} \right).
\end{align}
Under the change of variables $z_6 \to \sqrt{t_6}$ we uncover a higher-order pole, just as in eq.~\eqref{eq: LS_double_triangle_inf} for the double-triangle diagram. Therefore, we integrate the leading singularity between the roots of the Baikov polynomial, specified in this diagram by the end points $t_6=0$ and $t_6=\frac{q^2 (x^2-1)^2}{4x^2}$, leading to
\begin{equation}
\label{eq: LS_box_with_bubble}
\LS \left( \begin{tikzpicture}[baseline={([yshift=-0.1cm]current bounding box.center)}] 
	\node[] (a) at (0,0) {};
	\node[] (a1) at (0.8,0) {};
	\node[] (b) at (0,-0.8) {};
	\node[] (b1) at (0.8,-0.8) {};
	\node[] (p1) at ($(a)+(-0.2,0)$) {};
	\node[] (p2) at ($(b)+(-0.2,0)$) {};
	\node[] (p3) at ($(b1)+(0.2,0)$) {};
	\node[] (p4) at ($(a1)+(0.2,0)$) {};
	\draw[line width=0.15mm] (b.center) -- (a.center);
	\draw[line width=0.15mm] (b1.center) -- (a1.center);
	\draw[line width=0.5mm] (p1.center) -- (p4.center);
	\draw[line width=0.5mm] (p2.center) -- (p3.center);
	\draw[line width=0.15mm] (0.8,0) arc (0:180:0.4);
\end{tikzpicture} \right) \propto \frac{1}{|q|} \, \int_{\frac{q^2 (x^2-1)^2}{4x^2}}^0 \frac{x \, d t_6}{|q| \sqrt{q^2(x^2-1)^2-4x^2 t_6}} \propto \frac{x^2-1}{x},
\end{equation}
in agreement with refs.~\cite{Parra-Martinez:2020dzs,Herrmann:2021tct,DiVecchia:2021bdo}.

For the odd-parity contribution, we can add another factor of the ISP $z_6=2u_1 \cdot k_2$ in the numerator,
\begin{align}
\LS \left( \begin{tikzpicture}[baseline={([yshift=-0.1cm]current bounding box.center)}] 
	\node[] (a) at (0,0) {};
	\node[] (a1) at (0.8,0) {};
	\node[] (b) at (0,-0.8) {};
	\node[] (b1) at (0.8,-0.8) {};
	\node[] (p1) at ($(a)+(-0.2,0)$) {};
	\node[] (p2) at ($(b)+(-0.2,0)$) {};
	\node[] (p3) at ($(b1)+(0.2,0)$) {};
	\node[] (p4) at ($(a1)+(0.2,0)$) {};
	\draw[line width=0.15mm] (b.center) -- (a.center);
	\draw[line width=0.15mm] (b1.center) -- (a1.center);
	\draw[line width=0.5mm] (p1.center) -- (p4.center);
	\draw[line width=0.5mm] (p2.center) -- (p3.center);
	\draw[line width=0.15mm] (0.8,0) arc (0:180:0.4);
\end{tikzpicture} {\times} (2u_1 {\cdot} k_2) \right) & \propto \LS \left( \begin{tikzpicture}[baseline={([yshift=-0.1cm]current bounding box.center)}] 
	\node[] (a) at (0,0) {};
	\node[] (a1) at (0.8,0) {};
	\node[] (b) at (0,-0.8) {};
	\node[] (b1) at (0.8,-0.8) {};
	\node[] (p1) at ($(a)+(-0.2,0)$) {};
	\node[] (p2) at ($(b)+(-0.2,0)$) {};
	\node[] (p3) at ($(b1)+(0.2,0)$) {};
	\node[] (p4) at ($(a1)+(0.2,0)$) {};
	\draw[line width=0.15mm] (b.center) -- (a.center);
	\draw[line width=0.15mm] (b1.center) -- (a1.center);
	\draw[line width=0.5mm] (p1.center) -- (a.center);
	\draw[line width=0.5mm, dashed, dash phase=1pt, postaction={decorate}] (a.center) -- node[sloped, allow upside down, label={[xshift=0cm, yshift=-0.2cm]$2u_1 {\cdot} k_2$}] {\midarrow} (a1.center);
	\draw[line width=0.5mm] (a1.center) -- (p4.center);
	\draw[line width=0.5mm] (p2.center) -- node[sloped, allow upside down, label={[xshift=0cm, yshift=-0.85cm]\textcolor{white}{$2u_2 {\cdot} k_2$}}] {} (p3.center);
\end{tikzpicture} {\times} (2u_1 {\cdot} k_2)^2 \right) \nonumber \\
& \propto \LS \left( \int \frac{x \, z_6^2 \, d z_6}{|q| \sqrt{q^2(x^2-1)^2-4x^2 z_6^2}} \right) \nonumber \\[0.1cm]
& \propto \frac{|q| \, (x^2-1)^2}{x^2} \, \LS \left( \int d t_6 \frac{(t_6^2+1)^2}{t_6^3} \right) \nonumber \\
& \propto \frac{|q| \, (x^2-1)^2}{x^2},
\label{eq: LS_box_bubble_times_ISP}
\end{align}
where we rationalized the square root with the change of variables~\eqref{eq: change_of_variables_(z-r1)(z-r2)} to $t_6$.

\subsection{Two-loop non-Abelian diagrams}
\label{sec:two_loop_non-abelian}
Having studied the two-loop Abelian diagrams, let us now turn to the non-Abelian diagrams, that is, diagrams containing graviton self-interactions. In particular, after discarding all diagrams with zero master integrals in their sector and those that have a leading singularity that can be trivially calculated, we only have the H diagram, given by the last graph in fig.~\ref{fig: diagrams_indep_two_loop}.

Let us use the integration order $k_1 \to k_2$, where $k_1$ is the loop on top. This parametrization requires one extra Baikov variable $z_8=2u_1 \cdot k_2$, and leads to \cite{Parra-Martinez:2020dzs,Herrmann:2021tct,DiVecchia:2021bdo}
\begin{align}
\LS \left( \begin{tikzpicture}[baseline={([yshift=-0.1cm]current bounding box.center)}] 
	\node[] (a) at (0,0) {};
	\node[] (b) at (0,-0.5) {};
	\node[] (a1) at (0.5,0) {};
	\node[] (b1) at (0.5,-0.5) {};
	\node[] (c) at (0,-1) {};
	\node[] (c1) at (0.5,-1) {};
	\node[] (p1) at ($(a)+(-0.2,0)$) {};
	\node[] (p2) at ($(c)+(-0.2,0)$) {};
	\node[] (p3) at ($(c1)+(0.2,0)$) {};
	\node[] (p4) at ($(a1)+(0.2,0)$) {};
	\draw[line width=0.15mm] (a.center) -- (b.center);
	\draw[line width=0.15mm] (a1.center) -- (b1.center);
	\draw[line width=0.15mm] (b1.center) -- (b.center);
	\draw[line width=0.15mm] (b.center) -- (c.center);
	\draw[line width=0.15mm] (c1.center) -- (b1.center);
	\draw[line width=0.5mm] (a.center) -- (a1.center);
	\draw[line width=0.5mm] (c.center) -- (c1.center);
	\draw[line width=0.5mm] (p1.center) -- (a.center);
	\draw[line width=0.5mm] (a1.center) -- (p4.center);
	\draw[line width=0.5mm] (p2.center) -- (c.center);
	\draw[line width=0.5mm] (c1.center) -- (p3.center);
\end{tikzpicture} \right) & \propto \LS \left( \int \frac{x \, d z_8}{|q|^3 z_8 \sqrt{q^2(x^2-1)^2-4x^2 z_8^2}} \right) \nonumber \\[0.1cm]
& \propto \frac{x}{x^2-1} \, \LS \left(\int \frac{d t_8}{q^4 \, (1+t_8^2)} \right) \propto \frac{x}{q^4 (x^2-1)},
\end{align}
which we can rationalize using the change of variables~\eqref{eq: change_of_variables_(z-r1)(z-r2)} to $t_8$. We see that upon including the $(|q|^2)^2$ factors that come from dressing the graviton self-interactions, the leading singularity correctly scales as $|q|^0$.

The integral of odd parity can be obtained by adding one factor of the ISP $z_8=2u_1 \cdot k_2$ in the numerator, which using the same change of variables as before yields
\begin{align}
\LS \left( \begin{tikzpicture}[baseline={([yshift=-0.1cm]current bounding box.center)}] 
	\node[] (a) at (0,0) {};
	\node[] (b) at (0,-0.5) {};
	\node[] (a1) at (0.5,0) {};
	\node[] (b1) at (0.5,-0.5) {};
	\node[] (c) at (0,-1) {};
	\node[] (c1) at (0.5,-1) {};
	\node[] (p1) at ($(a)+(-0.2,0)$) {};
	\node[] (p2) at ($(c)+(-0.2,0)$) {};
	\node[] (p3) at ($(c1)+(0.2,0)$) {};
	\node[] (p4) at ($(a1)+(0.2,0)$) {};
	\draw[line width=0.15mm] (a.center) -- (b.center);
	\draw[line width=0.15mm] (a1.center) -- (b1.center);
	\draw[line width=0.15mm] (b1.center) -- (b.center);
	\draw[line width=0.15mm] (b.center) -- (c.center);
	\draw[line width=0.15mm] (c1.center) -- (b1.center);
	\draw[line width=0.5mm] (a.center) -- (a1.center);
	\draw[line width=0.5mm] (c.center) -- (c1.center);
	\draw[line width=0.5mm] (p1.center) -- (a.center);
	\draw[line width=0.5mm] (a1.center) -- (p4.center);
	\draw[line width=0.5mm] (p2.center) -- (c.center);
	\draw[line width=0.5mm] (c1.center) -- (p3.center);
\end{tikzpicture} {\times} (2u_1 {\cdot} k_2) \right) & \propto \LS \left( \int \frac{x \, d z_8}{|q|^3 \sqrt{q^2(x^2-1)^2-4x^2 z_8^2}} \right) \\
& \propto \frac{1}{|q|^3} \, \LS \left( \int \frac{d t_8}{t_8} \right) \propto \frac{1}{|q|^3}.
\end{align}

Together with the results for the two-loop Abelian subsectors in sec.~\ref{sec:two_loop_abelian}, this shows that all two-loop diagrams and their subsectors have a dlog form including both conservative and dissipative sectors. This is in agreement with explicit calculations~\cite{Bern:2019nnu,Bern:2019crd,Kalin:2020fhe,Parra-Martinez:2020dzs,Herrmann:2021tct,DiVecchia:2021bdo}, where it was found that the PM expansion is expressible in terms of polylogarithms at two loops.

\section{Three-loop diagrams}
\label{sec:three_loop}
\begin{figure}[t]
\centering
\parbox{\textwidth}{\begin{align*}
& \begin{tikzpicture}[baseline={([yshift=-0.1cm]current bounding box.center)}] 
	\node[] (a) at (0,0) {};
	\node[] (a1) at (0.8,0) {};
	\node[] (a2) at (1.2,0) {};
	\node[] (a3) at (2,0) {};
	\node[] (b) at (0.4,-0.8) {};
	\node[] (b1) at (1.6,-0.8) {};
	\node[] (p1) at ($(a)+(-0.2,0)$) {};
	\node[] (p2) at ($(b)+(-0.6,0)$) {};
	\node[] (p3) at ($(b1)+(0.6,0)$) {};
	\node[] (p4) at ($(a3)+(0.2,0)$) {};
	\draw[line width=0.15mm] (b.center) -- (a.center);
	\draw[line width=0.15mm] (b.center) -- (a1.center);
		\draw[line width=0.15mm] (b1.center) -- (a2.center);
	\draw[line width=0.15mm] (b1.center) -- (a3.center);
	\draw[line width=0.5mm] (p1.center) -- (p4.center);
	\draw[line width=0.5mm] (p2.center) -- (p3.center);
\end{tikzpicture} \qquad \begin{tikzpicture}[baseline={([yshift=-0.1cm]current bounding box.center)}] 
	\node[] (a) at (0,0) {};
	\node[] (a1) at (0.8,0) {};
	\node[] (a2) at (1.6,0) {};
	\node[] (b) at (0.4,-0.8) {};
	\node[] (b1) at (1.2,-0.8) {};
	\node[] (b2) at (2,-0.8) {};
	\node[] (p1) at ($(a)+(-0.2,0)$) {};
	\node[] (p2) at ($(b)+(-0.6,0)$) {};
	\node[] (p3) at ($(b2)+(0.2,0)$) {};
	\node[] (p4) at ($(a2)+(0.6,0)$) {};
	\draw[line width=0.15mm] (b.center) -- (a.center);
	\draw[line width=0.15mm] (b.center) -- (a1.center);
		\draw[line width=0.15mm] (b1.center) -- (a2.center);
	\draw[line width=0.15mm] (b2.center) -- (a2.center);
	\draw[line width=0.5mm] (p1.center) -- (p4.center);
	\draw[line width=0.5mm] (p2.center) -- (p3.center);
\end{tikzpicture} \qquad \begin{tikzpicture}[baseline={([yshift=-0.1cm]current bounding box.center)}] 
	\node[] (a) at (0,0) {};
	\node[] (a1) at (0.4,0) {};
	\node[] (a2) at (0.8,0) {};
	\node[] (a3) at (1.2,0) {};
	\node[] (b) at (0.6,-0.8) {};
	\node[] (p1) at ($(a)+(-0.2,0)$) {};
	\node[] (p2) at ($(b)+(-0.8,0)$) {};
	\node[] (p3) at ($(b)+(0.8,0)$) {};
	\node[] (p4) at ($(a3)+(0.2,0)$) {};
	\draw[line width=0.15mm] (b.center) -- (a.center);
	\draw[line width=0.15mm] (b.center) -- (a1.center);
	\draw[line width=0.15mm] (b.center) -- (a2.center);
	\draw[line width=0.15mm] (b.center) -- (a3.center);
	\draw[line width=0.5mm] (p1.center) -- (p4.center);
	\draw[line width=0.5mm] (p2.center) -- (p3.center);
\end{tikzpicture} \qquad \begin{tikzpicture}[baseline={([yshift=-0.1cm]current bounding box.center)}] 
	\node[] (a) at (0,0) {};
	\node[] (a1) at (0.4,0) {};
	\node[] (a2) at (0.8,0) {};
	\node[] (b) at (0.4,-0.8) {};
	\node[] (b1) at (1.2,-0.8) {};
	\node[] (p1) at ($(a)+(-0.2,0)$) {};
	\node[] (p2) at ($(b)+(-0.6,0)$) {};
	\node[] (p3) at ($(b1)+(0.2,0)$) {};
	\node[] (p4) at ($(a2)+(0.6,0)$) {};
	\draw[line width=0.15mm] (b.center) -- (a.center);
	\draw[line width=0.15mm] (b.center) -- (a1.center);
	\draw[line width=0.15mm] (b.center) -- (a2.center);
	\draw[line width=0.15mm] (b1.center) -- (a2.center);
	\draw[line width=0.5mm] (p1.center) -- (p4.center);
	\draw[line width=0.5mm] (p2.center) -- (p3.center);
\end{tikzpicture} \\[0.2cm]
& \enspace \begin{tikzpicture}[baseline={([yshift=-0.1cm]current bounding box.center)}] 
	\node[] (a) at (0,0) {};
	\node[] (a1) at (0.8,0) {};
	\node[] (a2) at (1.6,0) {};
	\node[] (b) at (0.4,-0.8) {};
	\node[] (b1) at (1.2,-0.8) {};
	\node[] (p1) at ($(a)+(-0.2,0)$) {};
	\node[] (p2) at ($(b)+(-0.6,0)$) {};
	\node[] (p3) at ($(b1)+(0.6,0)$) {};
	\node[] (p4) at ($(a2)+(0.2,0)$) {};
	\draw[line width=0.15mm] (b.center) -- (a.center);
	\draw[line width=0.15mm] (b.center) -- (a1.center);
	\draw[line width=0.15mm] (b1.center) -- (a1.center);
	\draw[line width=0.15mm] (b1.center) -- (a2.center);
	\draw[line width=0.5mm] (p1.center) -- (p4.center);
	\draw[line width=0.5mm] (p2.center) -- (p3.center);
\end{tikzpicture} \quad \qquad \enspace \begin{tikzpicture}[baseline={([yshift=-0.1cm]current bounding box.center)}] 
	\node[] (a) at (0,0) {};
	\node[] (a1) at (1,0) {};
	\node[] (b) at (0,-1) {};
	\node[] (b1) at (1,-1) {};
	\node[] (p1) at ($(a)+(-0.2,0)$) {};
	\node[] (p2) at ($(b)+(-0.2,0)$) {};
	\node[] (p3) at ($(b1)+(0.2,0)$) {};
	\node[] (p4) at ($(a1)+(0.2,0)$) {};
	\draw[line width=0.15mm] (b.center) -- (a.center);
	\draw[line width=0.15mm] (b1.center) -- (a1.center);
	\draw[line width=0.5mm] (p1.center) -- (p4.center);
	\draw[line width=0.5mm] (p2.center) -- (p3.center);
	\draw[line width=0.15mm] (0.5,0) arc (0:180:0.25);
	\draw[line width=0.15mm] (1,0) arc (0:180:0.25);
\end{tikzpicture} \quad \qquad \enspace \begin{tikzpicture}[baseline={([yshift=-0.1cm]current bounding box.center)}] 
	\node[] (a) at (0,0) {};
	\node[] (a1) at (1,0) {};
	\node[] (b) at (0,-1) {};
	\node[] (b1) at (1,-1) {};
	\node[] (p1) at ($(a)+(-0.2,0)$) {};
	\node[] (p2) at ($(b)+(-0.2,0)$) {};
	\node[] (p3) at ($(b1)+(0.2,0)$) {};
	\node[] (p4) at ($(a1)+(0.2,0)$) {};
	\draw[line width=0.15mm] (b.center) -- (a.center);
	\draw[line width=0.15mm] (b1.center) -- (a1.center);
	\draw[line width=0.15mm] (0.5,0) -- (b1.center);
	\draw[line width=0.5mm] (p1.center) -- (p4.center);
	\draw[line width=0.5mm] (p2.center) -- (p3.center);
	\draw[line width=0.15mm] (0.5,0) arc (0:180:0.25);
\end{tikzpicture} \qquad \enspace \raisebox{0.125cm}{\begin{tikzpicture}[baseline={([yshift=-0.1cm]current bounding box.center)}] 
	\node[] (a) at (0,0) {};
	\node[] (a1) at (1,0) {};
	\node[] (b) at (0,-1) {};
	\node[] (b1) at (1,-1) {};
	\node[] (p1) at ($(a)+(-0.2,0)$) {};
	\node[] (p2) at ($(b)+(-0.2,0)$) {};
	\node[] (p3) at ($(b1)+(0.2,0)$) {};
	\node[] (p4) at ($(a1)+(0.2,0)$) {};
	\draw[line width=0.15mm] (b.center) -- (a.center);
	\draw[line width=0.15mm] (b1.center) -- (a1.center);
	\draw[line width=0.15mm] (0.5,-1) -- (a.center);
	\draw[line width=0.5mm] (p1.center) -- (p4.center);
	\draw[line width=0.5mm] (p2.center) -- (p3.center);
	\draw[line width=0.15mm] (1,0) arc (0:180:0.5);
\end{tikzpicture}}
\end{align*}}
\caption{The 8 remaining three-loop Abelian diagrams after all diagrams with leading singularities that are trivial to calculate or that can be related to lower sectors have been discarded.}
\label{fig: diagrams_three_loop_Abelian}
\vspace{0.5cm}
\parbox{\textwidth}{\begin{equation*}
\begin{tikzpicture}[baseline={([yshift=-0.1cm]current bounding box.center)}] 
	\node[] (a) at (0,0) {};
	\node[] (a1) at (0.5,0) {};
	\node[] (a2) at (1,0) {};
	\node[] (b) at (0,-0.5) {};
	\node[] (c) at (0,-1) {};
	\node[] (c1) at (1,-1) {};
	\node[] (p1) at ($(a)+(-0.2,0)$) {};
	\node[] (p2) at ($(c)+(-0.2,0)$) {};
	\node[] (p3) at ($(c1)+(0.2,0)$) {};
	\node[] (p4) at ($(a2)+(0.2,0)$) {};
	\draw[line width=0.15mm] (b.center) -- (a.center);
	\draw[line width=0.15mm] (b.center) -- (a1.center);
	\draw[line width=0.15mm] (b.center) -- (a2.center);
	\draw[line width=0.15mm] (b.center) -- (c.center);
	\draw[line width=0.15mm] (c1.center) -- (a2.center);
	\draw[line width=0.5mm] (p1.center) -- (p4.center);
	\draw[line width=0.5mm] (p2.center) -- (p3.center);
\end{tikzpicture} \quad \begin{tikzpicture}[baseline={([yshift=-0.1cm]current bounding box.center)}] 
	\node[] (a) at (0,0) {};
	\node[] (a1) at (0.25,0) {};
	\node[] (a2) at (0.5,0) {};
	\node[] (a3) at (1,0) {};
	\node[] (b) at (0.25,-0.5) {};
	\node[] (c) at (0.5,-1) {};
	\node[] (p1) at ($(a)+(-0.2,0)$) {};
	\node[] (p2) at ($(c)+(-0.7,0)$) {};
	\node[] (p3) at ($(c)+(0.7,0)$) {};
	\node[] (p4) at ($(a3)+(0.2,0)$) {};
	\draw[line width=0.15mm] (a.center) -- (c.center);
	\draw[line width=0.15mm] (b.center) -- (a1.center);
	\draw[line width=0.15mm] (b.center) -- (a2.center);
	\draw[line width=0.15mm] (c.center) -- (a3.center);
	\draw[line width=0.5mm] (p1.center) -- (p4.center);
	\draw[line width=0.5mm] (p2.center) -- (p3.center);
\end{tikzpicture} \quad \begin{tikzpicture}[baseline={([yshift=-0.1cm]current bounding box.center)}] 
	\node[] (a) at (0,0) {};
	\node[] (a1) at (0.5,0) {};
	\node[] (a2) at (1,0) {};
	\node[] (b) at (0,-0.5) {};
	\node[] (b1) at (0.5,-0.5) {};
	\node[] (b2) at (1,-0.5) {};
	\node[] (c) at (0,-1) {};
	\node[] (c1) at (1,-1) {};
	\node[] (p1) at ($(a)+(-0.2,0)$) {};
	\node[] (p2) at ($(c)+(-0.2,0)$) {};
	\node[] (p3) at ($(c1)+(0.2,0)$) {};
	\node[] (p4) at ($(a2)+(0.2,0)$) {};
	\draw[line width=0.15mm] (b.center) -- (a.center);
	\draw[line width=0.15mm] (b1.center) -- (a1.center);
	\draw[line width=0.15mm] (b2.center) -- (a2.center);
	\draw[line width=0.15mm] (b2.center) -- (b.center);
	\draw[line width=0.15mm] (b.center) -- (c.center);
	\draw[line width=0.15mm] (c1.center) -- (b2.center);
	\draw[line width=0.5mm] (p1.center) -- (p4.center);
	\draw[line width=0.5mm] (p2.center) -- (p3.center);
\end{tikzpicture} \quad \begin{tikzpicture}[baseline={([yshift=-0.1cm]current bounding box.center)}] 
	\node[] (a) at (0,0) {};
	\node[] (a1) at (0.5,0) {};
	\node[] (a2) at (1,0) {};
	\node[] (b) at (0,-0.5) {};
	\node[] (b1) at (0.5,-0.5) {};
	\node[] (c) at (0,-1) {};
	\node[] (c1) at (1,-1) {};
	\node[] (p1) at ($(a)+(-0.2,0)$) {};
	\node[] (p2) at ($(c)+(-0.2,0)$) {};
	\node[] (p3) at ($(c1)+(0.2,0)$) {};
	\node[] (p4) at ($(a2)+(0.2,0)$) {};
	\draw[line width=0.15mm] (b.center) -- (a.center);
	\draw[line width=0.15mm] (b1.center) -- (a1.center);
	\draw[line width=0.15mm] (c1.center) -- (a2.center);
	\draw[line width=0.15mm] (b1.center) -- (b.center);
	\draw[line width=0.15mm] (b.center) -- (c.center);
	\draw[line width=0.15mm] (c1.center) -- (b1.center);
	\draw[line width=0.5mm] (p1.center) -- (p4.center);
	\draw[line width=0.5mm] (p2.center) -- (p3.center);
\end{tikzpicture} \quad  \begin{tikzpicture}[baseline={([yshift=-0.1cm]current bounding box.center)}] 
	\node[] (a) at (0,0) {};
	\node[] (a1) at (0.5,0) {};
	\node[] (a2) at (1,0) {};
	\node[] (b) at (0,-0.5) {};
	\node[] (b1) at (1,-0.5) {};
	\node[] (c) at (0,-1) {};
	\node[] (c1) at (1,-1) {};
	\node[] (p1) at ($(a)+(-0.2,0)$) {};
	\node[] (p2) at ($(c)+(-0.2,0)$) {};
	\node[] (p3) at ($(c1)+(0.2,0)$) {};
	\node[] (p4) at ($(a2)+(0.2,0)$) {};
	\draw[line width=0.15mm] (b.center) -- (a.center);
	\draw[line width=0.15mm] (b.center) -- (a1.center);
	\draw[line width=0.15mm] (b1.center) -- (a2.center);
	\draw[line width=0.15mm] (b1.center) -- (b.center);
	\draw[line width=0.15mm] (b.center) -- (c.center);
	\draw[line width=0.15mm] (c1.center) -- (b1.center);
	\draw[line width=0.5mm] (p1.center) -- (p4.center);
	\draw[line width=0.5mm] (p2.center) -- (p3.center);
\end{tikzpicture} \quad \begin{tikzpicture}[baseline={([yshift=-0.1cm]current bounding box.center)}] 
	\node[] (a) at (0,0) {};
	\node[] (a1) at (0.5,0) {};
	\node[] (a2) at (1,0) {};
	\node[] (b) at (0.5,-0.5) {};
	\node[] (c) at (0,-1) {};
	\node[] (c1) at (1,-1) {};
	\node[] (p1) at ($(a)+(-0.2,0)$) {};
	\node[] (p2) at ($(c)+(-0.2,0)$) {};
	\node[] (p3) at ($(c1)+(0.2,0)$) {};
	\node[] (p4) at ($(a2)+(0.2,0)$) {};
	\draw[line width=0.15mm] (c.center) -- (a.center);
	\draw[line width=0.15mm] (b.center) -- (a1.center);
	\draw[line width=0.15mm] (c1.center) -- (a2.center);
	\draw[line width=0.15mm] (b.center) -- (c.center);
	\draw[line width=0.15mm] (c1.center) -- (b.center);
	\draw[line width=0.5mm] (p1.center) -- (p4.center);
	\draw[line width=0.5mm] (p2.center) -- (p3.center);
\end{tikzpicture} \quad \begin{tikzpicture}[baseline={([yshift=-0.1cm]current bounding box.center)}] 
	\node[] (a) at (0,0) {};
	\node[] (a1) at (0.5,0) {};
	\node[] (a2) at (1,0) {};
	\node[] (b) at (0,-0.5) {};
	\node[] (c) at (0,-1) {};
	\node[] (c1) at (1,-1) {};
	\node[] (p1) at ($(a)+(-0.2,0)$) {};
	\node[] (p2) at ($(c)+(-0.2,0)$) {};
	\node[] (p3) at ($(c1)+(0.2,0)$) {};
	\node[] (p4) at ($(a2)+(0.2,0)$) {};
	\draw[line width=0.15mm] (b.center) -- (a.center);
	\draw[line width=0.15mm] (b.center) -- (a1.center);
	\draw[line width=0.15mm] (c1.center) -- (a2.center);
	\draw[line width=0.15mm] (b.center) -- (c.center);
	\draw[line width=0.15mm] (c1.center) -- (b.center);
	\draw[line width=0.5mm] (p1.center) -- (p4.center);
	\draw[line width=0.5mm] (p2.center) -- (p3.center);
\end{tikzpicture}
\end{equation*}}
\caption{The 7 remaining three-loop non-Abelian diagrams after all diagrams with leading singularities that are trivial to calculate or that can be related to lower sectors have been discarded.}
\label{fig: diagrams_three_loop_non_Abelian}
\end{figure}

In this section, we continue the analysis and classification of geometries to three loops. Having exemplified how to calculate the leading singularity up to two loops, we will thus omit the explicit calculations in the following, since they involve analogous steps.

At three loops there are 531 topologies, however, as can be seen in tab.~\ref{tab:indep_diagrams}, thanks to the relations gathered in sec.~\ref{sec:reduction_relations}, the number of relevant diagrams that we need to analyze is drastically reduced. Starting with the unraveling identity~\eqref{eq: unraveling_matter_props}, we can relate the leading singularity of all 214 non-planar diagrams to planar counterparts. Then, there are 263 diagrams that have zero master integrals since they contain bubble corrections with at least one cubic vertex (see fig.~\ref{fig: diagrams_zero_masters}(a)), and another 31 diagrams with triangle sub-loops that also have zero master integrals (see fig.~\ref{fig: diagrams_zero_masters}(b)). Moreover, there are 8 superclassical diagrams the leading singularity of which can be trivially computed with eq.~\eqref{eq: reduction_superclassical}. Omitting thus all of these diagrams, as well as those that completely factor into a product of lower-loop integrals, we find that at three loops there are 8 independent Abelian topologies, which we gather in fig.~\ref{fig: diagrams_three_loop_Abelian} and study in sec.~\ref{sec:three_loop_abelian}. On top of that, there are 7 independent non-Abelian topologies, which are gathered in fig.~\ref{fig: diagrams_three_loop_non_Abelian} and which we turn to in sec.~\ref{sec:three_loop_non-abelian} and sec.~\ref{sec:three_loop_K3}. The last of these diagrams, which we study in sec.~\ref{sec:three_loop_K3}, includes an integral over the first non-trivial geometry appearing in the PM expansion -- a K3 surface~\cite{Bern:2021dqo,Dlapa:2021npj,Dlapa:2022wdu}. 

With the exception of the second diagram of fig.~\ref{fig: diagrams_three_loop_non_Abelian}, the leading singularity of which can be trivially related to that of a one-loop diagram as shown below, we find that for the non-Abelian sector at three loops it suffices to limit our analysis of geometries to diagrams (and subsectors thereof) that pertain to the Mondrian family of diagrams\footnote{This notation goes back to the resemblance to paintings by Piet Mondrian, see e.g.\  ref.~\cite{Bern:2004kq}.}
\begin{equation}
\label{eq: diag_Mondrian}
\begin{tikzpicture}[baseline={([yshift=-0.1cm]current bounding box.center)}, scale=0.7] 
	\node[] (a) at (0,0) {};
	\node[] (a1) at (1,0) {};
	\node[] (a2) at (2,0) {};
	\node[] (a3) at (3,0) {};
	\node[] (a4) at (4,0) {};
	\node[] (a5) at (5,0) {};
	\node[] (a6) at (6,0) {};
	\node[] (b) at (0,-1) {};
	\node[] (b1) at (1,-1) {};
	\node[] (b2) at (2,-1) {};
	\node[] (b3) at (3,-1) {};
	\node[] (b4) at (4,-1) {};
	\node[] (b5) at (5,-1) {};
	\node[] (b6) at (6,-1) {};
	\node[] (c) at (0,-2) {};
	\node[] (c1) at (1,-2) {};
	\node[] (c2) at (2,-2) {};
	\node[] (c3) at (3,-2) {};
	\node[] (c4) at (4,-2) {};
	\node[] (c5) at (5,-2) {};
	\node[] (c6) at (6,-2) {};
	\node[] (p1) at ($(a)+(-0.3,0)$) {};
	\node[] (p2) at ($(c)+(-0.3,0)$) {};
	\node[] (p3) at ($(c6)+(0.3,0)$) {};
	\node[] (p4) at ($(a6)+(0.3,0)$) {};
	\draw[line width=0.15mm] (b.center) -- (a.center);
	\draw[line width=0.15mm] (b1.center) -- (a1.center);
	\draw[line width=0.15mm] (b3.center) -- (a3.center);
	\draw[line width=0.15mm] (b4.center) -- (a4.center);
	\draw[line width=0.15mm] (b6.center) -- (a6.center);
	\draw[line width=0.15mm] (b.center) -- (b6.center);
	\draw[line width=0.15mm] (b.center) -- (c.center);
	\draw[line width=0.15mm] (b2.center) -- (c2.center);
	\draw[line width=0.15mm] (b5.center) -- (c5.center);
	\draw[line width=0.15mm] (b6.center) -- (c6.center);
	\draw[line width=0.5mm] (p1.center) -- (p4.center);
	\draw[line width=0.5mm] (p2.center) -- (p3.center);
	\node at (5,-0.5)[circle,fill,inner sep=0.6pt]{};
	\node at (5.25,-0.5)[circle,fill,inner sep=0.6pt]{};
	\node at (4.75,-0.5)[circle,fill,inner sep=0.6pt]{};
	\node at (5.5,-1.5)[circle,fill,inner sep=0.6pt]{};
	\node at (5.25,-1.5)[circle,fill,inner sep=0.6pt]{};
	\node at (5.75,-1.5)[circle,fill,inner sep=0.6pt]{};
\end{tikzpicture},
\end{equation}
as well as to their superclassical iterations.

Starting at four loops, one also needs to include the non-planar variations of this family that cannot be related to planar counterparts by the unraveling of matter propagators~\eqref{eq: unraveling_matter_props}, i.e.\ by re-ordering the vertices at the matter lines. A pictorial way to obtain such non-planar topologies is for instance to perform Jacobi moves to the internal edges of the drawings, which amounts to replacing the corresponding $s$ channel by the $t$ or the $u$ channels~\cite{Bern:2008qj}. For instance, starting with the four-loop Mondrian diagram that contains two loops on top and two on the bottom, we can pictorially obtain two different topologies by doing a Jacobi move in the propagator highlighted in red:
\begin{equation}
\left\{ \begin{tikzpicture}[baseline={([yshift=-0.1cm]current bounding box.center)}] 
	\node[] (a) at (0,0) {};
	\node[] (a1) at (0.5,0) {};
	\node[] (a2) at (1.5,0) {};
	\node[] (b) at (0,-0.5) {};
	\node[] (b1) at (0.5,-0.5) {};
	\node[] (b2) at (1,-0.5) {};
	\node[] (b3) at (1.5,-0.5) {};
	\node[] (c) at (0,-1) {};
	\node[] (c1) at (1,-1) {};
	\node[] (c2) at (1.5,-1) {};
	\node[] (p1) at ($(a)+(-0.2,0)$) {};
	\node[] (p2) at ($(c)+(-0.2,0)$) {};
	\node[] (p3) at ($(c2)+(0.2,0)$) {};
	\node[] (p4) at ($(a2)+(0.2,0)$) {};
	\draw[line width=0.15mm] (b.center) -- (a.center);
	\draw[line width=0.15mm] (b1.center) -- (a1.center);
	\draw[line width=0.15mm] (b3.center) -- (a2.center);
	\draw[line width=0.15mm] (b2.center) -- (c1.center);
	\draw[line width=0.15mm] (b3.center) -- (b2.center);
	\draw[line width=0.15mm] (b1.center) -- (b.center);
	\draw[line width=0.3mm, red] (b1.center) -- (b2.center);
	\draw[line width=0.15mm] (b.center) -- (c.center);
	\draw[line width=0.15mm] (c2.center) -- (b3.center);
	\draw[line width=0.5mm] (p1.center) -- (p4.center);
	\draw[line width=0.5mm] (p2.center) -- (p3.center);
\end{tikzpicture}, \begin{tikzpicture}[baseline={([yshift=-0.1cm]current bounding box.center)}] 
	\node[] (a) at (0,0) {};
	\node[] (a1) at (1,0) {};
	\node[] (a2) at (1.5,0) {};
	\node[] (b) at (0,-0.5) {};
	\node[] (b1) at (0.5,-0.5) {};
	\node[] (b2) at (1,-0.5) {};
	\node[] (b3) at (1.5,-0.5) {};
	\node[] (c) at (0,-1) {};
	\node[] (c1) at (0.5,-1) {};
	\node[] (c2) at (1.5,-1) {};
	\node[] (p1) at ($(a)+(-0.2,0)$) {};
	\node[] (p2) at ($(c)+(-0.2,0)$) {};
	\node[] (p3) at ($(c2)+(0.2,0)$) {};
	\node[] (p4) at ($(a2)+(0.2,0)$) {};
	\draw[line width=0.15mm] (b.center) -- (a.center);
	\draw[line width=0.15mm] (b2.center) -- (a1.center);
	\draw[line width=0.15mm] (b3.center) -- (a2.center);
	\draw[line width=0.15mm] (b1.center) -- (c1.center);
	\draw[line width=0.15mm] (b3.center) -- (b2.center);
	\draw[line width=0.15mm] (b1.center) -- (b.center);
	\draw[line width=0.3mm, red] (b1.center) -- (b2.center);
	\draw[line width=0.15mm] (b.center) -- (c.center);
	\draw[line width=0.15mm] (c2.center) -- (b3.center);
	\draw[line width=0.5mm] (p1.center) -- (p4.center);
	\draw[line width=0.5mm] (p2.center) -- (p3.center);
\end{tikzpicture}, \raisebox{0.1cm}{\begin{tikzpicture}[baseline=(current bounding box.center), scale=0.25] 
	\node[] (a) at (0,0) {};
	\node[] (a1) at (3,0) {};
	\node[] (a2) at (6,0) {};
	\node[] (b) at (3.75,-1) {};
	\node[] (c) at (0,-2) {};
	\node[] (c1) at (2.5,-2) {};
	\node[] (c2) at (6,-2) {};
	\node[] (d) at (0,-4) {};
	\node[] (d1) at (3,-4) {};
	\node[] (d2) at (6,-4) {};
	\node[] (p1) at ($(a)+(-0.8,0)$) {};
	\node[] (p2) at ($(d)+(-0.8,0)$) {};
	\node[] (p3) at ($(d2)+(0.8,0)$) {};
	\node[] (p4) at ($(a2)+(0.8,0)$) {};
	\draw[line width=0.15mm] (c.center) -- (a.center);
	\draw[line width=0.15mm] (d.center) -- (c.center);
	\draw[line width=0.15mm] (a1.center) -- (b.center);
	\draw[line width=0.3mm, red] (b.center) -- (c1.center);
	\draw[line width=0.15mm] (c.center) -- (c1.center);
	\draw[line width=0.15mm] (a2.center) -- (c2.center);
	\draw[line width=0.15mm] (c2.center) -- (d2.center);
	\draw[line width=0.15mm] (c2.center) -- (4.4,-2);
	\draw[line width=0.15mm] (3.6,-2) -- (c1.center);
	\draw[line width=0.15mm] plot[smooth, tension=0.8] coordinates {(b.center) (4,-2) (3.75,-3) (d1.center)};
	\draw[line width=0.5mm] (a.center) -- (a1.center);
	\draw[line width=0.5mm] (a1.center) -- (a2.center);
	\draw[line width=0.5mm] (d.center) -- (d1.center);
	\draw[line width=0.5mm] (d1.center) -- (d2.center);
	\draw[line width=0.5mm] (p1.center) -- (p4.center);
	\draw[line width=0.5mm] (p2.center) -- (p3.center);
\end{tikzpicture}}\right\},
\end{equation}
where the last diagram is truly non-planar in the sense described above, and is studied in sec.~\ref{sec:four_loop_nonplanar}.

\subsection{Three-loop Abelian diagrams}
\label{sec:three_loop_abelian}

At three loops, we have the 8 independent Abelian subsectors of fig.~\ref{fig: diagrams_three_loop_Abelian}. The first two diagrams contain a superclassical box in the middle, but we cannot apply the reduction~\eqref{eq: reduction_superclassical_generic} since the momentum transfer for the box does not satisfy $k \cdot u_i=0$ for both matter lines at the maximal cut. However, we can re-use eq.~\eqref{eq: LS_triangle} for the leading singularity of the triangles on each side to obtain
\begin{equation}
\LS \left(
\begin{tikzpicture}[baseline={([yshift=-0.1cm]current bounding box.center)}] 
	\node[] (a) at (0,0) {};
	\node[] (a1) at (0.8,0) {};
	\node[] (a2) at (1.2,0) {};
	\node[] (a3) at (2,0) {};
	\node[] (b) at (0.4,-0.8) {};
	\node[] (b1) at (1.6,-0.8) {};
	\node[] (p1) at ($(a)+(-0.2,0)$) {};
	\node[] (p2) at ($(b)+(-0.6,0)$) {};
	\node[] (p3) at ($(b1)+(0.6,0)$) {};
	\node[] (p4) at ($(a3)+(0.2,0)$) {};
	\draw[line width=0.15mm] (b.center) -- (a.center);
	\draw[line width=0.15mm] (b.center) -- (a1.center);
		\draw[line width=0.15mm] (b1.center) -- (a2.center);
	\draw[line width=0.15mm] (b1.center) -- (a3.center);
	\draw[line width=0.5mm] (p1.center) -- (p4.center);
	\draw[line width=0.5mm] (p2.center) -- (p3.center);
\end{tikzpicture}
\right) \propto \LS \left(
\begin{tikzpicture}[baseline={([yshift=-0.1cm]current bounding box.center)}] 
	\node[] (a) at (0,0) {};
	\node[] (a1) at (0.8,0) {};
	\node[] (a2) at (1.6,0) {};
	\node[] (b) at (0.4,-0.8) {};
	\node[] (b1) at (1.2,-0.8) {};
	\node[] (b2) at (2,-0.8) {};
	\node[] (p1) at ($(a)+(-0.2,0)$) {};
	\node[] (p2) at ($(b)+(-0.6,0)$) {};
	\node[] (p3) at ($(b2)+(0.2,0)$) {};
	\node[] (p4) at ($(a2)+(0.6,0)$) {};
	\draw[line width=0.15mm] (b.center) -- (a.center);
	\draw[line width=0.15mm] (b.center) -- (a1.center);
		\draw[line width=0.15mm] (b1.center) -- (a2.center);
	\draw[line width=0.15mm] (b2.center) -- (a2.center);
	\draw[line width=0.5mm] (p1.center) -- (p4.center);
	\draw[line width=0.5mm] (p2.center) -- (p3.center);
\end{tikzpicture}
\right) \propto \frac{x}{x^2-1},
\end{equation}
while for the odd-parity sector we find
\begin{equation}
\LS \left(
\begin{tikzpicture}[baseline={([yshift=-0.1cm]current bounding box.center)}] 
	\node[] (a) at (0,0) {};
	\node[] (a1) at (0.8,0) {};
	\node[] (a2) at (1.2,0) {};
	\node[] (a3) at (2,0) {};
	\node[] (b) at (0.4,-0.8) {};
	\node[] (b1) at (1.6,-0.8) {};
	\node[] (p1) at ($(a)+(-0.2,0)$) {};
	\node[] (p2) at ($(b)+(-0.6,0)$) {};
	\node[] (p3) at ($(b1)+(0.6,0)$) {};
	\node[] (p4) at ($(a3)+(0.2,0)$) {};
	\draw[line width=0.15mm] (b.center) -- (a.center);
	\draw[line width=0.15mm] (b.center) -- (a1.center);
		\draw[line width=0.15mm] (b1.center) -- (a2.center);
	\draw[line width=0.15mm] (b1.center) -- (a3.center);
	\draw[line width=0.5mm] (p1.center) -- (p4.center);
	\draw[line width=0.5mm] (p2.center) -- (p3.center);
	\node at (1,0) [circle,fill,inner sep=1.5pt]{};
\end{tikzpicture}
\right) = \LS \left(
\begin{tikzpicture}[baseline={([yshift=-0.1cm]current bounding box.center)}] 
	\node[] (a) at (0,0) {};
	\node[] (a1) at (0.8,0) {};
	\node[] (a2) at (1.6,0) {};
	\node[] (b) at (0.4,-0.8) {};
	\node[] (b1) at (1.2,-0.8) {};
	\node[] (b2) at (2,-0.8) {};
	\node[] (p1) at ($(a)+(-0.2,0)$) {};
	\node[] (p2) at ($(b)+(-0.6,0)$) {};
	\node[] (p3) at ($(b2)+(0.2,0)$) {};
	\node[] (p4) at ($(a2)+(0.6,0)$) {};
	\draw[line width=0.15mm] (b.center) -- (a.center);
	\draw[line width=0.15mm] (b.center) -- (a1.center);
		\draw[line width=0.15mm] (b1.center) -- (a2.center);
	\draw[line width=0.15mm] (b2.center) -- (a2.center);
	\draw[line width=0.5mm] (p1.center) -- (p4.center);
	\draw[line width=0.5mm] (p2.center) -- (p3.center);
	\node at (1.2,0) [circle,fill,inner sep=1.5pt]{};
\end{tikzpicture}
\right) = 0.
\end{equation}

Next, for the third diagram in fig.~\ref{fig: diagrams_three_loop_Abelian} we can use eq.~\eqref{eq: LS_double_triangle_u1q0} to set $\LS \propto 1$ for the double triangle, since the momentum transfer $k_3$ also satisfies $2u_1 \cdot k_3=0$ at the maximal cut:
\begin{equation}
\LS \left(
\begin{tikzpicture}[baseline={([yshift=-0.1cm]current bounding box.center)}] 
	\node[] (a) at (0,0) {};
	\node[] (a1) at (0.4,0) {};
	\node[] (a2) at (0.8,0) {};
	\node[] (a3) at (1.2,0) {};
	\node[] (b) at (0.6,-0.8) {};
	\node[] (p1) at ($(a)+(-0.2,0)$) {};
	\node[] (p2) at ($(b)+(-0.8,0)$) {};
	\node[] (p3) at ($(b)+(0.8,0)$) {};
	\node[] (p4) at ($(a3)+(0.2,0)$) {};
	\draw[line width=0.15mm] (b.center) -- (a.center);
	\draw[line width=0.15mm] (b.center) -- (a1.center);
	\draw[line width=0.15mm] (b.center) -- (a2.center);
	\draw[line width=0.15mm] (b.center) -- (a3.center);
	\draw[line width=0.5mm] (p1.center) -- (p4.center);
	\draw[line width=0.5mm] (p2.center) -- (p3.center);
\end{tikzpicture}
\right) = \LS \left(
\begin{tikzpicture}[baseline={([yshift=-0.1cm]current bounding box.center)}] 
	\node[] (a) at (0,0) {};
	\node[] (a1) at (0.4,0) {};
	\node[] (a2) at (0.8,0) {};
	\node[] (b) at (0.4,-0.8) {};
	\node[] (p1) at ($(a)+(-0.2,0)$) {};
	\node[] (p2) at ($(b)+(-0.6,0)$) {};
	\node[] (p3) at ($(b)+(0.6,0)$) {};
	\node[] (p4) at ($(a2)+(0.2,0)$) {};
	\draw[line width=0.15mm] (b.center) -- (a.center);
	\draw[line width=0.15mm] (b.center) -- (a1.center);
	\draw[line width=0.15mm] (b.center) -- (a2.center);
	\draw[line width=0.15mm,-{Latex[length=2.2mm]}](1.1,-0.75) -- (1.1,-0.05);
	\node[label={[xshift=0.35cm, yshift=-0.45cm]$k_3$}] (k) at (1.1,-0.4) {};
	\draw[line width=0.5mm] (p1.center) -- (p4.center);
	\draw[line width=0.5mm] (p2.center) -- (p3.center);
\end{tikzpicture} \times \begin{tikzpicture}[baseline={([yshift=-0.1cm]current bounding box.center)}] 
	\node[] (a) at (0,0) {};
	\node[] (a1) at (0.8,0) {};
	\node[] (b) at (0.4,-0.8) {};
	\node[] (p1) at ($(a)+(-0.2,0)$) {};
	\node[] (p2) at ($(b)+(-0.6,0)$) {};
	\node[] (p3) at ($(b)+(0.6,0)$) {};
	\node[] (p4) at ($(a1)+(0.2,0)$) {};
	\draw[line width=0.15mm, dashed, postaction={decorate}] (b.center) -- node[sloped, allow upside down, label={[xshift=0.2cm, yshift=0.4cm]$k_3$}] {\midarrow} (a.center);
	\draw[line width=0.15mm] (b.center) -- (a1.center);
	\draw[line width=0.5mm] (p1.center) -- (p4.center);
	\draw[line width=0.5mm] (p2.center) -- (p3.center);
\end{tikzpicture}
\right) \propto \LS \left(
\begin{tikzpicture}[baseline={([yshift=-0.1cm]current bounding box.center)}] 
	\node[] (a) at (0,0) {};
	\node[] (a1) at (0.8,0) {};
	\node[] (b) at (0.4,-0.8) {};
	\node[] (p1) at ($(a)+(-0.2,0)$) {};
	\node[] (p2) at ($(b)+(-0.6,0)$) {};
	\node[] (p3) at ($(b)+(0.6,0)$) {};
	\node[] (p4) at ($(a1)+(0.2,0)$) {};
	\draw[line width=0.15mm, dashed, postaction={decorate}] (b.center) -- node[sloped, allow upside down, label={[xshift=0.2cm, yshift=0.4cm]$k_3$}] {\midarrow} (a.center);
	\draw[line width=0.15mm] (b.center) -- (a1.center);
	\draw[line width=0.5mm] (p1.center) -- (p4.center);
	\draw[line width=0.5mm] (p2.center) -- (p3.center);
\end{tikzpicture}
\right) \propto |q|.
\end{equation}
For the even-parity terms we again have vanishing leading singularity:
\begin{equation}
\LS \left(
\begin{tikzpicture}[baseline={([yshift=-0.1cm]current bounding box.center)}] 
	\node[] (a) at (0,0) {};
	\node[] (a1) at (0.4,0) {};
	\node[] (a2) at (0.8,0) {};
	\node[] (a3) at (1.2,0) {};
	\node[] (b) at (0.6,-0.8) {};
	\node[] (p1) at ($(a)+(-0.2,0)$) {};
	\node[] (p2) at ($(b)+(-0.8,0)$) {};
	\node[] (p3) at ($(b)+(0.8,0)$) {};
	\node[] (p4) at ($(a3)+(0.2,0)$) {};
	\draw[line width=0.15mm] (b.center) -- (a.center);
	\draw[line width=0.15mm] (b.center) -- (a1.center);
	\draw[line width=0.15mm] (b.center) -- (a2.center);
	\draw[line width=0.15mm] (b.center) -- (a3.center);
	\draw[line width=0.5mm] (p1.center) -- (p4.center);
	\draw[line width=0.5mm] (p2.center) -- (p3.center);
	\node at (0.6,0) [circle,fill,inner sep=1.5pt]{};
\end{tikzpicture}
\right) = 0.
\end{equation}

For the fourth and fifth diagrams in fig.~\ref{fig: diagrams_three_loop_Abelian}, we obtain that
\begin{align}
& \LS \left(
\begin{tikzpicture}[baseline={([yshift=-0.1cm]current bounding box.center)}] 
	\node[] (a) at (0,0) {};
	\node[] (a1) at (0.4,0) {};
	\node[] (a2) at (0.8,0) {};
	\node[] (b) at (0.4,-0.8) {};
	\node[] (b1) at (1.2,-0.8) {};
	\node[] (p1) at ($(a)+(-0.2,0)$) {};
	\node[] (p2) at ($(b)+(-0.6,0)$) {};
	\node[] (p3) at ($(b1)+(0.2,0)$) {};
	\node[] (p4) at ($(a2)+(0.6,0)$) {};
	\draw[line width=0.15mm] (b.center) -- (a.center);
	\draw[line width=0.15mm] (b.center) -- (a1.center);
	\draw[line width=0.15mm] (b.center) -- (a2.center);
	\draw[line width=0.15mm] (b1.center) -- (a2.center);
	\draw[line width=0.5mm] (p1.center) -- (p4.center);
	\draw[line width=0.5mm] (p2.center) -- (p3.center);
	\node at (0.6,-0.4) [circle,fill,inner sep=1.5pt]{};
\end{tikzpicture}
\right) \propto \LS \left(
\begin{tikzpicture}[baseline={([yshift=-0.1cm]current bounding box.center)}] 
	\node[] (a) at (0,0) {};
	\node[] (a1) at (0.8,0) {};
	\node[] (a2) at (1.6,0) {};
	\node[] (b) at (0.4,-0.8) {};
	\node[] (b1) at (1.2,-0.8) {};
	\node[] (p1) at ($(a)+(-0.2,0)$) {};
	\node[] (p2) at ($(b)+(-0.6,0)$) {};
	\node[] (p3) at ($(b1)+(0.6,0)$) {};
	\node[] (p4) at ($(a2)+(0.2,0)$) {};
	\draw[line width=0.15mm] (b.center) -- (a.center);
	\draw[line width=0.15mm] (b.center) -- (a1.center);
	\draw[line width=0.15mm] (b1.center) -- (a1.center);
	\draw[line width=0.15mm] (b1.center) -- (a2.center);
	\draw[line width=0.5mm] (p1.center) -- (p4.center);
	\draw[line width=0.5mm] (p2.center) -- (p3.center);
	\node at (1,-0.4) [circle,fill,inner sep=1.5pt]{};
\end{tikzpicture}
\right) \propto \frac{\varepsilon \, x}{|q| \, (x^2-1)}, \\[0.2cm]
& \LS \left(
\begin{tikzpicture}[baseline={([yshift=-0.1cm]current bounding box.center)}] 
	\node[] (a) at (0,0) {};
	\node[] (a1) at (0.4,0) {};
	\node[] (a2) at (0.8,0) {};
	\node[] (b) at (0.4,-0.8) {};
	\node[] (b1) at (1.2,-0.8) {};
	\node[] (p1) at ($(a)+(-0.2,0)$) {};
	\node[] (p2) at ($(b)+(-0.6,0)$) {};
	\node[] (p3) at ($(b1)+(0.2,0)$) {};
	\node[] (p4) at ($(a2)+(0.6,0)$) {};
	\draw[line width=0.15mm] (b.center) -- (a.center);
	\draw[line width=0.15mm] (b.center) -- (a1.center);
	\draw[line width=0.15mm] (b.center) -- (a2.center);
	\draw[line width=0.15mm] (b1.center) -- (a2.center);
	\draw[line width=0.5mm] (p1.center) -- (p4.center);
	\draw[line width=0.5mm] (p2.center) -- (p3.center);
	\node at (0.8,-0.8) [circle,fill,inner sep=1.5pt]{};
\end{tikzpicture}
\right) \propto \varepsilon, \qquad \qquad \LS \left(
\begin{tikzpicture}[baseline={([yshift=-0.1cm]current bounding box.center)}] 
	\node[] (a) at (0,0) {};
	\node[] (a1) at (0.8,0) {};
	\node[] (a2) at (1.6,0) {};
	\node[] (b) at (0.4,-0.8) {};
	\node[] (b1) at (1.2,-0.8) {};
	\node[] (p1) at ($(a)+(-0.2,0)$) {};
	\node[] (p2) at ($(b)+(-0.6,0)$) {};
	\node[] (p3) at ($(b1)+(0.6,0)$) {};
	\node[] (p4) at ($(a2)+(0.2,0)$) {};
	\draw[line width=0.15mm] (b.center) -- (a.center);
	\draw[line width=0.15mm] (b.center) -- (a1.center);
	\draw[line width=0.15mm] (b1.center) -- (a1.center);
	\draw[line width=0.15mm] (b1.center) -- (a2.center);
	\draw[line width=0.5mm] (p1.center) -- (p4.center);
	\draw[line width=0.5mm] (p2.center) -- (p3.center);
	\node at (1.2,0) [circle,fill,inner sep=1.5pt]{};
\end{tikzpicture}
\right) \propto \frac{\varepsilon \, x}{x^2-1},
\end{align}
where the dot on the graviton propagator does not modify the parity and simplifies the calculation. To obtain these results, we found it useful to employ the change of variables \eqref{eq: change_of_variables_(z-c_squared)}.

For the sixth diagram in fig.~\ref{fig: diagrams_three_loop_Abelian} we can first compute the leading singularities of the bubbles, each amounting to $2 u_1 \cdot k_3$, to find
\vspace{-0.2cm}
\begin{align}
\label{eq: LS_box_two_indep_bubbles}
\LS \left( \begin{tikzpicture}[baseline={([yshift=-0.1cm]current bounding box.center)}] 
	\node[] (a) at (0,0) {};
	\node[] (a1) at (1,0) {};
	\node[] (b) at (0,-1) {};
	\node[] (b1) at (1,-1) {};
	\node[] (p1) at ($(a)+(-0.2,0)$) {};
	\node[] (p2) at ($(b)+(-0.2,0)$) {};
	\node[] (p3) at ($(b1)+(0.2,0)$) {};
	\node[] (p4) at ($(a1)+(0.2,0)$) {};
	\draw[line width=0.15mm] (b.center) -- (a.center);
	\draw[line width=0.15mm] (b1.center) -- (a1.center);
	\draw[line width=0.5mm] (p1.center) -- (p4.center);
	\draw[line width=0.5mm] (p2.center) -- (p3.center);
	\draw[line width=0.15mm] (0.5,0) arc (0:180:0.25);
	\draw[line width=0.15mm] (1,0) arc (0:180:0.25);
\end{tikzpicture} \right) & \propto \LS \left( \begin{tikzpicture}[baseline={([yshift=-0.1cm]current bounding box.center)}] 
	\node[] (a) at (0,0) {};
	\node[] (a1) at (0.8,0) {};
	\node[] (b) at (0,-0.8) {};
	\node[] (b1) at (0.8,-0.8) {};
	\node[] (p1) at ($(a)+(-0.2,0)$) {};
	\node[] (p2) at ($(b)+(-0.2,0)$) {};
	\node[] (p3) at ($(b1)+(0.2,0)$) {};
	\node[] (p4) at ($(a1)+(0.2,0)$) {};
	\draw[line width=0.15mm] (b.center) -- (a.center);
	\draw[line width=0.15mm] (b1.center) -- (a1.center);
	\draw[line width=0.5mm] (p1.center) -- (a.center);
	\draw[line width=0.5mm, dashed, dash phase=1pt, postaction={decorate}] (a.center) -- node[sloped, allow upside down, label={[xshift=0cm, yshift=-0.2cm]$2u_1 {\cdot} k_3$}] {\midarrow} (a1.center);
	\draw[line width=0.5mm] (a1.center) -- (p4.center);
	\draw[line width=0.5mm] (p2.center) -- node[sloped, allow upside down, label={[xshift=0cm, yshift=-0.85cm]\textcolor{white}{$2u_2 {\cdot} k_3$}}] {} (p3.center);
\end{tikzpicture} {\times} (2u_1 {\cdot} k_3)^2 \right) \propto \frac{|q| \, (x^2-1)^2}{x^2}, \\
\LS \left( \begin{tikzpicture}[baseline={([yshift=-0.1cm]current bounding box.center)}] 
	\node[] (a) at (0,0) {};
	\node[] (a1) at (1,0) {};
	\node[] (b) at (0,-1) {};
	\node[] (b1) at (1,-1) {};
	\node[] (p1) at ($(a)+(-0.2,0)$) {};
	\node[] (p2) at ($(b)+(-0.2,0)$) {};
	\node[] (p3) at ($(b1)+(0.2,0)$) {};
	\node[] (p4) at ($(a1)+(0.2,0)$) {};
	\draw[line width=0.15mm] (b.center) -- (a.center);
	\draw[line width=0.15mm] (b1.center) -- (a1.center);
	\draw[line width=0.5mm] (p1.center) -- (p4.center);
	\draw[line width=0.5mm] (p2.center) -- (p3.center);
	\draw[line width=0.15mm] (0.5,0) arc (0:180:0.25);
	\draw[line width=0.15mm] (1,0) arc (0:180:0.25);
\end{tikzpicture} {\times} (2u_1 {\cdot} k_3) \right) & \propto \LS \left( \begin{tikzpicture}[baseline={([yshift=-0.1cm]current bounding box.center)}] 
	\node[] (a) at (0,0) {};
	\node[] (a1) at (0.8,0) {};
	\node[] (b) at (0,-0.8) {};
	\node[] (b1) at (0.8,-0.8) {};
	\node[] (p1) at ($(a)+(-0.2,0)$) {};
	\node[] (p2) at ($(b)+(-0.2,0)$) {};
	\node[] (p3) at ($(b1)+(0.2,0)$) {};
	\node[] (p4) at ($(a1)+(0.2,0)$) {};
	\draw[line width=0.15mm] (b.center) -- (a.center);
	\draw[line width=0.15mm] (b1.center) -- (a1.center);
	\draw[line width=0.5mm] (p1.center) -- (a.center);
	\draw[line width=0.5mm, dashed, dash phase=1pt, postaction={decorate}] (a.center) -- node[sloped, allow upside down, label={[xshift=0cm, yshift=-0.2cm]$2u_1 {\cdot} k_3$}] {\midarrow} (a1.center);
	\draw[line width=0.5mm] (a1.center) -- (p4.center);
	\draw[line width=0.5mm] (p2.center) -- node[sloped, allow upside down, label={[xshift=0cm, yshift=-0.85cm]\textcolor{white}{$2u_2 {\cdot} k_3$}}] {} (p3.center);
\end{tikzpicture} {\times} (2u_1 {\cdot} k_3)^3 \right),
\label{eq: LS_box_two_indep_bubbles_ISP}
\end{align}
where $k_3$ is the loop momentum parametrizing the box, and where we used the result in eq.~\eqref{eq: LS_box_bubble_times_ISP} to calculate the leading singularity~\eqref{eq: LS_box_two_indep_bubbles}. For eq.~\eqref{eq: LS_box_two_indep_bubbles_ISP}, we can notice that it pertains to the same parity sector in the differential equations as
\begin{equation}
\LS \left( \begin{tikzpicture}[baseline={([yshift=-0.1cm]current bounding box.center)}] 
	\node[] (a) at (0,0) {};
	\node[] (a1) at (0.8,0) {};
	\node[] (b) at (0,-0.8) {};
	\node[] (b1) at (0.8,-0.8) {};
	\node[] (p1) at ($(a)+(-0.2,0)$) {};
	\node[] (p2) at ($(b)+(-0.2,0)$) {};
	\node[] (p3) at ($(b1)+(0.2,0)$) {};
	\node[] (p4) at ($(a1)+(0.2,0)$) {};
	\draw[line width=0.15mm] (b.center) -- (a.center);
	\draw[line width=0.15mm] (b1.center) -- (a1.center);
	\draw[line width=0.5mm] (p1.center) -- (p4.center);
	\draw[line width=0.5mm] (p2.center) -- (p3.center);
	\draw[line width=0.15mm] (0.8,0) arc (0:180:0.4);
\end{tikzpicture} \right) \propto \LS \left( \begin{tikzpicture}[baseline={([yshift=-0.1cm]current bounding box.center)}] 
	\node[] (a) at (0,0) {};
	\node[] (a1) at (0.8,0) {};
	\node[] (b) at (0,-0.8) {};
	\node[] (b1) at (0.8,-0.8) {};
	\node[] (p1) at ($(a)+(-0.2,0)$) {};
	\node[] (p2) at ($(b)+(-0.2,0)$) {};
	\node[] (p3) at ($(b1)+(0.2,0)$) {};
	\node[] (p4) at ($(a1)+(0.2,0)$) {};
	\draw[line width=0.15mm] (b.center) -- (a.center);
	\draw[line width=0.15mm] (b1.center) -- (a1.center);
	\draw[line width=0.5mm] (p1.center) -- (a.center);
	\draw[line width=0.5mm, dashed, dash phase=1pt, postaction={decorate}] (a.center) -- node[sloped, allow upside down, label={[xshift=0cm, yshift=-0.2cm]$2u_1 {\cdot} k_2$}] {\midarrow} (a1.center);
	\draw[line width=0.5mm] (a1.center) -- (p4.center);
	\draw[line width=0.5mm] (p2.center) -- node[sloped, allow upside down, label={[xshift=0cm, yshift=-0.85cm]\textcolor{white}{$2u_2 {\cdot} k_2$}}] {} (p3.center);
\end{tikzpicture} {\times} (2u_1 {\cdot} k_2) \right) \propto \frac{x^2-1}{x},
\end{equation}
where we used the result in eq.~\eqref{eq: LS_box_with_bubble}. For the purpose of finding non-trivial geometries, since this leading singularity is algebraic, we can conclude that our original integral in eq.~\eqref{eq: LS_box_two_indep_bubbles_ISP} also has an algebraic leading singularity and thus a dlog form.

Finally, for the last two diagrams in fig.~\ref{fig: diagrams_three_loop_Abelian}, we find it easier to compute the leading singularity when we have a dotted graviton propagator in the bubble, which does not modify the parity:
\begin{align}
& \LS \left(
\begin{tikzpicture}[baseline={([yshift=-0.1cm]current bounding box.center)}] 
	\node[] (a) at (0,0) {};
	\node[] (a1) at (1,0) {};
	\node[] (b) at (0,-1) {};
	\node[] (b1) at (1,-1) {};
	\node[] (p1) at ($(a)+(-0.2,0)$) {};
	\node[] (p2) at ($(b)+(-0.2,0)$) {};
	\node[] (p3) at ($(b1)+(0.2,0)$) {};
	\node[] (p4) at ($(a1)+(0.2,0)$) {};
	\draw[line width=0.15mm] (b.center) -- (a.center);
	\draw[line width=0.15mm] (b1.center) -- (a1.center);
	\draw[line width=0.15mm] (0.5,0) -- (b1.center);
	\draw[line width=0.5mm] (p1.center) -- (p4.center);
	\draw[line width=0.5mm] (p2.center) -- (p3.center);
	\draw[line width=0.15mm] (0.5,0) arc (0:180:0.25);
	\node at (0.25,0.25) [circle,fill,inner sep=1.5pt]{};
\end{tikzpicture}
\right) \propto \LS \left(
\raisebox{0.125cm}{\begin{tikzpicture}[baseline={([yshift=-0.1cm]current bounding box.center)}] 
	\node[] (a) at (0,0) {};
	\node[] (a1) at (1,0) {};
	\node[] (b) at (0,-1) {};
	\node[] (b1) at (1,-1) {};
	\node[] (p1) at ($(a)+(-0.2,0)$) {};
	\node[] (p2) at ($(b)+(-0.2,0)$) {};
	\node[] (p3) at ($(b1)+(0.2,0)$) {};
	\node[] (p4) at ($(a1)+(0.2,0)$) {};
	\draw[line width=0.15mm] (b.center) -- (a.center);
	\draw[line width=0.15mm] (b1.center) -- (a1.center);
	\draw[line width=0.15mm] (0.5,-1) -- (a.center);
	\draw[line width=0.5mm] (p1.center) -- (p4.center);
	\draw[line width=0.5mm] (p2.center) -- (p3.center);
	\draw[line width=0.15mm] (1,0) arc (0:180:0.5);
	\node at (0.5,0.5) [circle,fill,inner sep=1.5pt]{};
\end{tikzpicture}}
\right) \propto \frac{x}{|q| \, (x^2-1)}, \\[0.2cm]
& \LS \left(
\begin{tikzpicture}[baseline={([yshift=-0.1cm]current bounding box.center)}] 
	\node[] (a) at (0,0) {};
	\node[] (a1) at (1,0) {};
	\node[] (b) at (0,-1) {};
	\node[] (b1) at (1,-1) {};
	\node[] (p1) at ($(a)+(-0.2,0)$) {};
	\node[] (p2) at ($(b)+(-0.2,0)$) {};
	\node[] (p3) at ($(b1)+(0.2,0)$) {};
	\node[] (p4) at ($(a1)+(0.2,0)$) {};
	\draw[line width=0.15mm] (b.center) -- (a.center);
	\draw[line width=0.15mm] (b1.center) -- (a1.center);
	\draw[line width=0.15mm] (0.5,0) -- (b1.center);
	\draw[line width=0.5mm] (p1.center) -- (p4.center);
	\draw[line width=0.5mm] (p2.center) -- (p3.center);
	\draw[line width=0.15mm] (0.5,0) arc (0:180:0.25);
	\node at (0.25,0.25) [circle,fill,inner sep=1.5pt]{};
\end{tikzpicture} {\times} (2u_1 {\cdot} k_3)
\right) \propto \frac{x}{x^2-1}, \qquad \quad \LS \left(
\raisebox{0.125cm}{\begin{tikzpicture}[baseline={([yshift=-0.1cm]current bounding box.center)}] 
	\node[] (a) at (0,0) {};
	\node[] (a1) at (1,0) {};
	\node[] (b) at (0,-1) {};
	\node[] (b1) at (1,-1) {};
	\node[] (p1) at ($(a)+(-0.2,0)$) {};
	\node[] (p2) at ($(b)+(-0.2,0)$) {};
	\node[] (p3) at ($(b1)+(0.2,0)$) {};
	\node[] (p4) at ($(a1)+(0.2,0)$) {};
	\draw[line width=0.15mm] (b.center) -- (a.center);
	\draw[line width=0.15mm] (b1.center) -- (a1.center);
	\draw[line width=0.15mm] (0.5,-1) -- (a.center);
	\draw[line width=0.5mm] (p1.center) -- (p4.center);
	\draw[line width=0.5mm] (p2.center) -- (p3.center);
	\draw[line width=0.15mm] (1,0) arc (0:180:0.5);
	\node at (0.5,0.5) [circle,fill,inner sep=1.5pt]{};
\end{tikzpicture}} {\times} (2u_1 {\cdot} k_3)
\right) \propto 1,
\end{align}
where the change of variables \eqref{eq: change_of_variables_(z-c_squared)} is used to obtain the results, and where in both diagrams $k_3$ parametrizes the loop below the bubble.

\subsection{Three-loop non-Abelian diagrams}
\label{sec:three_loop_non-abelian}

Turning to the non-Abelian sector, let us start with the first two diagrams in fig.~\ref{fig: diagrams_three_loop_non_Abelian}. Both diagrams contain a double-triangle sub-diagram with momentum transfer $k$ that satisfies $k^2 = 0$ at the maximal cut. Therefore, by virtue of eq.~\eqref{eq: LS_double_triangle_u1q_general}, we can set $\LS \propto 1$ for the double triangle, finding a simplification
\begin{align}
\LS \left( \begin{tikzpicture}[baseline={([yshift=-0.1cm]current bounding box.center)}] 
	\node[] (a) at (0,0) {};
	\node[] (a1) at (0.5,0) {};
	\node[] (a2) at (1,0) {};
	\node[] (b) at (0,-0.5) {};
	\node[] (c) at (0,-1) {};
	\node[] (c1) at (1,-1) {};
	\node[] (p1) at ($(a)+(-0.2,0)$) {};
	\node[] (p2) at ($(c)+(-0.2,0)$) {};
	\node[] (p3) at ($(c1)+(0.2,0)$) {};
	\node[] (p4) at ($(a2)+(0.2,0)$) {};
	\draw[line width=0.15mm] (b.center) -- (a.center);
	\draw[line width=0.15mm] (b.center) -- (a1.center);
	\draw[line width=0.15mm] (b.center) -- (a2.center);
	\draw[line width=0.15mm] (b.center) -- (c.center);
	\draw[line width=0.15mm] (c1.center) -- (a2.center);
	\draw[line width=0.5mm] (p1.center) -- (p4.center);
	\draw[line width=0.5mm] (p2.center) -- (p3.center);
\end{tikzpicture} \right) & \propto \LS \left( \begin{tikzpicture}[baseline={([yshift=-0.1cm]current bounding box.center)}] 
	\node[] (a) at (0,0) {};
	\node[] (a1) at (1,0) {};
	\node[] (b) at (0,-1) {};
	\node[] (b1) at (1,-1) {};
	\node[] (p1) at ($(a)+(-0.2,0)$) {};
	\node[] (p2) at ($(b)+(-0.2,0)$) {};
	\node[] (p3) at ($(b1)+(0.2,0)$) {};
	\node[] (p4) at ($(a1)+(0.2,0)$) {};
	\draw[line width=0.15mm] (b.center) -- (a.center);
	\draw[line width=0.15mm] (b1.center) -- (a1.center);
	\draw[line width=0.5mm] (p1.center) -- (a.center);
	\draw[line width=0.5mm, dashed, dash phase=3pt, postaction={decorate}] (a.center) -- node[sloped, allow upside down, label={[xshift=0cm, yshift=-0.2cm]$2u_1 {\cdot} k$}] {\midarrow} (a1.center);
	\draw[line width=0.5mm] (a1.center) -- (p4.center);
	\draw[line width=0.5mm] (p2.center) -- node[sloped, allow upside down, label={[xshift=0cm, yshift=-0.85cm]\textcolor{white}{$2u_2 {\cdot} k$}}] {} (p3.center);
\end{tikzpicture} \right) \propto \frac{1}{|q|}, \\
\LS \left( \begin{tikzpicture}[baseline={([yshift=-0.1cm]current bounding box.center)}] 
	\node[] (a) at (0,0) {};
	\node[] (a1) at (0.25,0) {};
	\node[] (a2) at (0.5,0) {};
	\node[] (a3) at (1,0) {};
	\node[] (b) at (0.25,-0.5) {};
	\node[] (c) at (0.5,-1) {};
	\node[] (p1) at ($(a)+(-0.2,0)$) {};
	\node[] (p2) at ($(c)+(-0.7,0)$) {};
	\node[] (p3) at ($(c)+(0.7,0)$) {};
	\node[] (p4) at ($(a3)+(0.2,0)$) {};
	\draw[line width=0.15mm] (a.center) -- (c.center);
	\draw[line width=0.15mm] (b.center) -- (a1.center);
	\draw[line width=0.15mm] (b.center) -- (a2.center);
	\draw[line width=0.15mm] (c.center) -- (a3.center);
	\draw[line width=0.5mm] (p1.center) -- (p4.center);
	\draw[line width=0.5mm] (p2.center) -- (p3.center);
\end{tikzpicture} \right) & \propto \LS \left( \begin{tikzpicture}[baseline={([yshift=-0.1cm]current bounding box.center)}] 
	\node[] (a) at (0,0) {};
	\node[] (a1) at (1,0) {};
	\node[] (b) at (0.5,-1) {};
	\node[] (p1) at ($(a)+(-0.2,0)$) {};
	\node[] (p2) at ($(b)+(-0.7,0)$) {};
	\node[] (p3) at ($(b)+(0.7,0)$) {};
	\node[] (p4) at ($(a1)+(0.2,0)$) {};
	\draw[line width=0.15mm] (a.center) -- (b.center);
	\draw[line width=0.15mm] (b.center) -- (a1.center);
	\draw[line width=0.5mm] (p1.center) -- (p4.center);
	\draw[line width=0.5mm] (p2.center) -- (p3.center);
\end{tikzpicture} \right) \propto \frac{1}{|q|},
\end{align}
and similarly for the integral of opposite parity. Therefore, these diagrams cannot contribute with new functions at three loops.

\begin{figure}[t]
\centering
\begin{tikzpicture}[baseline=(current bounding box.center)] 
	\node[] (a) at (0,0) {};
	\node[] (a1) at (2,0) {};
	\node[] (a2) at (4,0) {};
	\node[] (b) at (0,-2) {};
	\node[] (b1) at (2,-2) {};
	\node[] (b2) at (4,-2) {};
	\node[] (c) at (0,-4) {};
	\node[] (c1) at (4,-4) {};
	\node[label=left:{$\overline{p}_1{-}\frac{q}{2}$}] (p1) at ($(a)+(-0.5,0)$) {};
	\node[label=left:{$\overline{p}_2{+}\frac{q}{2}$}] (p2) at ($(c)+(-0.5,0)$) {};
	\node[label=right:{$\overline{p}_2{-}\frac{q}{2}$}] (p3) at ($(c1)+(0.5,0)$) {};
	\node[label=right:{$\overline{p}_1{+}\frac{q}{2}$}] (p4) at ($(a2)+(0.5,0)$) {};
	\draw[line width=0.15mm, postaction={decorate}] (b.center) -- node[sloped, allow upside down, label={[xshift=0.15cm, yshift=0cm]$k_1$}] {\midarrow} (a.center);
	\draw[line width=0.15mm, postaction={decorate}] (a1.center) -- node[sloped, allow upside down, label={[xshift=-0.15cm, yshift=0cm]$k_1{+}k_2$}] {\midarrow} (b1.center);
	\draw[line width=0.15mm, postaction={decorate}] (b2.center) -- node[sloped, allow upside down, label={[xshift=1.45cm, yshift=0cm]$k_2{+}q$}] {\midarrow} (a2.center);
	\draw[line width=0.15mm, postaction={decorate}] (b1.center) -- node[sloped, allow upside down, label={[xshift=0cm, yshift=0.9cm]$k_1{+}k_3$}] {\midarrow} (b.center);
	\draw[line width=0.15mm, postaction={decorate}] (b1.center) -- node[sloped, allow upside down, label={[xshift=0cm, yshift=-0.2cm]$k_2{-}k_3$}] {\midarrow} (b2.center);
	\draw[line width=0.15mm, postaction={decorate}] (b.center) -- node[sloped, allow upside down, label={[xshift=-0.9cm, yshift=0.05cm]$k_3$}] {\midarrow} (c.center);
	\draw[line width=0.15mm, postaction={decorate}] (c1.center) -- node[sloped, allow upside down, label={[xshift=1.45cm, yshift=0cm]$k_3{+}q$}] {\midarrow} (b2.center);
	\draw[line width=0.5mm, postaction={decorate}] (a.center) -- node[sloped, allow upside down, label={[xshift=0cm, yshift=-0.15cm]$2u_1 {\cdot} k_1$}] {\midarrow} (a1.center);
	\draw[line width=0.5mm, postaction={decorate}] (a1.center) -- node[sloped, allow upside down, label={[xshift=0cm, yshift=-0.15cm]$-2u_1 {\cdot} k_2$}] {\midarrow} (a2.center);
	\draw[line width=0.5mm, postaction={decorate}] (c.center) -- node[sloped, allow upside down, label={[xshift=0cm, yshift=-0.9cm]$2u_2 {\cdot} k_3$}] {\midarrow} (c1.center);
	\draw[line width=0.5mm, postaction={decorate}] (p1.center) -- node[sloped, allow upside down] {\midarrow} (a.center);
	\draw[line width=0.5mm, postaction={decorate}] (a2.center) -- node[sloped, allow upside down] {\midarrow} (p4.center);
	\draw[line width=0.5mm, postaction={decorate}] (p2.center) -- node[sloped, allow upside down] {\midarrow} (c.center);
	\draw[line width=0.5mm, postaction={decorate}] (c1.center) -- node[sloped, allow upside down] {\midarrow} (p3.center);
\end{tikzpicture}
\caption{Parametrization of the loop momenta for the three-loop Mondrian diagram.}
\label{fig: diag_half-tennis_court}
\end{figure}
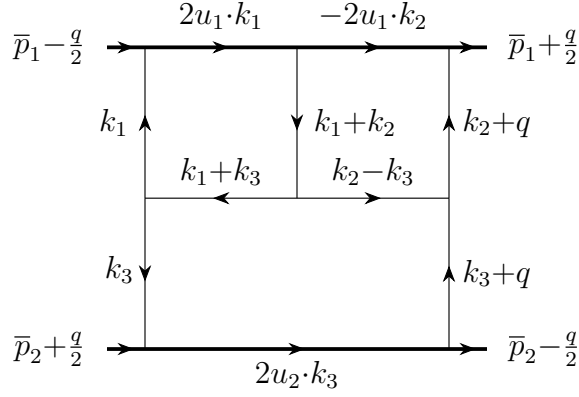

Let us then consider the three-loop Mondrian diagram, also known as the half tennis court diagram, as parametrized in fig.~\ref{fig: diag_half-tennis_court}. Using the integration order $k_1 \to k_2 \to k_3$ and introducing $z_{11}=k_2^2$ and $z_{12}=2u_1 \cdot k_3$ as extra Baikov variables, we obtain
\begin{align}
\begin{tikzpicture}[baseline={([yshift=-0.1cm]current bounding box.center)}] 
	\node[] (a) at (0,0) {};
	\node[] (a1) at (0.5,0) {};
	\node[] (a2) at (1,0) {};
	\node[] (b) at (0,-0.5) {};
	\node[] (b1) at (0.5,-0.5) {};
	\node[] (b2) at (1,-0.5) {};
	\node[] (c) at (0,-1) {};
	\node[] (c1) at (1,-1) {};
	\node[] (p1) at ($(a)+(-0.2,0)$) {};
	\node[] (p2) at ($(c)+(-0.2,0)$) {};
	\node[] (p3) at ($(c1)+(0.2,0)$) {};
	\node[] (p4) at ($(a2)+(0.2,0)$) {};
	\draw[line width=0.15mm] (b.center) -- (a.center);
	\draw[line width=0.15mm] (b1.center) -- (a1.center);
	\draw[line width=0.15mm] (b2.center) -- (a2.center);
	\draw[line width=0.15mm] (b2.center) -- (b.center);
	\draw[line width=0.15mm] (b.center) -- (c.center);
	\draw[line width=0.15mm] (c1.center) -- (b2.center);
	\draw[line width=0.5mm] (p1.center) -- (p4.center);
	\draw[line width=0.5mm] (p2.center) -- (p3.center);
\end{tikzpicture} & \propto \int \frac{d z_1 \cdots d z_{10}}{z_1 \cdots z_{10}} \frac{d z_{11} d z_{12}}{\sqrt{\det G(k_1,k_2,k_3,u_1)}}
\nonumber \\
& \quad \times \frac{1}{\sqrt{\det G(k_2,k_3,u_1,q)} \sqrt{\det G(k_3,u_1,u_2,q)}}, \\
\LS \left( \begin{tikzpicture}[baseline={([yshift=-0.1cm]current bounding box.center)}] 
	\node[] (a) at (0,0) {};
	\node[] (a1) at (0.5,0) {};
	\node[] (a2) at (1,0) {};
	\node[] (b) at (0,-0.5) {};
	\node[] (b1) at (0.5,-0.5) {};
	\node[] (b2) at (1,-0.5) {};
	\node[] (c) at (0,-1) {};
	\node[] (c1) at (1,-1) {};
	\node[] (p1) at ($(a)+(-0.2,0)$) {};
	\node[] (p2) at ($(c)+(-0.2,0)$) {};
	\node[] (p3) at ($(c1)+(0.2,0)$) {};
	\node[] (p4) at ($(a2)+(0.2,0)$) {};
	\draw[line width=0.15mm] (b.center) -- (a.center);
	\draw[line width=0.15mm] (b1.center) -- (a1.center);
	\draw[line width=0.15mm] (b2.center) -- (a2.center);
	\draw[line width=0.15mm] (b2.center) -- (b.center);
	\draw[line width=0.15mm] (b.center) -- (c.center);
	\draw[line width=0.15mm] (c1.center) -- (b2.center);
	\draw[line width=0.5mm] (p1.center) -- (p4.center);
	\draw[line width=0.5mm] (p2.center) -- (p3.center);
\end{tikzpicture} \right) & \propto \LS \left( \int \frac{x \, d z_{11} d z_{12}}{|q| (q^2- z_{11}) z_{11} z_{12}^2 \sqrt{q^2(x^2-1)^2-4x^2 z_{12}^2}} \right),
\end{align}
where the leading singularity contains a double pole for $z_{12}$. As discussed earlier in sec.~\ref{sec:methodology_systematics}, this can be remedied by adding a numerator factor, which amounts to choosing a different master integral with powers of an ISP in the numerator. In this case, we can take it to be $z_{12}^2$ to avoid modifying the parity of the diagram, leading to
\begin{equation}
\LS \left( \begin{tikzpicture}[baseline={([yshift=-0.1cm]current bounding box.center)}] 
	\node[] (a) at (0,0) {};
	\node[] (a1) at (0.5,0) {};
	\node[] (a2) at (1,0) {};
	\node[] (b) at (0,-0.5) {};
	\node[] (b1) at (0.5,-0.5) {};
	\node[] (b2) at (1,-0.5) {};
	\node[] (c) at (0,-1) {};
	\node[] (c1) at (1,-1) {};
	\node[] (p1) at ($(a)+(-0.2,0)$) {};
	\node[] (p2) at ($(c)+(-0.2,0)$) {};
	\node[] (p3) at ($(c1)+(0.2,0)$) {};
	\node[] (p4) at ($(a2)+(0.2,0)$) {};
	\draw[line width=0.15mm] (b.center) -- (a.center);
	\draw[line width=0.15mm] (b1.center) -- (a1.center);
	\draw[line width=0.15mm] (b2.center) -- (a2.center);
	\draw[line width=0.15mm] (b2.center) -- (b.center);
	\draw[line width=0.15mm] (b.center) -- (c.center);
	\draw[line width=0.15mm] (c1.center) -- (b2.center);
	\draw[line width=0.5mm] (p1.center) -- (p4.center);
	\draw[line width=0.5mm] (p2.center) -- (p3.center);
\end{tikzpicture} {\times} (2 u_1 {\cdot} k_3)^2 \right) \propto \frac{1}{|q|^3},
\end{equation}
while the even-parity integral can be obtained instead with
\begin{equation}
\LS \left( \begin{tikzpicture}[baseline={([yshift=-0.1cm]current bounding box.center)}] 
	\node[] (a) at (0,0) {};
	\node[] (a1) at (0.5,0) {};
	\node[] (a2) at (1,0) {};
	\node[] (b) at (0,-0.5) {};
	\node[] (b1) at (0.5,-0.5) {};
	\node[] (b2) at (1,-0.5) {};
	\node[] (c) at (0,-1) {};
	\node[] (c1) at (1,-1) {};
	\node[] (p1) at ($(a)+(-0.2,0)$) {};
	\node[] (p2) at ($(c)+(-0.2,0)$) {};
	\node[] (p3) at ($(c1)+(0.2,0)$) {};
	\node[] (p4) at ($(a2)+(0.2,0)$) {};
	\draw[line width=0.15mm] (b.center) -- (a.center);
	\draw[line width=0.15mm] (b1.center) -- (a1.center);
	\draw[line width=0.15mm] (b2.center) -- (a2.center);
	\draw[line width=0.15mm] (b2.center) -- (b.center);
	\draw[line width=0.15mm] (b.center) -- (c.center);
	\draw[line width=0.15mm] (c1.center) -- (b2.center);
	\draw[line width=0.5mm] (p1.center) -- (p4.center);
	\draw[line width=0.5mm] (p2.center) -- (p3.center);
\end{tikzpicture} {\times} (2 u_1 {\cdot} k_3) \right) \propto \frac{x}{q^4 (x^2-1)}.
\end{equation}

After removing all cases with trivial leading singularity or that can be related to lower sectors, for this diagram there are 4 relevant subsectors, which we gather in fig.~\ref{fig: diagrams_three_loop_non_Abelian}. For the odd sector of the first three diagrams, we find:
\begin{equation}
\LS \left( \begin{tikzpicture}[baseline={([yshift=-0.1cm]current bounding box.center)}] 
	\node[] (a) at (0,0) {};
	\node[] (a1) at (0.5,0) {};
	\node[] (a2) at (1,0) {};
	\node[] (b) at (0,-0.5) {};
	\node[] (b1) at (0.5,-0.5) {};
	\node[] (c) at (0,-1) {};
	\node[] (c1) at (1,-1) {};
	\node[] (p1) at ($(a)+(-0.2,0)$) {};
	\node[] (p2) at ($(c)+(-0.2,0)$) {};
	\node[] (p3) at ($(c1)+(0.2,0)$) {};
	\node[] (p4) at ($(a2)+(0.2,0)$) {};
	\draw[line width=0.15mm] (b.center) -- (a.center);
	\draw[line width=0.15mm] (b1.center) -- (a1.center);
	\draw[line width=0.15mm] (c1.center) -- (a2.center);
	\draw[line width=0.15mm] (b1.center) -- (b.center);
	\draw[line width=0.15mm] (b.center) -- (c.center);
	\draw[line width=0.15mm] (c1.center) -- (b1.center);
	\draw[line width=0.5mm] (p1.center) -- (p4.center);
	\draw[line width=0.5mm] (p2.center) -- (p3.center);
\end{tikzpicture} \times k_2^2, \, \begin{tikzpicture}[baseline={([yshift=-0.1cm]current bounding box.center)}] 
	\node[] (a) at (0,0) {};
	\node[] (a1) at (0.5,0) {};
	\node[] (a2) at (1,0) {};
	\node[] (b) at (0.5,-0.5) {};
	\node[] (c) at (0,-1) {};
	\node[] (c1) at (1,-1) {};
	\node[] (p1) at ($(a)+(-0.2,0)$) {};
	\node[] (p2) at ($(c)+(-0.2,0)$) {};
	\node[] (p3) at ($(c1)+(0.2,0)$) {};
	\node[] (p4) at ($(a2)+(0.2,0)$) {};
	\draw[line width=0.15mm] (c.center) -- (a.center);
	\draw[line width=0.15mm] (b.center) -- (a1.center);
	\draw[line width=0.15mm] (c1.center) -- (a2.center);
	\draw[line width=0.15mm] (b.center) -- (c.center);
	\draw[line width=0.15mm] (c1.center) -- (b.center);
	\draw[line width=0.5mm] (p1.center) -- (p4.center);
	\draw[line width=0.5mm] (p2.center) -- (p3.center);
\end{tikzpicture} \right) \propto \frac{x}{|q| \, (x^2-1)},
\end{equation}
\begin{equation}
\LS \left( \begin{tikzpicture}[baseline={([yshift=-0.1cm]current bounding box.center)}] 
	\node[] (a) at (0,0) {};
	\node[] (a1) at (0.5,0) {};
	\node[] (a2) at (1,0) {};
	\node[] (b) at (0,-0.5) {};
	\node[] (b1) at (1,-0.5) {};
	\node[] (c) at (0,-1) {};
	\node[] (c1) at (1,-1) {};
	\node[] (p1) at ($(a)+(-0.2,0)$) {};
	\node[] (p2) at ($(c)+(-0.2,0)$) {};
	\node[] (p3) at ($(c1)+(0.2,0)$) {};
	\node[] (p4) at ($(a2)+(0.2,0)$) {};
	\draw[line width=0.15mm] (b.center) -- (a.center);
	\draw[line width=0.15mm] (b.center) -- (a1.center);
	\draw[line width=0.15mm] (b1.center) -- (a2.center);
	\draw[line width=0.15mm] (b1.center) -- (b.center);
	\draw[line width=0.15mm] (b.center) -- (c.center);
	\draw[line width=0.15mm] (c1.center) -- (b1.center);
	\draw[line width=0.5mm] (p1.center) -- (p4.center);
	\draw[line width=0.5mm] (p2.center) -- (p3.center);
\end{tikzpicture} \right) \propto \frac{x}{|q|^3 \, (x^2-1)},
\vspace{0.2cm}
\end{equation}
where we parametrize the loop momenta following the same convention as in fig.~\ref{fig: diag_half-tennis_court}. Notice that for one of the subsectors we need to add an ISP in the numerator in order to avoid higher-order poles.

For the even-parity integrals, we have
\begin{equation}
\LS \left( \begin{tikzpicture}[baseline={([yshift=-0.1cm]current bounding box.center)}] 
	\node[] (a) at (0,0) {};
	\node[] (a1) at (0.5,0) {};
	\node[] (a2) at (1,0) {};
	\node[] (b) at (0,-0.5) {};
	\node[] (b1) at (0.5,-0.5) {};
	\node[] (c) at (0,-1) {};
	\node[] (c1) at (1,-1) {};
	\node[] (p1) at ($(a)+(-0.2,0)$) {};
	\node[] (p2) at ($(c)+(-0.2,0)$) {};
	\node[] (p3) at ($(c1)+(0.2,0)$) {};
	\node[] (p4) at ($(a2)+(0.2,0)$) {};
	\draw[line width=0.15mm] (b.center) -- (a.center);
	\draw[line width=0.15mm] (b1.center) -- (a1.center);
	\draw[line width=0.15mm] (c1.center) -- (a2.center);
	\draw[line width=0.15mm] (b1.center) -- (b.center);
	\draw[line width=0.15mm] (b.center) -- (c.center);
	\draw[line width=0.15mm] (c1.center) -- (b1.center);
	\draw[line width=0.5mm] (p1.center) -- (p4.center);
	\draw[line width=0.5mm] (p2.center) -- (p3.center);
\end{tikzpicture} {\times} (2u_1 {\cdot} k_3) \right) \propto \frac{x}{q^2 (x^2-1)}, \quad
\LS \left( \begin{tikzpicture}[baseline={([yshift=-0.1cm]current bounding box.center)}] 
	\node[] (a) at (0,0) {};
	\node[] (a1) at (0.5,0) {};
	\node[] (a2) at (1,0) {};
	\node[] (b) at (0,-0.5) {};
	\node[] (b1) at (1,-0.5) {};
	\node[] (c) at (0,-1) {};
	\node[] (c1) at (1,-1) {};
	\node[] (p1) at ($(a)+(-0.2,0)$) {};
	\node[] (p2) at ($(c)+(-0.2,0)$) {};
	\node[] (p3) at ($(c1)+(0.2,0)$) {};
	\node[] (p4) at ($(a2)+(0.2,0)$) {};
	\draw[line width=0.15mm] (b.center) -- (a.center);
	\draw[line width=0.15mm] (b.center) -- (a1.center);
	\draw[line width=0.15mm] (b1.center) -- (a2.center);
	\draw[line width=0.15mm] (b1.center) -- (b.center);
	\draw[line width=0.15mm] (b.center) -- (c.center);
	\draw[line width=0.15mm] (c1.center) -- (b1.center);
	\draw[line width=0.5mm] (p1.center) -- (p4.center);
	\draw[line width=0.5mm] (p2.center) -- (p3.center);
\end{tikzpicture} {\times} (2u_1 {\cdot} k_3) \right) \propto \frac{1}{q^2},
\end{equation}
\begin{equation}
\LS \left( \begin{tikzpicture}[baseline={([yshift=-0.1cm]current bounding box.center)}] 
	\node[] (a) at (0,0) {};
	\node[] (a1) at (0.5,0) {};
	\node[] (a2) at (1,0) {};
	\node[] (b) at (0.5,-0.5) {};
	\node[] (c) at (0,-1) {};
	\node[] (c1) at (1,-1) {};
	\node[] (p1) at ($(a)+(-0.2,0)$) {};
	\node[] (p2) at ($(c)+(-0.2,0)$) {};
	\node[] (p3) at ($(c1)+(0.2,0)$) {};
	\node[] (p4) at ($(a2)+(0.2,0)$) {};
	\draw[line width=0.15mm] (c.center) -- (a.center);
	\draw[line width=0.15mm] (b.center) -- (a1.center);
	\draw[line width=0.15mm] (c1.center) -- (a2.center);
	\draw[line width=0.15mm] (b.center) -- (c.center);
	\draw[line width=0.15mm] (c1.center) -- (b.center);
	\draw[line width=0.5mm] (p1.center) -- (p4.center);
	\draw[line width=0.5mm] (p2.center) -- (p3.center);
\end{tikzpicture} {\times} (2u_2 {\cdot} k_1) \right) \propto 1.
\end{equation}

As can be seen, the leading singularities of the majority of non-Abelian diagrams are algebraic, and all integrals in the three-loop PM expansion up to this point would therefore have a dlog form. This is however not the case for the last diagram in fig.~\ref{fig: diagrams_three_loop_non_Abelian}, as we shall see in the following.

\subsection{A K3 surface at three loops}
\label{sec:three_loop_K3}

\begin{figure}[t]
\centering
\begin{tikzpicture}[baseline=(current bounding box.center)] 
	\node[] (a) at (0,0) {};
	\node[] (a1) at (2,0) {};
	\node[] (a2) at (4,0) {};
	\node[] (b) at (0,-2) {};
	\node[] (c) at (0,-4) {};
	\node[] (c1) at (4,-4) {};
	\node[label=left:{$\overline{p}_1{-}\frac{q}{2}$}] (p1) at ($(a)+(-0.5,0)$) {};
	\node[label=left:{$\overline{p}_2{+}\frac{q}{2}$}] (p2) at ($(c)+(-0.5,0)$) {};
	\node[label=right:{$\overline{p}_2{-}\frac{q}{2}$}] (p3) at ($(c1)+(0.5,0)$) {};
	\node[label=right:{$\overline{p}_1{+}\frac{q}{2}$}] (p4) at ($(a2)+(0.5,0)$) {};
	\draw[line width=0.15mm, postaction={decorate}] (b.center) -- node[sloped, allow upside down, label={[xshift=0.2cm, yshift=0cm]$k_1$}] {\midarrow} (a.center);
	\draw[line width=0.15mm, postaction={decorate}] (a1.center) -- node[sloped, allow upside down, label={[xshift=0cm, yshift=0.4cm]$k_1{+}k_2$}] {\midarrow} (b.center);
	\draw[line width=0.15mm, postaction={decorate}] (c1.center) -- node[sloped, allow upside down, label={[xshift=1.45cm, yshift=0cm]$k_2{+}q$}] {\midarrow} (a2.center);
	\draw[line width=0.15mm, postaction={decorate}] (b.center) -- node[sloped, allow upside down, label={[xshift=0cm, yshift=-0.35cm]$k_2{-}k_3$}] {\midarrow} (c1.center);
	\draw[line width=0.15mm, postaction={decorate}] (b.center) -- node[sloped, allow upside down, label={[xshift=-0.9cm, yshift=0.05cm]$k_3$}] {\midarrow} (c.center);
	\draw[line width=0.5mm, postaction={decorate}] (a.center) -- node[sloped, allow upside down, label={[xshift=0cm, yshift=-0.15cm]$2u_1 {\cdot} k_1$}] {\midarrow} (a1.center);
	\draw[line width=0.5mm, postaction={decorate}] (a1.center) -- node[sloped, allow upside down, label={[xshift=0cm, yshift=-0.15cm]$-2u_1 {\cdot} k_2$}] {\midarrow} (a2.center);
	\draw[line width=0.5mm, postaction={decorate}] (c.center) -- node[sloped, allow upside down, label={[xshift=0cm, yshift=-0.9cm]$2u_2 {\cdot} k_3$}] {\midarrow} (c1.center);
	\draw[line width=0.5mm, postaction={decorate}] (p1.center) -- node[sloped, allow upside down] {\midarrow} (a.center);
	\draw[line width=0.5mm, postaction={decorate}] (a2.center) -- node[sloped, allow upside down] {\midarrow} (p4.center);
	\draw[line width=0.5mm, postaction={decorate}] (p2.center) -- node[sloped, allow upside down] {\midarrow} (c.center);
	\draw[line width=0.5mm, postaction={decorate}] (c1.center) -- node[sloped, allow upside down] {\midarrow} (p3.center);
\end{tikzpicture}
\caption{Parametrization of the loop momenta for the last three-loop classical non-Abelian subsector.}
\label{fig: diag_3-loop_elliptic}
\end{figure}
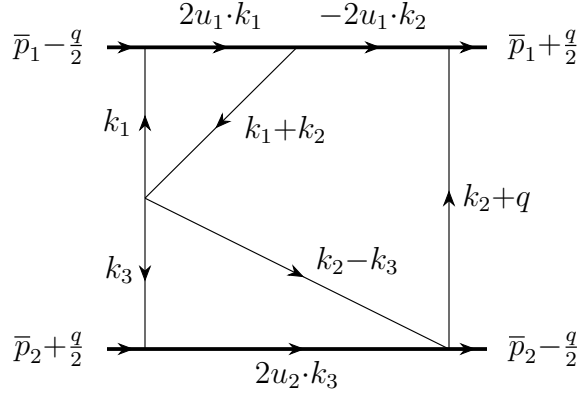

Let us now study the last non-Abelian subsector, which we parametrize as shown in fig.~\ref{fig: diag_3-loop_elliptic}. Using the integration order $k_1 \to k_3 \to k_2$, and introducing $z_9=k_2^2$ and $z_{10}=2u_2 \cdot k_2$, we obtain
\begin{align}
& \begin{tikzpicture}[baseline={([yshift=-0.1cm]current bounding box.center)}] 
	\node[] (a) at (0,0) {};
	\node[] (a1) at (0.5,0) {};
	\node[] (a2) at (1,0) {};
	\node[] (b) at (0,-0.5) {};
	\node[] (c) at (0,-1) {};
	\node[] (c1) at (1,-1) {};
	\node[] (p1) at ($(a)+(-0.2,0)$) {};
	\node[] (p2) at ($(c)+(-0.2,0)$) {};
	\node[] (p3) at ($(c1)+(0.2,0)$) {};
	\node[] (p4) at ($(a2)+(0.2,0)$) {};
	\draw[line width=0.15mm] (b.center) -- (a.center);
	\draw[line width=0.15mm] (b.center) -- (a1.center);
	\draw[line width=0.15mm] (c1.center) -- (a2.center);
	\draw[line width=0.15mm] (b.center) -- (c.center);
	\draw[line width=0.15mm] (c1.center) -- (b.center);
	\draw[line width=0.5mm] (p1.center) -- (p4.center);
	\draw[line width=0.5mm] (p2.center) -- (p3.center);
\end{tikzpicture} \propto \int \frac{d z_1 \cdots d z_{8}}{z_1 \cdots z_{8}} \frac{d z_9 d z_{10}}{\sqrt{\det G(k_2,u_1)} \sqrt{\det G(k_2,u_2)} \sqrt{\det G(k_2,u_1,u_2,q)}}, \\[0.2cm]
& \LS \left( \begin{tikzpicture}[baseline={([yshift=-0.1cm]current bounding box.center)}] 
	\node[] (a) at (0,0) {};
	\node[] (a1) at (0.5,0) {};
	\node[] (a2) at (1,0) {};
	\node[] (b) at (0,-0.5) {};
	\node[] (c) at (0,-1) {};
	\node[] (c1) at (1,-1) {};
	\node[] (p1) at ($(a)+(-0.2,0)$) {};
	\node[] (p2) at ($(c)+(-0.2,0)$) {};
	\node[] (p3) at ($(c1)+(0.2,0)$) {};
	\node[] (p4) at ($(a2)+(0.2,0)$) {};
	\draw[line width=0.15mm] (b.center) -- (a.center);
	\draw[line width=0.15mm] (b.center) -- (a1.center);
	\draw[line width=0.15mm] (c1.center) -- (a2.center);
	\draw[line width=0.15mm] (b.center) -- (c.center);
	\draw[line width=0.15mm] (c1.center) -- (b.center);
	\draw[line width=0.5mm] (p1.center) -- (p4.center);
	\draw[line width=0.5mm] (p2.center) -- (p3.center);
\end{tikzpicture} \right) \propto \LS \left( \int \frac{x \, d z_9 d z_{10}}{\sqrt{z_9} \sqrt{4z_9-z_{10}^2} \sqrt{(x^2-1)^2(q^2-z_9)^2-4q^2x^2z_{10}^2}} \right).
\label{eq: LS_diag_3-loop_elliptic_pre}
\end{align}
At this point, we need to rationalize at least two of the square roots simultaneously. However, doing so does not reveal any further residues. To make the result more transparent, we can first introduce the change of variables $z_9=q^2 \, t_9^2$, which rationalizes the first square root. The leading singularity becomes
\begin{align}
\LS \left( \begin{tikzpicture}[baseline={([yshift=-0.1cm]current bounding box.center)}] 
	\node[] (a) at (0,0) {};
	\node[] (a1) at (0.5,0) {};
	\node[] (a2) at (1,0) {};
	\node[] (b) at (0,-0.5) {};
	\node[] (c) at (0,-1) {};
	\node[] (c1) at (1,-1) {};
	\node[] (p1) at ($(a)+(-0.2,0)$) {};
	\node[] (p2) at ($(c)+(-0.2,0)$) {};
	\node[] (p3) at ($(c1)+(0.2,0)$) {};
	\node[] (p4) at ($(a2)+(0.2,0)$) {};
	\draw[line width=0.15mm] (b.center) -- (a.center);
	\draw[line width=0.15mm] (b.center) -- (a1.center);
	\draw[line width=0.15mm] (c1.center) -- (a2.center);
	\draw[line width=0.15mm] (b.center) -- (c.center);
	\draw[line width=0.15mm] (c1.center) -- (b.center);
	\draw[line width=0.5mm] (p1.center) -- (p4.center);
	\draw[line width=0.5mm] (p2.center) -- (p3.center);
\end{tikzpicture} \right) & \propto \LS \left( \int \frac{x \, d t_9 d z_{10}}{\sqrt{4 q^2 t_9^2 -z_{10}^2} \sqrt{q^2 (x^2-1)^2(t_9^2-1)^2-4x^2 z_{10}^2}} \right) \nonumber \\
& \propto \LS \left( \int \frac{x \, d t_1 d t_2}{|q| \, \sqrt{t_2^2 (t_1^2-1)^2 (x^2-1)^2 - 4 x^2 t_1^2 (t_2^2+1)^2}} \right),
\label{eq: LS_diag_3-loop_elliptic}
\end{align}
where in the second step we used the change of variables~\eqref{eq: change_of_variables_(z-r1)(z-r2)} to $t_{10}$ to rationalize the first square root, and where we relabeled $t_9 \to t_1$ and $t_{10} \to t_2$. We observe the appearance of a square root of a polynomial $P_6(t_1,t_2)$ of total degree six in two variables that is of degree four in both variables individually. Homogenizing this polynomial to $\widetilde{P}_6(t_1,t_2,t_3)=P_6(t_1/t_3,t_2/t_3)t_3^6$, we obtain a homogeneous equation of degree six in weighted projective space $[t_1,t_2,t_3,t_4]\sim [\lambda^1 t_1,\lambda^1 t_2, \lambda^1 t_3, \lambda^3 t_4]\in\mathbb{WP}^{1,1,1,3}$: 
\begin{equation}
 t_4^2-\widetilde{P}_6(t_1,t_2,t_3)=0.
\end{equation}
Since the sum of the projective weights is equal to the degree, its solution in general defines a Calabi-Yau two-fold,\footnote{Strictly speaking, this defines a Calabi-Yau two-variety, since the geometries in Feynman integrals typically have singularities. While it is possible to desingularize it to arrive at a Calabi-Yau manifold, this is not necessary for calculating the Feynman integral by bringing its differential equation into canonical form. We thus use the words Calabi-Yau variety and Calabi-Yau manifold interchangeably.} or K3 surface \cite{Hubsch:1992nu,Bourjaily:2019hmc}. Thus this diagram is the first instance in our analysis that depends on a non-trivial geometry! 

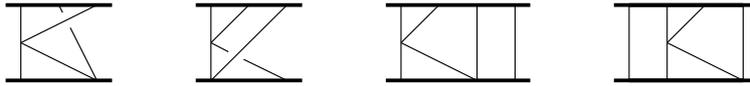
\begin{figure}[t]
\centering
\parbox{\textwidth}{\begin{equation*}
\begin{tikzpicture}[baseline={([yshift=-0.1cm]current bounding box.center)}] 
	\node[] (a) at (0,0) {};
	\node[] (a1) at (0.5,0) {};
	\node[] (a2) at (1,0) {};
	\node[] (b) at (0,-0.5) {};
	\node[] (c) at (0,-1) {};
	\node[] (c1) at (1,-1) {};
	\node[] (p1) at ($(a)+(-0.2,0)$) {};
	\node[] (p2) at ($(c)+(-0.2,0)$) {};
	\node[] (p3) at ($(c1)+(0.2,0)$) {};
	\node[] (p4) at ($(a2)+(0.2,0)$) {};
	\draw[line width=0.15mm] (b.center) -- (a.center);
	\draw[line width=0.15mm] (b.center) -- (a2.center);
	\draw[line width=0.15mm] (c1.center) -- (0.65,-0.3);
	\draw[line width=0.15mm] (0.55,-0.1) -- (a1.center);
	\draw[line width=0.15mm] (b.center) -- (c.center);
	\draw[line width=0.15mm] (c1.center) -- (b.center);
	\draw[line width=0.5mm] (p1.center) -- (p4.center);
	\draw[line width=0.5mm] (p2.center) -- (p3.center);
\end{tikzpicture} \qquad \begin{tikzpicture}[baseline={([yshift=-0.1cm]current bounding box.center)}] 
	\node[] (a) at (0,0) {};
	\node[] (a1) at (0.5,0) {};
	\node[] (a2) at (1,0) {};
	\node[] (b) at (0,-0.5) {};
	\node[] (c) at (0,-1) {};
	\node[] (c1) at (1,-1) {};
	\node[] (p1) at ($(a)+(-0.2,0)$) {};
	\node[] (p2) at ($(c)+(-0.2,0)$) {};
	\node[] (p3) at ($(c1)+(0.2,0)$) {};
	\node[] (p4) at ($(a2)+(0.2,0)$) {};
	\draw[line width=0.15mm] (b.center) -- (a.center);
	\draw[line width=0.15mm] (b.center) -- (a1.center);
	\draw[line width=0.15mm] (c.center) -- (a2.center);
	\draw[line width=0.15mm] (b.center) -- (c.center);
	\draw[line width=0.15mm] (c1.center) -- (0.43,-0.72);
	\draw[line width=0.15mm] (0.23,-0.62) -- (b.center);
	\draw[line width=0.5mm] (p1.center) -- (p4.center);
	\draw[line width=0.5mm] (p2.center) -- (p3.center);
\end{tikzpicture} \qquad \begin{tikzpicture}[baseline={([yshift=-0.1cm]current bounding box.center)}] 
	\node[] (a) at (0,0) {};
	\node[] (a1) at (0.5,0) {};
	\node[] (a2) at (1,0) {};
	\node[] (a3) at (1.5,0) {};
	\node[] (b) at (0,-0.5) {};
	\node[] (c) at (0,-1) {};
	\node[] (c1) at (1,-1) {};
	\node[] (c2) at (1.5,-1) {};
	\node[] (p1) at ($(a)+(-0.2,0)$) {};
	\node[] (p2) at ($(c)+(-0.2,0)$) {};
	\node[] (p3) at ($(c2)+(0.2,0)$) {};
	\node[] (p4) at ($(a3)+(0.2,0)$) {};
	\draw[line width=0.15mm] (b.center) -- (a.center);
	\draw[line width=0.15mm] (b.center) -- (a1.center);
	\draw[line width=0.15mm] (c1.center) -- (a2.center);
	\draw[line width=0.15mm] (b.center) -- (c.center);
	\draw[line width=0.15mm] (c1.center) -- (b.center);
	\draw[line width=0.15mm] (a3.center) -- (c2.center);
	\draw[line width=0.5mm] (p1.center) -- (p4.center);
	\draw[line width=0.5mm] (p2.center) -- (p3.center);
\end{tikzpicture} \qquad \begin{tikzpicture}[baseline={([yshift=-0.1cm]current bounding box.center)}] 
	\node[] (a) at (0,0) {};
	\node[] (a1) at (0.5,0) {};
	\node[] (a2) at (1,0) {};
	\node[] (a3) at (1.5,0) {};
	\node[] (b) at (0.5,-0.5) {};
	\node[] (c) at (0,-1) {};
	\node[] (c1) at (0.5,-1) {};
	\node[] (c2) at (1.5,-1) {};
	\node[] (p1) at ($(a)+(-0.2,0)$) {};
	\node[] (p2) at ($(c)+(-0.2,0)$) {};
	\node[] (p3) at ($(c2)+(0.2,0)$) {};
	\node[] (p4) at ($(a3)+(0.2,0)$) {};
	\draw[line width=0.15mm] (c.center) -- (a.center);
	\draw[line width=0.15mm] (b.center) -- (a1.center);
	\draw[line width=0.15mm] (a2.center) -- (b.center);
	\draw[line width=0.15mm] (c2.center) -- (a3.center);
	\draw[line width=0.15mm] (b.center) -- (c2.center);
	\draw[line width=0.15mm] (b.center) -- (c1.center);
	\draw[line width=0.5mm] (a.center) -- (a1.center);
	\draw[line width=0.5mm] (a1.center) -- (a2.center);
	\draw[line width=0.5mm] (a2.center) -- (a3.center);
	\draw[line width=0.5mm] (c1.center) -- (c2.center);
	\draw[line width=0.5mm] (c.center) -- (c1.center);
	\draw[line width=0.5mm] (p1.center) -- (p4.center);
	\draw[line width=0.5mm] (p2.center) -- (p3.center);
\end{tikzpicture}
\end{equation*}}
\caption{Examples of three and four-loop diagrams that trivially depend on the same K3 surface as the diagram in fig.~\ref{fig: diag_3-loop_elliptic}.}
\label{fig: examples_same_K3}
\end{figure}

From the perspective of the differential equation, one obtains a Picard-Fuchs operator of order 3 \cite{Ruf:2021egk,Dlapa:2022wdu}:
\begin{equation}
\label{eq: Picard_Fuchs_3-loop}
\mathcal{L}_3 =
 \frac{\partial^3}{\partial x^3} - \frac{6 x}{1 - x^2} \frac{\partial^2}{\partial x^2} + \frac{1 - 4 x^2 + 7 x^4}{x^2 (1 - x^2)^2} \frac{\partial}{\partial x} - \frac{1 + x^2}{x^3 (1 - x^2)},
\end{equation}
which analogously corresponds to a K3 surface. However, for univariate cases such as the post-Minkowskian expansion which depends only on the parameter $x$, the Picard-Fuchs operator of a K3 surface is necessarily related through an operation known as a \textit{symmetric square}~\cite{Joyce:1972,Joyce:1973,Verrill:1996,Doran:1998hm} to a differential operator of order 2, which defines instead an elliptic curve. Then, the solutions of $\mathcal{L}_3 \Psi =0$ become products of the solutions of the order 2 operator. In the present case, it becomes the symmetric square of~\cite{Ruf:2021egk,Dlapa:2022wdu}
\begin{equation}
\mathcal{L}_2 =
 \frac{\partial^2}{\partial x^2} + \frac{2x}{x^2 -1} \frac{\partial}{\partial x} + \frac{1}{4 x^2},
\end{equation}
whose solutions are complete elliptic integrals. As already pointed out in ref.~\cite{Ruf:2021egk,Dlapa:2022wdu}, this explains the proliferation of products of complete elliptic integrals in the three-loop conservative results computed in the literature~\cite{Bern:2021dqo,Dlapa:2021npj,Bern:2021yeh,Dlapa:2021vgp,Dlapa:2022lmu,Dlapa:2023hsl,Jakobsen:2023ndj,Jakobsen:2023pvx,Jakobsen:2023hig}, where this diagram (plus reflections and non-planar versions) was already highlighted as their source~\cite{Ruf:2021egk,Bern:2022jvn}. In our analysis, the three-loop non-planar variations can be trivially seen to depend on the same K3 surface as the planar diagram since they are related to it by the unraveling of matter propagators. Similarly, superclassical higher-loop box iterations of these diagrams also depend on the same K3 surface, which is in agreement with explicit four-loop calculations~\cite{Klemm:2024wtd}; see fig.~\ref{fig: examples_same_K3} for some examples.

Let us now turn to the even-parity case. Concretely, let us add a factor of the ISP $z_{10}=2u_2\cdot k_2$ in the numerator,
\begin{align}
& \LS \left( \begin{tikzpicture}[baseline={([yshift=-0.1cm]current bounding box.center)}] 
	\node[] (a) at (0,0) {};
	\node[] (a1) at (0.5,0) {};
	\node[] (a2) at (1,0) {};
	\node[] (b) at (0,-0.5) {};
	\node[] (c) at (0,-1) {};
	\node[] (c1) at (1,-1) {};
	\node[] (p1) at ($(a)+(-0.2,0)$) {};
	\node[] (p2) at ($(c)+(-0.2,0)$) {};
	\node[] (p3) at ($(c1)+(0.2,0)$) {};
	\node[] (p4) at ($(a2)+(0.2,0)$) {};
	\draw[line width=0.15mm] (b.center) -- (a.center);
	\draw[line width=0.15mm] (b.center) -- (a1.center);
	\draw[line width=0.15mm] (c1.center) -- (a2.center);
	\draw[line width=0.15mm] (b.center) -- (c.center);
	\draw[line width=0.15mm] (c1.center) -- (b.center);
	\draw[line width=0.5mm] (p1.center) -- (p4.center);
	\draw[line width=0.5mm] (p2.center) -- (p3.center);
\end{tikzpicture} {\times} (2u_2 {\cdot} k_2) \right)  \nonumber \\
& \qquad \qquad \qquad \propto \LS \left( \int \frac{x \, z_{10} \, d z_9 d z_{10}}{\sqrt{z_9} \sqrt{4z_9-z_{10}^2} \sqrt{(x^2-1)^2(q^2-z_9)^2-4q^2x^2z_{10}^2}} \right).
\end{align}
As it turns out, this numerator factor drastically changes the result, as it allows us now to change variables to $z_{10} \to \sqrt{z'_{10}}$, such that the Jacobian cancels the extra factor of $z_{10}$ in the numerator. Subsequently, we can simultaneously rationalize the two square roots, now linear in $z'_{10}$, with the change of variables~\eqref{eq: change_of_variables_(z-r1)(z-r2)},
\begin{align}
& \LS \left( \begin{tikzpicture}[baseline={([yshift=-0.1cm]current bounding box.center)}] 
	\node[] (a) at (0,0) {};
	\node[] (a1) at (0.5,0) {};
	\node[] (a2) at (1,0) {};
	\node[] (b) at (0,-0.5) {};
	\node[] (c) at (0,-1) {};
	\node[] (c1) at (1,-1) {};
	\node[] (p1) at ($(a)+(-0.2,0)$) {};
	\node[] (p2) at ($(c)+(-0.2,0)$) {};
	\node[] (p3) at ($(c1)+(0.2,0)$) {};
	\node[] (p4) at ($(a2)+(0.2,0)$) {};
	\draw[line width=0.15mm] (b.center) -- (a.center);
	\draw[line width=0.15mm] (b.center) -- (a1.center);
	\draw[line width=0.15mm] (c1.center) -- (a2.center);
	\draw[line width=0.15mm] (b.center) -- (c.center);
	\draw[line width=0.15mm] (c1.center) -- (b.center);
	\draw[line width=0.5mm] (p1.center) -- (p4.center);
	\draw[line width=0.5mm] (p2.center) -- (p3.center);
\end{tikzpicture} {\times} (2u_2 {\cdot} k_2) \right)  \nonumber \\
& \qquad \qquad \qquad \propto \LS \left( \int \frac{x \, d z_9 d z'_{10}}{\sqrt{z_9} \sqrt{4z_9-z'_{10}} \sqrt{(x^2-1)^2(q^2-z_9)^2-4q^2x^2z'_{10}}} \right) \nonumber \\
& \qquad \qquad \qquad \propto \frac{1}{|q|} \, \LS \left( \int \frac{d z_9 d t_{10}}{t_{10} \sqrt{z_9}}  \right) \propto \frac{1}{|q|} \, \int_{\frac{q^2 (x-1)^2}{(x+1)^2}}^{\frac{q^2 (x+1)^2}{(x-1)^2}} \frac{dz_9}{\sqrt{z_9}} \propto \frac{x}{x^2-1},
\end{align}
where we integrate the leading singularity between the roots of the Baikov polynomial.

Therefore, the sector of even parity does no longer involve a K3 surface and products of elliptic integrals, but its leading singularity is actually algebraic. Hence, we find that all integrals contributing to the 4PM dissipative sector have a dlog form, in agreement with explicit calculations~\cite{Bern:2021yeh,Dlapa:2022lmu,Jakobsen:2023hig,Dlapa:2023hsl}.

\section{A four-loop non-planar diagram}
\label{sec:four_loop_nonplanar}

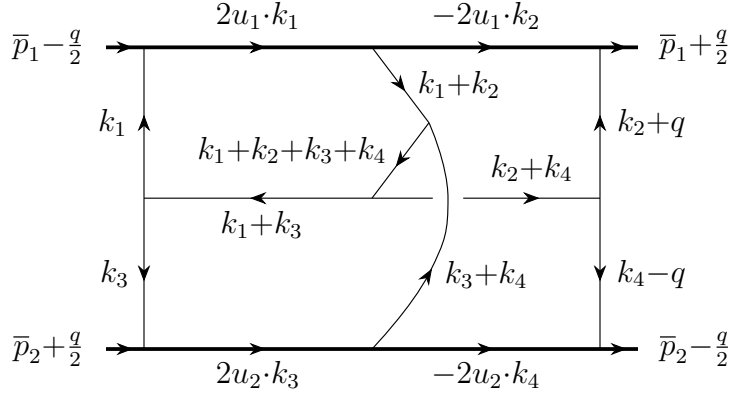
\begin{figure}[t]
\centering
\begin{tikzpicture}[baseline=(current bounding box.center)] 
	\node[] (a) at (0,0) {};
	\node[] (a1) at (3,0) {};
	\node[] (a2) at (6,0) {};
	\node[] (b) at (3.75,-1) {};
	\node[] (c) at (0,-2) {};
	\node[] (c1) at (3,-2) {};
	\node[] (c2) at (6,-2) {};
	\node[] (d) at (0,-4) {};
	\node[] (d1) at (3,-4) {};
	\node[] (d2) at (6,-4) {};
	\node[label=left:{$\overline{p}_1{-}\frac{q}{2}$}] (p1) at ($(a)+(-0.5,0)$) {};
	\node[label=left:{$\overline{p}_2{+}\frac{q}{2}$}] (p2) at ($(d)+(-0.5,0)$) {};
	\node[label=right:{$\overline{p}_2{-}\frac{q}{2}$}] (p3) at ($(d2)+(0.5,0)$) {};
	\node[label=right:{$\overline{p}_1{+}\frac{q}{2}$}] (p4) at ($(a2)+(0.5,0)$) {};
	\draw[line width=0.15mm, postaction={decorate}] (c.center) -- node[sloped, allow upside down, label={[xshift=0.15cm, yshift=0cm]$k_1$}] {\midarrow} (a.center);
	\draw[line width=0.15mm, postaction={decorate}] (c.center) -- node[sloped, allow upside down, label={[xshift=-0.95cm, yshift=0cm]$k_3$}] {\midarrow} (d.center);
	\draw[line width=0.15mm, postaction={decorate}] (a1.center) -- node[sloped, allow upside down, label={[xshift=-0.1cm, yshift=-0.45cm]$k_1{+}k_2$}] {\midarrow} (b.center);
	\draw[line width=0.15mm, postaction={decorate}] (b.center) -- node[sloped, allow upside down, label={[xshift=-3cm, yshift=0.6cm]$k_1{+}k_2{+}k_3{+}k_4$}] {\midarrow} (c1.center);
	\draw[line width=0.15mm, postaction={decorate}] (c1.center) -- node[sloped, allow upside down, label={[xshift=0.05cm, yshift=0.2cm]$k_1{+}k_3$}] {\midarrow} (c.center);
	\draw[line width=0.15mm, postaction={decorate}] (c2.center) -- node[sloped, allow upside down, label={[xshift=1.5cm, yshift=0cm]$k_2{+}q$}] {\midarrow} (a2.center);
	\draw[line width=0.15mm, postaction={decorate}] (c2.center) -- node[sloped, allow upside down, label={[xshift=-0.15cm, yshift=0cm]$k_4{-}q$}] {\midarrow} (d2.center);
	\draw[line width=0.15mm, postaction={decorate}] (4.2,-2) -- node[sloped, allow upside down, label={[xshift=0cm, yshift=-0.15cm]$k_2{+}k_4$}] {\midarrow} (c2.center);
	\draw[line width=0.15mm] (3.8,-2) -- (c1.center);
	\draw[line width=0.15mm] plot[smooth, tension=0.8] coordinates {(b.center) (4,-2) (3.75,-3) (d1.center)};
	\draw[line width=0.15mm, postaction={decorate}] (3.72,-3.05) -- node[sloped, allow upside down, label={[xshift=1.65cm, yshift=-0.425cm]$k_3{+}k_4$}] {\midarrow} (3.75,-3);
	\draw[line width=0.5mm, postaction={decorate}] (a.center) -- node[sloped, allow upside down, label={[xshift=0cm, yshift=-0.15cm]$2u_1 {\cdot} k_1$}] {\midarrow} (a1.center);
	\draw[line width=0.5mm, postaction={decorate}] (a1.center) -- node[sloped, allow upside down, label={[xshift=0cm, yshift=-0.15cm]$-2u_1 {\cdot} k_2$}] {\midarrow} (a2.center);
	\draw[line width=0.5mm, postaction={decorate}] (d.center) -- node[sloped, allow upside down, label={[xshift=0cm, yshift=-0.9cm]$2u_2 {\cdot} k_3$}] {\midarrow} (d1.center);
	\draw[line width=0.5mm, postaction={decorate}] (d1.center) -- node[sloped, allow upside down, label={[xshift=0cm, yshift=-0.9cm]$-2u_2 {\cdot} k_4$}] {\midarrow} (d2.center);
	\draw[line width=0.5mm, postaction={decorate}] (p1.center) -- node[sloped, allow upside down] {\midarrow} (a.center);
	\draw[line width=0.5mm, postaction={decorate}] (a2.center) -- node[sloped, allow upside down] {\midarrow} (p4.center);
	\draw[line width=0.5mm, postaction={decorate}] (p2.center) -- node[sloped, allow upside down] {\midarrow} (d.center);
	\draw[line width=0.5mm, postaction={decorate}] (d2.center) -- node[sloped, allow upside down] {\midarrow} (p3.center);
\end{tikzpicture}
\caption{Parametrization of the loop momenta for the four-loop non-planar top topology.}
\label{fig: diag_4-loop_nonplanar}
\end{figure}

Let us finally study the four-loop non-planar top topology, with a parametrization shown in fig.~\ref{fig: diag_4-loop_nonplanar}. It is the most complicated sector at four loops with respect to IBP identities. This diagram is truly non-planar, in the sense that its leading singularity cannot be related to a planar counterpart using the unraveling of matter propagators~\eqref{eq: unraveling_matter_props}. Moreover, it is the first diagram we consider that contains pentagon sub-loops that cannot be avoided. The fact that we can nevertheless easily calculate its leading singularity clearly demonstrates the potential of our approach.

Using the integration order $k_1 \to k_3 \to k_2 \to k_4$, and introducing the six extra Baikov variables $z_{14}=(k_2+k_3)^2$, $z_{15}=k_2^2$, $z_{16}=k_4^2$, $z_{17}=2u_1 {\cdot} k_3$, $z_{18}=2u_2 {\cdot} k_2$ and $z_{19}=2u_1 {\cdot} k_4$, we obtain in generic $D$ dimensions
\begin{align}
\raisebox{0.1cm}{\begin{tikzpicture}[baseline=(current bounding box.center), scale=0.25] 
	\node[] (a) at (0,0) {};
	\node[] (a1) at (3,0) {};
	\node[] (a2) at (6,0) {};
	\node[] (b) at (3.75,-1) {};
	\node[] (c) at (0,-2) {};
	\node[] (c1) at (2.5,-2) {};
	\node[] (c2) at (6,-2) {};
	\node[] (d) at (0,-4) {};
	\node[] (d1) at (3,-4) {};
	\node[] (d2) at (6,-4) {};
	\node[] (p1) at ($(a)+(-0.8,0)$) {};
	\node[] (p2) at ($(d)+(-0.8,0)$) {};
	\node[] (p3) at ($(d2)+(0.8,0)$) {};
	\node[] (p4) at ($(a2)+(0.8,0)$) {};
	\draw[line width=0.15mm] (c.center) -- (a.center);
	\draw[line width=0.15mm] (d.center) -- (c.center);
	\draw[line width=0.15mm] (a1.center) -- (b.center);
	\draw[line width=0.15mm] (b.center) -- (c1.center);
	\draw[line width=0.15mm] (c.center) -- (c1.center);
	\draw[line width=0.15mm] (a2.center) -- (c2.center);
	\draw[line width=0.15mm] (c2.center) -- (d2.center);
	\draw[line width=0.15mm] (c2.center) -- (4.4,-2);
	\draw[line width=0.15mm] (3.6,-2) -- (c1.center);
	\draw[line width=0.15mm] plot[smooth, tension=0.8] coordinates {(b.center) (4,-2) (3.75,-3) (d1.center)};
	\draw[line width=0.5mm] (a.center) -- (a1.center);
	\draw[line width=0.5mm] (a1.center) -- (a2.center);
	\draw[line width=0.5mm] (d.center) -- (d1.center);
	\draw[line width=0.5mm] (d1.center) -- (d2.center);
	\draw[line width=0.5mm] (p1.center) -- (p4.center);
	\draw[line width=0.5mm] (p2.center) -- (p3.center);
\end{tikzpicture}} \propto & \, \frac{1}{{\Gamma(\frac{D-4}{2})}^3 \, \Gamma(\frac{D-3}{2})} \int \frac{d z_1 \cdots d z_{13}}{z_1 \cdots z_{13}} \, dz_{14} \cdots dz_{19} \,{\det G(k_1,k_2,k_3,k_4,u_1)}^{\frac{D-6}{2}} \nonumber \\
& \times {\det G(k_2,k_3,k_4,u_1,u_2)}^{\frac{D-6}{2}} {\det G(k_2,k_4,u_1,u_2,q)}^{\frac{D-6}{2}} \label{eq: diag_4-loop_nonplanar} \\[0.1cm]
& \times {\det G(k_2,k_3,k_4,u_1)}^{\frac{5-D}{2}} {\det G(k_2,k_4,u_1,u_2)}^{\frac{5-D}{2}} {\det G(u_1,u_2,q)}^{\frac{4-D}{2}},
\nonumber
\end{align}
where there are two instances of the Baikov polynomial $\det G(k_4,u_1,u_2,q)$ that cancel each other.

First of all, we notice that in $D=4-2\varepsilon$, the prefactor vanishes in dimensional regularization, as it becomes of order $\mathcal{O}(\varepsilon^3)$. As explained in sec.~\ref{sec:methodology_systematics}, this is an artifact of dimensional regularisation, which arises due to the presence of three pentagon sub-loops during the loop-by-loop parametrization. While the explicit $\varepsilon$-dependence is compensated at the end by $\varepsilon^{-1}$ divergences when integrating over the Baikov contour, to avoid the vanishing prefactor, we can alternatively work in $D=6-2\varepsilon$. Since the six-dimensional result can be related to the four-dimensional one by dimension-shift identities~\cite{Tarasov:1996br, Lee:2012te}, both results contain the same geometry. On top of that, three of the Gram determinants have exponents that vanish in six dimensions, such that we find a much simpler result:
\begin{equation}
\raisebox{0.1cm}{\begin{tikzpicture}[baseline=(current bounding box.center), scale=0.25] 
	\node[] (a) at (0,0) {};
	\node[] (a1) at (3,0) {};
	\node[] (a2) at (6,0) {};
	\node[] (b) at (3.75,-1) {};
	\node[] (c) at (0,-2) {};
	\node[] (c1) at (2.5,-2) {};
	\node[] (c2) at (6,-2) {};
	\node[] (d) at (0,-4) {};
	\node[] (d1) at (3,-4) {};
	\node[] (d2) at (6,-4) {};
	\node[] (p1) at ($(a)+(-0.8,0)$) {};
	\node[] (p2) at ($(d)+(-0.8,0)$) {};
	\node[] (p3) at ($(d2)+(0.8,0)$) {};
	\node[] (p4) at ($(a2)+(0.8,0)$) {};
	\draw[line width=0.15mm] (c.center) -- (a.center);
	\draw[line width=0.15mm] (d.center) -- (c.center);
	\draw[line width=0.15mm] (a1.center) -- (b.center);
	\draw[line width=0.15mm] (b.center) -- (c1.center);
	\draw[line width=0.15mm] (c.center) -- (c1.center);
	\draw[line width=0.15mm] (a2.center) -- (c2.center);
	\draw[line width=0.15mm] (c2.center) -- (d2.center);
	\draw[line width=0.15mm] (c2.center) -- (4.4,-2);
	\draw[line width=0.15mm] (3.6,-2) -- (c1.center);
	\draw[line width=0.15mm] plot[smooth, tension=0.8] coordinates {(b.center) (4,-2) (3.75,-3) (d1.center)};
	\draw[line width=0.5mm] (a.center) -- (a1.center);
	\draw[line width=0.5mm] (a1.center) -- (a2.center);
	\draw[line width=0.5mm] (d.center) -- (d1.center);
	\draw[line width=0.5mm] (d1.center) -- (d2.center);
	\draw[line width=0.5mm] (p1.center) -- (p4.center);
	\draw[line width=0.5mm] (p2.center) -- (p3.center);
\end{tikzpicture}} \propto \int \frac{d z_1 \cdots d z_{13}}{z_1 \cdots z_{13}} \frac{x^2 \, d z_{14} \cdots d z_{19}}{q^2 (x^2-1)^2 \sqrt{\det G(k_2,k_3,k_4,u_1)} \sqrt{\det G(k_2,k_4,u_1,u_2)}},
\end{equation}
\begin{align}
& \LS \left( \raisebox{0.1cm}{\begin{tikzpicture}[baseline=(current bounding box.center), scale=0.25] 
	\node[] (a) at (0,0) {};
	\node[] (a1) at (3,0) {};
	\node[] (a2) at (6,0) {};
	\node[] (b) at (3.75,-1) {};
	\node[] (c) at (0,-2) {};
	\node[] (c1) at (2.5,-2) {};
	\node[] (c2) at (6,-2) {};
	\node[] (d) at (0,-4) {};
	\node[] (d1) at (3,-4) {};
	\node[] (d2) at (6,-4) {};
	\node[] (p1) at ($(a)+(-0.8,0)$) {};
	\node[] (p2) at ($(d)+(-0.8,0)$) {};
	\node[] (p3) at ($(d2)+(0.8,0)$) {};
	\node[] (p4) at ($(a2)+(0.8,0)$) {};
	\draw[line width=0.15mm] (c.center) -- (a.center);
	\draw[line width=0.15mm] (d.center) -- (c.center);
	\draw[line width=0.15mm] (a1.center) -- (b.center);
	\draw[line width=0.15mm] (b.center) -- (c1.center);
	\draw[line width=0.15mm] (c.center) -- (c1.center);
	\draw[line width=0.15mm] (a2.center) -- (c2.center);
	\draw[line width=0.15mm] (c2.center) -- (d2.center);
	\draw[line width=0.15mm] (c2.center) -- (4.4,-2);
	\draw[line width=0.15mm] (3.6,-2) -- (c1.center);
	\draw[line width=0.15mm] plot[smooth, tension=0.8] coordinates {(b.center) (4,-2) (3.75,-3) (d1.center)};
	\draw[line width=0.5mm] (a.center) -- (a1.center);
	\draw[line width=0.5mm] (a1.center) -- (a2.center);
	\draw[line width=0.5mm] (d.center) -- (d1.center);
	\draw[line width=0.5mm] (d1.center) -- (d2.center);
	\draw[line width=0.5mm] (p1.center) -- (p4.center);
	\draw[line width=0.5mm] (p2.center) -- (p3.center);
\end{tikzpicture}} \right) \propto \LS \left( \int \frac{x^3 \, d z_{14} \cdots d z_{19}}{q^2 (x^2-1)^2} \Big[ -4 z_{16} (z_{14} - 2 z_{15}) (z_{14} - z_{15} - z_{16}) \right. \nonumber \\
& + z_{17}^2 (z_{15} - z_{16})^2 + z_{17} \Big( 2 z_{14} z_{19}(z_{15} + z_{16}) - 2 z_{15} z_{17} (z_{15} + 3 z_{16}) + z_{19} (z_{14} - z_{15})^2 \Big) \Big]^{-\frac{1}{2}} \nonumber \\
& \times \Big[ (x^2-1)^2 (z_{15} - z_{16})^2 + x z_{18} z_{19} (2 (x^2+1) (z_{15} + z_{16}) + x z_{18} z_{19}) \nonumber \\
& \left. -4 x^2 z_{16} z_{18}^2 - 4 x^2 z_{15} z_{19}^2 \Big]^{-\frac{1}{2}} \right) .
\end{align}

Since the first and second square roots are at most quadratic in $z_{17}$ and $z_{18}$, respectively, we can use the change of variables~\eqref{eq: change_of_variables_(z-r1)(z-r2)} to rationalize them, leading to
\begin{equation}
\LS \left( \raisebox{0.1cm}{\begin{tikzpicture}[baseline=(current bounding box.center), scale=0.25] 
	\node[] (a) at (0,0) {};
	\node[] (a1) at (3,0) {};
	\node[] (a2) at (6,0) {};
	\node[] (b) at (3.75,-1) {};
	\node[] (c) at (0,-2) {};
	\node[] (c1) at (2.5,-2) {};
	\node[] (c2) at (6,-2) {};
	\node[] (d) at (0,-4) {};
	\node[] (d1) at (3,-4) {};
	\node[] (d2) at (6,-4) {};
	\node[] (p1) at ($(a)+(-0.8,0)$) {};
	\node[] (p2) at ($(d)+(-0.8,0)$) {};
	\node[] (p3) at ($(d2)+(0.8,0)$) {};
	\node[] (p4) at ($(a2)+(0.8,0)$) {};
	\draw[line width=0.15mm] (c.center) -- (a.center);
	\draw[line width=0.15mm] (d.center) -- (c.center);
	\draw[line width=0.15mm] (a1.center) -- (b.center);
	\draw[line width=0.15mm] (b.center) -- (c1.center);
	\draw[line width=0.15mm] (c.center) -- (c1.center);
	\draw[line width=0.15mm] (a2.center) -- (c2.center);
	\draw[line width=0.15mm] (c2.center) -- (d2.center);
	\draw[line width=0.15mm] (c2.center) -- (4.4,-2);
	\draw[line width=0.15mm] (3.6,-2) -- (c1.center);
	\draw[line width=0.15mm] plot[smooth, tension=0.8] coordinates {(b.center) (4,-2) (3.75,-3) (d1.center)};
	\draw[line width=0.5mm] (a.center) -- (a1.center);
	\draw[line width=0.5mm] (a1.center) -- (a2.center);
	\draw[line width=0.5mm] (d.center) -- (d1.center);
	\draw[line width=0.5mm] (d1.center) -- (d2.center);
	\draw[line width=0.5mm] (p1.center) -- (p4.center);
	\draw[line width=0.5mm] (p2.center) -- (p3.center);
\end{tikzpicture}} \right) \propto \frac{x^2}{q^2 (x^2-1)^2} \, \LS \left( \int \frac{d z_{14} d z_{15} d z_{16} d t_{17} d t_{18} d z_{19}}{t_{17} t_{18} (z_{15}-z_{16}) \sqrt{4 z_{16}-z_{19}^2}} \right) .
\end{equation}
We can now shift $z_{15} \to z_{15} + z_{16}$, and use the same change of variables to rationalize the square root with respect to $z_{19}$,
\begin{equation}
\LS \left( \raisebox{0.1cm}{\begin{tikzpicture}[baseline=(current bounding box.center), scale=0.25] 
	\node[] (a) at (0,0) {};
	\node[] (a1) at (3,0) {};
	\node[] (a2) at (6,0) {};
	\node[] (b) at (3.75,-1) {};
	\node[] (c) at (0,-2) {};
	\node[] (c1) at (2.5,-2) {};
	\node[] (c2) at (6,-2) {};
	\node[] (d) at (0,-4) {};
	\node[] (d1) at (3,-4) {};
	\node[] (d2) at (6,-4) {};
	\node[] (p1) at ($(a)+(-0.8,0)$) {};
	\node[] (p2) at ($(d)+(-0.8,0)$) {};
	\node[] (p3) at ($(d2)+(0.8,0)$) {};
	\node[] (p4) at ($(a2)+(0.8,0)$) {};
	\draw[line width=0.15mm] (c.center) -- (a.center);
	\draw[line width=0.15mm] (d.center) -- (c.center);
	\draw[line width=0.15mm] (a1.center) -- (b.center);
	\draw[line width=0.15mm] (b.center) -- (c1.center);
	\draw[line width=0.15mm] (c.center) -- (c1.center);
	\draw[line width=0.15mm] (a2.center) -- (c2.center);
	\draw[line width=0.15mm] (c2.center) -- (d2.center);
	\draw[line width=0.15mm] (c2.center) -- (4.4,-2);
	\draw[line width=0.15mm] (3.6,-2) -- (c1.center);
	\draw[line width=0.15mm] plot[smooth, tension=0.8] coordinates {(b.center) (4,-2) (3.75,-3) (d1.center)};
	\draw[line width=0.5mm] (a.center) -- (a1.center);
	\draw[line width=0.5mm] (a1.center) -- (a2.center);
	\draw[line width=0.5mm] (d.center) -- (d1.center);
	\draw[line width=0.5mm] (d1.center) -- (d2.center);
	\draw[line width=0.5mm] (p1.center) -- (p4.center);
	\draw[line width=0.5mm] (p2.center) -- (p3.center);
\end{tikzpicture}} \right) \propto \frac{x^2}{q^2 (x^2-1)^2} \, \LS \left( \int \frac{d z_{14} d z_{15} d z_{16} d t_{17} d t_{18} d t_{19}}{z_{15} t_{17} t_{18} t_{19}} \right) .
\end{equation}
Having rationalized all square roots, we may now take residues at the simple poles. However, as can be seen, not all variables have a pole; thus, we also need to integrate some of them over the contour $\mathfrak{C}$, specified by the roots of the Baikov polynomials in eq.~\eqref{eq: diag_4-loop_nonplanar}. In this case, we can immediately take the residue at $t_{17}=t_{18}=0$, while we need to integrate the remaining variables, leading to
\begin{align}
\LS \left( \raisebox{0.1cm}{\begin{tikzpicture}[baseline=(current bounding box.center), scale=0.25] 
	\node[] (a) at (0,0) {};
	\node[] (a1) at (3,0) {};
	\node[] (a2) at (6,0) {};
	\node[] (b) at (3.75,-1) {};
	\node[] (c) at (0,-2) {};
	\node[] (c1) at (2.5,-2) {};
	\node[] (c2) at (6,-2) {};
	\node[] (d) at (0,-4) {};
	\node[] (d1) at (3,-4) {};
	\node[] (d2) at (6,-4) {};
	\node[] (p1) at ($(a)+(-0.8,0)$) {};
	\node[] (p2) at ($(d)+(-0.8,0)$) {};
	\node[] (p3) at ($(d2)+(0.8,0)$) {};
	\node[] (p4) at ($(a2)+(0.8,0)$) {};
	\draw[line width=0.15mm] (c.center) -- (a.center);
	\draw[line width=0.15mm] (d.center) -- (c.center);
	\draw[line width=0.15mm] (a1.center) -- (b.center);
	\draw[line width=0.15mm] (b.center) -- (c1.center);
	\draw[line width=0.15mm] (c.center) -- (c1.center);
	\draw[line width=0.15mm] (a2.center) -- (c2.center);
	\draw[line width=0.15mm] (c2.center) -- (d2.center);
	\draw[line width=0.15mm] (c2.center) -- (4.4,-2);
	\draw[line width=0.15mm] (3.6,-2) -- (c1.center);
	\draw[line width=0.15mm] plot[smooth, tension=0.8] coordinates {(b.center) (4,-2) (3.75,-3) (d1.center)};
	\draw[line width=0.5mm] (a.center) -- (a1.center);
	\draw[line width=0.5mm] (a1.center) -- (a2.center);
	\draw[line width=0.5mm] (d.center) -- (d1.center);
	\draw[line width=0.5mm] (d1.center) -- (d2.center);
	\draw[line width=0.5mm] (p1.center) -- (p4.center);
	\draw[line width=0.5mm] (p2.center) -- (p3.center);
\end{tikzpicture}} \right) & \propto \frac{x^2}{q^2 (x^2-1)^2} \, \int_{\frac{i}{x}}^{ix} \frac{d t_{19}}{t_{19}} \int_{0}^{-\frac{q^2}{t_{19}^2}} d z_{16} \int_{-z_{16}-z_{16} t_{19}^2}^{-z_{16}} \! \frac{d z_{15}}{z_{15}} \int_{z_{15}+2z_{16}}^{2z_{15}+2z_{16}} \! d z_{14} \nonumber \\[0.1cm]
& \propto \frac{q^2 (x^2+1)}{x^2-1}.
\end{align}

Turning to the odd-parity case, we add one factor of the ISP $z_{19}=2u_1 \cdot k_4$ in the numerator. Following the same steps as before, we similarly find
\begin{align}
& \LS \left( \raisebox{0.1cm}{\begin{tikzpicture}[baseline=(current bounding box.center), scale=0.25] 
	\node[] (a) at (0,0) {};
	\node[] (a1) at (3,0) {};
	\node[] (a2) at (6,0) {};
	\node[] (b) at (3.75,-1) {};
	\node[] (c) at (0,-2) {};
	\node[] (c1) at (2.5,-2) {};
	\node[] (c2) at (6,-2) {};
	\node[] (d) at (0,-4) {};
	\node[] (d1) at (3,-4) {};
	\node[] (d2) at (6,-4) {};
	\node[] (p1) at ($(a)+(-0.8,0)$) {};
	\node[] (p2) at ($(d)+(-0.8,0)$) {};
	\node[] (p3) at ($(d2)+(0.8,0)$) {};
	\node[] (p4) at ($(a2)+(0.8,0)$) {};
	\draw[line width=0.15mm] (c.center) -- (a.center);
	\draw[line width=0.15mm] (d.center) -- (c.center);
	\draw[line width=0.15mm] (a1.center) -- (b.center);
	\draw[line width=0.15mm] (b.center) -- (c1.center);
	\draw[line width=0.15mm] (c.center) -- (c1.center);
	\draw[line width=0.15mm] (a2.center) -- (c2.center);
	\draw[line width=0.15mm] (c2.center) -- (d2.center);
	\draw[line width=0.15mm] (c2.center) -- (4.4,-2);
	\draw[line width=0.15mm] (3.6,-2) -- (c1.center);
	\draw[line width=0.15mm] plot[smooth, tension=0.8] coordinates {(b.center) (4,-2) (3.75,-3) (d1.center)};
	\draw[line width=0.5mm] (a.center) -- (a1.center);
	\draw[line width=0.5mm] (a1.center) -- (a2.center);
	\draw[line width=0.5mm] (d.center) -- (d1.center);
	\draw[line width=0.5mm] (d1.center) -- (d2.center);
	\draw[line width=0.5mm] (p1.center) -- (p4.center);
	\draw[line width=0.5mm] (p2.center) -- (p3.center);
\end{tikzpicture}} {\times} (2u_1 {\cdot} k_4) \right) \nonumber \\
& \qquad \propto \frac{x^2}{q^2 (x^2-1)^2} \, \LS \left( \int \frac{\sqrt{z_{16}} \,  (t_{19}^2+1) \, d z_{14} d z_{15} d z_{16} d t_{17} d t_{18} d t_{19}}{z_{15} t_{17} t_{18} t_{19}^2} \right) \nonumber \\
& \qquad \propto \frac{x^2}{q^2 (x^2-1)^2} \, \int_{\frac{i}{x}}^{ix} \frac{t_{19}^2+1}{t_{19}^2} \, d t_{19} \int_{0}^{-\frac{q^2}{t_{19}^2}} \sqrt{z_{16}} \, d z_{16} \int_{-z_{16}-z_{16} t_{19}^2}^{-z_{16}} \! \frac{d z_{15}}{z_{15}} \int_{z_{15}+2z_{16}}^{2z_{15}+2z_{16}} \! d z_{14} \nonumber \\[0.1cm]
& \qquad \propto \frac{|q|^3 (x^4-1)}{x^2}.
\end{align}

Since the leading singularities are algebraic for both parities, the four-loop non-planar top sector has a dlog form at the maximal cut. As a complementary check, we have calculated the differential equation for this integral family via \FIRE~\cite{Smirnov:2023yhb}, 
which involves 16 master integrals for the even-parity sector and 12 master integrals for the odd sector. We have verified that the corresponding Picard-Fuchs operators factorize into a product of differential operators of order one for both parity sectors, which can be done by using the {\texttt{\textup{DFactor}}} command implemented in {\Maple}.
As a consequence, a dependence on non-trivial geometries can only arise through its subsectors.

\section{Conclusions and further directions}
\label{sec:conclusion}
In this work, we have analyzed the geometries appearing in Feynman integrals within the post-Minkowskian expansion of general relativity, which contribute to the emission of gravitational waves through the classical two-body dynamics. We have shown that the loop-by-loop Baikov representation can be used as a tool to efficiently compute the leading singularity of PM integrals and to classify their associated geometries. As a proof of principle, we have performed a systematic analysis of the diagrams up to three loops including conservative and dissipative effects, reproducing all known results in the literature. 

Moreover, using the loop-by-loop Baikov representation we have found identities relating the leading singularities of different diagrams -- namely the unraveling of matter propagators~\eqref{eq: unraveling_matter_props} and the reduction of superclassical diagrams with box iterations~\eqref{eq: reduction_superclassical}--\eqref{eq: reduction_superclassical_generic}. Moreover, we also found that certain diagrams containing bubble corrections and triangles with cubic vertices are reducible to lower sectors, see fig.\ \ref{fig: diagrams_zero_masters}. Besides allowing us to relate the geometries of diagrams across different sectors and loop orders, these identities also greatly simplify the analysis, as they significantly reduce the number of diagrams which need to be studied independently at a given loop order, see for example tab.~\ref{tab:indep_diagrams}. For instance, we can disregard non-planar diagrams whose non-planarity relies on the ordering of the vertices in the matter lines, as their geometry is related to that of planar counterparts. At four loops, however, truly non-planar diagrams appear. As we have shown, the most complicated of those diagrams, namely the four-loop non-planar top topology, actually has an algebraic leading singularity. Therefore, it can only depend on non-trivial geometries through its subsectors.

For the purpose of classifying geometries, we focussed on calculating the leading singularity of one integral per parity sector. However, using the same techniques it is similarly possible to calculate the leading singularity for a full basis of master integrals, which is helpful for bringing the differential equation into canonical form~\cite{Henn:2013pwa,Henn:2014qga, Primo:2016ebd, WasserMSc,Dlapa:2021qsl}.

As has already been identified in refs.~\cite{Frellesvig:2023bbf,Klemm:2024wtd}, at four loops new non-trivial geometries start to appear, in particular a novel Calabi-Yau three-fold. We provide a full classification of the geometries appearing at four loops in upcoming work \cite{Brammer:toapp}, to serve as a guide for the development and understanding of the function spaces these integrals evaluate to. Moreover, it would be interesting to study whether the K3 surface at three loops (discussed in sec.~\ref{sec:three_loop_K3}) and the Calabi-Yau three-fold found at four loops belong to a family of Calabi-Yau ($L-1$)-folds at $L$-loop order in the PM expansion, as it has been found for other integral families~\cite{Broadhurst:1993mw,Bourjaily:2018ycu,Bourjaily:2018yfy,Bonisch:2021yfw,Broedel:2021zij,Duhr:2022pch,Lairez:2022zkj,Pogel:2022vat,Duhr:2022dxb,Cao:2023tpx}.  

It would also be interesting to apply our methodology to diagrams contributing to the waveform, which has recently received increasing attention \cite{Brandhuber:2023hhy,Herderschee:2023fxh,Elkhidir:2023dco,Georgoudis:2023lgf,Caron-Huot:2023vxl,Georgoudis:2023ozp,Bohnenblust:2023qmy,Adamo:2024oxy,Brunello:2024ibk}. More broadly, the tools developed in this work could be used for a similar classification of geometries in other Feynman integrals, such as for QCD integrals relevant for particle phenomenology. We leave these interesting questions to future work.

\subsection*{Acknowledgments}

We are grateful to Zvi Bern, Enrico Herrmann, Zhengwen Liu and Michael Ruf for many enlightening discussions. We further thank Daniel Brammer, Christoph Dlapa, Gustav Jakobsen, Andres Luna, Andrew McLeod, Jan Plefka, Sebastian Pögel, Rafael Porto, Cristian Vergu, Matt von Hippel and Stefan Weinzierl for fruitful discussions, Daniel Brammer, Christoph Dlapa, Zhengwen Liu, Rafael Porto and Michael Ruf for comments on the manuscript, as well as Alexander Smirnov for communication. This work was supported by the research grant 00025445 from Villum Fonden. HF is moreover supported by a Carlsberg Foundation Reintegration Fellowship, and has received funding from the European Union's Horizon 2020 research and innovation program under the Marie Sk{\l}odowska-Curie grant agreement No. 847523 `INTERACTIONS'. RM is also grateful for the hospitality and support received from the Mani L. Bhaumik Institute for Theoretical Physics during part of this project.

\newpage
\appendix

\section{Unraveling of matter propagators}
\label{app: Nonplanar_unraveling}

In this appendix, we show that matter propagators can be unraveled, thus proving eq.~\eqref{eq: unraveling_matter_props}. This allows us to relate the leading singularity of many non-planar diagrams to planar counterparts. Only starting at four loops, are there truly non-planar diagrams in the PM expansion (see section \ref{sec:four_loop_nonplanar}), whose leading singularity cannot be related to planar counterparts by unraveling transformations.
\begin{figure}[t]
\centering
\begin{tikzpicture}[baseline=(current bounding box.center)] 
	\node[] (a) at (0,0) {};
	\node[] (b) at (0,-2) {};
	\node[] (a1) at (2,0) {};
	\node[] (b1) at (2,-2) {};
	\node[] (a2) at (4,0) {};
	\node[] (b2) at (4,-2) {};
	\node[label=left:{$\overline{p}_1{-}\frac{q}{2}$}] (p1) at ($(a)+(-0.5,0)$) {};
	\node[label=left:{$\overline{p}_2{+}\frac{q}{2}$}] (p2) at ($(b)+(-0.5,0)$) {};
	\node[label=right:{$\overline{p}_2{-}\frac{q}{2}$}] (p3) at ($(b2)+(0.5,0)$) {};
	\node[label=right:{$\overline{p}_1{+}\frac{q}{2}$}] (p4) at ($(a2)+(0.5,0)$) {};
	\draw[line width=0.15mm, postaction={decorate}] (b.center) -- node[sloped, allow upside down, label={[xshift=0.2cm, yshift=0cm]$k_1$}] {\midarrow} (a.center);
	\draw[line width=0.15mm, postaction={decorate}] (3,-1) -- node[sloped, allow upside down, label={[xshift=-1.6cm, yshift=0.6cm]$k_1{-}k_2$}] {\midarrow} (b1.center);
	\draw[line width=0.15mm] (3,-1) -- (a2.center);
	\draw[line width=0.15mm, postaction={decorate}] (a1.center) -- (2.9,-0.9);
	\draw[line width=0.15mm, postaction={decorate}] (3.1,-1.1) -- node[sloped, allow upside down, label={[xshift=-0.15cm, yshift=-0.25cm]$k_2{-}q$}] {\midarrow} (b2.center);
	\draw[line width=0.5mm, postaction={decorate}] (a.center) -- node[sloped, allow upside down, label={[xshift=0cm, yshift=-0.15cm]$2u_1 {\cdot} k_1$}] {\midarrow} (a1.center);
		\draw[line width=0.5mm, postaction={decorate}] (a1.center) -- node[sloped, allow upside down, label={[xshift=0cm, yshift=-0.15cm]$2u_1 {\cdot} (k_1{-}k_2)$}] {\midarrow} (a2.center);
	\draw[line width=0.5mm, postaction={decorate}] (b.center) -- node[sloped, allow upside down, label={[xshift=0cm, yshift=-0.9cm]$-2u_2 {\cdot} k_1$}] {\midarrow} (b1.center);
	\draw[line width=0.5mm, postaction={decorate}] (b1.center) -- node[sloped, allow upside down, label={[xshift=0cm, yshift=-0.9cm]$-2u_2 {\cdot} k_2$}] {\midarrow} (b2.center);
	\draw[line width=0.5mm, postaction={decorate}] (p1.center) -- node[sloped, allow upside down] {\midarrow} (a.center);
	\draw[line width=0.5mm, postaction={decorate}] (a2.center) -- node[sloped, allow upside down] {\midarrow} (p4.center);
	\draw[line width=0.5mm, postaction={decorate}] (p2.center) -- node[sloped, allow upside down] {\midarrow} (b.center);
	\draw[line width=0.5mm, postaction={decorate}] (b2.center) -- node[sloped, allow upside down] {\midarrow} (p3.center);
\end{tikzpicture}
\caption{Parametrization of the loop momenta for the crossed double-box diagram.}
\label{fig: app_crossed_double_box_param}
\end{figure}

Before studying the generic case, let us first look at the crossed double-box diagram for pedagogical purposes, which we parametrize in fig.~\ref{fig: app_crossed_double_box_param}. We see that the $k_1$-loop has a propagator $2u_1 {\cdot} k_1 - 2u_1 {\cdot} k_2$, where the second term effectively acts as a constant. Using the loop-by-loop Baikov representation, this means that for the $k_1$-loop we have $E=3$ and thus $N_{\text{V}}=4<\nint$, so we are in the special case in the Baikov representation where the propagators are not independent. We can then obtain the loop-by-loop Baikov representation for the $k_1$-loop via~\cite{Weinzierl:2022eaz}
\begin{align}
\label{eq: app_proof_unraveling_Baikov}
& \text{Baikov} \left( \begin{tikzpicture}[baseline={([yshift=-0.1cm]current bounding box.center)}] 
	\node[] (a) at (0,0) {};
	\node[] (b) at (0,-2) {};
	\node[] (a1) at (2,0) {};
	\node[] (b1) at (2,-2) {};
	\node[] (a2) at (4,0) {};
	\node[] (b2) at (4,-2) {};
	\node[] (p1) at ($(a)+(-0.5,0)$) {};
	\node[] (p2) at ($(b)+(-0.5,0)$) {};
	\node[] (p3) at ($(b1)+(0.5,0)$) {};
	\node[] (p4) at ($(a2)+(0.5,0)$) {};
	\draw[line width=0.15mm, postaction={decorate}] (b.center) -- node[sloped, allow upside down, label={[xshift=0.15cm, yshift=0cm]$k_1$}] {\midarrow} (a.center);
	\draw[line width=0.15mm, postaction={decorate}] (a1.center) -- node[sloped, allow upside down] {\midarrow} (2.5,-0.5);
	\draw[line width=0.15mm, postaction={decorate}] (3,-1) -- node[sloped, allow upside down, label={[xshift=-1.6cm, yshift=0.6cm]$k_1{-}k_2$}] {\midarrow} (b1.center);
	\draw[line width=0.15mm] (3,-1) -- (a2.center);
	\draw[line width=0.5mm, postaction={decorate}] (a.center) -- node[sloped, allow upside down, label={[xshift=0cm, yshift=-0.15cm]$2u_1{\cdot} k_1$}] {\midarrow} (a1.center);
		\draw[line width=0.5mm, postaction={decorate}] (a1.center) -- node[sloped, allow upside down, label={[xshift=0cm, yshift=-0.15cm]$2u_1 {\cdot}(k_1{-}k_2)$}] {\midarrow} (a2.center);
	\draw[line width=0.5mm, postaction={decorate}] (b.center) -- node[sloped, allow upside down, label={[xshift=0cm, yshift=-0.9cm]$-2u_2{\cdot} k_1$}] {\midarrow} (b1.center);
	\draw[line width=0.5mm, postaction={decorate}] (p1.center) -- node[sloped, allow upside down] {\midarrow} (a.center);
	\draw[line width=0.5mm, postaction={decorate}] (a2.center) -- node[sloped, allow upside down] {\midarrow} (p4.center);
	\draw[line width=0.5mm, postaction={decorate}] (p2.center) -- node[sloped, allow upside down] {\midarrow} (b.center);
	\draw[line width=0.5mm, postaction={decorate}] (b1.center) -- node[sloped, allow upside down] {\midarrow} (p3.center);
\end{tikzpicture} \right) \nonumber \\
& \qquad \qquad \qquad = 
\text{Baikov} \left( \begin{tikzpicture}[baseline={([yshift=-0.1cm]current bounding box.center)}] 
	\node[] (a) at (0,0) {};
	\node[] (b) at (0,-2) {};
	\node[] (a1) at (2,0) {};
	\node[] (b1) at (2,-2) {};
	\node[] (p1) at ($(a)+(-0.5,0)$) {};
	\node[] (p2) at ($(b)+(-0.5,0)$) {};
	\node[] (p3) at ($(b1)+(0.5,0)$) {};
	\node[] (p4) at ($(a1)+(0.5,0)$) {};
	\draw[line width=0.15mm, postaction={decorate}] (b.center) -- node[sloped, allow upside down, label={[xshift=0.15cm, yshift=0cm]$k_1$}] {\midarrow} (a.center);
	\draw[line width=0.15mm, postaction={decorate}] (a1.center) -- node[sloped, allow upside down] {\midarrow} (2.5,-0.5);
	\draw[line width=0.15mm, postaction={decorate}] (a1.center) -- node[sloped, allow upside down, label={[xshift=-1.6cm, yshift=0cm]$k_1{-}k_2$}] {\midarrow} (b1.center);
	\draw[line width=0.5mm, postaction={decorate}] (a.center) -- node[sloped, allow upside down, label={[xshift=0cm, yshift=-0.15cm]$2u_1{\cdot} k_1$}] {\midarrow} (a1.center);
	\draw[line width=0.5mm, postaction={decorate}] (b.center) -- node[sloped, allow upside down, label={[xshift=0cm, yshift=-0.9cm]$-2u_2{\cdot} k_1$}] {\midarrow} (b1.center);
	\draw[line width=0.5mm, postaction={decorate}] (p1.center) -- node[sloped, allow upside down] {\midarrow} (a.center);
	\draw[line width=0.5mm, postaction={decorate}] (a1.center) -- node[sloped, allow upside down] {\midarrow} (p4.center);
	\draw[line width=0.5mm, postaction={decorate}] (p2.center) -- node[sloped, allow upside down] {\midarrow} (b.center);
	\draw[line width=0.5mm, postaction={decorate}] (b1.center) -- node[sloped, allow upside down] {\midarrow} (p3.center);
\end{tikzpicture} \right) \times \frac{1}{2u_1 {\cdot} k_1 - 2u_1 {\cdot} k_2}\Bigg|_{2u_1 \cdot k_1 \to z},
\end{align}
where we replace the corresponding Baikov variable in the last propagator.

Since the Baikov variable $z=2u_1 \cdot k_1$ vanishes at the maximal cut, the propagator $\frac{1}{- 2u_1 \cdot k_2}$ becomes exactly that of the planar double box (up to a minus sign) for the $k_2$-loop. Hence, this proves that the leading singularity is equivalent to that of the planar double box. Repeating the process but choosing instead the propagator $\frac{1}{2u_1 \cdot (k_1{-}k_2)}$ as the independent one, we finally obtain the relation
\begin{equation}
\label{eq: app_LS_double_box_relation}
\LS \left( \begin{tikzpicture}[baseline={([yshift=-0.1cm]current bounding box.center)}] 
	\node[] (a) at (0,0) {};
	\node[] (a1) at (0.8,0) {};
	\node[] (a2) at (1.6,0) {};
	\node[] (b) at (0,-0.8) {};
	\node[] (b1) at (0.8,-0.8) {};
	\node[] (b2) at (1.6,-0.8) {};
	\node[] (p1) at ($(a)+(-0.2,0)$) {};
	\node[] (p2) at ($(b)+(-0.2,0)$) {};
	\node[] (p3) at ($(b2)+(0.2,0)$) {};
	\node[] (p4) at ($(a2)+(0.2,0)$) {};
	\draw[line width=0.15mm] (b.center) -- (a.center);
	\draw[line width=0.15mm] (b1.center) -- (a1.center);
	\draw[line width=0.15mm] (b2.center) -- (a2.center);
	\draw[line width=0.5mm] (p1.center) -- (p4.center);
	\draw[line width=0.5mm] (p2.center) -- (p3.center);
\end{tikzpicture} \right) \propto \LS \left( \begin{tikzpicture}[baseline={([yshift=-0.1cm]current bounding box.center)}] 
	\node[] (a) at (0,0) {};
	\node[] (a1) at (0.8,0) {};
	\node[] (a2) at (1.6,0) {};
	\node[] (b) at (0,-0.8) {};
	\node[] (b1) at (0.8,-0.8) {};
	\node[] (b2) at (1.6,-0.8) {};
	\node[] (p1) at ($(a)+(-0.2,0)$) {};
	\node[] (p2) at ($(b)+(-0.2,0)$) {};
	\node[] (p3) at ($(b2)+(0.2,0)$) {};
	\node[] (p4) at ($(a2)+(0.2,0)$) {};
	\draw[line width=0.15mm] (b.center) -- (a.center);
	\draw[line width=0.15mm] (b1.center) -- (a2.center);
	\draw[line width=0.15mm] (b2.center) -- (1.3,-0.5);
	\draw[line width=0.15mm] (1.1,-0.3) -- (a1.center);
	\draw[line width=0.5mm] (p1.center) -- (p4.center);
	\draw[line width=0.5mm] (p2.center) -- (p3.center);
\end{tikzpicture} \right) \propto \LS \left( \begin{tikzpicture}[baseline={([yshift=-0.1cm]current bounding box.center)}] 
	\node[] (a) at (2.4,0) {};
	\node[] (a1) at (0.8,0) {};
	\node[] (a2) at (1.6,0) {};
	\node[] (b) at (2.4,-0.8) {};
	\node[] (b1) at (0.8,-0.8) {};
	\node[] (b2) at (1.6,-0.8) {};
	\node[] (p1) at ($(a1)+(-0.2,0)$) {};
	\node[] (p2) at ($(b1)+(-0.2,0)$) {};
	\node[] (p3) at ($(b)+(0.2,0)$) {};
	\node[] (p4) at ($(a)+(0.2,0)$) {};
	\draw[line width=0.15mm] (b.center) -- (a.center);
	\draw[line width=0.15mm] (b1.center) -- (a2.center);
	\draw[line width=0.15mm] (b2.center) -- (1.3,-0.5);
	\draw[line width=0.15mm] (1.1,-0.3) -- (a1.center);
	\draw[line width=0.5mm] (p1.center) -- (p4.center);
	\draw[line width=0.5mm] (p2.center) -- (p3.center);
\end{tikzpicture} \right).
\end{equation}
The equivalence of these leading singularities can be related to the fact that, using partial fraction on the linearized propagators, we can write
\begin{equation}
\label{eq: app_LS_double_box_PF}
\begin{tikzpicture}[baseline={([yshift=-0.1cm]current bounding box.center)}] 
	\node[] (a) at (0,0) {};
	\node[] (a1) at (0.8,0) {};
	\node[] (a2) at (1.6,0) {};
	\node[] (b) at (0,-0.8) {};
	\node[] (b1) at (0.8,-0.8) {};
	\node[] (b2) at (1.6,-0.8) {};
	\node[] (p1) at ($(a)+(-0.2,0)$) {};
	\node[] (p2) at ($(b)+(-0.2,0)$) {};
	\node[] (p3) at ($(b2)+(0.2,0)$) {};
	\node[] (p4) at ($(a2)+(0.2,0)$) {};
	\draw[line width=0.15mm] (b.center) -- (a.center);
	\draw[line width=0.15mm] (b1.center) -- (a1.center);
	\draw[line width=0.15mm] (b2.center) -- (a2.center);
	\draw[line width=0.5mm] (p1.center) -- (p4.center);
	\draw[line width=0.5mm] (p2.center) -- (p3.center);
\end{tikzpicture} + \begin{tikzpicture}[baseline={([yshift=-0.1cm]current bounding box.center)}] 
	\node[] (a) at (0,0) {};
	\node[] (a1) at (0.8,0) {};
	\node[] (a2) at (1.6,0) {};
	\node[] (b) at (0,-0.8) {};
	\node[] (b1) at (0.8,-0.8) {};
	\node[] (b2) at (1.6,-0.8) {};
	\node[] (p1) at ($(a)+(-0.2,0)$) {};
	\node[] (p2) at ($(b)+(-0.2,0)$) {};
	\node[] (p3) at ($(b2)+(0.2,0)$) {};
	\node[] (p4) at ($(a2)+(0.2,0)$) {};
	\draw[line width=0.15mm] (b.center) -- (a.center);
	\draw[line width=0.15mm] (b1.center) -- (a2.center);
	\draw[line width=0.15mm] (b2.center) -- (1.3,-0.5);
	\draw[line width=0.15mm] (1.1,-0.3) -- (a1.center);
	\draw[line width=0.5mm] (p1.center) -- (p4.center);
	\draw[line width=0.5mm] (p2.center) -- (p3.center);
\end{tikzpicture} + \begin{tikzpicture}[baseline={([yshift=-0.1cm]current bounding box.center)}] 
	\node[] (a) at (2.4,0) {};
	\node[] (a1) at (0.8,0) {};
	\node[] (a2) at (1.6,0) {};
	\node[] (b) at (2.4,-0.8) {};
	\node[] (b1) at (0.8,-0.8) {};
	\node[] (b2) at (1.6,-0.8) {};
	\node[] (p1) at ($(a1)+(-0.2,0)$) {};
	\node[] (p2) at ($(b1)+(-0.2,0)$) {};
	\node[] (p3) at ($(b)+(0.2,0)$) {};
	\node[] (p4) at ($(a)+(0.2,0)$) {};
	\draw[line width=0.15mm] (b.center) -- (a.center);
	\draw[line width=0.15mm] (b1.center) -- (a2.center);
	\draw[line width=0.15mm] (b2.center) -- (1.3,-0.5);
	\draw[line width=0.15mm] (1.1,-0.3) -- (a1.center);
	\draw[line width=0.5mm] (p1.center) -- (p4.center);
	\draw[line width=0.5mm] (p2.center) -- (p3.center);
\end{tikzpicture} = 0.
\end{equation}
Using the Baikov representation, we can conclude that the same geometry appears in all the terms in the partial fraction expansion. However, at first glance it would seem as if eq.~\eqref{eq: app_LS_double_box_relation} is not compatible with the partial fraction in eq.~\eqref{eq: app_LS_double_box_PF}. As we shall see in the following, this is due to the ambiguity inherent to the definition of the leading singularity, which is dependent on the contour chosen to compute the residues.

First of all, we can parametrize the three diagrams as in fig.~\ref{fig: app_crossed_double_box_param}, such that the only difference resides in the linearized propagators for the $u_1$ matter line. Therefore, using the standard Baikov representation~\eqref{eq: standard_Baikov_ISP}, the three diagrams can be written as integrals over the same Baikov polynomial $\mathcal{B}(z)$, but with different linearized propagators:
\begin{align}
\begin{tikzpicture}[baseline={([yshift=-0.1cm]current bounding box.center)}] 
	\node[] (a) at (0,0) {};
	\node[] (a1) at (0.8,0) {};
	\node[] (a2) at (1.6,0) {};
	\node[] (b) at (0,-0.8) {};
	\node[] (b1) at (0.8,-0.8) {};
	\node[] (b2) at (1.6,-0.8) {};
	\node[] (p1) at ($(a)+(-0.2,0)$) {};
	\node[] (p2) at ($(b)+(-0.2,0)$) {};
	\node[] (p3) at ($(b2)+(0.2,0)$) {};
	\node[] (p4) at ($(a2)+(0.2,0)$) {};
	\draw[line width=0.15mm] (b.center) -- (a.center);
	\draw[line width=0.15mm] (b1.center) -- (a1.center);
	\draw[line width=0.15mm] (b2.center) -- (a2.center);
	\draw[line width=0.5mm] (p1.center) -- (p4.center);
	\draw[line width=0.5mm] (p2.center) -- (p3.center);
\end{tikzpicture} & \propto \int \frac{d z_1 \cdots d z_9}{z_1 z_2 z_3 \cdots z_7} \ {\mathcal{B}(z_1,z_2,z_3,\dots,z_9)}^{-1}, \\
\begin{tikzpicture}[baseline={([yshift=-0.1cm]current bounding box.center)}] 
	\node[] (a) at (0,0) {};
	\node[] (a1) at (0.8,0) {};
	\node[] (a2) at (1.6,0) {};
	\node[] (b) at (0,-0.8) {};
	\node[] (b1) at (0.8,-0.8) {};
	\node[] (b2) at (1.6,-0.8) {};
	\node[] (p1) at ($(a)+(-0.2,0)$) {};
	\node[] (p2) at ($(b)+(-0.2,0)$) {};
	\node[] (p3) at ($(b2)+(0.2,0)$) {};
	\node[] (p4) at ($(a2)+(0.2,0)$) {};
	\draw[line width=0.15mm] (b.center) -- (a.center);
	\draw[line width=0.15mm] (b1.center) -- (a2.center);
	\draw[line width=0.15mm] (b2.center) -- (1.3,-0.5);
	\draw[line width=0.15mm] (1.1,-0.3) -- (a1.center);
	\draw[line width=0.5mm] (p1.center) -- (p4.center);
	\draw[line width=0.5mm] (p2.center) -- (p3.center);
\end{tikzpicture} & \propto \int \frac{d z_1 \cdots d z_9}{z_1 (z_1 - z_2) z_3 \cdots z_7} \ {\mathcal{B}(z_1,z_1 -z_2,z_3,\dots,z_9)}^{-1}, \\
\begin{tikzpicture}[baseline={([yshift=-0.1cm]current bounding box.center)}] 
	\node[] (a) at (2.4,0) {};
	\node[] (a1) at (0.8,0) {};
	\node[] (a2) at (1.6,0) {};
	\node[] (b) at (2.4,-0.8) {};
	\node[] (b1) at (0.8,-0.8) {};
	\node[] (b2) at (1.6,-0.8) {};
	\node[] (p1) at ($(a1)+(-0.2,0)$) {};
	\node[] (p2) at ($(b1)+(-0.2,0)$) {};
	\node[] (p3) at ($(b)+(0.2,0)$) {};
	\node[] (p4) at ($(a)+(0.2,0)$) {};
	\draw[line width=0.15mm] (b.center) -- (a.center);
	\draw[line width=0.15mm] (b1.center) -- (a2.center);
	\draw[line width=0.15mm] (b2.center) -- (1.3,-0.5);
	\draw[line width=0.15mm] (1.1,-0.3) -- (a1.center);
	\draw[line width=0.5mm] (p1.center) -- (p4.center);
	\draw[line width=0.5mm] (p2.center) -- (p3.center);
\end{tikzpicture} & \propto \int \frac{d z_1 \cdots d z_9}{(z_2-z_1) z_2 z_3 \cdots z_7} \ {\mathcal{B}(z_2-z_1,z_2,z_3,\dots,z_9)}^{-1},
\end{align}
with $z_1 = 2u_1 \cdot k_1$ and $z_2 = 2 u_1 \cdot k_2$, where $z_3,\dots,z_7$ are the remaining Baikov variables for the propagators, and where we drop the common constant prefactors. For the three diagrams we also require the extra Baikov variables $z_8=k_2^2$ and $z_9=(k_1{-}q)^2$. As can be seen, these expressions indeed satisfy the partial fraction relation~\eqref{eq: app_LS_double_box_PF}.

At this point, we would like to calculate the residue around the poles $z_1=z_2=0$ when we cut the matter propagators for the $u_1$ matter line in the planar double box. However, this accidentally implies that we also have $z_1 - z_2 = 0$ for the non-planar diagrams, thus also cutting their propagators. To separate the poles, we can rescale $z_2 \to z_2 z_1$ in all diagrams (an operation known as a \textit{blow-up}), leading to
\begin{align}
\begin{tikzpicture}[baseline={([yshift=-0.1cm]current bounding box.center)}] 
	\node[] (a) at (0,0) {};
	\node[] (a1) at (0.8,0) {};
	\node[] (a2) at (1.6,0) {};
	\node[] (b) at (0,-0.8) {};
	\node[] (b1) at (0.8,-0.8) {};
	\node[] (b2) at (1.6,-0.8) {};
	\node[] (p1) at ($(a)+(-0.2,0)$) {};
	\node[] (p2) at ($(b)+(-0.2,0)$) {};
	\node[] (p3) at ($(b2)+(0.2,0)$) {};
	\node[] (p4) at ($(a2)+(0.2,0)$) {};
	\draw[line width=0.15mm] (b.center) -- (a.center);
	\draw[line width=0.15mm] (b1.center) -- (a1.center);
	\draw[line width=0.15mm] (b2.center) -- (a2.center);
	\draw[line width=0.5mm] (p1.center) -- (p4.center);
	\draw[line width=0.5mm] (p2.center) -- (p3.center);
\end{tikzpicture} & \propto \int \frac{d z_1 \cdots d z_9}{z_1 z_2 z_3 \cdots z_7} \ {\mathcal{B}(z_1,z_1 z_2,z_3,\dots,z_9)}^{-1}, \\
\begin{tikzpicture}[baseline={([yshift=-0.1cm]current bounding box.center)}] 
	\node[] (a) at (0,0) {};
	\node[] (a1) at (0.8,0) {};
	\node[] (a2) at (1.6,0) {};
	\node[] (b) at (0,-0.8) {};
	\node[] (b1) at (0.8,-0.8) {};
	\node[] (b2) at (1.6,-0.8) {};
	\node[] (p1) at ($(a)+(-0.2,0)$) {};
	\node[] (p2) at ($(b)+(-0.2,0)$) {};
	\node[] (p3) at ($(b2)+(0.2,0)$) {};
	\node[] (p4) at ($(a2)+(0.2,0)$) {};
	\draw[line width=0.15mm] (b.center) -- (a.center);
	\draw[line width=0.15mm] (b1.center) -- (a2.center);
	\draw[line width=0.15mm] (b2.center) -- (1.3,-0.5);
	\draw[line width=0.15mm] (1.1,-0.3) -- (a1.center);
	\draw[line width=0.5mm] (p1.center) -- (p4.center);
	\draw[line width=0.5mm] (p2.center) -- (p3.center);
\end{tikzpicture} & \propto \int \frac{d z_1 \cdots d z_9}{z_1 (1 - z_2) z_3 \cdots z_7} \ {\mathcal{B}(z_1,z_1(1 -z_2),z_3,\dots,z_9)}^{-1}, \\
\begin{tikzpicture}[baseline={([yshift=-0.1cm]current bounding box.center)}] 
	\node[] (a) at (2.4,0) {};
	\node[] (a1) at (0.8,0) {};
	\node[] (a2) at (1.6,0) {};
	\node[] (b) at (2.4,-0.8) {};
	\node[] (b1) at (0.8,-0.8) {};
	\node[] (b2) at (1.6,-0.8) {};
	\node[] (p1) at ($(a1)+(-0.2,0)$) {};
	\node[] (p2) at ($(b1)+(-0.2,0)$) {};
	\node[] (p3) at ($(b)+(0.2,0)$) {};
	\node[] (p4) at ($(a)+(0.2,0)$) {};
	\draw[line width=0.15mm] (b.center) -- (a.center);
	\draw[line width=0.15mm] (b1.center) -- (a2.center);
	\draw[line width=0.15mm] (b2.center) -- (1.3,-0.5);
	\draw[line width=0.15mm] (1.1,-0.3) -- (a1.center);
	\draw[line width=0.5mm] (p1.center) -- (p4.center);
	\draw[line width=0.5mm] (p2.center) -- (p3.center);
\end{tikzpicture} & \propto \int \frac{d z_1 \cdots d z_9}{z_1(z_2-1) z_2 z_3 \cdots z_7} \ {\mathcal{B}(z_1(z_2-1),z_1 z_2,z_3,\dots,z_9)}^{-1}.
\end{align}
Now, we can see that the maximal cuts are actually related pair-wise in the partial fraction identity~\eqref{eq: app_LS_double_box_PF}. For example, taking the maximal cut as the residue around the poles $z_1=\dots=z_7=0$ yields
\begin{equation}
\LS \left( \begin{tikzpicture}[baseline={([yshift=-0.1cm]current bounding box.center)}] 
	\node[] (a) at (0,0) {};
	\node[] (a1) at (0.8,0) {};
	\node[] (a2) at (1.6,0) {};
	\node[] (b) at (0,-0.8) {};
	\node[] (b1) at (0.8,-0.8) {};
	\node[] (b2) at (1.6,-0.8) {};
	\node[] (p1) at ($(a)+(-0.2,0)$) {};
	\node[] (p2) at ($(b)+(-0.2,0)$) {};
	\node[] (p3) at ($(b2)+(0.2,0)$) {};
	\node[] (p4) at ($(a2)+(0.2,0)$) {};
	\draw[line width=0.15mm] (b.center) -- (a.center);
	\draw[line width=0.15mm] (b1.center) -- (a1.center);
	\draw[line width=0.15mm] (b2.center) -- (a2.center);
	\draw[line width=0.5mm] (p1.center) -- (p4.center);
	\draw[line width=0.5mm] (p2.center) -- (p3.center);
\end{tikzpicture} \right) = - \, \LS \left( \begin{tikzpicture}[baseline={([yshift=-0.1cm]current bounding box.center)}] 
	\node[] (a) at (2.4,0) {};
	\node[] (a1) at (0.8,0) {};
	\node[] (a2) at (1.6,0) {};
	\node[] (b) at (2.4,-0.8) {};
	\node[] (b1) at (0.8,-0.8) {};
	\node[] (b2) at (1.6,-0.8) {};
	\node[] (p1) at ($(a1)+(-0.2,0)$) {};
	\node[] (p2) at ($(b1)+(-0.2,0)$) {};
	\node[] (p3) at ($(b)+(0.2,0)$) {};
	\node[] (p4) at ($(a)+(0.2,0)$) {};
	\draw[line width=0.15mm] (b.center) -- (a.center);
	\draw[line width=0.15mm] (b1.center) -- (a2.center);
	\draw[line width=0.15mm] (b2.center) -- (1.3,-0.5);
	\draw[line width=0.15mm] (1.1,-0.3) -- (a1.center);
	\draw[line width=0.5mm] (p1.center) -- (p4.center);
	\draw[line width=0.5mm] (p2.center) -- (p3.center);
\end{tikzpicture} \right) \propto \LS \left( \int \frac{d z_8 d z_9}{\mathcal{B}(0,\dots,0,z_8,z_9)} \right),
\end{equation}
while the residue for the second diagram vanishes, thus satisfying eq.~\eqref{eq: app_LS_double_box_PF}. Similarly, for $z_2=1$ and $z_1=z_3=\dots=z_7=0$,
\begin{equation}
\LS \left( \begin{tikzpicture}[baseline={([yshift=-0.1cm]current bounding box.center)}] 
	\node[] (a) at (2.4,0) {};
	\node[] (a1) at (0.8,0) {};
	\node[] (a2) at (1.6,0) {};
	\node[] (b) at (2.4,-0.8) {};
	\node[] (b1) at (0.8,-0.8) {};
	\node[] (b2) at (1.6,-0.8) {};
	\node[] (p1) at ($(a1)+(-0.2,0)$) {};
	\node[] (p2) at ($(b1)+(-0.2,0)$) {};
	\node[] (p3) at ($(b)+(0.2,0)$) {};
	\node[] (p4) at ($(a)+(0.2,0)$) {};
	\draw[line width=0.15mm] (b.center) -- (a.center);
	\draw[line width=0.15mm] (b1.center) -- (a2.center);
	\draw[line width=0.15mm] (b2.center) -- (1.3,-0.5);
	\draw[line width=0.15mm] (1.1,-0.3) -- (a1.center);
	\draw[line width=0.5mm] (p1.center) -- (p4.center);
	\draw[line width=0.5mm] (p2.center) -- (p3.center);
\end{tikzpicture} \right) = - \, \LS \left( \begin{tikzpicture}[baseline={([yshift=-0.1cm]current bounding box.center)}] 
	\node[] (a) at (0,0) {};
	\node[] (a1) at (0.8,0) {};
	\node[] (a2) at (1.6,0) {};
	\node[] (b) at (0,-0.8) {};
	\node[] (b1) at (0.8,-0.8) {};
	\node[] (b2) at (1.6,-0.8) {};
	\node[] (p1) at ($(a)+(-0.2,0)$) {};
	\node[] (p2) at ($(b)+(-0.2,0)$) {};
	\node[] (p3) at ($(b2)+(0.2,0)$) {};
	\node[] (p4) at ($(a2)+(0.2,0)$) {};
	\draw[line width=0.15mm] (b.center) -- (a.center);
	\draw[line width=0.15mm] (b1.center) -- (a2.center);
	\draw[line width=0.15mm] (b2.center) -- (1.3,-0.5);
	\draw[line width=0.15mm] (1.1,-0.3) -- (a1.center);
	\draw[line width=0.5mm] (p1.center) -- (p4.center);
	\draw[line width=0.5mm] (p2.center) -- (p3.center);
\end{tikzpicture} \right) \propto \LS \left( \int \frac{d z_8 d z_9}{\mathcal{B}(0,\dots,0,z_8,z_9)} \right),
\end{equation}
while the residue for the first diagram vanishes, again in accordance with eq.~\eqref{eq: app_LS_double_box_PF}. Performing instead a blow-up $z_1 \to z_1 z_2$ in the three diagrams, and taking the maximal cut as the residue around the poles $z_1=\dots=z_7=0$, we obtain
\begin{equation}
\LS \left( \begin{tikzpicture}[baseline={([yshift=-0.1cm]current bounding box.center)}] 
	\node[] (a) at (0,0) {};
	\node[] (a1) at (0.8,0) {};
	\node[] (a2) at (1.6,0) {};
	\node[] (b) at (0,-0.8) {};
	\node[] (b1) at (0.8,-0.8) {};
	\node[] (b2) at (1.6,-0.8) {};
	\node[] (p1) at ($(a)+(-0.2,0)$) {};
	\node[] (p2) at ($(b)+(-0.2,0)$) {};
	\node[] (p3) at ($(b2)+(0.2,0)$) {};
	\node[] (p4) at ($(a2)+(0.2,0)$) {};
	\draw[line width=0.15mm] (b.center) -- (a.center);
	\draw[line width=0.15mm] (b1.center) -- (a1.center);
	\draw[line width=0.15mm] (b2.center) -- (a2.center);
	\draw[line width=0.5mm] (p1.center) -- (p4.center);
	\draw[line width=0.5mm] (p2.center) -- (p3.center);
\end{tikzpicture} \right) = - \, \LS \left( \begin{tikzpicture}[baseline={([yshift=-0.1cm]current bounding box.center)}] 
	\node[] (a) at (0,0) {};
	\node[] (a1) at (0.8,0) {};
	\node[] (a2) at (1.6,0) {};
	\node[] (b) at (0,-0.8) {};
	\node[] (b1) at (0.8,-0.8) {};
	\node[] (b2) at (1.6,-0.8) {};
	\node[] (p1) at ($(a)+(-0.2,0)$) {};
	\node[] (p2) at ($(b)+(-0.2,0)$) {};
	\node[] (p3) at ($(b2)+(0.2,0)$) {};
	\node[] (p4) at ($(a2)+(0.2,0)$) {};
	\draw[line width=0.15mm] (b.center) -- (a.center);
	\draw[line width=0.15mm] (b1.center) -- (a2.center);
	\draw[line width=0.15mm] (b2.center) -- (1.3,-0.5);
	\draw[line width=0.15mm] (1.1,-0.3) -- (a1.center);
	\draw[line width=0.5mm] (p1.center) -- (p4.center);
	\draw[line width=0.5mm] (p2.center) -- (p3.center);
\end{tikzpicture} \right) \propto \LS \left( \int \frac{d z_8 d z_9}{\mathcal{B}(0,\dots,0,z_8,z_9)} \right),
\end{equation}
while the residue for the third diagram vanishes, again satisfying eq.~\eqref{eq: app_LS_double_box_PF}. As can be seen, the result of the leading singularity is the same for the three diagrams (up to a sign), while the contour chosen to compute the residues depends on the specific diagram.

Generalizing the previous observations, the equivalence between these leading singularities only relied on having two adjacent linearized propagators, not on the diagram itself, and is also applicable beyond cubic vertices. Therefore, the same argument holds for all non-planar diagrams with two adjacent matter propagators. In the worst case, the non-planar loop appears in the middle of a diagram, so that we have instead three adjacent matter propagators.\footnote{For higher-loop diagrams we can always re-route the remaining loop momenta around the three adjacent matter propagators, where they act as a constant in all of them, thus not modifying the argument.} However, we are always able to find a parametrization such that the following unraveling (to be understood as being part of a bigger diagram) is allowed:

\begin{align}
& \LS \left(
\begin{tikzpicture}[baseline=(current bounding box.center)] 
	\node[] (a) at (0,0) {};
	\node[] (b) at (-0.3,-1) {};
	\node[] (bmod) at (0.3,-1) {};
	\node[] (a1) at (2,0) {};
	\node[] (a2) at (4.5,0) {};
	\node[] (b1) at (1.7,-1) {};
	\node[] (b1mod) at (2.3,-1) {};
	\node[] (p1) at ($(a)+(-1.75,0)$) {};
	\draw[line width=0.15mm] (b.center) -- (a.center);
	\draw[line width=0.15mm] (bmod.center) -- (a.center);
	\draw[line width=0.15mm,-{Latex[length=2.2mm]}](-0.4,-0.9) -- (-0.4,-0.2);
	\node[label={[xshift=-0.5cm, yshift=-0.5cm]$k_2$}] (label1) at (-0.2,-0.5) {};
	\node at (-0.15,-0.9)[circle,fill,inner sep=0.65pt]{};
	\node at (0,-0.9)[circle,fill,inner sep=0.65pt]{};
	\node at (0.15,-0.9)[circle,fill,inner sep=0.65pt]{};
	\draw[line width=0.15mm] (b1.center) -- (a1.center);
	\draw[line width=0.15mm] (b1mod.center) -- (a1.center);
	\draw[line width=0.15mm,-{Latex[length=2.2mm]}](2.4,-0.9) -- (2.4,-0.2);
	\node[label={[xshift=1.65cm, yshift=-0.5cm]$k_3{-}k_1{-}k_2$}] (label2) at (1.8,-0.5) {};
	\node at (1.85,-0.9)[circle,fill,inner sep=0.65pt]{};
	\node at (2,-0.9)[circle,fill,inner sep=0.65pt]{};
	\node at (2.15,-0.9)[circle,fill,inner sep=0.65pt]{};
	\draw[line width=0.5mm, postaction={decorate}] (p1.center) -- node[sloped, allow upside down, label={[xshift=0cm, yshift=-0.15cm]$2u_1{\cdot} k_1$}] {\midarrow} (a.center);
	\draw[line width=0.5mm, postaction={decorate}] (a.center) -- node[sloped, allow upside down, label={[xshift=0cm, yshift=-0.15cm]$2u_1{\cdot} (k_1{+}k_2)$}] {\midarrow} (a1.center);
		\draw[line width=0.5mm, postaction={decorate}] (a1.center) -- node[sloped, allow upside down, label={[xshift=0cm, yshift=-0.15cm]$2u_1{\cdot} k_3$}] {\midarrow} (a2.center);
\end{tikzpicture} \right) \nonumber \\
& \qquad \qquad  = \LS \left(
\begin{tikzpicture}[baseline=(current bounding box.center)] 
	\node[] (a) at (0,0) {};
	\node[] (b) at (-0.9,-1) {};
	\node[] (bmod) at (-0.3,-1) {};
	\node[] (a1) at (2,0) {};
	\node[] (a2) at (3,0) {};
	\node[] (b1) at (0.3,-1) {};
	\node[] (b1mod) at (0.9,-1) {};
	\node[] (p1) at ($(a)+(-1.75,0)$) {};
	\draw[line width=0.15mm] (b.center) -- (a.center);
	\draw[line width=0.15mm] (bmod.center) -- (a.center);
	\draw[line width=0.15mm,-{Latex[length=2.2mm]}](-1.1,-0.9) -- (-0.4,-0.2);
	\node[label={[xshift=-0.5cm, yshift=-0.4cm]$k_2$}] (label1) at (-0.6,-0.5) {};
	\node at (-0.7,-0.9)[circle,fill,inner sep=0.65pt]{};
	\node at (-0.55,-0.9)[circle,fill,inner sep=0.65pt]{};
	\node at (-0.4,-0.9)[circle,fill,inner sep=0.65pt]{};
	\draw[line width=0.15mm] (b1.center) -- (a.center);
	\draw[line width=0.15mm] (b1mod.center) -- (a.center);
	\draw[line width=0.15mm,-{Latex[length=2.2mm]}](1.1,-0.9) -- (0.4,-0.2);
	\node[label={[xshift=0.8cm, yshift=-0.4cm]$k_3{-}k_1{-}k_2$}] (label2) at (1,-0.5) {};
	\node at (0.4,-0.9)[circle,fill,inner sep=0.65pt]{};
	\node at (0.55,-0.9)[circle,fill,inner sep=0.65pt]{};
	\node at (0.7,-0.9)[circle,fill,inner sep=0.65pt]{};		
	\draw[line width=0.5mm, postaction={decorate}] (p1.center) -- node[sloped, allow upside down, label={[xshift=0cm, yshift=-0.15cm]$2u_1{\cdot} k_1$}] {\midarrow} (a.center);
	\draw[line width=0.5mm, postaction={decorate}] (a.center) -- node[sloped, allow upside down, label={[xshift=0cm, yshift=-0.15cm]$2u_1{\cdot} k_3$}] {\midarrow} (a2.center);
\end{tikzpicture} \times \frac{-1}{\underbrace{2u_1{\cdot} k_3}_{0}-\underbrace{2u_1{\cdot} k_1}_{0} - 2u_1{\cdot} k_2} \right) \nonumber \\
& \qquad \qquad = - \ \LS \left(
\begin{tikzpicture}[baseline=(current bounding box.center)] 
	\node[] (a) at (0,0) {};
	\node[] (b) at (-0.4,-1) {};
	\node[] (bmod) at (0.4,-1) {};
	\node[] (a1) at (2,0) {};
	\node[] (a2) at (4.5,0) {};
	\node[] (b1) at (1.6,-1) {};
	\node[] (b1mod) at (2.4,-1) {};
	\node[] (p1) at ($(a)+(-1.75,0)$) {};
	\draw[line width=0.15mm] (b.center) -- (a1.center);
	\draw[line width=0.15mm] (bmod.center) -- (a1.center);
	\draw[line width=0.15mm,-{Latex[length=2.2mm]}](-0.6,-0.9) -- (0.3,-0.5);
	\node[label={[xshift=0cm, yshift=-0.5cm]$k_2$}] (label1) at (-0.6,-0.5) {};
	\node at (0.05,-0.9)[circle,fill,inner sep=0.65pt]{};
	\node at (0.2,-0.9)[circle,fill,inner sep=0.65pt]{};
	\node at (0.35,-0.9)[circle,fill,inner sep=0.65pt]{};
	\draw[line width=0.15mm] (b1.center) -- (1.08,-0.68);
	\draw[line width=0.15mm] (0.72,-0.45) -- (a.center);
	\draw[line width=0.15mm] (b1mod.center) -- (1.3,-0.54);
	\draw[line width=0.15mm] (0.9,-0.37) -- (a.center);
	\draw[line width=0.15mm,-{Latex[length=2.2mm]}](2.6,-0.9) -- (1.7,-0.5);
	\node[label={[xshift=0.7cm, yshift=-0.5cm]$k_3{-}k_1{-}k_2$}] (label2) at (2.6,-0.5) {};
	\node at (1.65,-0.9)[circle,fill,inner sep=0.65pt]{};
	\node at (1.8,-0.9)[circle,fill,inner sep=0.65pt]{};
	\node at (1.95,-0.9)[circle,fill,inner sep=0.65pt]{};
	\draw[line width=0.5mm, postaction={decorate}] (p1.center) -- node[sloped, allow upside down, label={[xshift=0cm, yshift=-0.15cm]$2u_1{\cdot} k_1$}] {\midarrow} (a.center);
	\draw[line width=0.5mm, postaction={decorate}] (a.center) -- node[sloped, allow upside down, label={[xshift=0cm, yshift=-0.15cm]$2u_1{\cdot} (k_3{-}k_2)$}] {\midarrow} (a1.center);
		\draw[line width=0.5mm, postaction={decorate}] (a1.center) -- node[sloped, allow upside down, label={[xshift=0cm, yshift=-0.15cm]$2u_1{\cdot} k_3$}] {\midarrow} (a2.center);
\end{tikzpicture} \right),
\end{align}
where we removed $2u_1{\cdot} k_1$ and added $2u_1{\cdot} k_3$, since both vanish on the maximal cut, which leads to relation~\eqref{eq: unraveling_matter_props}. Alternatively, one can prove the unraveling of matter propagators using partial fraction identities as we have shown explicitly above, since these partial fraction identities always arise when two or more matter propagators are adjacent.

\section{IBP reductions for bubble and triangle subdiagrams}
\label{app: IBPs_bubbles_triangles}

In this appendix, we study the integration-by-parts (IBP) identities of diagrams containing certain bubble and triangle subdiagrams on the matter lines. In particular, in sec.~\ref{app: IBPs_bubbles} we show that if a diagram contains a bubble subdiagram on a matter propagator and at least one of its vertices is cubic, as represented in fig.~\ref{fig: diagrams_zero_masters}(a), then the diagram has zero master integrals in both parity sectors. In addition, as we explicitly show at two loops, together with partial fractioning this also allows us to discard from our analysis all diagrams with homologous bubble corrections that go over more than one matter propagator. Analogously, in sec.~\ref{app: IBPs_triangles} we find that diagrams containing a triangle sub-loop attached to a matter line, where at least two of its vertices are cubic and one of them is the graviton self-interaction vertex (see fig.~\ref{fig: diagrams_zero_masters}(b)), also have zero master integrals in their sector, which generalizes the previous argument of ref.~\cite{Frellesvig:2023bbf}. This proves that these diagrams are fully expressible in terms of integrals in lower sectors, and thus can be discarded from our analysis.

\subsection{IBPs for bubble subdiagrams}
\label{app: IBPs_bubbles}

Let us begin by studying the family of integrals for a generic bubble correction to a matter propagator
\begin{equation}
\mathcal{I}(\nu_1,\nu_2) = \int \frac{d^D k}{{[ k^2 ]}^{\nu_1} {[ 2u_1 \cdot(k+l) ]}^{\nu_2}},
\end{equation}
where $l$ is a different loop momentum flowing through the bubble. This integral family represents subdiagrams such as
\begin{equation}
\begin{tikzpicture}[baseline={([yshift=-0.1cm]current bounding box.center)}, scale=0.53]
	\node[] (a) at (0,-0.3) {};
	\node[] (a1) at (1.1,-0.3) {};
	\node[] (a2) at (2.7,-0.3) {};
	\node[] (p1) at (0.69,-0.3) {};
	\node[] (p4) at (3.31,-0.3) {};
	\fill[gray!20] (2,0.2) circle (2.1);
	\fill[white] (2,0.2) circle (1.4);
	\draw[line width=0.15mm] (p4.center) arc (0:180:1.31);
	\draw[line width=0.5mm] (p1.center) -- (p4.center);
\end{tikzpicture},
\end{equation}
where the grey zone indicates that it should be understood as being part of a bigger diagram. For this integral family, we can consider IBPs of the form
\begin{equation}
0 = \int d^D k \; \frac{\partial}{\partial k^{\mu}} \frac{v^{\mu}}{{[ k^2 ]}^{\nu_1} {[ 2u_1 \cdot(k+l) ]}^{\nu_2}},
\end{equation}
where $v^\mu$ can be either $k^\mu$ or $u_1^\mu$. In particular, one obtains a system of equations
\begin{align}
& 0= (D-2 \nu_1 - \nu_2) \, \mathcal{I}(\nu_1,\nu_2) + 2 (u_1 \cdot l) \, \nu_2 \, \mathcal{I}(\nu_1,\nu_2+1), \\
& 0= - \nu_1 \, \mathcal{I}(\nu_1+1,\nu_2-1) + 2 (u_1 \cdot l) \, \nu_1 \, \mathcal{I}(\nu_1+1,\nu_2)-2 \nu_2 \, \mathcal{I}(\nu_1,\nu_2+1).
\end{align}
As can be easily seen, one cannot isolate the integral $\mathcal{I}(\nu_1,\nu_2)$ solely in terms of subsectors. Therefore, a generic bubble correction has master integrals in its sector (unless a reduction is possible using IBPs with respect to other loop momenta).

However, we notice that the second term in the first IBP relation would reduce the power of a hypothetical propagator $2 u_1 \cdot l$ if it were present, thus allowing for a reduction into lower sectors. Consequently, let us consider a new family of integrals
\begin{equation}
\mathcal{I}(\nu_1,\nu_2,\nu_3) = \int \frac{d^D k}{{[ k^2 ]}^{\nu_1} {[ 2u_1 \cdot(k+l) ]}^{\nu_2} {[ 2u_1 \cdot l ]}^{\nu_3}}.
\end{equation}
The third propagator should be understood as being integrated over in a subsequent integral over the loop momentum $l$. This integral family denotes a bubble correction to a matter propagator with one of its vertices being cubic, such as
\begin{equation}
\begin{tikzpicture}[baseline={([yshift=-0.1cm]current bounding box.center)}, scale=0.53]
	\node[] (a) at (0,-0.3) {};
	\node[] (a1) at (1.1,-0.3) {};
	\node[] (a2) at (2.7,-0.3) {};
	\node[] (p1) at (0.69,-0.3) {};
	\node[] (p4) at (3.31,-0.3) {};
	\fill[gray!20] (2,0.2) circle (2.1);
	\fill[white] (2,0.2) circle (1.4);
	\draw[line width=0.15mm] (p4.center) arc (0:180:0.8);
	\draw[line width=0.5mm] (p1.center) -- (p4.center);
\end{tikzpicture}.
\end{equation}

Repeating the IBP identities for $v^\mu = k^\mu$, we obtain
\begin{equation}
\mathcal{I}(\nu_1,\nu_2,\nu_3) = - \frac{\nu_2}{D - 2 \nu_1 - \nu_2} \, \mathcal{I}(\nu_1,\nu_2+1,\nu_3-1).
\end{equation}
Therefore, iterating this IBP relation one can reduce all integrals in this family to  $\mathcal{I}(\nu_1,\nu_2+\nu_3,0)$ where the third propagator is absent. This shows that all diagrams containing bubble corrections to matter propagators where at least one of its vertices is cubic have zero master integrals in their sector.

For instance, we find
\begin{align}
\begin{tikzpicture}[baseline={([yshift=-0.1cm]current bounding box.center)}, scale=0.53]
	\node[] (a) at (0,-0.3) {};
	\node[] (a1) at (1.1,-0.3) {};
	\node[] (a2) at (2.7,-0.3) {};
	\node[] (p1) at (0.69,-0.3) {};
	\node[] (p4) at (3.31,-0.3) {};
	\fill[gray!20] (2,0.2) circle (2.1);
	\fill[white] (2,0.2) circle (1.4);
	\draw[line width=0.15mm] (p4.center) arc (0:180:0.8);
	\draw[line width=0.5mm] (p1.center) -- (p4.center);
\end{tikzpicture} & = \frac{-1}{D-3} \begin{tikzpicture}[baseline={([yshift=-0.1cm]current bounding box.center)}, scale=0.53]
	\node[] (a) at (0,-0.3) {};
	\node[] (a1) at (1.1,-0.3) {};
	\node[] (a2) at (2.7,-0.3) {};
	\node[] (p1) at (0.69,-0.3) {};
	\node[] (p4) at (3.31,-0.3) {};
	\fill[gray!20] (2,0.2) circle (2.1);
	\fill[white] (2,0.2) circle (1.4);
	\draw[line width=0.15mm] (p4.center) arc (0:180:1.31);
	\draw[line width=0.5mm] (p1.center) -- (p4.center);
	\node at (2,-0.3) [circle,fill,inner sep=1.5pt]{};
\end{tikzpicture}, \\[0.2cm] \begin{tikzpicture}[baseline={([yshift=-0.1cm]current bounding box.center)}, scale=0.53]
	\node[] (a) at (0,-0.3) {};
	\node[] (a1) at (1.1,-0.3) {};
	\node[] (a2) at (2.7,-0.3) {};
	\node[] (p1) at (0.69,-0.3) {};
	\node[] (p4) at (3.31,-0.3) {};
	\fill[gray!20] (2,0.2) circle (2.1);
	\fill[white] (2,0.2) circle (1.4);
	\draw[line width=0.15mm] (p4.center) arc (0:180:0.8);
	\draw[line width=0.5mm] (p1.center) -- (p4.center);
	\node at (2.51,0.5) [circle,fill,inner sep=1.5pt]{};
\end{tikzpicture} &= \frac{-1}{D-5} \begin{tikzpicture}[baseline={([yshift=-0.1cm]current bounding box.center)}, scale=0.53]
	\node[] (a) at (0,-0.3) {};
	\node[] (a1) at (1.1,-0.3) {};
	\node[] (a2) at (2.7,-0.3) {};
	\node[] (p1) at (0.69,-0.3) {};
	\node[] (p4) at (3.31,-0.3) {};
	\fill[gray!20] (2,0.2) circle (2.1);
	\fill[white] (2,0.2) circle (1.4);
	\draw[line width=0.15mm] (p4.center) arc (0:180:1.31);
	\draw[line width=0.5mm] (p1.center) -- (p4.center);
	\node at (2,-0.3) [circle,fill,inner sep=1.5pt]{};
	\node at (2,1.01) [circle,fill,inner sep=1.5pt]{};
\end{tikzpicture}, \\[0.2cm]
\begin{tikzpicture}[baseline={([yshift=-0.1cm]current bounding box.center)}, scale=0.53]
	\node[] (a) at (0,-0.3) {};
	\node[] (a1) at (1.1,-0.3) {};
	\node[] (a2) at (2.7,-0.3) {};
	\node[] (p1) at (0.69,-0.3) {};
	\node[] (p4) at (3.31,-0.3) {};
	\fill[gray!20] (2,0.2) circle (2.1);
	\fill[white] (2,0.2) circle (1.4);
	\draw[line width=0.15mm] (p4.center) arc (0:180:0.8);
	\draw[line width=0.5mm] (p1.center) -- (p4.center);
	\node at (1.2,-0.3) [circle,fill,inner sep=1.5pt]{};
\end{tikzpicture} &= \frac{2}{(D-3)(D-4)} \begin{tikzpicture}[baseline={([yshift=-0.1cm]current bounding box.center)}, scale=0.53]
	\node[] (a) at (0,-0.3) {};
	\node[] (a1) at (1.1,-0.3) {};
	\node[] (a2) at (2.7,-0.3) {};
	\node[] (p1) at (0.69,-0.3) {};
	\node[] (p4) at (3.31,-0.3) {};
	\fill[gray!20] (2,0.2) circle (2.1);
	\fill[white] (2,0.2) circle (1.4);
	\draw[line width=0.15mm] (p4.center) arc (0:180:1.31);
	\draw[line width=0.5mm] (p1.center) -- (p4.center);
	\node at (1.56,-0.3) [circle,fill,inner sep=1.5pt]{};
	\node at (2.43,-0.3) [circle,fill,inner sep=1.5pt]{};
\end{tikzpicture}.
\end{align}

In fact, we can exploit the unraveling of matter propagators to relate all diagrams containing bubble corrections with at least one cubic vertex to the cases above, even if they go over more than one matter propagator. Let us exemplify it at two loops, where we have five possibilities:
\begin{equation}
\raisebox{0.15cm}{\begin{tikzpicture}[baseline={([yshift=-0.1cm]current bounding box.center)}, scale=0.8] 
	\node[] (a) at (0,0) {};
	\node[] (a1) at (1,0) {};
	\node[] (b) at (0,-1) {};
	\node[] (b1) at (1,-1) {};
	\node[] (p1) at ($(a)+(-0.45,0)$) {};
	\node[] (p4) at ($(a1)+(0.45,0)$) {};
	\node[] (p2) at ($(b)+(-0.45,0)$) {};
	\node[] (p3) at ($(b1)+(0.45,0)$) {};
	\draw[line width=0.15mm] (b.center) -- (a.center);
	\draw[line width=0.15mm] (a1.center) -- (b1.center);
	\draw[line width=0.5mm] (p1.center) -- (p4.center);
	\draw[line width=0.5mm] (p2.center) -- (p3.center);
	\draw[line width=0.15mm] (1.2,0) arc (0:180:0.7);
\end{tikzpicture}}, \enspace \begin{tikzpicture}[baseline={([yshift=-0.1cm]current bounding box.center)}, scale=0.8] 
	\node[] (a) at (0,0) {};
	\node[] (a1) at (1,0) {};
	\node[] (b) at (0,-1) {};
	\node[] (b1) at (1,-1) {};
	\node[] (p1) at ($(a)+(-0.25,0)$) {};
	\node[] (p4) at ($(a1)+(0.25,0)$) {};
	\node[] (p2) at ($(b)+(-0.25,0)$) {};
	\node[] (p3) at ($(b1)+(0.25,0)$) {};
	\draw[line width=0.15mm] (b.center) -- (a.center);
	\draw[line width=0.15mm] (b1.center) -- (a1.center);
	\draw[line width=0.5mm] (p1.center) -- (p4.center);
	\draw[line width=0.5mm] (p2.center) -- (p3.center);
	\draw[line width=0.15mm] (0.8,0) arc (0:180:0.3);
\end{tikzpicture}, \enspace
\begin{tikzpicture}[baseline={([yshift=-0.1cm]current bounding box.center)}, scale=0.8] 
	\node[] (a) at (0,0) {};
	\node[] (a1) at (1,0) {};
	\node[] (b) at (0,-1) {};
	\node[] (b1) at (1,-1) {};
	\node[] (p1) at ($(a)+(-0.55,0)$) {};
	\node[] (p4) at ($(a1)+(0.25,0)$) {};
	\node[] (p2) at ($(b)+(-0.55,0)$) {};
	\node[] (p3) at ($(b1)+(0.25,0)$) {};
	\draw[line width=0.15mm] (b.center) -- (a.center);
	\draw[line width=0.15mm] (b1.center) -- (a1.center);
	\draw[line width=0.5mm] (p1.center) -- (p4.center);
	\draw[line width=0.5mm] (p2.center) -- (p3.center);
	\draw[line width=0.15mm] (0.3,0) arc (0:180:0.3);
\end{tikzpicture}, \enspace \raisebox{0.175cm}{\begin{tikzpicture}[baseline={([yshift=-0.1cm]current bounding box.center)}, scale=0.8] 
	\node[] (a) at (0,0) {};
	\node[] (a1) at (1,0) {};
	\node[] (b) at (0,-1) {};
	\node[] (b1) at (1,-1) {};
	\node[] (p1) at ($(a)+(-0.85,0)$) {};
	\node[] (p4) at ($(a1)+(0.25,0)$) {};
	\node[] (p2) at ($(b)+(-0.85,0)$) {};
	\node[] (p3) at ($(b1)+(0.25,0)$) {};
	\draw[line width=0.15mm] (b.center) -- (a.center);
	\draw[line width=0.15mm] (a1.center) -- (b1.center);
	\draw[line width=0.5mm] (p1.center) -- (p4.center);
	\draw[line width=0.5mm] (p2.center) -- (p3.center);
	\draw[line width=0.15mm] (1,0) arc (0:180:0.8);
\end{tikzpicture}}, \enspace \begin{tikzpicture}[baseline={([yshift=-0.1cm]current bounding box.center)}, scale=0.8] 
	\node[] (a) at (0,0) {};
	\node[] (a1) at (1.2,0) {};
	\node[] (b) at (0,-1) {};
	\node[] (b1) at (1.2,-1) {};
	\node[] (p1) at ($(a)+(-0.25,0)$) {};
	\node[] (p4) at ($(a1)+(0.25,0)$) {};
	\node[] (p2) at ($(b)+(-0.25,0)$) {};
	\node[] (p3) at ($(b1)+(0.25,0)$) {};
	\draw[line width=0.15mm] (b.center) -- (a.center);
	\draw[line width=0.15mm] (b1.center) -- (a1.center);
	\draw[line width=0.5mm] (p1.center) -- (p4.center);
	\draw[line width=0.5mm] (p2.center) -- (p3.center);
	\draw[line width=0.15mm] (1.2,0) arc (0:180:0.3);
\end{tikzpicture}.
\end{equation}
Notice that the last diagram has a bubble that contains a single matter line; thus, it has zero master integrals as we showed above. Using partial fraction, for the first diagram we find
\begin{equation}
\raisebox{0.15cm}{\begin{tikzpicture}[baseline={([yshift=-0.1cm]current bounding box.center)}, scale=0.8] 
	\node[] (a) at (0,0) {};
	\node[] (a1) at (1,0) {};
	\node[] (b) at (0,-1) {};
	\node[] (b1) at (1,-1) {};
	\node[] (p1) at ($(a)+(-0.45,0)$) {};
	\node[] (p4) at ($(a1)+(0.45,0)$) {};
	\node[] (p2) at ($(b)+(-0.45,0)$) {};
	\node[] (p3) at ($(b1)+(0.45,0)$) {};
	\draw[line width=0.15mm] (b.center) -- (a.center);
	\draw[line width=0.15mm] (a1.center) -- (b1.center);
	\draw[line width=0.5mm] (p1.center) -- (p4.center);
	\draw[line width=0.5mm] (p2.center) -- (p3.center);
	\draw[line width=0.15mm] (1.2,0) arc (0:180:0.7);
\end{tikzpicture}} = \raisebox{0.175cm}{\begin{tikzpicture}[baseline={([yshift=-0.1cm]current bounding box.center)}, scale=0.8] 
	\node[] (a) at (0,0) {};
	\node[] (a1) at (1,0) {};
	\node[] (b) at (0,-1) {};
	\node[] (b1) at (1,-1) {};
	\node[] (p1) at ($(a)+(-0.85,0)$) {};
	\node[] (p4) at ($(a1)+(0.25,0)$) {};
	\node[] (p2) at ($(b)+(-0.85,0)$) {};
	\node[] (p3) at ($(b1)+(0.25,0)$) {};
	\draw[line width=0.15mm] (b.center) -- (a.center);
	\draw[line width=0.15mm] (a1.center) -- (b1.center);
	\draw[line width=0.5mm] (p1.center) -- (p4.center);
	\draw[line width=0.5mm] (p2.center) -- (p3.center);
	\draw[line width=0.15mm] (1,0) arc (0:180:0.8);
	\node at (-0.3,0) [circle,fill,inner sep=1.5pt]{};
\end{tikzpicture}} = - \begin{tikzpicture}[baseline={([yshift=-0.1cm]current bounding box.center)}, scale=0.8] 
	\node[] (a) at (0,0) {};
	\node[] (a1) at (1,0) {};
	\node[] (b) at (0,-1) {};
	\node[] (b1) at (1,-1) {};
	\node[] (p1) at ($(a)+(-0.85,0)$) {};
	\node[] (p4) at ($(a1)+(0.25,0)$) {};
	\node[] (p2) at ($(b)+(-0.85,0)$) {};
	\node[] (p3) at ($(b1)+(0.25,0)$) {};
	\draw[line width=0.15mm] (b.center) -- (a.center);
	\draw[line width=0.15mm] (b1.center) -- (a1.center);
	\draw[line width=0.5mm] (p1.center) -- (p4.center);
	\draw[line width=0.5mm] (p2.center) -- (p3.center);
	\draw[line width=0.15mm] (0,0) arc (0:180:0.3);
	\node at (0.5,0) [circle,fill,inner sep=1.5pt]{};
\end{tikzpicture} + \begin{tikzpicture}[baseline={([yshift=-0.1cm]current bounding box.center)}, scale=0.8] 
	\node[] (a) at (0,0) {};
	\node[] (a1) at (1,0) {};
	\node[] (b) at (0,-1) {};
	\node[] (b1) at (1,-1) {};
	\node[] (p1) at ($(a)+(-0.85,0)$) {};
	\node[] (p4) at ($(a1)+(0.25,0)$) {};
	\node[] (p2) at ($(b)+(-0.85,0)$) {};
	\node[] (p3) at ($(b1)+(0.25,0)$) {};
	\draw[line width=0.15mm] (b.center) -- (a.center);
	\draw[line width=0.15mm] (b1.center) -- (a1.center);
	\draw[line width=0.5mm] (p1.center) -- (p4.center);
	\draw[line width=0.5mm] (p2.center) -- (p3.center);
	\draw[line width=0.15mm] (0,0) arc (0:180:0.3);
	\node at (-0.3,0) [circle,fill,inner sep=1.5pt]{};
\end{tikzpicture} + \begin{tikzpicture}[baseline={([yshift=-0.1cm]current bounding box.center)}, scale=0.8] 
	\node[] (a) at (0,0) {};
	\node[] (a1) at (1.2,0) {};
	\node[] (b) at (0,-1) {};
	\node[] (b1) at (1.2,-1) {};
	\node[] (p1) at ($(a)+(-0.25,0)$) {};
	\node[] (p4) at ($(a1)+(0.25,0)$) {};
	\node[] (p2) at ($(b)+(-0.25,0)$) {};
	\node[] (p3) at ($(b1)+(0.25,0)$) {};
	\draw[line width=0.15mm] (b.center) -- (a.center);
	\draw[line width=0.15mm] (b1.center) -- (a1.center);
	\draw[line width=0.5mm] (p1.center) -- (p4.center);
	\draw[line width=0.5mm] (p2.center) -- (p3.center);
	\draw[line width=0.15mm] (0.6,0) arc (0:180:0.3);
	\node at (0.9,0) [circle,fill,inner sep=1.5pt]{};
\end{tikzpicture}.
\end{equation}
As can be seen, employing partial fraction we reduce the diagram to a linear combination of diagrams with bubble corrections with one cubic vertex and that contain only one matter line, which as we have proven above have zero master integrals. We can similarly proceed for the remaining diagrams to find
\begin{align}
\begin{tikzpicture}[baseline={([yshift=-0.1cm]current bounding box.center)}, scale=0.8] 
	\node[] (a) at (0,0) {};
	\node[] (a1) at (1,0) {};
	\node[] (b) at (0,-1) {};
	\node[] (b1) at (1,-1) {};
	\node[] (p1) at ($(a)+(-0.25,0)$) {};
	\node[] (p4) at ($(a1)+(0.25,0)$) {};
	\node[] (p2) at ($(b)+(-0.25,0)$) {};
	\node[] (p3) at ($(b1)+(0.25,0)$) {};
	\draw[line width=0.15mm] (b.center) -- (a.center);
	\draw[line width=0.15mm] (b1.center) -- (a1.center);
	\draw[line width=0.5mm] (p1.center) -- (p4.center);
	\draw[line width=0.5mm] (p2.center) -- (p3.center);
	\draw[line width=0.15mm] (0.8,0) arc (0:180:0.3);
\end{tikzpicture} & = \begin{tikzpicture}[baseline={([yshift=-0.1cm]current bounding box.center)}, scale=0.8] 
	\node[] (a) at (0,0) {};
	\node[] (a1) at (1.2,0) {};
	\node[] (b) at (0,-1) {};
	\node[] (b1) at (1.2,-1) {};
	\node[] (p1) at ($(a)+(-0.25,0)$) {};
	\node[] (p4) at ($(a1)+(0.25,0)$) {};
	\node[] (p2) at ($(b)+(-0.25,0)$) {};
	\node[] (p3) at ($(b1)+(0.25,0)$) {};
	\draw[line width=0.15mm] (b.center) -- (a.center);
	\draw[line width=0.15mm] (b1.center) -- (a1.center);
	\draw[line width=0.5mm] (p1.center) -- (p4.center);
	\draw[line width=0.5mm] (p2.center) -- (p3.center);
	\draw[line width=0.15mm] (0.6,0) arc (0:180:0.3);
	\node at (0.9,0) [circle,fill,inner sep=1.5pt]{};
\end{tikzpicture}, \\[0.1cm]
\begin{tikzpicture}[baseline={([yshift=-0.1cm]current bounding box.center)}, scale=0.8] 
	\node[] (a) at (0,0) {};
	\node[] (a1) at (1,0) {};
	\node[] (b) at (0,-1) {};
	\node[] (b1) at (1,-1) {};
	\node[] (p1) at ($(a)+(-0.55,0)$) {};
	\node[] (p4) at ($(a1)+(0.25,0)$) {};
	\node[] (p2) at ($(b)+(-0.55,0)$) {};
	\node[] (p3) at ($(b1)+(0.25,0)$) {};
	\draw[line width=0.15mm] (b.center) -- (a.center);
	\draw[line width=0.15mm] (b1.center) -- (a1.center);
	\draw[line width=0.5mm] (p1.center) -- (p4.center);
	\draw[line width=0.5mm] (p2.center) -- (p3.center);
	\draw[line width=0.15mm] (0.3,0) arc (0:180:0.3);
\end{tikzpicture} & = \begin{tikzpicture}[baseline={([yshift=-0.1cm]current bounding box.center)}, scale=0.8] 
	\node[] (a) at (0,0) {};
	\node[] (a1) at (1,0) {};
	\node[] (b) at (0,-1) {};
	\node[] (b1) at (1,-1) {};
	\node[] (p1) at ($(a)+(-0.85,0)$) {};
	\node[] (p4) at ($(a1)+(0.25,0)$) {};
	\node[] (p2) at ($(b)+(-0.85,0)$) {};
	\node[] (p3) at ($(b1)+(0.25,0)$) {};
	\draw[line width=0.15mm] (b.center) -- (a.center);
	\draw[line width=0.15mm] (b1.center) -- (a1.center);
	\draw[line width=0.5mm] (p1.center) -- (p4.center);
	\draw[line width=0.5mm] (p2.center) -- (p3.center);
	\draw[line width=0.15mm] (0,0) arc (0:180:0.3);
	\node at (0.5,0) [circle,fill,inner sep=1.5pt]{};
\end{tikzpicture} - \begin{tikzpicture}[baseline={([yshift=-0.1cm]current bounding box.center)}, scale=0.8] 
	\node[] (a) at (0,0) {};
	\node[] (a1) at (1.2,0) {};
	\node[] (b) at (0,-1) {};
	\node[] (b1) at (1.2,-1) {};
	\node[] (p1) at ($(a)+(-0.25,0)$) {};
	\node[] (p4) at ($(a1)+(0.25,0)$) {};
	\node[] (p2) at ($(b)+(-0.25,0)$) {};
	\node[] (p3) at ($(b1)+(0.25,0)$) {};
	\draw[line width=0.15mm] (b.center) -- (a.center);
	\draw[line width=0.15mm] (b1.center) -- (a1.center);
	\draw[line width=0.5mm] (p1.center) -- (p4.center);
	\draw[line width=0.5mm] (p2.center) -- (p3.center);
	\draw[line width=0.15mm] (0.6,0) arc (0:180:0.3);
	\node at (0.9,0) [circle,fill,inner sep=1.5pt]{};
\end{tikzpicture}, \\
\raisebox{0.175cm}{\begin{tikzpicture}[baseline={([yshift=-0.1cm]current bounding box.center)}, scale=0.8] 
	\node[] (a) at (0,0) {};
	\node[] (a1) at (1,0) {};
	\node[] (b) at (0,-1) {};
	\node[] (b1) at (1,-1) {};
	\node[] (p1) at ($(a)+(-0.85,0)$) {};
	\node[] (p4) at ($(a1)+(0.25,0)$) {};
	\node[] (p2) at ($(b)+(-0.85,0)$) {};
	\node[] (p3) at ($(b1)+(0.25,0)$) {};
	\draw[line width=0.15mm] (b.center) -- (a.center);
	\draw[line width=0.15mm] (a1.center) -- (b1.center);
	\draw[line width=0.5mm] (p1.center) -- (p4.center);
	\draw[line width=0.5mm] (p2.center) -- (p3.center);
	\draw[line width=0.15mm] (1,0) arc (0:180:0.8);
\end{tikzpicture}} & = \begin{tikzpicture}[baseline={([yshift=-0.1cm]current bounding box.center)}, scale=0.8] 
	\node[] (a) at (0,0) {};
	\node[] (a1) at (1,0) {};
	\node[] (b) at (0,-1) {};
	\node[] (b1) at (1,-1) {};
	\node[] (p1) at ($(a)+(-0.85,0)$) {};
	\node[] (p4) at ($(a1)+(0.25,0)$) {};
	\node[] (p2) at ($(b)+(-0.85,0)$) {};
	\node[] (p3) at ($(b1)+(0.25,0)$) {};
	\draw[line width=0.15mm] (b.center) -- (a.center);
	\draw[line width=0.15mm] (b1.center) -- (a1.center);
	\draw[line width=0.5mm] (p1.center) -- (p4.center);
	\draw[line width=0.5mm] (p2.center) -- (p3.center);
	\draw[line width=0.15mm] (0,0) arc (0:180:0.3);
\end{tikzpicture} - \begin{tikzpicture}[baseline={([yshift=-0.1cm]current bounding box.center)}, scale=0.8] 
	\node[] (a) at (0,0) {};
	\node[] (a1) at (1.2,0) {};
	\node[] (b) at (0,-1) {};
	\node[] (b1) at (1.2,-1) {};
	\node[] (p1) at ($(a)+(-0.25,0)$) {};
	\node[] (p4) at ($(a1)+(0.25,0)$) {};
	\node[] (p2) at ($(b)+(-0.25,0)$) {};
	\node[] (p3) at ($(b1)+(0.25,0)$) {};
	\draw[line width=0.15mm] (b.center) -- (a.center);
	\draw[line width=0.15mm] (b1.center) -- (a1.center);
	\draw[line width=0.5mm] (p1.center) -- (p4.center);
	\draw[line width=0.5mm] (p2.center) -- (p3.center);
	\draw[line width=0.15mm] (1.2,0) arc (0:180:0.3);
\end{tikzpicture},
\end{align}
where all diagrams in the right-hand side have zero master integrals in their sectors. Following analogous steps one can show that the integrals in the opposite parity sector also have zero master integrals in their sector.
As a consequence, at two loops there is only one relevant diagram containing a bubble sub-loop, since neither of its vertices are cubic:
\begin{equation}
\begin{tikzpicture}[baseline={([yshift=-0.1cm]current bounding box.center)}] 
	\node[] (a) at (0,0) {};
	\node[] (a1) at (0.8,0) {};
	\node[] (b) at (0,-0.8) {};
	\node[] (b1) at (0.8,-0.8) {};
	\node[] (p1) at ($(a)+(-0.2,0)$) {};
	\node[] (p4) at ($(a1)+(0.2,0)$) {};
	\node[] (p2) at ($(b)+(-0.2,0)$) {};
	\node[] (p3) at ($(b1)+(0.2,0)$) {};
	\draw[line width=0.15mm] (b.center) -- (a.center);
	\draw[line width=0.15mm] (b1.center) -- (a1.center);
	\draw[line width=0.5mm] (p1.center) -- (p4.center);
	\draw[line width=0.5mm] (p2.center) -- (p3.center);
	\draw[line width=0.15mm] (0.8,0) arc (0:180:0.4);
\end{tikzpicture}.
\end{equation}

Notice that by following the same steps, the same conclusion holds if these box diagrams with bubble corrections appear as part of a bigger diagram. In general, if a bubble correction contains two or more adjacent matter propagators, we can use partial fraction to relate it to diagrams where the bubble contains only a single matter line. As we have explicitly shown at two loops, if at least one of the vertices in the bubble is cubic, the diagram is expressible in terms of lower sectors. Therefore, these diagrams can be totally omitted from our analysis.

\subsection{IBPs for triangles subdiagrams}
\label{app: IBPs_triangles}

Let us now study the family of integrals 
\begin{equation}
\mathcal{I}(\nu_1, \nu_2, \nu_3, \nu_4) = \int \frac{d^D k}{{[ k^2 ]}^{\nu_1} {[ (k+l)^2 ]}^{\nu_2} {[ 2u_1 \cdot k ]}^{\nu_3} {[ l^2 ]}^{\nu_4}}
\end{equation}
for a generic triangle sub-loop attached to a $u_1$ matter line at one end of the diagram, with a different loop momentum $l$ flowing through the triangle, such as
\begin{equation}
\begin{tikzpicture}[baseline={([yshift=-0.1cm]current bounding box.center)}] 
	\node[] (a) at (0,0) {};
	\node[] (a1) at (1.1,0) {};
	\node[] (a2) at (1.75,0) {};
	\node[] (b) at (1.2,-0.35) {};
	\node[] (c) at (0.8,-0.6) {};
	\node[] (d) at (0,-1.2) {};
	\node[] (d1) at (1.75,-1.2) {};
	\node[] (p1) at ($(a)+(-0.3,0)$) {};
	\node[] (p2) at ($(d)+(-0.3,0)$) {};
	\node[] (p3) at ($(d1)+(0.3,0)$) {};
	\node[] (p4) at ($(a2)+(0.3,0)$) {};
	\draw[line width=0.15mm] (c.center) -- (a2.center);
	\draw[line width=0.15mm] (b.center) -- (0.6,0);
	\draw[line width=0.5mm] (p1.center) -- (p4.center);
	\draw[line width=0.5mm] (p2.center) -- (p3.center);
	\fill[gray!50] (0.4,-0.6) ellipse (0.4 and 0.7);
	\draw (0.4,-0.6) ellipse (0.4 and 0.7);
\end{tikzpicture}.
\end{equation}
For this integral family we can consider IBPs of the form
\begin{equation}
0 = \int d^D k \; \frac{\partial}{\partial k^{\mu}} \frac{v^{\mu}}{{[ k^2 ]}^{\nu_1} {[ (k+l)^2 ]}^{\nu_2} {[ 2u_1 \cdot k ]}^{\nu_3} {[ l^2 ]}^{\nu_4}},
\end{equation}
where $v^\mu$ can be either $k^\mu$, $l^\mu$ or $u_1^\mu$. Particularly, for $v^\mu = k^\mu$ one obtains that
\begin{equation}
\mathcal{I}(\nu_1,\nu_2,\nu_3,\nu_4) = \frac{\nu_2 \Big( \mathcal{I}(\nu_1-1,\nu_2+1,\nu_3,\nu_4) - \mathcal{I}(\nu_1,\nu_2+1,\nu_3,\nu_4-1) \Big)}{D - 2 \nu_1 - \nu_2 - \nu_3}.
\end{equation}
Therefore, iterating this IBP relation one can reduce all integrals in this family to subsectors $\mathcal{I}(0,\nu'_2,\nu'_3,\nu'_4)$ and $\mathcal{I}(\nu'_1,\nu'_2,\nu'_3,0)$ for some $\nu'_i$, where the first and fourth propagators are respectively absent. For example,
\begin{align}
\begin{tikzpicture}[baseline={([yshift=-0.1cm]current bounding box.center)}] 
	\node[] (a) at (0,0) {};
	\node[] (a1) at (1.1,0) {};
	\node[] (a2) at (1.75,0) {};
	\node[] (b) at (1.2,-0.35) {};
	\node[] (c) at (0.8,-0.6) {};
	\node[] (d) at (0,-1.2) {};
	\node[] (d1) at (1.75,-1.2) {};
	\node[] (p1) at ($(a)+(-0.3,0)$) {};
	\node[] (p2) at ($(d)+(-0.3,0)$) {};
	\node[] (p3) at ($(d1)+(0.3,0)$) {};
	\node[] (p4) at ($(a2)+(0.3,0)$) {};
	\draw[line width=0.15mm] (c.center) -- (a2.center);
	\draw[line width=0.15mm] (b.center) -- (0.6,0);
	\draw[line width=0.5mm] (p1.center) -- (p4.center);
	\draw[line width=0.5mm] (p2.center) -- (p3.center);
	\fill[gray!50] (0.4,-0.6) ellipse (0.4 and 0.7);
	\draw (0.4,-0.6) ellipse (0.4 and 0.7);
\end{tikzpicture} &= \frac{1}{D-4} \left( \raisebox{0.2cm}{\begin{tikzpicture}[baseline={([yshift=-0.1cm]current bounding box.center)}] 
	\node[] (a) at (0,0) {};
	\node[] (a1) at (1.1,0) {};
	\node[] (a2) at (1.6,0) {};
	\node[] (b) at (1.2,-0.35) {};
	\node[] (c) at (0.8,-0.6) {};
	\node[] (d) at (0,-1.2) {};
	\node[] (d1) at (1.75,-1.2) {};
	\node[] (p1) at ($(a)+(-0.3,0)$) {};
	\node[] (p2) at ($(d)+(-0.3,0)$) {};
	\node[] (p3) at ($(d1)+(0.15,0)$) {};
	\node[] (p4) at ($(a2)+(0.3,0)$) {};
	\draw[line width=0.15mm] (c.center) -- (a2.center);
	\draw[line width=0.5mm] (p1.center) -- (p4.center);
	\draw[line width=0.5mm] (p2.center) -- (p3.center);
	\draw[line width=0.15mm] (a2.center) arc (0:180:0.5);
	\fill[gray!50] (0.4,-0.6) ellipse (0.4 and 0.7);
	\draw (0.4,-0.6) ellipse (0.4 and 0.7);
	\node at (1.1,0.5) [circle,fill,inner sep=1.5pt]{};
\end{tikzpicture}} - \begin{tikzpicture}[baseline={([yshift=-0.1cm]current bounding box.center)}] 
	\node[] (a) at (0,0) {};
	\node[] (a1) at (1.1,0) {};
	\node[] (a2) at (1.75,0) {};
	\node[] (b) at (1.2,-0.35) {};
	\node[] (c) at (0.8,-0.6) {};
	\node[] (d) at (0,-1.2) {};
	\node[] (d1) at (1.75,-1.2) {};
	\node[] (p1) at ($(a)+(-0.3,0)$) {};
	\node[] (p2) at ($(d)+(-0.3,0)$) {};
	\node[] (p3) at ($(d1)+(0.3,0)$) {};
	\node[] (p4) at ($(a2)+(0.3,0)$) {};
	\draw[line width=0.15mm] (c.center) -- (a2.center);
	\draw[line width=0.5mm] (p1.center) -- (p4.center);
	\draw[line width=0.5mm] (p2.center) -- (p3.center);
	\fill[gray!50] (0.4,-0.6) ellipse (0.4 and 0.7);
	\draw (0.4,-0.6) ellipse (0.4 and 0.7);
	\draw[line width=0.15mm] plot[smooth, tension=1] coordinates {(0.8,-0.6) (1,-0.25) (0.6,0)};
	\node at (1,-0.25) [circle,fill,inner sep=1.5pt]{};
\end{tikzpicture} \right), \\
\begin{tikzpicture}[baseline={([yshift=-0.1cm]current bounding box.center)}] 
	\node[] (a) at (0,0) {};
	\node[] (a1) at (1.1,0) {};
	\node[] (a2) at (1.75,0) {};
	\node[] (b) at (1.2,-0.35) {};
	\node[] (c) at (0.8,-0.6) {};
	\node[] (d) at (0,-1.2) {};
	\node[] (d1) at (1.75,-1.2) {};
	\node[] (p1) at ($(a)+(-0.3,0)$) {};
	\node[] (p2) at ($(d)+(-0.3,0)$) {};
	\node[] (p3) at ($(d1)+(0.3,0)$) {};
	\node[] (p4) at ($(a2)+(0.3,0)$) {};
	\draw[line width=0.15mm] (c.center) -- (a2.center);
	\draw[line width=0.15mm] (b.center) -- (0.6,0);
	\draw[line width=0.5mm] (p1.center) -- (p4.center);
	\draw[line width=0.5mm] (p2.center) -- (p3.center);
	\fill[gray!50] (0.4,-0.6) ellipse (0.4 and 0.7);
	\draw (0.4,-0.6) ellipse (0.4 and 0.7);
	\node at (1.2,0) [circle,fill,inner sep=1.5pt]{};
\end{tikzpicture} &= \frac{1}{D-5} \left( \raisebox{0.2cm}{\begin{tikzpicture}[baseline={([yshift=-0.1cm]current bounding box.center)}] 
	\node[] (a) at (0,0) {};
	\node[] (a1) at (1.1,0) {};
	\node[] (a2) at (1.6,0) {};
	\node[] (b) at (1.2,-0.35) {};
	\node[] (c) at (0.8,-0.6) {};
	\node[] (d) at (0,-1.2) {};
	\node[] (d1) at (1.75,-1.2) {};
	\node[] (p1) at ($(a)+(-0.3,0)$) {};
	\node[] (p2) at ($(d)+(-0.3,0)$) {};
	\node[] (p3) at ($(d1)+(0.15,0)$) {};
	\node[] (p4) at ($(a2)+(0.3,0)$) {};
	\draw[line width=0.15mm] (c.center) -- (a2.center);
	\draw[line width=0.5mm] (p1.center) -- (p4.center);
	\draw[line width=0.5mm] (p2.center) -- (p3.center);
	\draw[line width=0.15mm] (a2.center) arc (0:180:0.5);
	\fill[gray!50] (0.4,-0.6) ellipse (0.4 and 0.7);
	\draw (0.4,-0.6) ellipse (0.4 and 0.7);
	\node at (1.1,0.5) [circle,fill,inner sep=1.5pt]{};
	\node at (1.1,0) [circle,fill,inner sep=1.5pt]{};
\end{tikzpicture}} - \begin{tikzpicture}[baseline={([yshift=-0.1cm]current bounding box.center)}] 
	\node[] (a) at (0,0) {};
	\node[] (a1) at (1.1,0) {};
	\node[] (a2) at (1.75,0) {};
	\node[] (b) at (1.2,-0.35) {};
	\node[] (c) at (0.8,-0.6) {};
	\node[] (d) at (0,-1.2) {};
	\node[] (d1) at (1.75,-1.2) {};
	\node[] (p1) at ($(a)+(-0.3,0)$) {};
	\node[] (p2) at ($(d)+(-0.3,0)$) {};
	\node[] (p3) at ($(d1)+(0.3,0)$) {};
	\node[] (p4) at ($(a2)+(0.3,0)$) {};
	\draw[line width=0.15mm] (c.center) -- (a2.center);
	\draw[line width=0.5mm] (p1.center) -- (p4.center);
	\draw[line width=0.5mm] (p2.center) -- (p3.center);
	\fill[gray!50] (0.4,-0.6) ellipse (0.4 and 0.7);
	\draw (0.4,-0.6) ellipse (0.4 and 0.7);
	\draw[line width=0.15mm] plot[smooth, tension=1] coordinates {(0.8,-0.6) (1,-0.25) (0.6,0)};
	\node at (1,-0.25) [circle,fill,inner sep=1.5pt]{};
	\node at (1.2,0) [circle,fill,inner sep=1.5pt]{};
\end{tikzpicture} \right).
\end{align}

We can similarly consider the case where the triangle sub-loop appears in the middle of a diagram, such as
\begin{equation}
\begin{tikzpicture}[baseline={([yshift=-0.1cm]current bounding box.center)}, scale=0.7]
	\node[] (a) at (1.6,0.3) {};
	\node[] (a1) at (2.4,0.3) {};
	\node[] (b) at (2,-1.2) {};
	\node[] (p1) at (1.13,0.3) {};
	\node[] (p4) at (2.87,0.3) {};
	\fill[gray!20] (2,-0.2) circle (1.6);
	\fill[white] (2,-0.2) circle (1);
	\draw[line width=0.15mm] (p1.center) -- (2,-0.4);
	\draw[line width=0.15mm] (p4.center) -- (2,-0.4);
	\draw[line width=0.15mm] (b.center) -- (2,-0.4);
	\draw[line width=0.5mm] (p1.center) -- (p4.center);
\end{tikzpicture} \, ,
\end{equation}
which is characterized by a family of integrals 
\begin{equation}
\mathcal{I}(\nu_1, \nu_2, \nu_3, \nu_4) = \int \frac{d^D k}{{[ k^2 ]}^{\nu_1} {[ (k+l_1)^2 ]}^{\nu_2} {[ 2u_1 \cdot (k+l_2) ]}^{\nu_3} {[ l_1^2 ]}^{\nu_4}},
\end{equation}
with two different loop momenta $l_1$ and $l_2$ flowing through the triangle.
If we consider the IBP
\begin{equation}
0 = \int d^D k \; \frac{\partial}{\partial k^{\mu}} \frac{v^{\mu}}{{[ k^2 ]}^{\nu_1} {[ (k+l_1)^2 ]}^{\nu_2} {[ 2u_1 \cdot (k+l_2) ]}^{\nu_3} {[ l_1^2 ]}^{\nu_4}},
\end{equation}
for $v^\mu = k^\mu$, we get that
\begin{align}
0 = & \, (D - 2 \nu_1 - \nu_2 - \nu_3) \, \mathcal{I}(\nu_1,\nu_2,\nu_3,\nu_4) - \nu_2 \, \mathcal{I}(\nu_1-1,\nu_2+1,\nu_3,\nu_4) \nonumber \\
& \, + \nu_2 \, \mathcal{I}(\nu_1,\nu_2+1,\nu_3,\nu_4-1) + 2 (u_1 \cdot l_2) \, \nu_3 \, \mathcal{I}(\nu_1,\nu_2,\nu_3+1,\nu_4).
\end{align}
Just as it happened in sec.~\ref{app: IBPs_bubbles} for the bubble without cubic matter vertices, the IBP relations for this integral family do not yield a reduction to lower sectors, even taking into account the remaining IBP identities for $v^\mu=l_1^\mu$ and $v^\mu=u_1^\mu$. However, we can achieve a reduction to lower sectors if we include a fifth propagator ${[2 u_1 \cdot l_2]}^{\nu_5}$ in the integral family, which means that at least one of the vertices of the triangle on the matter line is cubic. Doing so, we obtain
\begin{align}
\mathcal{I}(\nu_1,\nu_2,\nu_3,\nu_4,\nu_5) = & \, \frac{1}{D - 2 \nu_1 - \nu_2 - \nu_3} \Big( \nu_2 \, \mathcal{I}(\nu_1-1,\nu_2+1,\nu_3,\nu_4,\nu_5) \nonumber \\
& - \nu_2 \, \mathcal{I}(\nu_1,\nu_2+1,\nu_3,\nu_4-1,\nu_5) - \nu_3 \, \mathcal{I}(\nu_1,\nu_2,\nu_3+1,\nu_4,\nu_5-1) \Big),
\end{align}
which is always reducible to the subsectors $\mathcal{I}(0,\nu'_2,\nu'_3,\nu'_4,\nu'_5)$, $\mathcal{I}(\nu'_1,\nu'_2,\nu'_3,0,\nu'_5)$ and $\mathcal{I}(\nu'_1,\nu'_2,\nu'_3,\nu'_4,0)$ for some $\nu'_i$. For example,
\begin{align}
\begin{tikzpicture}[baseline={([yshift=-0.1cm]current bounding box.center)}, scale=0.7]
	\node[] (a) at (1.6,0.3) {};
	\node[] (a1) at (2.4,0.3) {};
	\node[] (b) at (2,-1.2) {};
	\node[] (p1) at (1.13,0.3) {};
	\node[] (p4) at (2.87,0.3) {};
	\fill[gray!20] (2,-0.2) circle (1.6);
	\fill[white] (2,-0.2) circle (1);
	\draw[line width=0.15mm] (p1.center) -- (2,-0.4);
	\draw[line width=0.15mm] (a1.center) -- (2,-0.4);
	\draw[line width=0.15mm] (b.center) -- (2,-0.4);
	\draw[line width=0.5mm] (p1.center) -- (p4.center);
\end{tikzpicture} &= \frac{1}{D-4} \left( \begin{tikzpicture}[baseline={([yshift=-0.1cm]current bounding box.center)}, scale=0.7]
	\node[] (a) at (1.6,0.3) {};
	\node[] (a1) at (2.4,0.3) {};
	\node[] (b) at (2,-1.2) {};
	\node[] (p1) at (1.13,0.3) {};
	\node[] (p4) at (2.87,0.3) {};
	\fill[gray!20] (2,-0.2) circle (1.6);
	\fill[white] (2,-0.2) circle (1);
	\draw[line width=0.15mm] (2.4,-1.2) -- (a1.center);
	\draw[line width=0.5mm] (p1.center) -- (p4.center);
	\draw[line width=0.15mm] (a1.center) arc (360:180:0.635);
	\node at (1.765,-0.335) [circle,fill,inner sep=1.5pt]{};
\end{tikzpicture} - \begin{tikzpicture}[baseline={([yshift=-0.1cm]current bounding box.center)}, scale=0.7]
	\node[] (a) at (1.6,0.3) {};
	\node[] (a1) at (2.4,0.3) {};
	\node[] (b) at (2,-1.2) {};
	\node[] (p1) at (1.13,0.3) {};
	\node[] (p4) at (2.87,0.3) {};
	\fill[gray!20] (2,-0.2) circle (1.6);
	\fill[white] (2,-0.2) circle (1);
	\draw[line width=0.15mm] (p1.center) -- (b.center);
	\draw[line width=0.15mm] (a1.center) -- (b.center);
	\draw[line width=0.5mm] (p1.center) -- (p4.center);
	\node at (1.565,-0.45) [circle,fill,inner sep=1.5pt]{};
\end{tikzpicture} - \begin{tikzpicture}[baseline={([yshift=-0.1cm]current bounding box.center)}, scale=0.7]
	\node[] (a) at (1.6,0.3) {};
	\node[] (a1) at (2.4,0.3) {};
	\node[] (b) at (2,-1.2) {};
	\node[] (p1) at (1.13,0.3) {};
	\node[] (p4) at (2.87,0.3) {};
	\fill[gray!20] (2,-0.2) circle (1.6);
	\fill[white] (2,-0.2) circle (1);
	\draw[line width=0.15mm] (p1.center) -- (2,-0.4);
	\draw[line width=0.15mm] (2,-0.4) -- (p4.center);
	\draw[line width=0.15mm] (b.center) -- (2,-0.4);
	\draw[line width=0.5mm] (p1.center) -- (p4.center);
	\node at (2,0.3) [circle,fill,inner sep=1.5pt]{};
\end{tikzpicture} \right), \\[0.2cm]
\begin{tikzpicture}[baseline={([yshift=-0.1cm]current bounding box.center)}, scale=0.7]
	\node[] (a) at (1.6,0.3) {};
	\node[] (a1) at (2.4,0.3) {};
	\node[] (b) at (2,-1.2) {};
	\node[] (p1) at (1.13,0.3) {};
	\node[] (p4) at (2.87,0.3) {};
	\fill[gray!20] (2,-0.2) circle (1.6);
	\fill[white] (2,-0.2) circle (1);
	\draw[line width=0.15mm] (p1.center) -- (2,-0.4);
	\draw[line width=0.15mm] (a1.center) -- (2,-0.4);
	\draw[line width=0.15mm] (b.center) -- (2,-0.4);
	\draw[line width=0.5mm] (p1.center) -- (p4.center);
	\node at (1.565,-0.05) [circle,fill,inner sep=1.5pt]{};
\end{tikzpicture} &= \frac{1}{D-5} \left( 2 \  \begin{tikzpicture}[baseline={([yshift=-0.1cm]current bounding box.center)}, scale=0.7]
	\node[] (a) at (1.6,0.3) {};
	\node[] (a1) at (2.4,0.3) {};
	\node[] (b) at (2,-1.2) {};
	\node[] (p1) at (1.13,0.3) {};
	\node[] (p4) at (2.87,0.3) {};
	\fill[gray!20] (2,-0.2) circle (1.6);
	\fill[white] (2,-0.2) circle (1);
	\draw[line width=0.15mm] (2.4,-1.2) -- (a1.center);
	\draw[line width=0.5mm] (p1.center) -- (p4.center);
	\draw[line width=0.15mm] (a1.center) arc (360:180:0.635);
	\node at (1.53,-0.275) [circle,fill,inner sep=1.5pt]{};
	\node at (2,-0.275) [circle,fill,inner sep=1.5pt]{};
\end{tikzpicture} - 2 \  \begin{tikzpicture}[baseline={([yshift=-0.1cm]current bounding box.center)}, scale=0.7]
	\node[] (a) at (1.6,0.3) {};
	\node[] (a1) at (2.4,0.3) {};
	\node[] (b) at (2,-1.2) {};
	\node[] (p1) at (1.13,0.3) {};
	\node[] (p4) at (2.87,0.3) {};
	\fill[gray!20] (2,-0.2) circle (1.6);
	\fill[white] (2,-0.2) circle (1);
	\draw[line width=0.15mm] (p1.center) -- (b.center);
	\draw[line width=0.15mm] (a1.center) -- (b.center);
	\draw[line width=0.5mm] (p1.center) -- (p4.center);
	\node at (1.42,-0.2) [circle,fill,inner sep=1.5pt]{};
	\node at (1.71,-0.7) [circle,fill,inner sep=1.5pt]{};
\end{tikzpicture} - \begin{tikzpicture}[baseline={([yshift=-0.1cm]current bounding box.center)}, scale=0.7]
	\node[] (a) at (1.6,0.3) {};
	\node[] (a1) at (2.4,0.3) {};
	\node[] (b) at (2,-1.2) {};
	\node[] (p1) at (1.13,0.3) {};
	\node[] (p4) at (2.87,0.3) {};
	\fill[gray!20] (2,-0.2) circle (1.6);
	\fill[white] (2,-0.2) circle (1);
	\draw[line width=0.15mm] (p1.center) -- (2,-0.4);
	\draw[line width=0.15mm] (2,-0.4) -- (p4.center);
	\draw[line width=0.15mm] (b.center) -- (2,-0.4);
	\draw[line width=0.5mm] (p1.center) -- (p4.center);
	\node at (2,0.3) [circle,fill,inner sep=1.5pt]{};
	\node at (1.565,-0.05) [circle,fill,inner sep=1.5pt]{};
\end{tikzpicture} \right).
\end{align}

In conclusion, regardless of whether they are located at one end of the diagram or in the middle, all triangle sub-loops with at least two cubic vertices and one of them being the graviton self-interaction vertex are expressible in terms of lower sectors. In general, this reduction allows us to disregard from the analysis top topologies containing such triangles, since their subsectors either have zero master integrals if they still contain the triangle or a bubble correction, or they have already been included in the analysis of different diagrams at the same loop order.

\section{Leading singularity of box iterations}
\label{app: Superclassical_reduction}

In this appendix, we show that the leading singularity of superclassical diagrams containing box iterations at one end can be reduced to lower-loop leading singularities, hence proving eqs.~\eqref{eq: reduction_superclassical} and~\eqref{eq: reduction_superclassical_generic}.

Let us first study the double-box diagram for pedagogic purposes. The loop-by-loop Baikov representation yields a leading singularity given by
\begin{align}
\LS \left(
\begin{tikzpicture}[baseline={([yshift=-0.1cm]current bounding box.center)}] 
	\node[] (a) at (0,0) {};
	\node[] (a1) at (0.8,0) {};
	\node[] (a2) at (1.6,0) {};
	\node[] (b) at (0,-0.8) {};
	\node[] (b1) at (0.8,-0.8) {};
	\node[] (b2) at (1.6,-0.8) {};
	\node[] (p1) at ($(a)+(-0.2,0)$) {};
	\node[] (p2) at ($(b)+(-0.2,0)$) {};
	\node[] (p3) at ($(b2)+(0.2,0)$) {};
	\node[] (p4) at ($(a2)+(0.2,0)$) {};
	\draw[line width=0.15mm] (b.center) -- (a.center);
	\draw[line width=0.15mm] (b1.center) -- (a1.center);
	\draw[line width=0.15mm] (b2.center) -- (a2.center);
	\draw[line width=0.5mm] (p1.center) -- (p4.center);
	\draw[line width=0.5mm] (p2.center) -- (p3.center);
\end{tikzpicture}
\right) & = \LS \left(
\begin{tikzpicture}[baseline={([yshift=-0.1cm]current bounding box.center)}] 
	\node[] (a) at (0,0) {};
	\node[] (a1) at (0.8,0) {};
	\node[] (b) at (0,-0.8) {};
	\node[] (b1) at (0.8,-0.8) {};
	\node[] (p1) at ($(a)+(-0.2,0)$) {};
	\node[] (p2) at ($(b)+(-0.2,0)$) {};
	\node[] (p3) at ($(b1)+(0.2,0)$) {};
	\node[] (p4) at ($(a1)+(0.2,0)$) {};
	\draw[line width=0.15mm] (b.center) -- (a.center);
	\draw[line width=0.15mm] (b1.center) -- (a1.center);
	\draw[line width=0.15mm,-{Latex[length=2.2mm]}](1.1,-0.75) -- (1.1,-0.05);
	\node[label={[xshift=0.35cm, yshift=-0.45cm]$k_2$}] (k) at (1.1,-0.4) {};
	\draw[line width=0.5mm] (p1.center) -- (p4.center);
	\draw[line width=0.5mm] (p2.center) -- (p3.center);
\end{tikzpicture} \times
\begin{tikzpicture}[baseline={([yshift=-0.1cm]current bounding box.center)}] 
	\node[] (a) at (0,0) {};
	\node[] (a1) at (0.8,0) {};
	\node[] (b) at (0,-0.8) {};
	\node[] (b1) at (0.8,-0.8) {};
	\node[] (p1) at ($(a)+(-0.2,0)$) {};
	\node[] (p2) at ($(b)+(-0.2,0)$) {};
	\node[] (p3) at ($(b1)+(0.2,0)$) {};
	\node[] (p4) at ($(a1)+(0.2,0)$) {};
	\draw[line width=0.15mm, dashed, postaction={decorate}] (b.center) -- node[sloped, allow upside down, label={[xshift=0.2cm, yshift=0cm]$k_2$}] {\midarrow} (a.center);
	\draw[line width=0.15mm] (b1.center) -- (a1.center);
	\draw[line width=0.5mm] (p1.center) -- (p4.center);
	\draw[line width=0.5mm] (p2.center) -- (p3.center);
\end{tikzpicture}
\right) \nonumber \\
& \propto \frac{x^2}{(x^2-1)^2} \, \LS \left( \int \frac{d z_8}{z_8 (q^2-z_8)} \right),
\end{align}
with $z_8=k_2^2$. Since at the maximal cut we have $2u_i \cdot k_2 = 0$ and $(k_1{+}k_2)^2 = 0$, the only dependence of the $k_1$-loop leading singularity on $k_2$ is via $k_2^2$; in this case we had $\frac{x}{k_2^2 \, (x^2-1)}$. On the other hand, the $k_2$-loop generates $\frac{x}{x^2-1} \, \int \frac{d z_8}{(q^2-z_8)}$, which reveals a new pole at $z_8 = q^2$. Therefore, upon taking the residue at the new pole $k_2^2 = q^2$, it brings us back to the leading singularity of the initial $k_1$-loop, now with momentum transfer $q$, times a factor of $\frac{x}{x^2-1}$.

The significance of this approach is that it does not rely on the specific leading singularity of the previous loop. Therefore, we find the following reduction for all box iterations,
\begin{align}
\LS \left(
\begin{tikzpicture}[baseline={([yshift=-0.1cm]current bounding box.center)}] 
	\node[] (a) at (0,0) {};
	\node[] (a1) at (1.5,0) {};
	\node[] (b) at (0,-1.2) {};
	\node[] (b1) at (1.5,-1.2) {};
	\node[] (p1) at ($(a)+(-0.3,0)$) {};
	\node[] (p2) at ($(b)+(-0.3,0)$) {};
	\node[] (p3) at ($(b1)+(0.3,0)$) {};
	\node[] (p4) at ($(a1)+(0.3,0)$) {};
	\draw[line width=0.15mm] (b1.center) -- (a1.center);
	\draw[line width=0.5mm] (p1.center) -- (0.4,0);
	\draw[line width=0.5mm] (0.4,0) -- (p4.center);
	\draw[line width=0.5mm] (p2.center) -- (0.4,-1.2);
	\draw[line width=0.5mm] (0.4,-1.2) -- (p3.center);
	\fill[gray!50] (0.4,-0.6) ellipse (0.4 and 0.7);
	\draw (0.4,-0.6) ellipse (0.4 and 0.7);
\end{tikzpicture} \right) & = 
\LS \left( 
\begin{tikzpicture}[baseline={([yshift=-0.1cm]current bounding box.center)}] 
	\node[] (a) at (0,0) {};
	\node[] (a1) at (0.8,0) {};
	\node[] (b) at (0,-1.2) {};
	\node[] (b1) at (0.8,-1.2) {};
	\node[] (p1) at ($(a)+(-0.3,0)$) {};
	\node[] (p2) at ($(b)+(-0.3,0)$) {};
	\node[] (p3) at ($(b1)+(0.3,0)$) {};
	\node[] (p4) at ($(a1)+(0.3,0)$) {};
	\draw[line width=0.5mm] (p1.center) -- (0.4,0);
	\draw[line width=0.5mm] (0.4,0) -- (p4.center);
	\draw[line width=0.5mm] (p2.center) -- (0.4,-1.2);
	\draw[line width=0.5mm] (0.4,-1.2) -- (p3.center);
	\fill[gray!50] (0.4,-0.6) ellipse (0.4 and 0.7);
	\draw (0.4,-0.6) ellipse (0.4 and 0.7);
	\draw[line width=0.15mm,-{Latex[length=2.2mm]}](1,-1) -- (1,-0.2);
	\node[label={[xshift=0.35cm, yshift=-0.45cm]$k_L$}] (k) at (1,-0.6) {};
\end{tikzpicture} \times
\begin{tikzpicture}[baseline={([yshift=-0.1cm]current bounding box.center)}] 
	\node[] (a) at (0,0) {};
	\node[] (a1) at (1.2,0) {};
	\node[] (b) at (0,-1.2) {};
	\node[] (b1) at (1.2,-1.2) {};
	\node[] (p1) at ($(a)+(-0.3,0)$) {};
	\node[] (p2) at ($(b)+(-0.3,0)$) {};
	\node[] (p3) at ($(b1)+(0.3,0)$) {};
	\node[] (p4) at ($(a1)+(0.3,0)$) {};
	\draw[line width=0.15mm, dashed, postaction={decorate}] (b.center) -- node[sloped, allow upside down, label={[xshift=0.2cm, yshift=0cm]$k_L$}] {\midarrow} (a.center);	
	\draw[line width=0.15mm] (b1.center) -- (a1.center);
	\draw[line width=0.5mm] (p1.center) -- (p4.center);
	\draw[line width=0.5mm] (p2.center) -- (p3.center);
\end{tikzpicture}
\right) \nonumber \\
& \propto \frac{x}{x^2-1} \, \LS \left( \left. \int \frac{d z}{q^2-z}
\begin{tikzpicture}[baseline={([yshift=-0.1cm]current bounding box.center)}] 
	\node[] (a) at (0,0) {};
	\node[] (a1) at (0.8,0) {};
	\node[] (b) at (0,-1.2) {};
	\node[] (b1) at (0.8,-1.2) {};
	\node[] (p1) at ($(a)+(-0.3,0)$) {};
	\node[] (p2) at ($(b)+(-0.3,0)$) {};
	\node[] (p3) at ($(b1)+(0.3,0)$) {};
	\node[] (p4) at ($(a1)+(0.3,0)$) {};
	\draw[line width=0.5mm] (p1.center) -- (0.4,0);
	\draw[line width=0.5mm] (0.4,0) -- (p4.center);
	\draw[line width=0.5mm] (p2.center) -- (0.4,-1.2);
	\draw[line width=0.5mm] (0.4,-1.2) -- (p3.center);
	\fill[gray!50] (0.4,-0.6) ellipse (0.4 and 0.7);
	\draw (0.4,-0.6) ellipse (0.4 and 0.7);
	\draw[line width=0.15mm,-{Latex[length=2.2mm]}](1,-1) -- (1,-0.2);
	\node[label={[xshift=0.35cm, yshift=-0.45cm]$k_L$}] (k) at (1,-0.6) {};
\end{tikzpicture} \right|_{k_L^2 \to \, z} \, \right) \nonumber \\
& = \frac{x}{x^2-1} \, \LS \left(
\begin{tikzpicture}[baseline={([yshift=-0.1cm]current bounding box.center)}] 
	\node[] (a) at (0,0) {};
	\node[] (a1) at (0.8,0) {};
	\node[] (b) at (0,-1.2) {};
	\node[] (b1) at (0.8,-1.2) {};
	\node[] (p1) at ($(a)+(-0.3,0)$) {};
	\node[] (p2) at ($(b)+(-0.3,0)$) {};
	\node[] (p3) at ($(b1)+(0.3,0)$) {};
	\node[] (p4) at ($(a1)+(0.3,0)$) {};
	\draw[line width=0.5mm] (p1.center) -- (0.4,0);
	\draw[line width=0.5mm] (0.4,0) -- (p4.center);
	\draw[line width=0.5mm] (p2.center) -- (0.4,-1.2);
	\draw[line width=0.5mm] (0.4,-1.2) -- (p3.center);
	\fill[gray!50] (0.4,-0.6) ellipse (0.4 and 0.7);
	\draw (0.4,-0.6) ellipse (0.4 and 0.7);
	\draw[line width=0.15mm,-{Latex[length=2.2mm]}](1,-1) -- (1,-0.2);
	\node[label={[xshift=0.25cm, yshift=-0.45cm]$q$}] (k) at (1,-0.6) {};
\end{tikzpicture} \right) = \frac{x}{x^2-1} \, \LS \left(
\begin{tikzpicture}[baseline={([yshift=-0.1cm]current bounding box.center)}] 
	\node[] (a) at (0,0) {};
	\node[] (a1) at (0.8,0) {};
	\node[] (b) at (0,-1.2) {};
	\node[] (b1) at (0.8,-1.2) {};
	\node[] (p1) at ($(a)+(-0.3,0)$) {};
	\node[] (p2) at ($(b)+(-0.3,0)$) {};
	\node[] (p3) at ($(b1)+(0.3,0)$) {};
	\node[] (p4) at ($(a1)+(0.3,0)$) {};
	\draw[line width=0.5mm] (p1.center) -- (0.4,0);
	\draw[line width=0.5mm] (0.4,0) -- (p4.center);
	\draw[line width=0.5mm] (p2.center) -- (0.4,-1.2);
	\draw[line width=0.5mm] (0.4,-1.2) -- (p3.center);
	\fill[gray!50] (0.4,-0.6) ellipse (0.4 and 0.7);
	\draw (0.4,-0.6) ellipse (0.4 and 0.7);
\end{tikzpicture} \right).
\label{eq: proof_reduction_superclassical}
\end{align}
Since the previous reduction is obtained loop-by-loop, we can embed it in a bigger diagram as long as the momentum transfer satisfies $k \cdot u_i = 0$ at the maximal cut, finding a second reduction of box iterations:
\begin{equation}
\LS \left(
\begin{tikzpicture}[baseline={([yshift=-0.1cm]current bounding box.center)}] 
	\node[] (a) at (0,0) {};
	\node[] (a1) at (0.4,0) {};
	\node[] (a2) at (1.5,0) {};
	\node[] (a3) at (2.6,0) {};
	\node[] (a4) at (3,0) {};
	\node[] (b) at (0,-1.2) {};
	\node[] (b1) at (0.4,-1.2) {};
	\node[] (b2) at (1.5,-1.2) {};
	\node[] (b3) at (2.6,-1.2) {};
	\node[] (b4) at (3,-1.2) {};
	\node[] (p1) at ($(a)+(-0.3,0)$) {};
	\node[] (p2) at ($(b)+(-0.3,0)$) {};
	\node[] (p3) at ($(b4)+(0.3,0)$) {};
	\node[] (p4) at ($(a4)+(0.3,0)$) {};
	\draw[line width=0.15mm] (a2.center) -- (b2.center);
	\draw[line width=0.5mm] (p1.center) -- (0.4,0);
	\draw[line width=0.5mm] (0.4,0) -- (2.6,0);
	\draw[line width=0.5mm] (2.6,0) -- (p4.center);
	\draw[line width=0.5mm] (p2.center) -- (0.4,-1.2);
	\draw[line width=0.5mm] (0.4,-1.2) -- (2.6,-1.2);
	\draw[line width=0.5mm] (2.6,-1.2) -- (p3.center);
	\fill[gray!50] ($(a1)+(0,-0.6)$) ellipse (0.4 and 0.7);
	\draw ($(a1)+(0,-0.6)$) ellipse (0.4 and 0.7);
	\fill[gray!50] ($(a3)+(0,-0.6)$) ellipse (0.4 and 0.7);
	\draw ($(a3)+(0,-0.6)$) ellipse (0.4 and 0.7);
\end{tikzpicture} \right) \propto \frac{x}{x^2-1} \, \LS \left(
\begin{tikzpicture}[baseline={([yshift=-0.1cm]current bounding box.center)}] 
	\node[] (a) at (0,0) {};
	\node[] (a1) at (0.4,0) {};
	\node[] (a2) at (2.4,0) {};
	\node[] (a3) at (2.8,0) {};
	\node[] (b) at (0,-1.2) {};
	\node[] (b1) at (0.4,-1.2) {};
	\node[] (b2) at (2.4,-1.2) {};
	\node[] (b3) at (2.8,-1.2) {};
	\node[] (p1) at ($(a)+(-0.3,0)$) {};
	\node[] (p2) at ($(b)+(-0.3,0)$) {};
	\node[] (p3) at ($(b3)+(0.3,0)$) {};
	\node[] (p4) at ($(a3)+(0.3,0)$) {};
	\draw[line width=0.5mm] (p1.center) -- (0.4,0);
	\draw[line width=0.5mm] (0.4,0) -- (2.6,0);
	\draw[line width=0.5mm] (2.6,0) -- (p4.center);
	\draw[line width=0.5mm] (p2.center) -- (0.4,-1.2);
	\draw[line width=0.5mm] (0.4,-1.2) -- (2.6,-1.2);
	\draw[line width=0.5mm] (2.6,-1.2) -- (p3.center);
	\fill[gray!50] ($(a1)+(0,-0.6)$) ellipse (0.4 and 0.7);
	\draw ($(a1)+(0,-0.6)$) ellipse (0.4 and 0.7);
	\fill[gray!50] ($(a2)+(0,-0.6)$) ellipse (0.4 and 0.7);
	\draw ($(a2)+(0,-0.6)$) ellipse (0.4 and 0.7);
\end{tikzpicture} \right).
\end{equation}

These reduction relations are applicable regardless of the parities of the blobs. Nonetheless, if we dot one of the matter propagators inside the superclassical loop we find instead:
\begin{align}
& \LS \left( \begin{tikzpicture}[baseline={([yshift=-0.1cm]current bounding box.center)}] 
	\node[] (a) at (0,0) {};
	\node[] (a1) at (1.2,0) {};
	\node[] (b) at (0,-1.2) {};
	\node[] (b1) at (1.2,-1.2) {};
	\node[] (p1) at ($(a)+(-0.3,0)$) {};
	\node[] (p2) at ($(b)+(-0.3,0)$) {};
	\node[] (p3) at ($(b1)+(0.3,0)$) {};
	\node[] (p4) at ($(a1)+(0.3,0)$) {};
	\draw[line width=0.15mm, dashed, postaction={decorate}] (b.center) -- node[sloped, allow upside down, label={[xshift=0.2cm, yshift=0cm]$k_L$}] {\midarrow} (a.center);	
	\draw[line width=0.15mm] (b1.center) -- (a1.center);
	\draw[line width=0.5mm] (p1.center) -- (p4.center);
	\draw[line width=0.5mm] (p2.center) -- (p3.center);
	\node at (0.6,0) [circle,fill,inner sep=1.5pt]{};
\end{tikzpicture} \right)  \nonumber \\
& \propto \int dz \frac{(u_1 \cdot q -y \, (u_2 \cdot q )) (q^2 + z)}{\sqrt[3]{(y^2 - 1) (q^2 - z)^2 - 4z ((u_1 \cdot q)^2 + (u_2 \cdot q)^2 - 2 y \, (u_1 \cdot q) (u_2 \cdot q)) }} \, .
\end{align}
Therefore, if the momentum transfer $q$ satisfies $u_i \cdot q = 0$, the corresponding leading singularity vanishes:
\begin{equation}
\LS \left(
\begin{tikzpicture}[baseline={([yshift=-0.1cm]current bounding box.center)}] 
	\node[] (a) at (0,0) {};
	\node[] (a1) at (1.5,0) {};
	\node[] (b) at (0,-1.2) {};
	\node[] (b1) at (1.5,-1.2) {};
	\node[] (p1) at ($(a)+(-0.3,0)$) {};
	\node[] (p2) at ($(b)+(-0.3,0)$) {};
	\node[] (p3) at ($(b1)+(0.3,0)$) {};
	\node[] (p4) at ($(a1)+(0.3,0)$) {};
	\draw[line width=0.15mm] (b1.center) -- (a1.center);
	\draw[line width=0.5mm] (p1.center) -- (0.4,0);
	\draw[line width=0.5mm] (0.4,0) -- (p4.center);
	\draw[line width=0.5mm] (p2.center) -- (0.4,-1.2);
	\draw[line width=0.5mm] (0.4,-1.2) -- (p3.center);
	\fill[gray!50] (0.4,-0.6) ellipse (0.4 and 0.7);
	\draw (0.4,-0.6) ellipse (0.4 and 0.7);
	\node at (1,0) [circle,fill,inner sep=1.5pt]{};
\end{tikzpicture} \right) = \LS \left(
\begin{tikzpicture}[baseline={([yshift=-0.1cm]current bounding box.center)}] 
	\node[] (a) at (0,0) {};
	\node[] (a1) at (0.4,0) {};
	\node[] (a2) at (1.5,0) {};
	\node[] (a3) at (2.6,0) {};
	\node[] (a4) at (3,0) {};
	\node[] (b) at (0,-1.2) {};
	\node[] (b1) at (0.4,-1.2) {};
	\node[] (b2) at (1.5,-1.2) {};
	\node[] (b3) at (2.6,-1.2) {};
	\node[] (b4) at (3,-1.2) {};
	\node[] (p1) at ($(a)+(-0.3,0)$) {};
	\node[] (p2) at ($(b)+(-0.3,0)$) {};
	\node[] (p3) at ($(b4)+(0.3,0)$) {};
	\node[] (p4) at ($(a4)+(0.3,0)$) {};
	\draw[line width=0.15mm] (a2.center) -- (b2.center);
	\draw[line width=0.5mm] (p1.center) -- (0.4,0);
	\draw[line width=0.5mm] (0.4,0) -- (2.6,0);
	\draw[line width=0.5mm] (2.6,0) -- (p4.center);
	\draw[line width=0.5mm] (p2.center) -- (0.4,-1.2);
	\draw[line width=0.5mm] (0.4,-1.2) -- (2.6,-1.2);
	\draw[line width=0.5mm] (2.6,-1.2) -- (p3.center);
	\fill[gray!50] ($(a1)+(0,-0.6)$) ellipse (0.4 and 0.7);
	\draw ($(a1)+(0,-0.6)$) ellipse (0.4 and 0.7);
	\fill[gray!50] ($(a3)+(0,-0.6)$) ellipse (0.4 and 0.7);
	\draw ($(a3)+(0,-0.6)$) ellipse (0.4 and 0.7);
	\node at (1,0) [circle,fill,inner sep=1.5pt]{};
\end{tikzpicture} \right) = 0.
\end{equation}
Therefore, the opposite-parity contributions are directly expressible in terms of lower sectors.

\section{Parity of post-Minkowskian Feynman integrals}
\label{app:parity}

In this appendix, we provide more details on the parity of post-Minkowskian Feynman integrals and prove the parity splitting in the corresponding differential equations. As an example, we illustrate the decoupling of the differential equations for the integral family involving the K3 surface discussed in sec.~\ref{sec:three_loop_K3}.

PM integrals contain propagators of two different kinds: Graviton propagators which are quadratic functions of the loop momenta $k_i$ and the momentum transfer $q$, e.g. of the form $k_i^2$, $(k_i-k_j)^2$, or $(k_i-q)^2$, and matter propagators which are linear in $k_i$ and are given by scalar products $2 u_j \cdot k_i$.

Let us now define two operators $\mathcal{P}_1$ and $\mathcal{P}_2$. $\mathcal{P}_1$ has the effect of (under the integral sign) changing the sign of all linearized propagators and ISPs involving $u_1$, and likewise for $\mathcal{P}_2$ and $u_2$. Thus $\mathcal{P}_i$ corresponds to changing the sign of $u_i$ while also conjugating the Feynman $i \epsilon$ in the corresponding propagators. In particular, we have\footnote{Note that in the worldline formulation, $L$ linearized propagators (odd under parity) are replaced by $L$ delta functions (even under parity); thus, changing the action of parity. Translating it to our formulation, the integrals would change under parity by a sign $(-1)^{a_1 + a_2 - L}$, such that the new even integrals are always of $b$-type in the momentum impulse, while odd integrals are always of $v$-type~\cite{Mogull:2020sak}.}
\begin{align}
\mathcal{P}_i \, \mathcal{I}_{a_1,a_2} = (-1)^{a_i} \, \mathcal{I}_{a_1,a_2},
\end{align}
where the $a_i$ denote the sums of the powers of the linearized propagators involving $u_i$.

Let us first study the combined transformation $\mathcal{P} := \mathcal{P}_1 \mathcal{P}_2$, which we refer to as the \textit{parity} of the integral. The behavior under the parity transformation sorts the integrals into two groups, \textit{even} and \textit{odd}, depending on whether the sum of powers $a_1{+}a_2$ is even or odd. Since parity is fixed by the number of scalar products $2 u_j \cdot k_i$ appearing either as propagators or as ISPs in the numerator (both of which are linear in $|q|$, while all other factors in PM integrals depend on $|q|^2$), we may also view the parity as a separation between even and odd-in-$|q|$ integrals.

As it turns out, using parity we can deduce various properties of the integrals. Let us first look at linear relations of the form often derived using IBP relations. We may take an integral $\mathcal{J}$ and write it as a linear combination of $N_{\text{basis}}$ master integrals $\mathcal{I}_i$,
\begin{align}
\mathcal{J} = \sum_{i=1}^{N_{\text{basis}}} \! c_i \, \mathcal{I}_i.
\label{eq:IBPparity}
\end{align}
The coefficients $c_i$ are functions only of $y = u_1 \cdot u_2$ and the dimension $D$, both of which are invariant under parity; thus, the $c_i$ are even. This means that if $\mathcal{J}$ is of even (odd) parity, then all master integrals $\mathcal{I}_i$ must also be even (odd). In other words, if an $\mathcal{I}_i$ has the opposite parity to $\mathcal{J}$, then the corresponding $c_i$ must vanish. This shows that the even and odd integrals decouple, since they are not connected through linear relations. 

The same property can be found for the differential equations obeyed by the master integrals, which take the form
\begin{align}
\partial_{y} \mathcal{I}_i = \sum_{j=1}^{N_{\text{basis}}} \! A_{ij} \, \mathcal{I}_j.
\label{eq:difeqparity}
\end{align}
Since the differential operator $\partial_y$ is even, and the matrix entries $A_{ij}$ are functions only of $y$ and $D$, we conclude that any $A_{ij}$ coupling an $\mathcal{I}_i$ and $\mathcal{I}_j$ of opposite parity have to be zero. Thus, the differential equations show a decoupling in parity. A similar decoupling is also found for the third type of linear relations between Feynman integrals, the dimension-shift identities~\cite{Tarasov:1996br, Lee:2012te}.

We may also study the behavior under the individual $\mathcal{P}_i$ (without the simultaneous application of the other $\mathcal{P}_j$), which can be used to infer further limitations on the forms of the $c_i$ and $A_{ij}$. Let us define the \textit{sub-parity} of an integral $\mathcal{P}_{\text{sub}} := \mathcal{P}_1$. Unlike the case of $\mathcal{P}$, the kinematic variable $y$ changes sign under $\mathcal{P}_{\text{sub}}$.
Consequently, for those $\mathcal{I}_i$ with the same (opposite) sub-parity as $\mathcal{J}$, we can deduce that the corresponding coefficients $c_i$ of eq.~\eqref{eq:IBPparity} have to be even (odd) functions of $y$. Since the differential operator $\partial_{y}$ is odd under $\mathcal{P}_{\text{sub}}$, we can similarly conclude that the matrix entries $A_{ij}$ in eq.~\eqref{eq:difeqparity} that couple integrals of the same (opposite) sub-parity have to be odd (even) functions of $y$.

\begin{figure}[t]
\centering
\begin{tikzpicture}[baseline=(current bounding box.center), scale=0.8] 
	\node[] (a) at (0,0) {};
	\node[] (a1) at (2,0) {};
	\node[] (a2) at (4,0) {};
	\node[] (b) at (0,-2) {};
	\node[] (c) at (0,-4) {};
	\node[] (c1) at (4,-4) {};
	\node[] (p1) at ($(a)+(-0.5,0)$) {};
	\node[] (p2) at ($(c)+(-0.5,0)$) {};
	\node[] (p3) at ($(c1)+(0.5,0)$) {};
	\node[] (p4) at ($(a2)+(0.5,0)$) {};
	\draw[line width=0.15mm] (b.center) -- (0,-1.3);
	\node[label={[xshift=0cm, yshift=-0.4cm]$1$}] (l1) at (0,-1) {};
	\draw[line width=0.15mm] (0,-0.7) -- (a.center);
	\draw[line width=0.15mm] (a1.center) -- (1.2,-0.8);
	\node[label={[xshift=0cm, yshift=-0.4cm]$2$}] (l2) at (1,-1) {};
	\draw[line width=0.15mm] (0.8,-1.2) -- (b.center);
	\draw[line width=0.15mm] (c1.center) -- (4,-2.3);
	\node[label={[xshift=0cm, yshift=-0.4cm]$5$}] (l5) at (4,-2) {};
	\draw[line width=0.15mm] (4,-1.7) -- (a2.center);
	\draw[line width=0.15mm] (b.center) -- (1.6,-2.8);
	\node[label={[xshift=0cm, yshift=-0.4cm]$4$}] (l4) at (2,-3) {};
	\draw[line width=0.15mm] (2.4,-3.2) -- (c1.center);
	\draw[line width=0.15mm] (b.center) -- (0,-2.7);
	\node[label={[xshift=0cm, yshift=-0.4cm]$3$}] (l3) at (0,-3) {};
	\draw[line width=0.15mm] (0,-3.3) -- (c.center);
	\draw[line width=0.5mm] (a.center) -- (0.7,0);
	\node[label={[xshift=0cm, yshift=-0.4cm]$6$}] (l6) at (1,0) {};
	\draw[line width=0.5mm] (1.3,0) -- (a1.center);
	\draw[line width=0.5mm] (a1.center) -- (2.7,0);
	\node[label={[xshift=0cm, yshift=-0.4cm]$7$}] (l7) at (3,0) {};
	\draw[line width=0.5mm] (3.3,0) -- (a2.center);
	\draw[line width=0.5mm] (c.center) -- (1.7,-4);
	\node[label={[xshift=0cm, yshift=-0.4cm]$8$}] (l8) at (2,-4) {};
	\draw[line width=0.5mm] (2.3,-4) -- (c1.center);
	\draw[line width=0.5mm] (p1.center) -- (a.center);
	\draw[line width=0.5mm] (a2.center) -- (p4.center);
	\draw[line width=0.5mm] (p2.center) -- (c.center);
	\draw[line width=0.5mm] (c1.center) -- (p3.center);
\end{tikzpicture}
\caption{Propagators for the integral family giving rise to the K3 surface of sec.~\ref{sec:three_loop_K3}.}
\label{fig: diag_propagators_K3}
\end{figure}
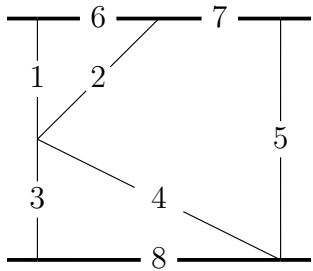

As an example, let us look at the three-loop integral of fig.~\ref{fig: diag_propagators_K3} giving rise to the K3 surface discussed in sec.~\ref{sec:three_loop_K3}. We consider generic integrals of the form
\begin{equation}
\mathcal{I}_{\nu_1,\nu_2,\dots,\nu_8}=\int \frac{d^D k_1 \, d^D k_2 \, d^D k_3}{\rho_1^{\nu_1} \rho_2^{\nu_2} \cdots \rho_8^{\nu_8}},
\end{equation}
where the propagators are defined using the same convention as in fig.~\ref{fig: diag_3-loop_elliptic},
\begin{equation}
\begin{aligned}
& \rho_1=k_1^2, \qquad \qquad \enspace \rho_2=(k_1{+}k_2)^2, \qquad \rho_3=k_3^2, \qquad \qquad \quad \rho_4=(k_2{-}k_3)^2, \\
& \rho_5=(k_2{+}q)^2, \qquad \, \rho_6=2u_1 \cdot k_1, \qquad \enspace \, \rho_7=-2u_1 \cdot k_2, \qquad \rho_8=2u_2 \cdot k_3.
\end{aligned}
\end{equation}

The top-sector of this integral family contains five master integrals, which may be chosen as
\begin{equation}
\begin{aligned}
\mathcal{I}_1 &= \mathcal{I}_{11111111} \;, & \mathcal{I}_2 &= \mathcal{I}_{11211111} \;, & \mathcal{I}_3 &= \mathcal{I}_{11121111} \;, \\
\mathcal{I}_4 &= \mathcal{I}_{11111112} \;, & \mathcal{I}_5 &= \mathcal{I}_{11111121} \;.
\end{aligned}
\end{equation}
The system of differential equations can then be written as in eq.~\eqref{eq:difeqparity}:
\begin{align}
\partial_{y} \mathcal{I}_i = \sum_{j=1}^{5} \! A_{ij}^{(5 \times 5)} \, \mathcal{I}_j.
\end{align}
The three integrals $\mathcal{I}_1$--$\mathcal{I}_3$ have odd parity, while the last two have even parity. By the arguments above, we conclude that the differential equation matrix splits in two independent blocks as
\begin{align}
A^{(5 \times 5)} = \left( \begin{array}{cc} A^{(3 \times 3)}_{\text{odd}} & 0^{(3 \times 2)} \\ 0^{(2 \times 3)} & A^{(2 \times 2)}_{\text{even}} \end{array} \right).
\end{align}
The K3 surface of sec.~\ref{sec:three_loop_K3} is present in the odd sector only, the two integrals in the even sector have instead a dlog form.

Regarding the sub-parity, we see that $\mathcal{I}_5$ has odd sub-parity, while the remaining four master integrals have even. This means that of the non-zero matrix elements, $A_{45}$ and $A_{54}$ have to be even functions of $y$, and the remaining elements all have to be odd functions of $y$.

\bibliography{References}
\bibliographystyle{JHEP}

\end{document}